\documentclass[11pt,a4paper]{article}
\pdfoutput=1
\usepackage{jcappub}
\usepackage{aasmacros}

\usepackage{float}

\usepackage{graphicx}
\usepackage{amsmath}
\usepackage{amssymb}
\usepackage{rotating} 
\usepackage{xspace}
\usepackage{color}
\usepackage{booktabs}
\usepackage[normalem]{ulem}

\newcommand{\Msun}{{\rm M}_\odot}
\newcommand{\FeH}{$[{\rm Fe}/{\rm H}]$}
\newcommand{\FeHone}{$[{\rm Fe}/{\rm H}]<-1$}
\newcommand{\FeHtwo}{$[{\rm Fe}/{\rm H}]<-2$}
\newcommand{\FeHthree}{$[{\rm Fe}/{\rm H}]<-3$}
\newcommand{\aFe}{$[\alpha/{\rm Fe}]$}
\newcommand{\aFetwo}{$[\alpha/{\rm Fe}]>0.2$}

\newcommand{\auriga}         {\small Auriga }

\newcommand{\kpc}            {\,{\rm kpc}}

\newcommand{\feh}            {{\rm [Fe/H]}}

\title{On the correlation between the local dark matter and stellar velocities}

\author[a]{Nassim Bozorgnia,}
\author[b]{Azadeh Fattahi,}
\author[a]{David G.~Cerde\~no,}
\author[b]{Carlos S. Frenk,}
\author[c, d]{Facundo A. G\'{o}mez,}
\author[e]{Robert J. J. Grand,}
\author[f]{Federico Marinacci,}
\author[e]{and R\"{u}diger Pakmor}

\affiliation[a]{Institute for Particle Physics Phenomenology, Department of Physics,\\
Durham University, Durham DH1 3LE, UK} 
\affiliation[b]{Institute for Computational Cosmology, Durham University,\\
South Road, Durham DH1 3LE, UK}
\affiliation[c]{Instituto de Investigaci\'on Multidisciplinar en Ciencia y Tecnolog\'ia, \\
Universidad de La Serena, Ra\'ul Bitr\'an 1305, La Serena, Chile}
\affiliation[d]{Departamento de F\'isica y Astronom\'ia, Universidad de La Serena,\\
Av. Juan Cisternas 1200 Norte, La Serena, Chile}
\affiliation[e]{Max-Planck-Institut f\"{u}r Astrophysik,\\
Karl-Schwarzschild-Str. 1, D-85748, Garching, Germany}
\affiliation[f]{Institute for Theory and Computation, Harvard-Smithsonian Center for Astrophysics, \\
  60 Garden Street, Cambridge, MA 02138, USA}

\emailAdd{nassim.bozorgnia@durham.ac.uk}

\abstract{
The dark matter velocity distribution in the Solar neighbourhood is an important astrophysical input which enters in the predicted event rate of dark matter direct detection experiments. It has been recently suggested that the local dark matter velocity distribution can be inferred from that of old or metal-poor stars in the Milky Way. We investigate this potential relation using six high resolution magneto-hydrodynamical simulations of Milky Way-like galaxies of the Auriga project. We do not find any correlation between the velocity distributions of dark matter and old stars in the Solar neighbourhood. Likewise, there are no strong correlations between the local velocity distributions of dark matter and metal-poor stars selected by applying reasonable cuts on metallicity. In some simulated galaxies, extremely metal-poor stars have a velocity distribution that is statistically consistent with that of the dark matter, but the sample of such stars is so small that we cannot draw any strong conclusions.}

\arxivnumber{IPPP/18/97}

\begin{document}
\maketitle

\section{Introduction}
\label{sec:introduction}

Discovering the identity of dark matter (DM) is one of the main goals of particle astrophysics~\cite{Bertone:2010zza,Jungman:1995df,Bergstrom00,Bertone05}. Many direct detection experiments are currently operating around the world, searching for the recoil of a nucleus in an underground detector after a collision with a DM particle. An important input which enters the calculations of direct detection event rates is the DM abundance, and its velocity distribution in the Solar neighbourhood. Variations in the astrophysical parameters that define this distribution as well as its functional form lead to large uncertainties in the interpretation of direct detection data. 

In the analysis of direct detection data, usually the Standard Halo Model (SHM)~\cite{Drukier:1986tm} is assumed for the DM distribution. In the SHM, the DM is distributed in an isothermal sphere and has a Maxwell-Boltzmann velocity distribution with the peak speed equal to the local circular speed, usually taken to be 220~km$/$s. The DM velocity distribution could, however, be different from the SHM and this could alter the exclusion limits derived from direct detection \cite{Green:2000jg,Green:2002ht,Vogelsberger:2008qb,Kuhlen:2009vh}. Departures from the SHM do not affect all the experiments in the same way, since, depending on the nuclear target and on the range of recoil energies that are analysed, different detectors probe different regions of the velocity distribution function. Therefore, an accurate description of the velocity distribution function is crucial to interpret current and future experimental results \cite{Belli:2002yt,Green:2003yh,Bottino:2005qj,Fairbairn:2008gz,McCabe:2010zh}.

An insightful way to obtain information on the DM velocity distribution is to use cosmological simulations of Milky Way (MW)-like galaxies. High resolution hydrodynamic simulations of galaxy formation including both DM and baryons have recently become possible and have achieved significant agreement with observations. Recently, The EAGLE/APOSTLE, MaGICC, and the Sloane {\it et al.} hydrodynamic simulations studied the DM velocity distribution in the Solar neighbourhood in MW-like galaxies and found that the Maxwellian velocity distribution provides a good fit to the DM velocity distributions of MW-like halos~\cite{Bozorgnia:2016ogo, Kelso:2016qqj, Sloane:2016kyi, Bozorgnia:2017brl}. The peak speed of the best fit Maxwellian distribution can however be different from the local circular speed (see figure 1 of ref.~\cite{Bozorgnia:2017brl}).

An alternative way to obtain information on the DM velocity distribution is to use both hydrodynamic simulations and observations of stellar velocities. Recently, it was suggested that old metal-poor stars in  the Eris simulation of a MW-like galaxy with a quiet merger history, trace the local DM velocity distribution, due to their common origin with DM~\cite{Herzog-Arbeitman:2017fte}. Using the SDSS-{\it Gaia} DR2 data, ref.~\cite{Necib:2018iwb} produced an empirical DM velocity distribution function which was inferred from the observed velocity distribution of metal-poor and intermediate metallicity stars belonging to the halo and a substructure population, respectively. The DM distribution inferred in this way diverges substantially from the SHM. This result depends on the assumption that there is a strong correlation between the local velocity distribution of DM particles and a population of old metal-poor stars. This has been confirmed in a single halo with a quiet merger history in one simulation~\cite{Herzog-Arbeitman:2017fte}\footnote{Strong correlations have also been observed between the local velocity distribution of stars and DM accreted from {\it luminous satellites} in a recent analysis of  two MW mass halos in the Latte suite of FIRE-2 simulations~\cite{Necib:2018igl}.}. Such correlations need to be tested with larger samples and also with different hydrodynamical simulations using different galaxy formation models in order to be generalised.

In addition, one should bear in mind a number of uncertainties involved in the comparison of simulated and observed metallicities. Incompleteness of the observational data, particularly at very low metallicities, as well as assumptions and approximations made in the {\it subgrid} physics in hydrodynamical simulations, make the comparison difficult. We discuss this further in section~\ref{sec:age-metal}, where we show a comparison of the age-metallicity relation in simulations and observations.

In this paper, we study possible correlations between the velocity distributions of old and metal-poor stars with those of DM particles in the Solar neighbourhood in six high resolution magneto-hydrodynamical simulations of MW-like galaxies within the Auriga project. In section~\ref{sec:sims} we discuss the details of the Auriga simulations relevant for this work. In section~\ref{sec:age-metal} we show the age-metallicity relation in Auriga halos. We present the velocity distributions in section~\ref{sec:velocities}, and in section~\ref{sec:halointegral} we present the halo integrals which are the relevant functions for studying DM direct detection implications. We present the density profiles in section~\ref{sec:density}. Finally, we present our concluding remarks in section~\ref{sec:conclusions}. In appendix~\ref{sec:Diffmetal} we discuss the sensitivity of our results to the metallicity cut, and in appendix \ref{sec:volume} we investigate how our results change if we consider a larger Solar neighbourhood region for the stars.

\section{Simulations}
\label{sec:sims}

The Auriga project \citep{Grand:2016mgo} is a suite of cosmological magneto-hydrodynamical simulations of 30~MW halos, performed by the Tree-PM, moving mesh code, {\small Arepo} \citep{Springel2010}. The MW-like halos were selected from the $100^3~{\rm Mpc}^3$ cosmological, periodic box of the {\small EAGLE} project \cite{Schaye2015,Crain2015}, with the requirements to have virial\footnote{Virial quantities are defined as those corresponding to a sphere with mean enclosed density of 200 times the critical density of the Universe.} mass of the order of $10^{12}~\Msun$, and be relatively isolated at z=0. The halos were then resimulated using the {\it zoom-in} technique \citep{Frenk1996,Jenkins2013} at three different levels of resolution with full hydrodynamics and a comprehensive subgrid galaxy formation model. 

In summary, the galaxy formation model includes primordial and metal-line cooling with self-shielding enabled, star formation, stellar evolution and supernovae feedback, X-ray/UV ionising background radiation, and supermassive black hole growth and feedback. We refer the reader to ref.~\cite{Grand:2016mgo} and references therein for details of the model.

While all 30 halos have been resimulated at the fiducial resolution, labeled level 4 with $5\times10^4~\Msun$ per DM particle, in this study, we use 6 halos (Au6, Au16, Au21, Au23, Au24, Au27) at the highest resolution, level 3. The level 3 runs have $m_{\rm DM}=4 \times 10^4~\Msun$ per DM particle,  typical mass of $m_b=6 \times 10^3~\Msun$ per baryonic element, and a maximum softening length of $\epsilon =184$~pc.

DM halos and bound structures in the simulations are defined using the Friends of Friends (FoF) algorithm and {\small SUBFIND}, respectively \citep{Davis1985,Springel2001a}. MW analogs are referred to the central bound structure of the main FoF group. The position and velocity of the centre of MW analogs are calculated using the {\it shrinking sphere} method on DM particles, where we start by computing the centre of mass of particles within the virial radius and shrink the radius iteratively by $5\%$ at each step until 1000 particles is reached. The disc of the MW analogs are defined to be perpendicular to the net angular momentum of bound star particles within $10$ kpc.

For our analysis, we consider only star particles bound to the MW analogs (not to the existing satellites), while all DM particles are included regardless of whether they are bound to the MW analogs or to subhalos. The reason being stars belonging to known satellites are removed in observations, while DM detection experiments are sensitive to all DM particles, particularly those bound to numerous dark subhalos. Notice that there are no luminous satellites in the Solar neighbourhood region (defined in section \ref{sec:velocities} as a cylindrical shell aligned with the stellar disc with a radius of $7-9$~kpc from the Galactic centre and a height of $|z| \leq 2$~kpc from the Galactic plane) of the simulated halos, which is the region we consider for the main analyses of this work. Luminous satellites, however, exist in the inner 30~kpc of the halos, which is the Galactocentric distance we use to present the density profiles (in section~\ref{sec:density}). We have checked that including the stars bound to satellites in the analysis of the density profiles does not change our results.


\section{Age-metallicity relation in Auriga}
\label{sec:age-metal}

Star particles in the \auriga simulations are formed stochastically from gas cells satisfying the starformation density criterion, and represent a single stellar population with Chabrier Initial Mass Function \citep{Chabrier2003}. As a star particle evolves, mass loss and metal deposition due to supernovae-Ia and Asymptotic Giant Branch stars are calculated and distributed to the neighbouring cells. To model the feedback and metal enrichment from supernovae-II, the code creates a wind particle instead of a star particle, which is then expelled from the star-forming gas. Nine chemical elements are tracked self-consistently in this process: H, He, C, O, N, Ne, Mg, Si, Fe~\cite{Vogelsberger:2013eka}.

The relative abundance of two elements $X$ and $Y$ in a star is defined with respect to their relative Solar abundances,
\begin{equation}
\left[\frac{X}{Y}\right]=\log_{10}\left(\frac{N_X}{N_Y}\right) - \log_{10}\left(\frac{N_X}{N_Y}\right)_\odot.
\end{equation}
Here $N_X$ and $N_Y$ are the number density of elements $X$ and $Y$, respectively. 

We use the mean abundance of available $\alpha$-elements in \auriga (O, Mg, and Si) for calculating \aFe. In particular, we take take the average of $[{\rm O}/{\rm Fe}]$, $[{\rm Mg}/{\rm Fe}]$, and $[{\rm Si}/{\rm Fe}]$. We adopt the following Solar abundances $A_{Fe}=7.5$, $A_{O}=8.69$, $A_{Mg}=7.60$, $A_{Si}=7.51$ from table 1 of ref.~\citep{Asplund2009}, where $A_{X}=\log_{10}(N_X/N_H)+12$.

\begin{figure}[t]
  \hspace{6pt}\includegraphics[width=0.48\textwidth]{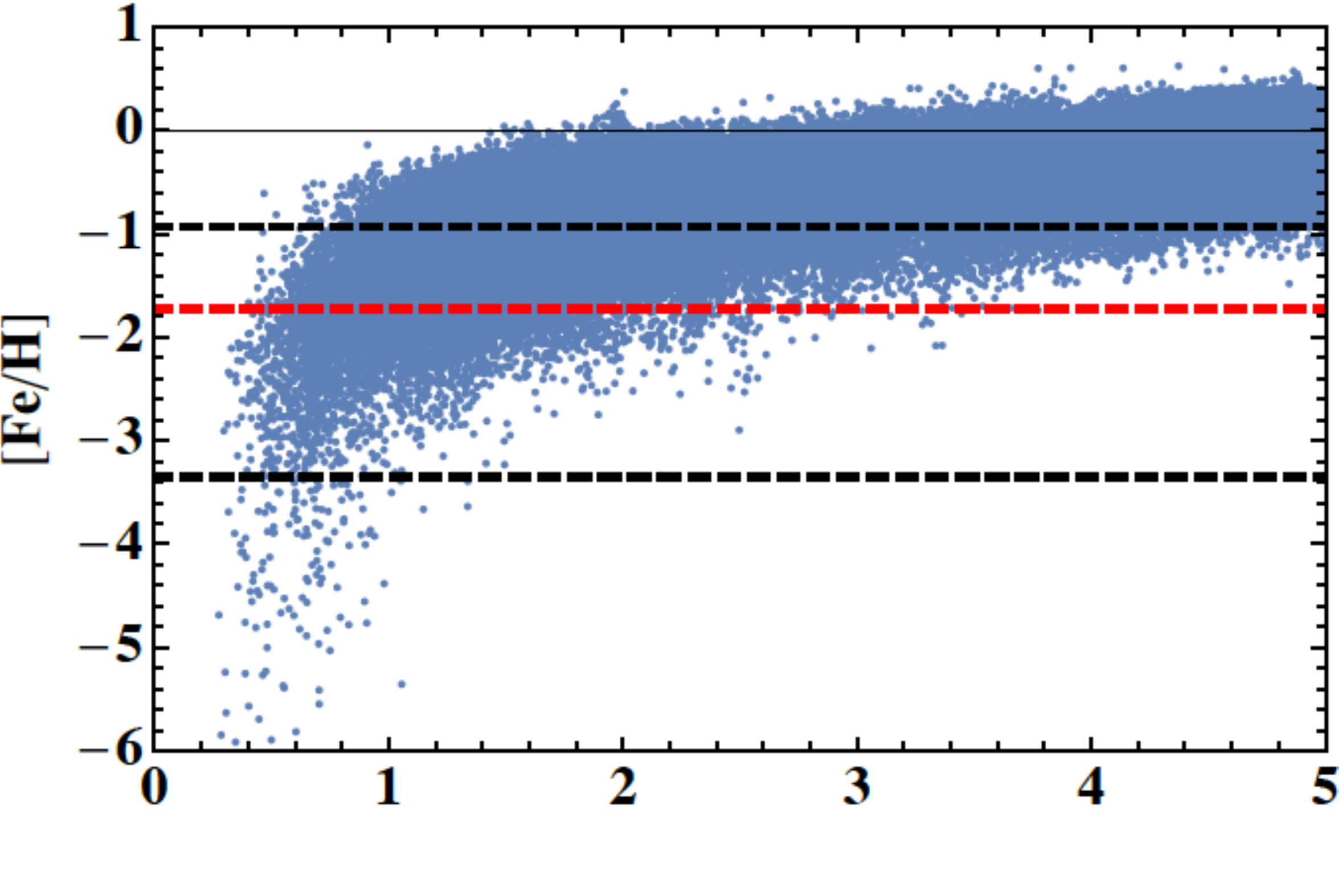}\\
  \includegraphics[width=0.495\textwidth]{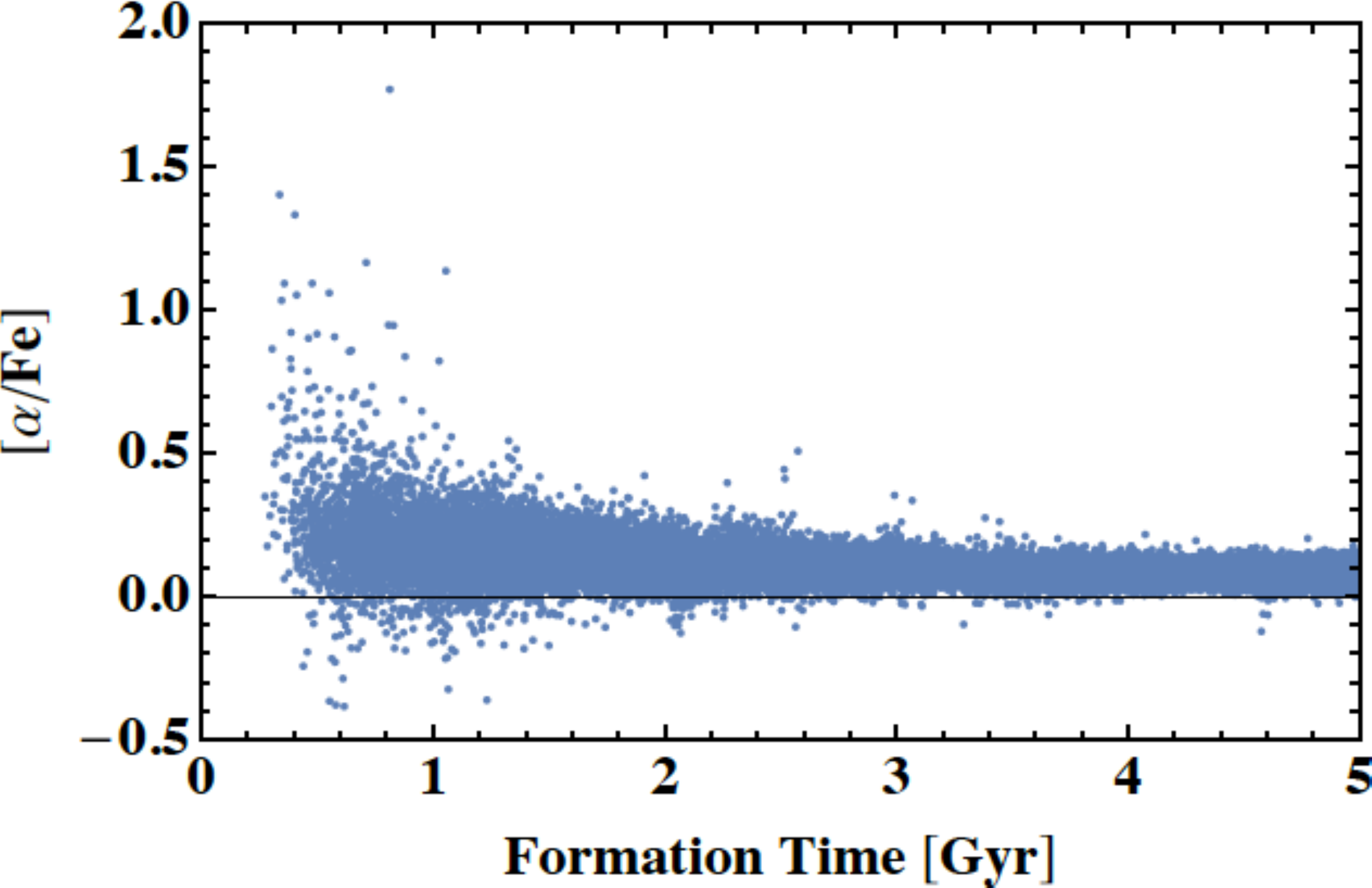}
    \includegraphics[width=0.465\textwidth]{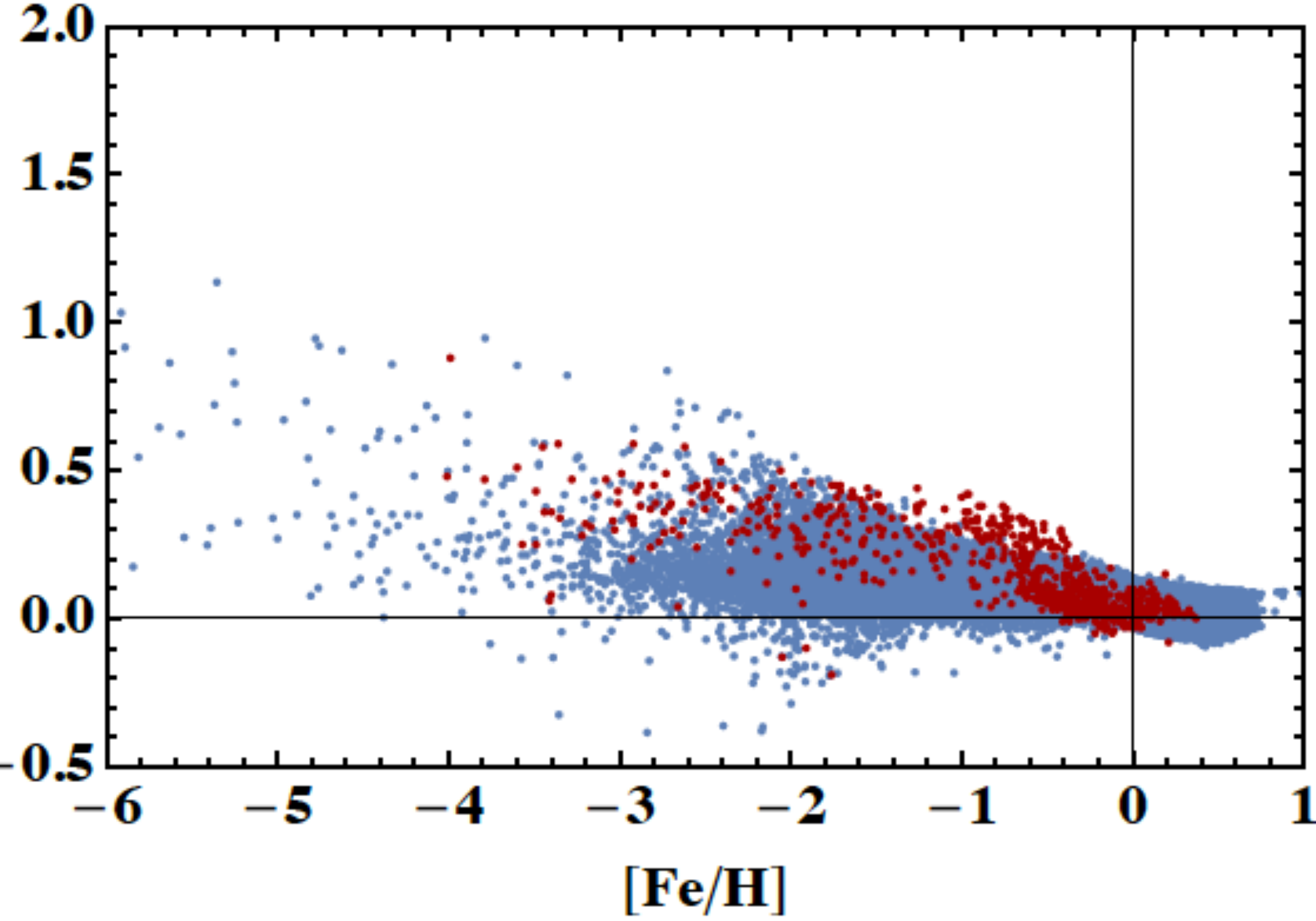}
\caption{\FeH~(top panel) and \aFe~(bottom left panel) as a function of their formation time, and \aFe~as a function of \FeH~(bottom right panel) for all stars in a cylindrical shell with radial distance $7 \leq \rho \leq 9$~kpc from the Galactic centre for halo Au6. The red data points in the bottom right panel are the observed metallicity of stars in the Solar neighbourhood, extracted from ref.~\cite{Venn:2004hk}. The black dashed lines in the top panel show the 5th and 95th percentile range of the \FeH~values for stars formed less than 1 Gyr after the Big Bang, while the red dashed line specifies the median of the distribution.}
\label{fig:MetalAge}
\end{figure}

Figure~\ref{fig:MetalAge} shows \FeH~and \aFe~versus formation time (since the Big Bang), as well \aFe~versus \FeH, for all stars in a cylindrical shell around the Solar circle, with radial distance $7 \leq \rho \leq 9$~kpc, from the centre of the galaxy for Au6. For comparison, we also show the observed metallicities of stars in the Solar neighbourhood as red data points in the bottom right panel. These observational data are extracted from table 2 of ref.~\cite{Venn:2004hk}, and show a reasonable agreement with the metallicities of Solar circle stars in the simulated halo. However, one can see that the data are sparse at very low metallicities, making a detailed comparison with simulations difficult. One can also see that there are many more stars with low \aFe~at $-3<\FeH<-1$ in the simulation compared to the observational data. This can be due to the incompleteness of the observational data at low metallicities, as well as uncertainties in the simulations due to subgrid modelling of the metals. Nevertheless, we show in appendix~\ref{sec:Diffmetal} that our results are not particularly sensitive to the \aFe~cut, when choosing the metal-poor stars in the simulations.

We obtain similar results for the other Auriga halos. As expected, stars formed at later times have higher metallicities on average (approaching Solar values at $\sim$ 6 Gyr after the Big Bang), while the oldest stars formed right after the big bang are very metal poor. The relation, however, has a large scatter particularly for the oldest stars; those which formed in the first Gyr after the big bang can have $\feh$ values spanning a large range from $-3.35$ to $-0.93$ (5 to 95 percentile range) with a median of -1.72.

\section{Velocity distributions}
\label{sec:velocities}

In this section, we will set different cuts on the age and metallicity of stars in the Solar neighbourhood and study the correlations of their velocities with the DM velocity distribution.

To describe the velocity vector of DM and star simulation particles, we define a reference frame with the origin at the Galactic centre, $z$-axis perpendicular to the stellar disc, $\rho$ in the radial direction, and $\phi$ in the azimuthal direction. We define the {\it Solar neighbourhood} region for extracting the DM and stellar velocity distributions, as a cylindrical shell with radius $7 \leq \rho \leq 9$~kpc and height $|z| \leq 2$~kpc, such that the particles are constrained to the disc. We then compute the vertical ($v_z$), radial ($v_\rho$), and azimuthal ($v_\phi$) components of the DM and stellar  velocity distributions. These three components of the velocity distribution are individually normalised to unity, such that $\int dv_i f(v_i)=1$ for $i=z,\, \rho ,\, \phi$.

Notice that we have chosen the same Solar neighbourhood region for DM and stars, such that their distributions can be compared in the same region of the Galaxy. This is different from the approach of ref.~\cite{Herzog-Arbeitman:2017fte} which does not restrict the Solar neighbourhood stellar distributions to the disc. In principle, velocity distributions can vary with height above the Galactic plane, and the same Solar neighbourhood region should be used to compare the DM and stellar velocity distributions. However, to directly compare our results with those of ref.~\cite{Herzog-Arbeitman:2017fte}, in appendix~\ref{sec:volume} we present the DM and stellar velocity distributions when the Solar neighbourhood star particles are not constrained to the disc.

\begin{figure}[H]
\begin{center}
   \includegraphics[width=0.31\textwidth]{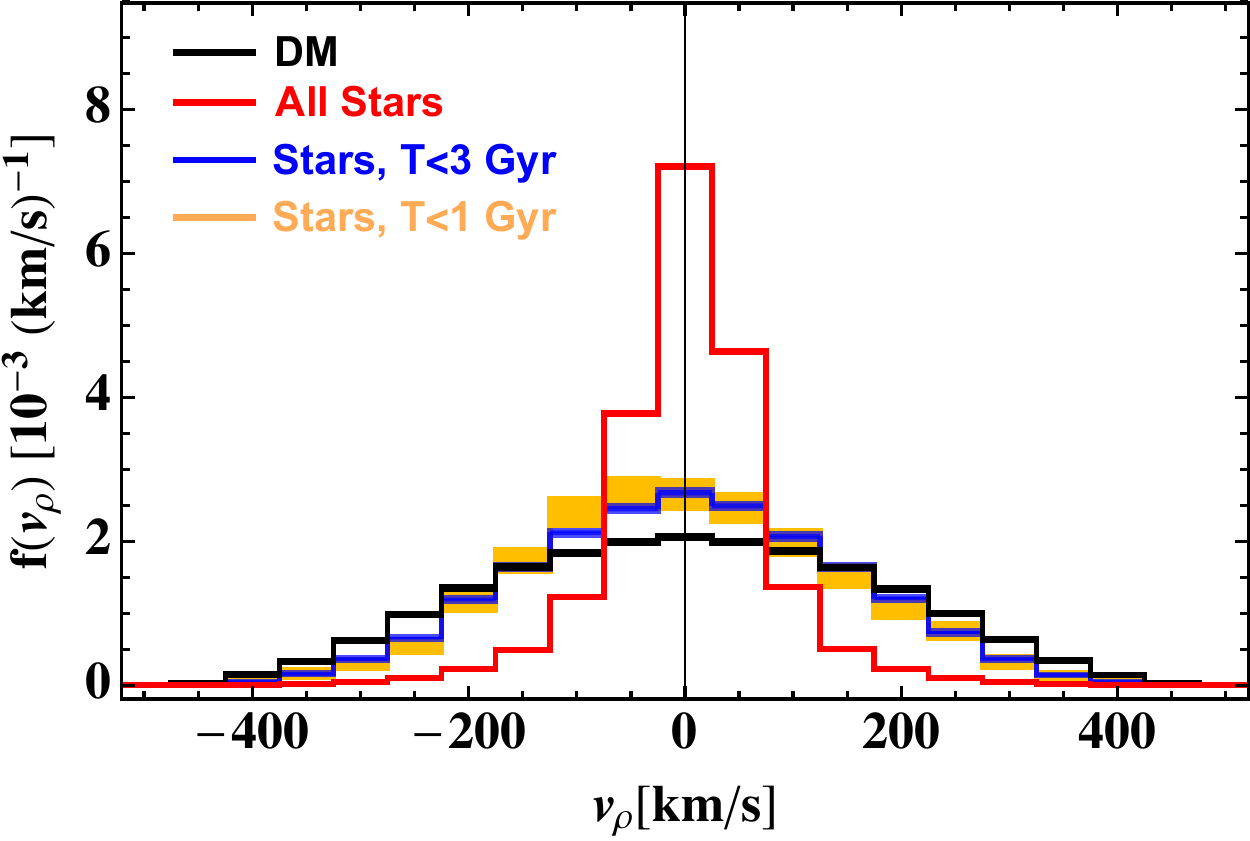}
   \includegraphics[width=0.31\textwidth]{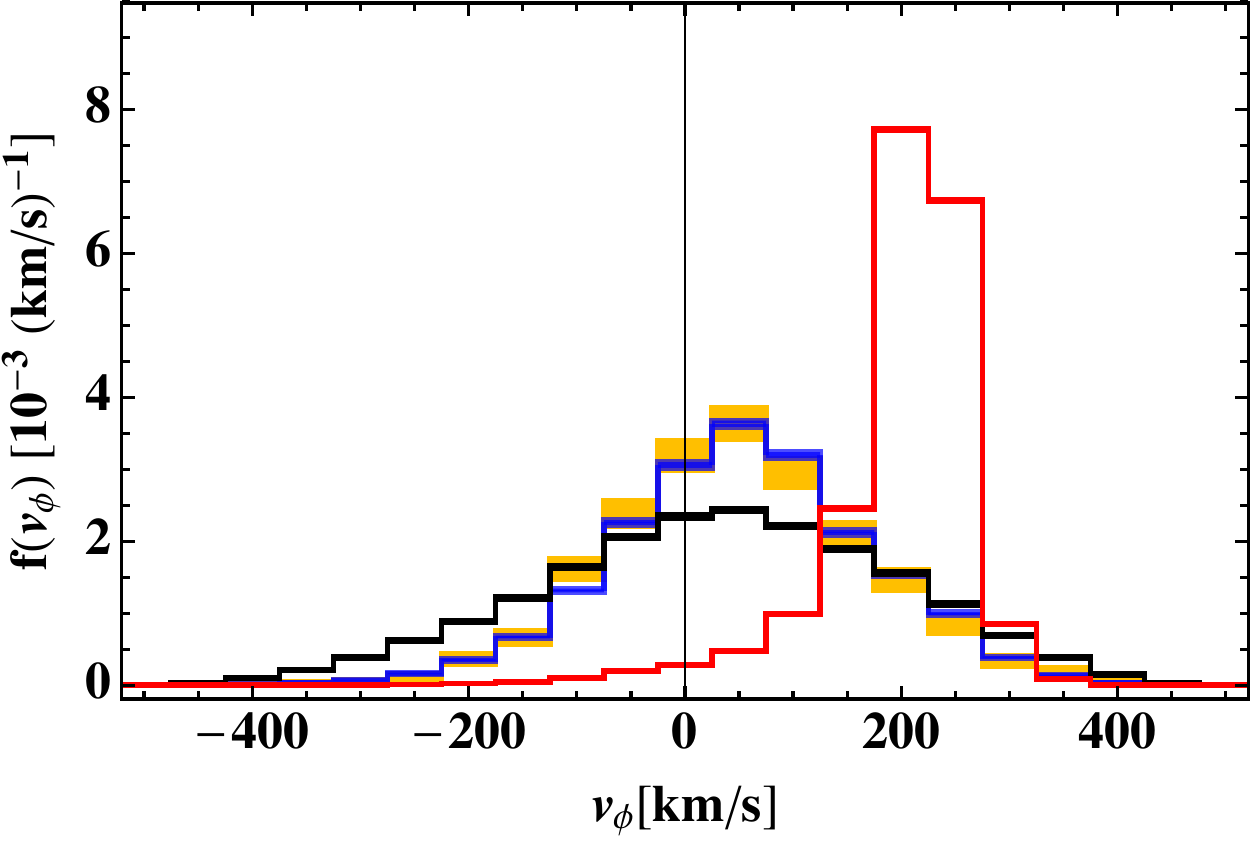}
   \includegraphics[width=0.31\textwidth]{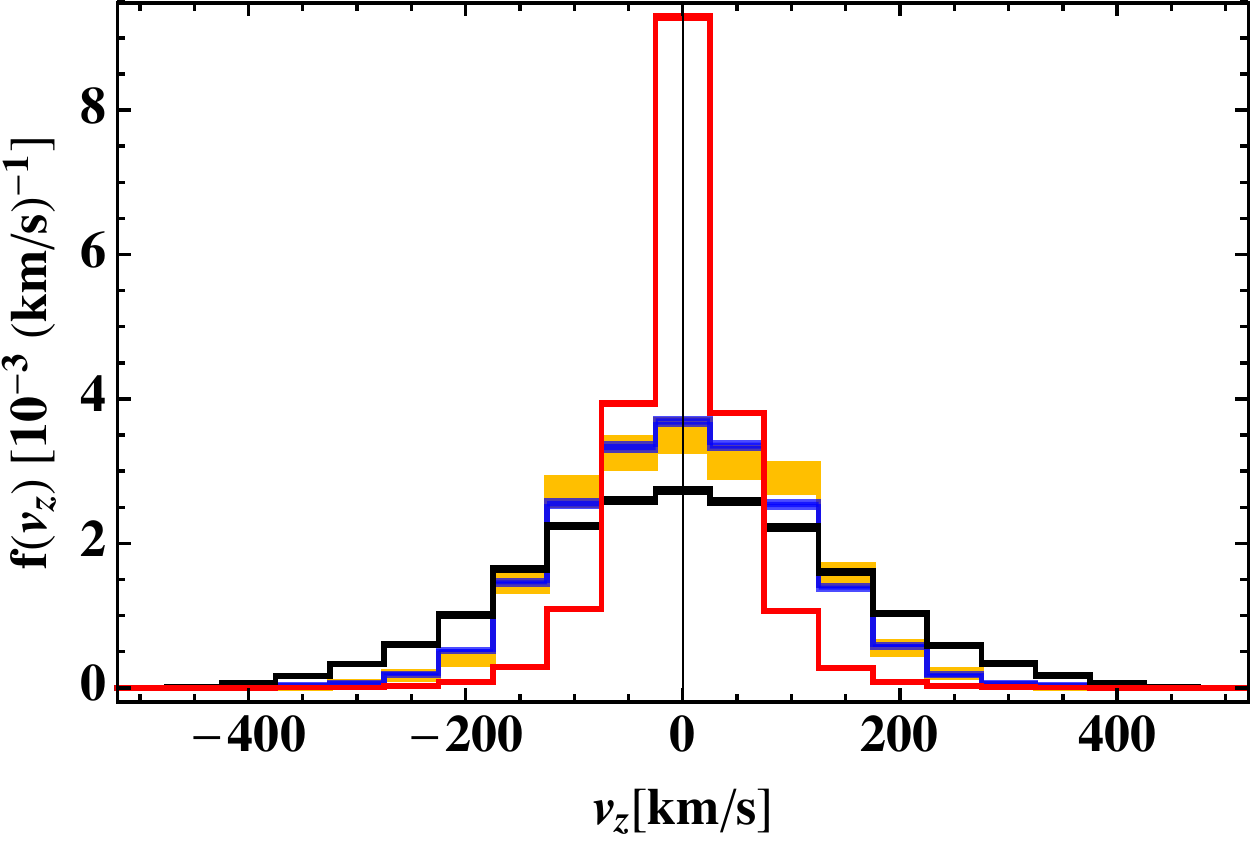}\\
   \includegraphics[width=0.31\textwidth]{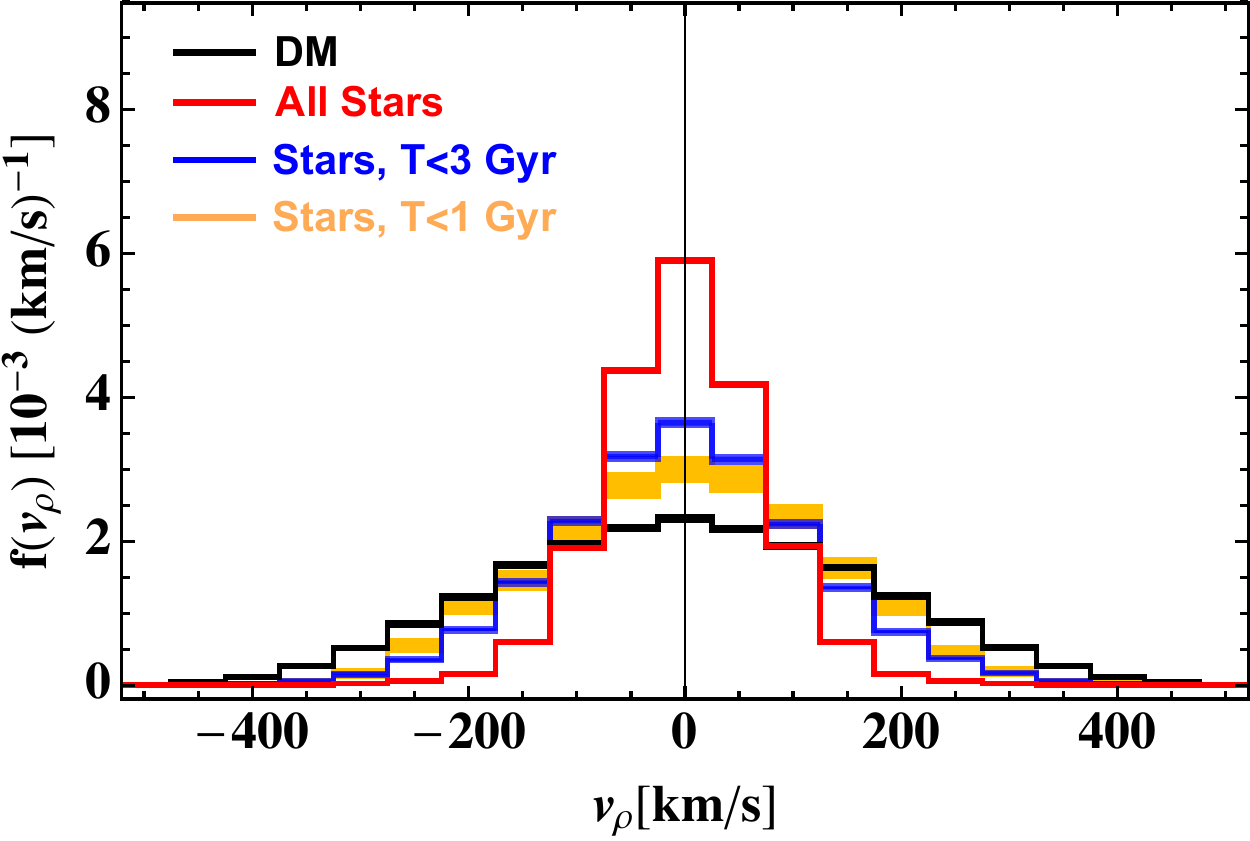}
   \includegraphics[width=0.31\textwidth]{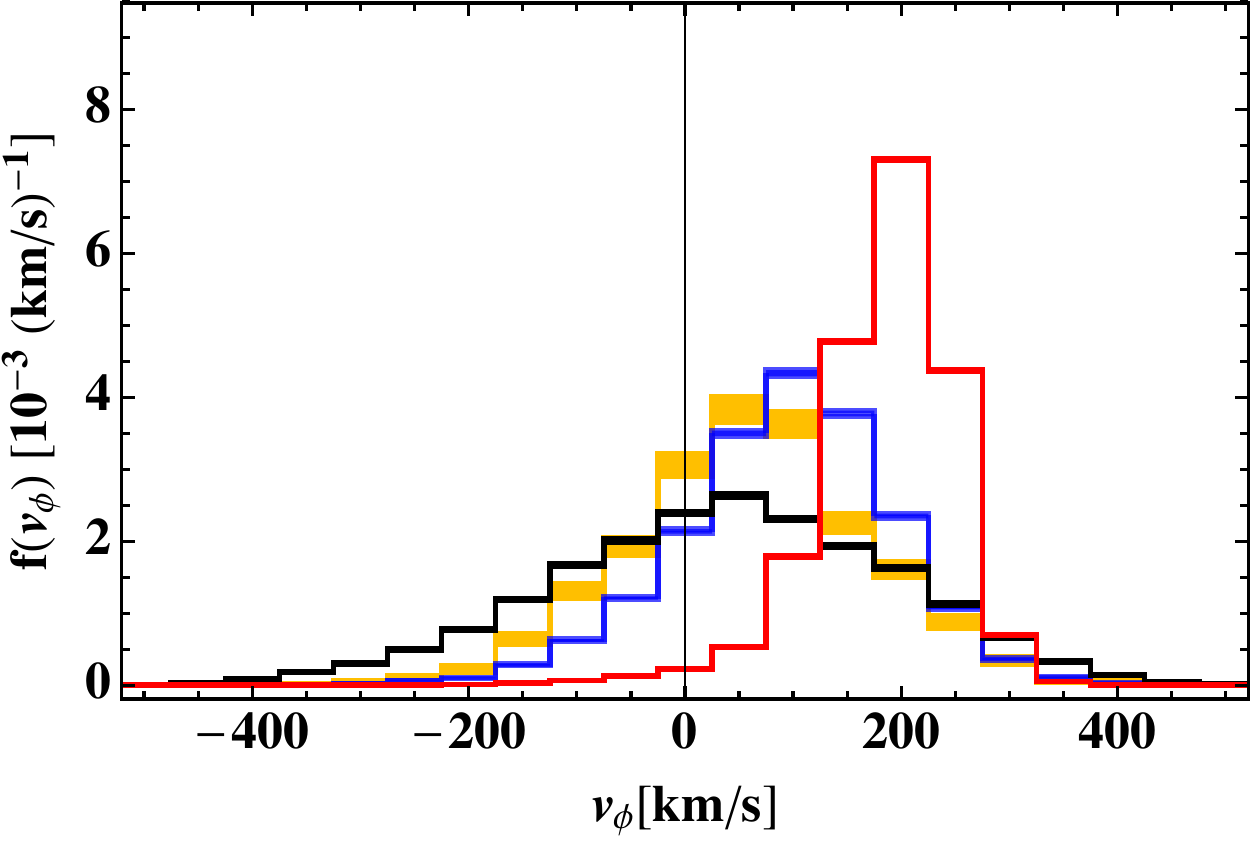}
   \includegraphics[width=0.31\textwidth]{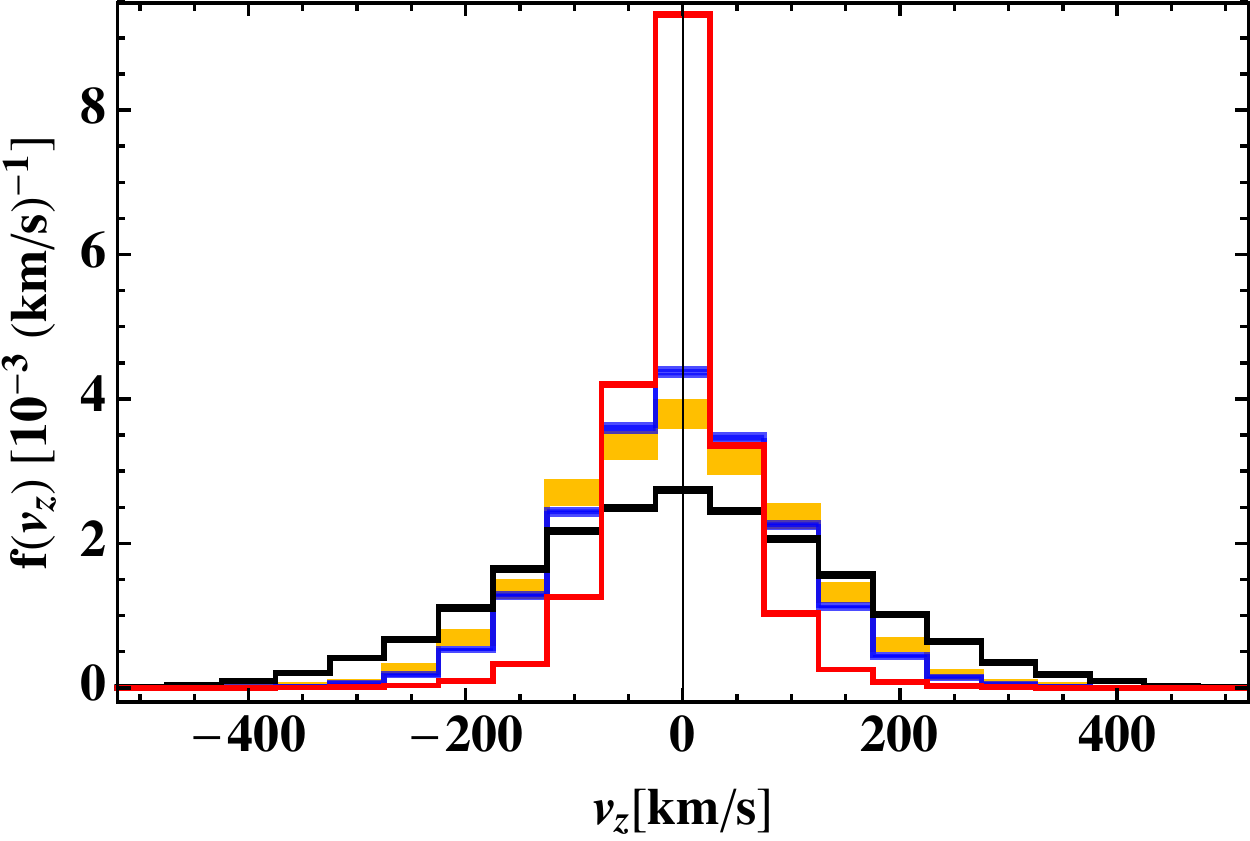}\\
   \includegraphics[width=0.31\textwidth]{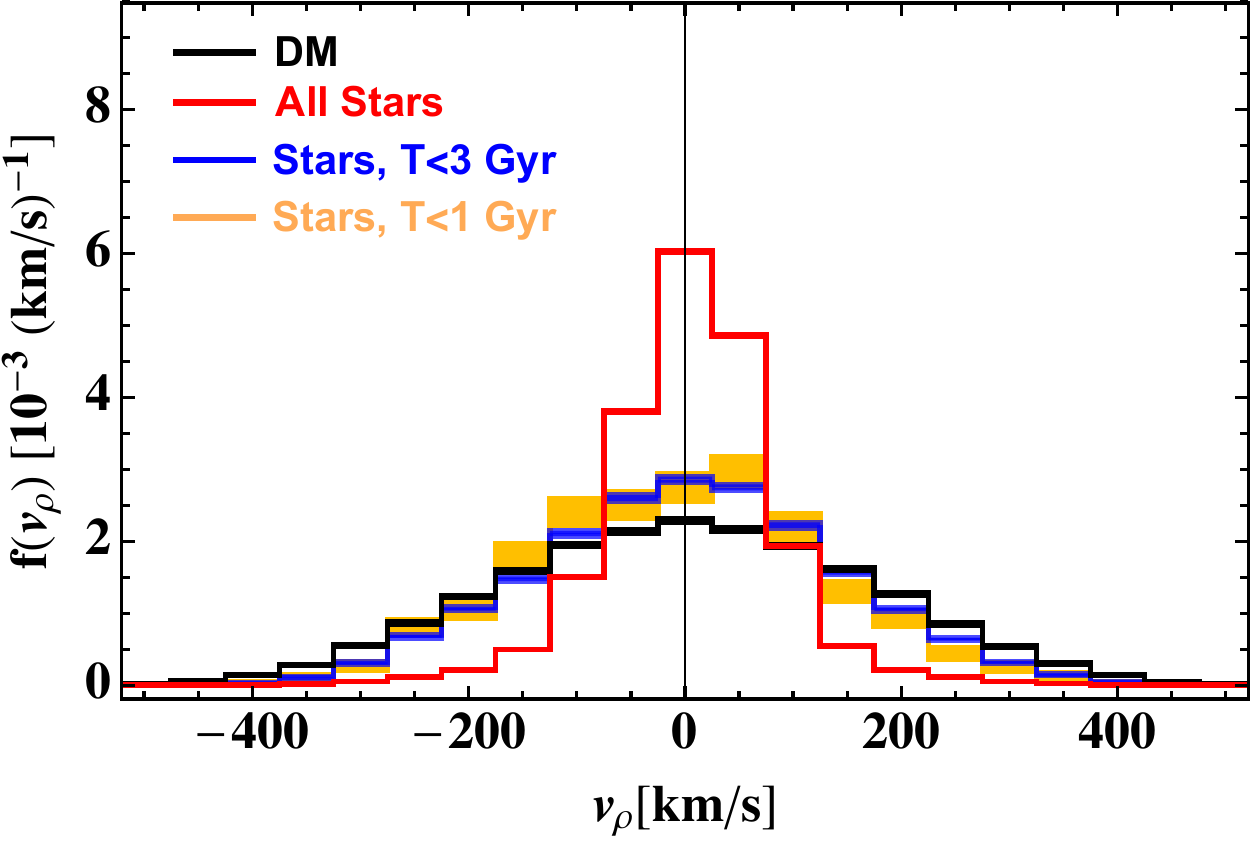}
   \includegraphics[width=0.31\textwidth]{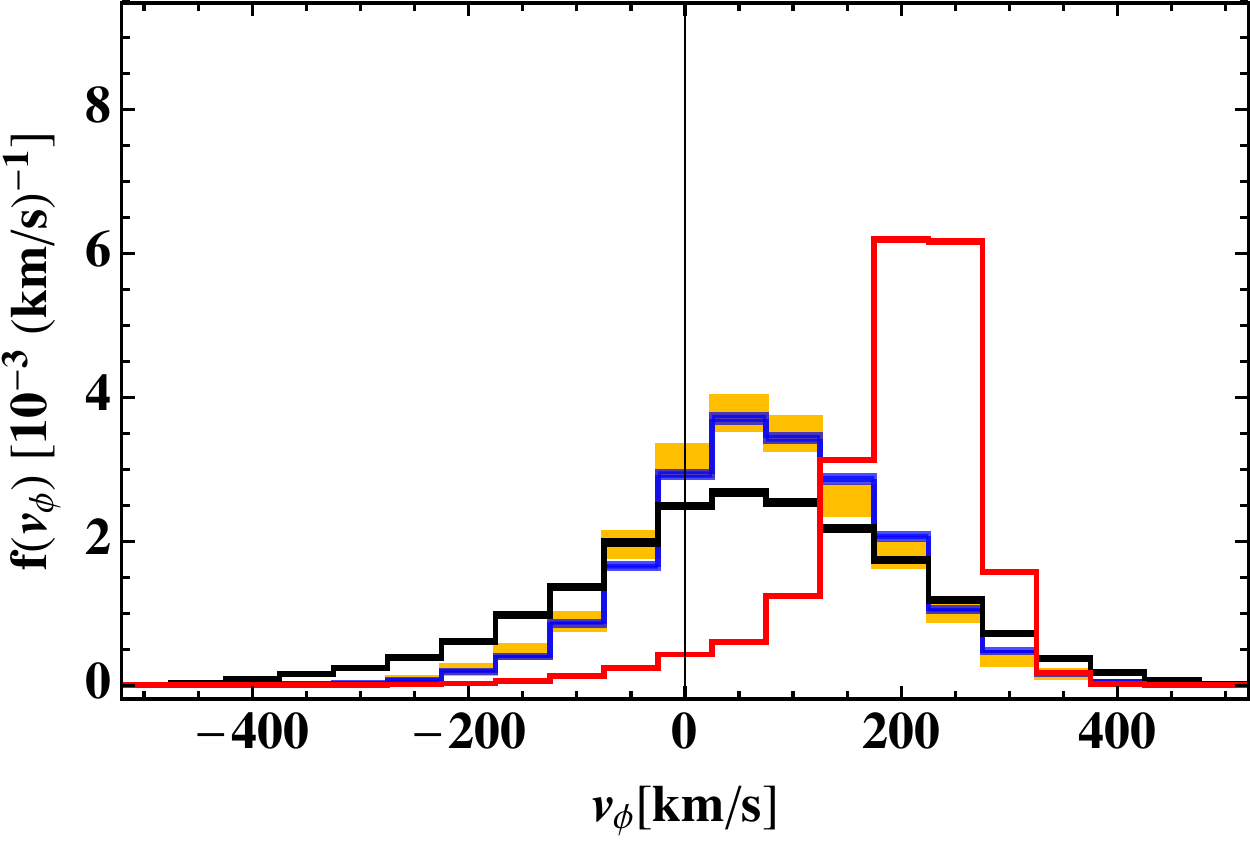}
   \includegraphics[width=0.31\textwidth]{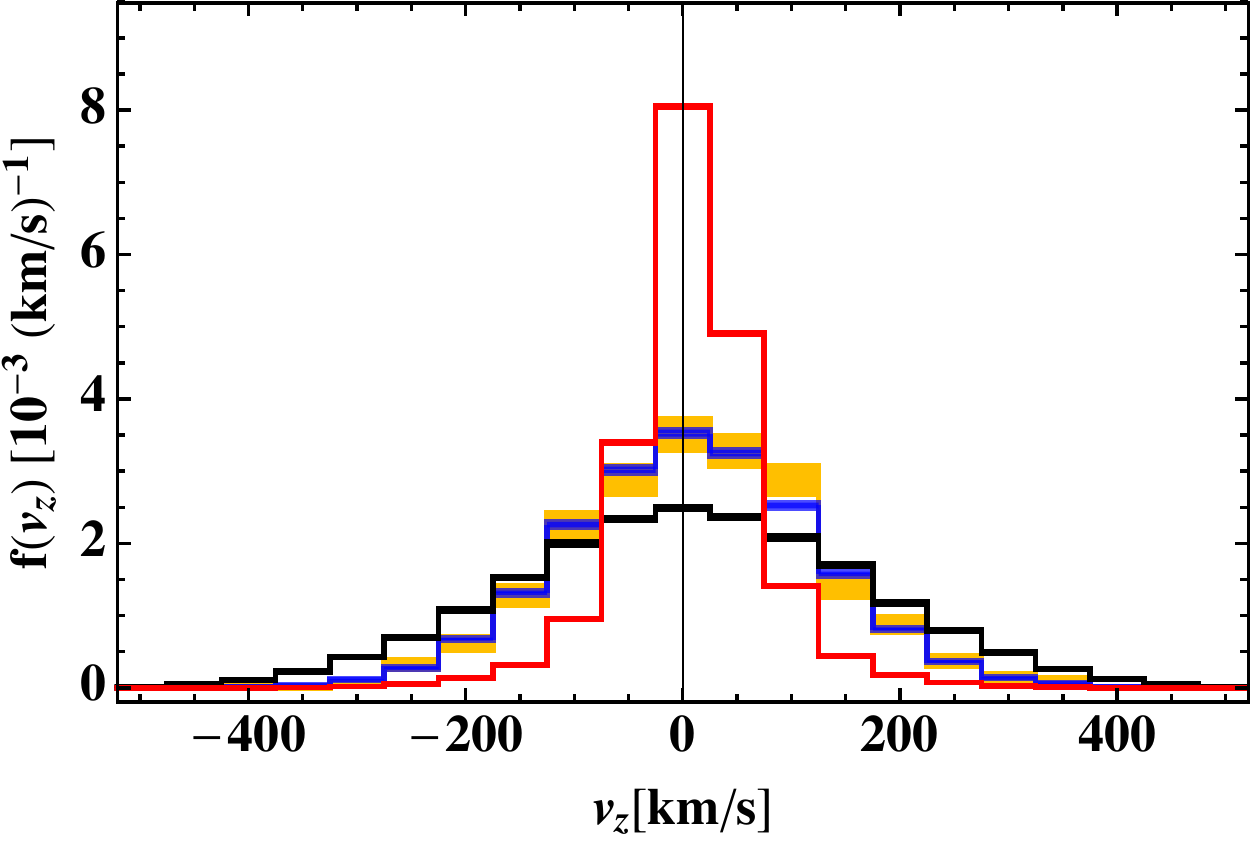}\\
   \includegraphics[width=0.31\textwidth]{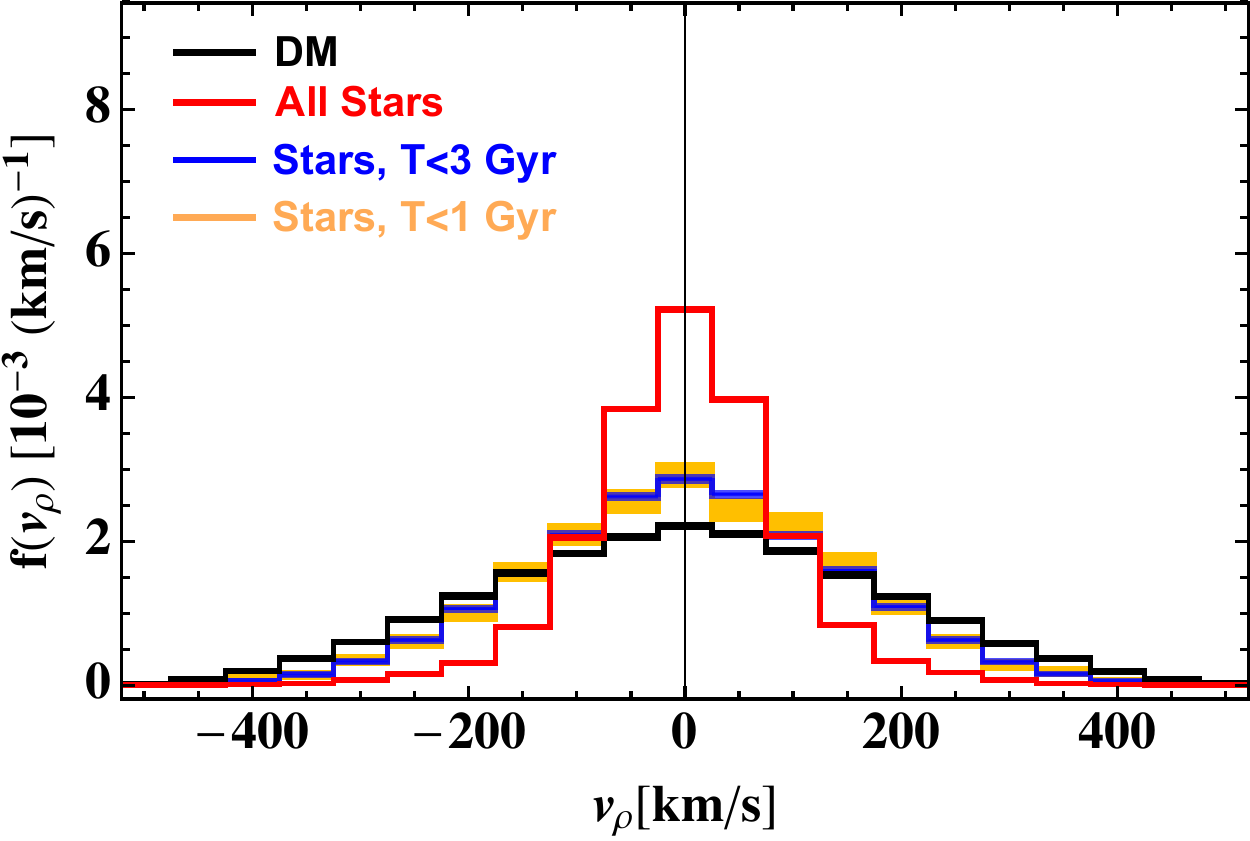}
   \includegraphics[width=0.31\textwidth]{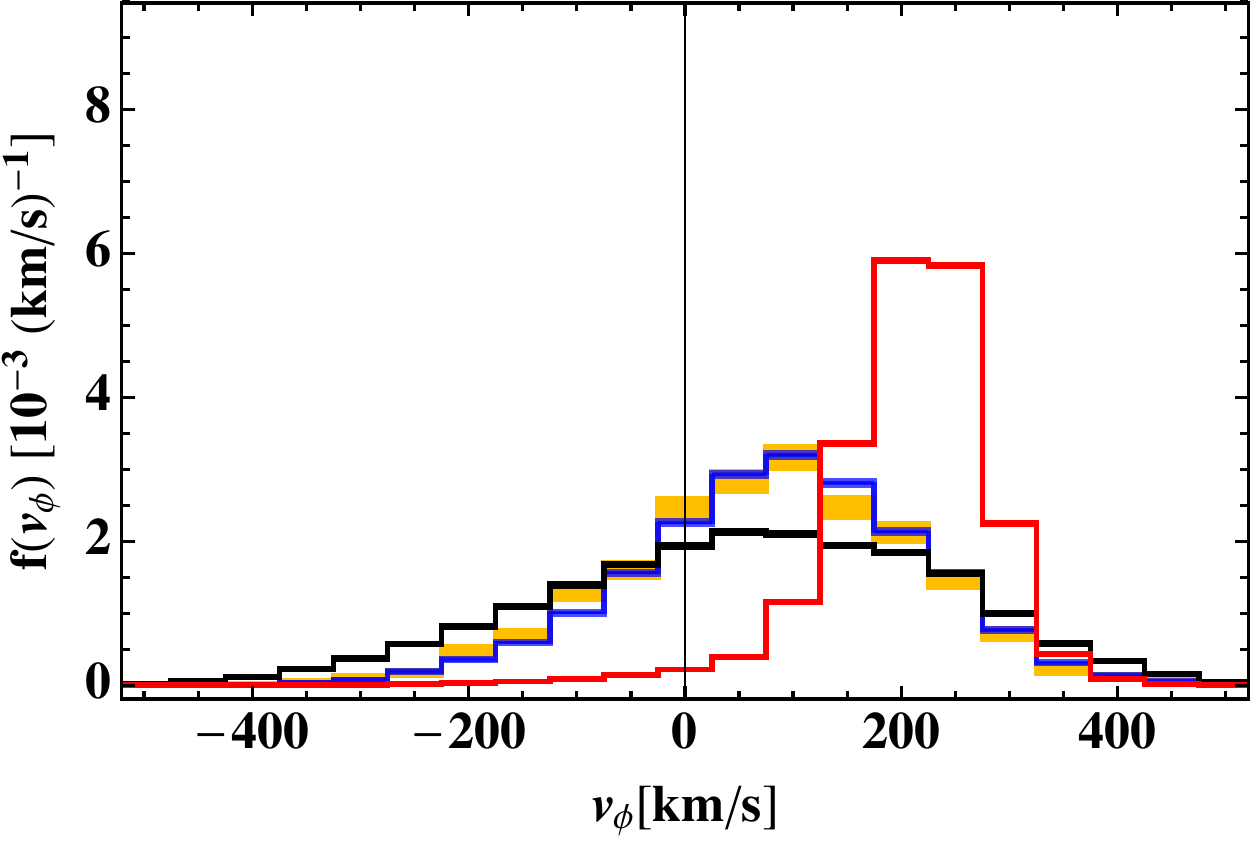}
   \includegraphics[width=0.31\textwidth]{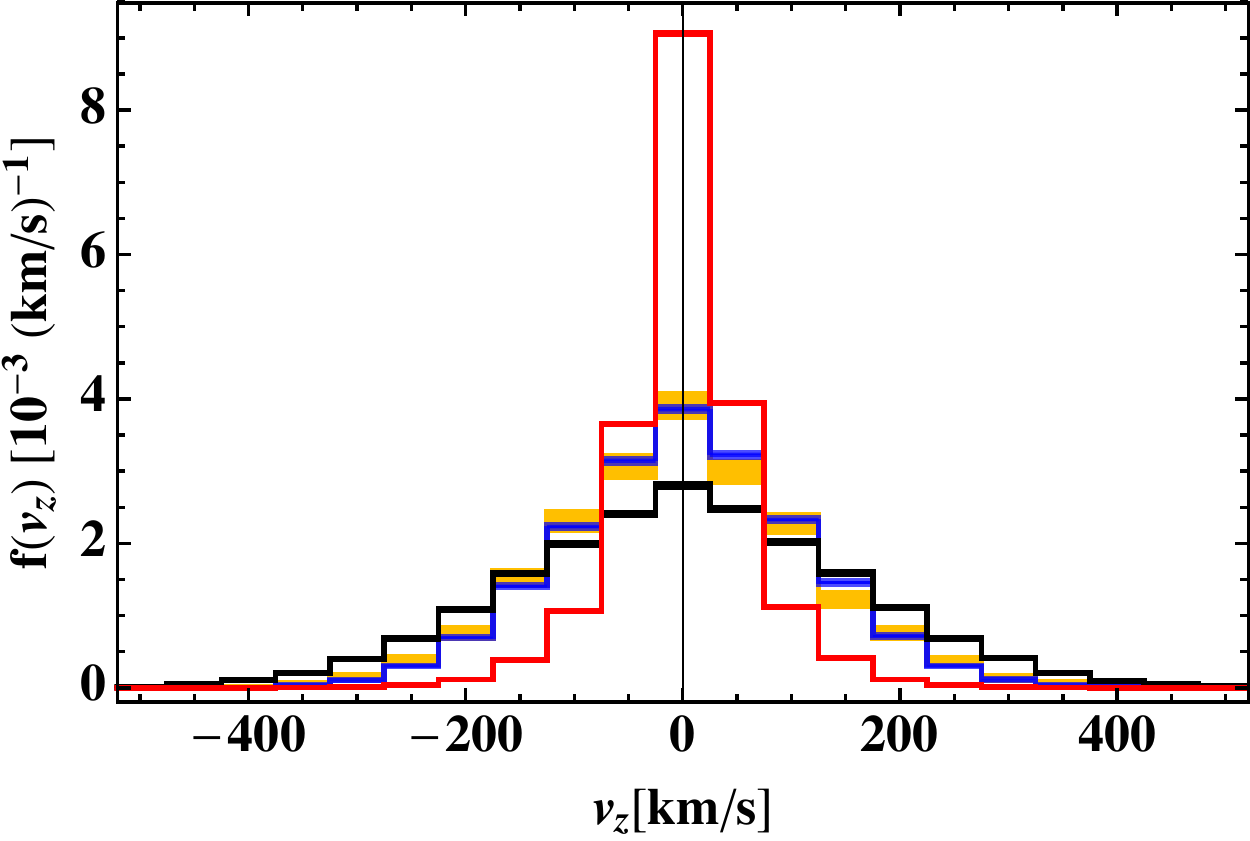}\\
   \includegraphics[width=0.31\textwidth]{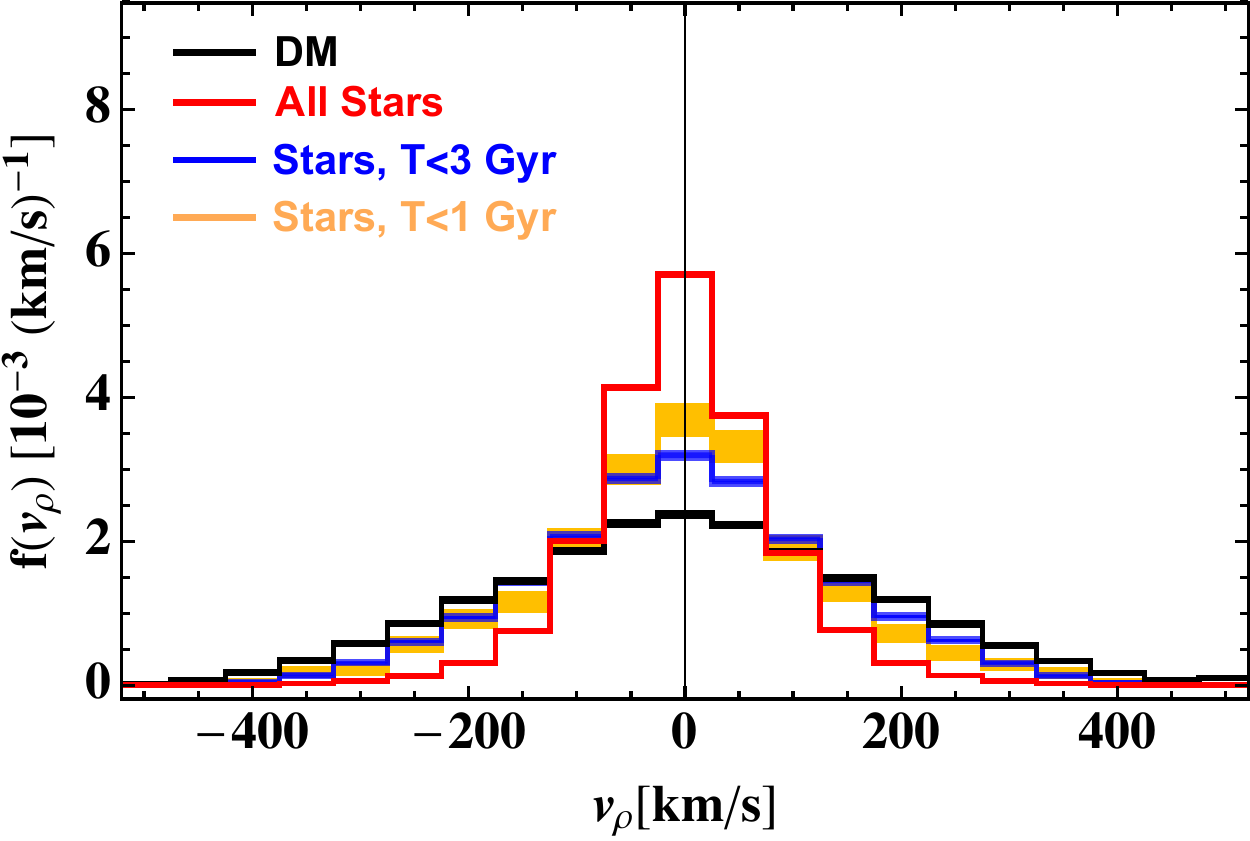}
   \includegraphics[width=0.31\textwidth]{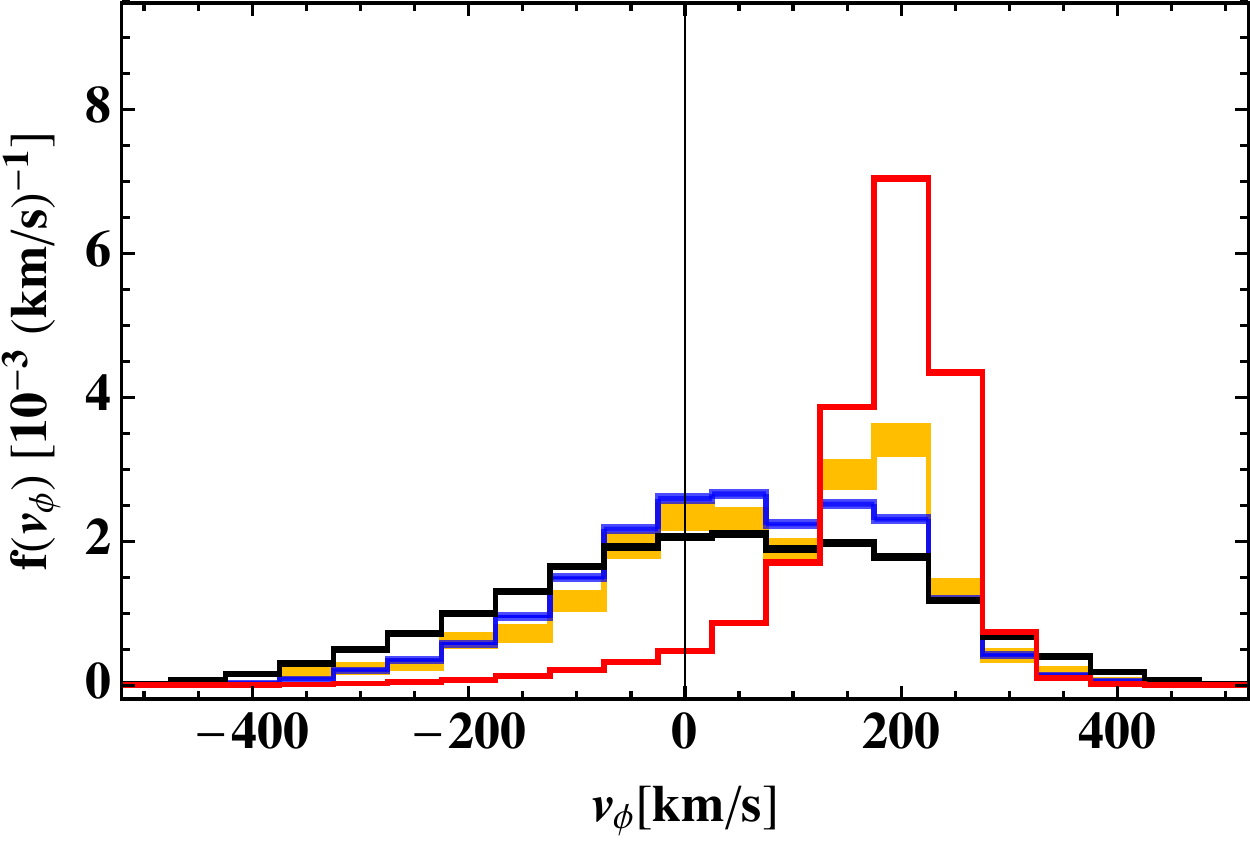}
   \includegraphics[width=0.31\textwidth]{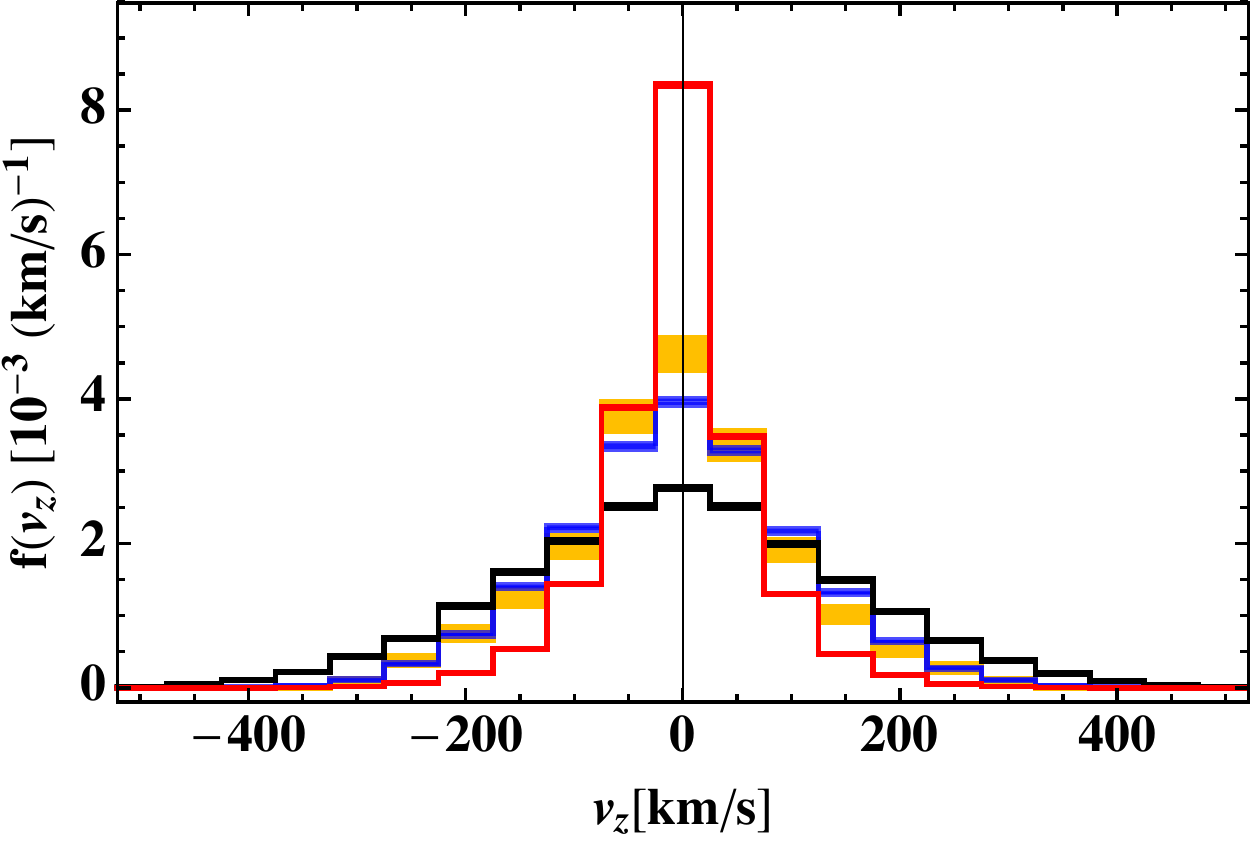}\\
   \includegraphics[width=0.31\textwidth]{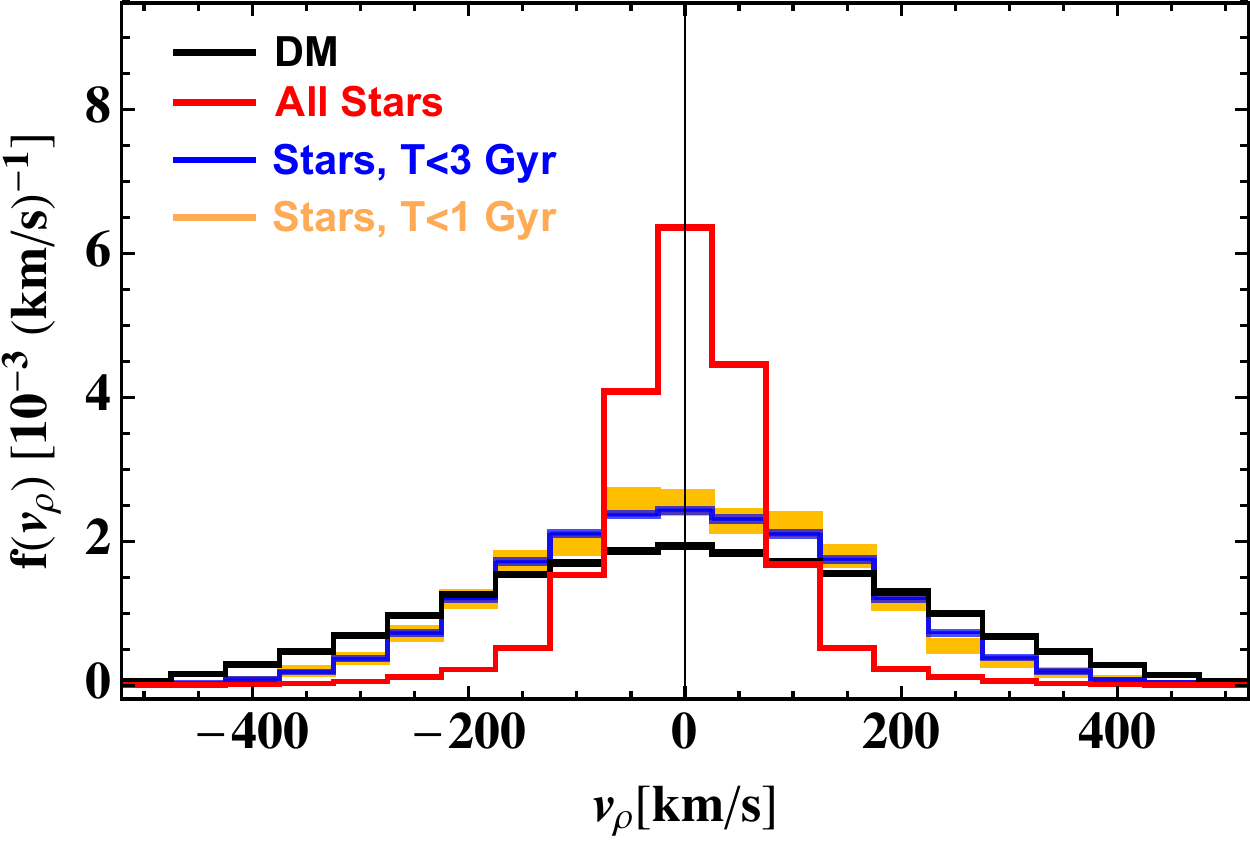}
   \includegraphics[width=0.31\textwidth]{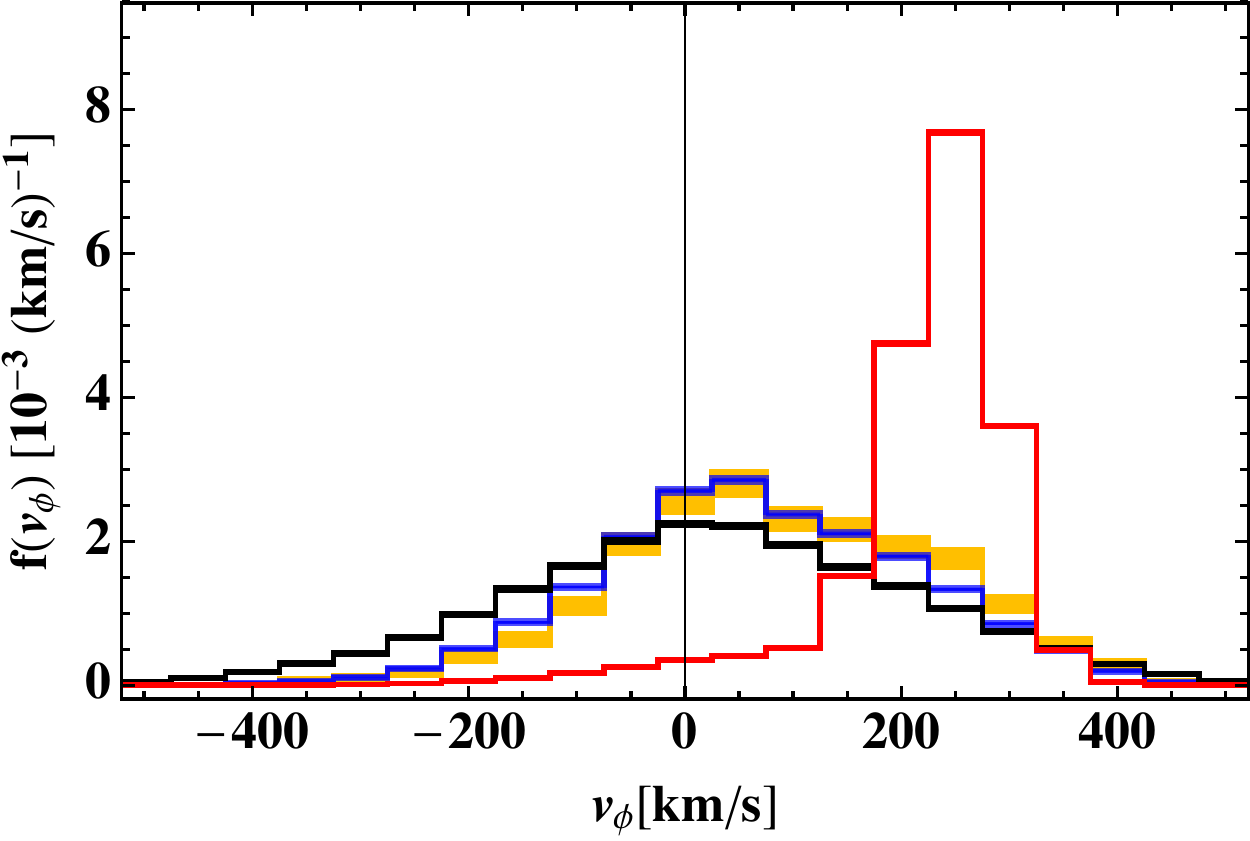}
   \includegraphics[width=0.31\textwidth]{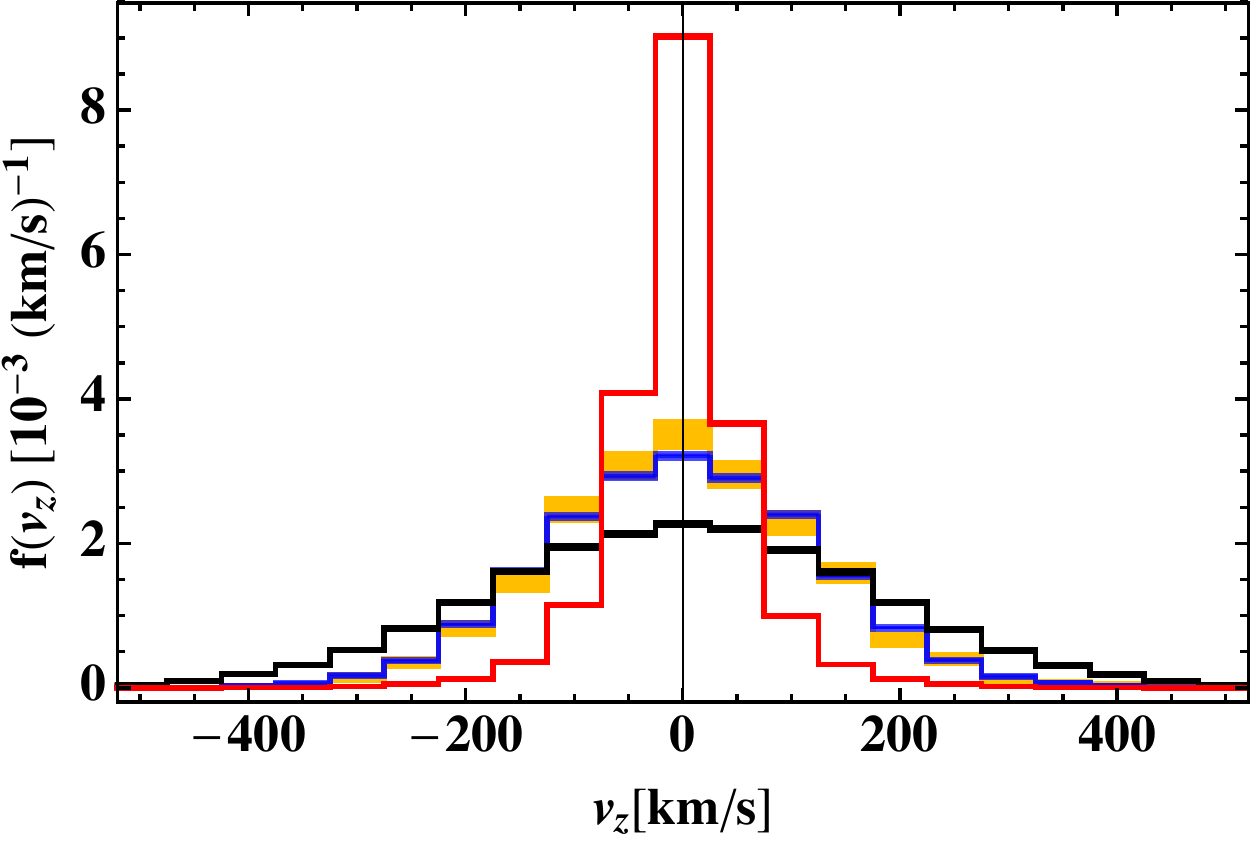}
\caption{The radial (left column), azimuthal (middle column), and vertical (right column) components of the DM (black) and stellar velocity distributions in the Galactic reference frame for six Auriga halos: Au6, Au16, Au21, Au23, Au24, and Au27 in rows one to six, respectively. The shaded regions specify the $1\sigma$  Poisson error on the data points. The velocity distribution of all stars (red) is different from those of old stars with formation time less than 3 Gyr (blue) and less than 1 Gyr (yellow) after the Big Bang. Both the DM and stellar velocity distributions are shown in the Solar neighbourhood, $7 \leq \rho \leq 9$~kpc and $|z|\leq 2$~kpc.}
\label{fig:TCuts}
\end{center}
\end{figure}

In figure~\ref{fig:TCuts} we show the radial, azimuthal, and vertical components of the DM and stellar velocity distributions in the Solar neighbourhood for six Level 3 Auriga halos: Au6, Au16, Au21, Au23, Au24, and Au27. Additionally, we show the velocity distributions of stars which have formed less than 3 Gyr and less than 1 Gyr after the Big Bang. The shaded regions specify the $1\sigma$  Poisson error on the data points. Due to the lower number of old stars, their Poisson errors are larger compared to all stars or DM particles. In the Solar neighbourhood region $7 \leq \rho \leq 9$~kpc and $|z| \leq 2$~kpc, the total number of stars is in the range of $[0.87 - 1.76] \times 10^6$, depending on the halo. The number of stars with formation time $T<3$~Gyr and $T<1$~Gyr, is $[3.2 - 7.6] \times 10^4$ and $[1.5 - 2.9] \times 10^3$, respectively. The number of DM particles in the same region is $[1.2 - 1.6] \times 10^5$, depending on the halo.

We can see from figure~\ref{fig:TCuts} that the velocity distribution of very old stars formed less than 1 Gyr after the Big Bang is similar to the distribution of stars which formed less than 3 Gyr after the Big Bang. Both of these distributions, however, significantly differ from the DM velocity distributions, and have smaller velocity dispersions. It is clear that there is no correlation between the velocity distribution of old stars and DM.

A number of other interesting features can be recognised from figure~\ref{fig:TCuts}. The stellar velocity distribution in the azimuthal direction has a peak speed of $\sim 200$~km$/$s, clearly indicating the presence of the stellar disc. This is particularly visible for the full stellar distribution, and old stars formed within $T<3$~Gyr do not present a strong disc kinematics. This implies that the younger stars are more likely found in the disc or that the disc assembled progressively. Notice also that, as expected, the DM distribution does not show this strong disc kinematic signature. There is, however, a small shift in the peak speed of the azimuthal component of the DM velocity distribution, which may be an indication of the presence of a disrupted substructure, remnant of a satellite merger\footnote{See e.g.~ref.~\cite{Gomez:2017yzr} which shows that co-planar mergers are the progenitors of ex-situ stellar discs in some Auriga simulated galaxies. Remnants of such mergers may also cause a shift in the azimuthal DM velocity distribution.}, or some other transient effect in the Solar neighbourhood region we considered.

\begin{figure}[ht!]
\begin{center}
   \includegraphics[width=0.31\textwidth]{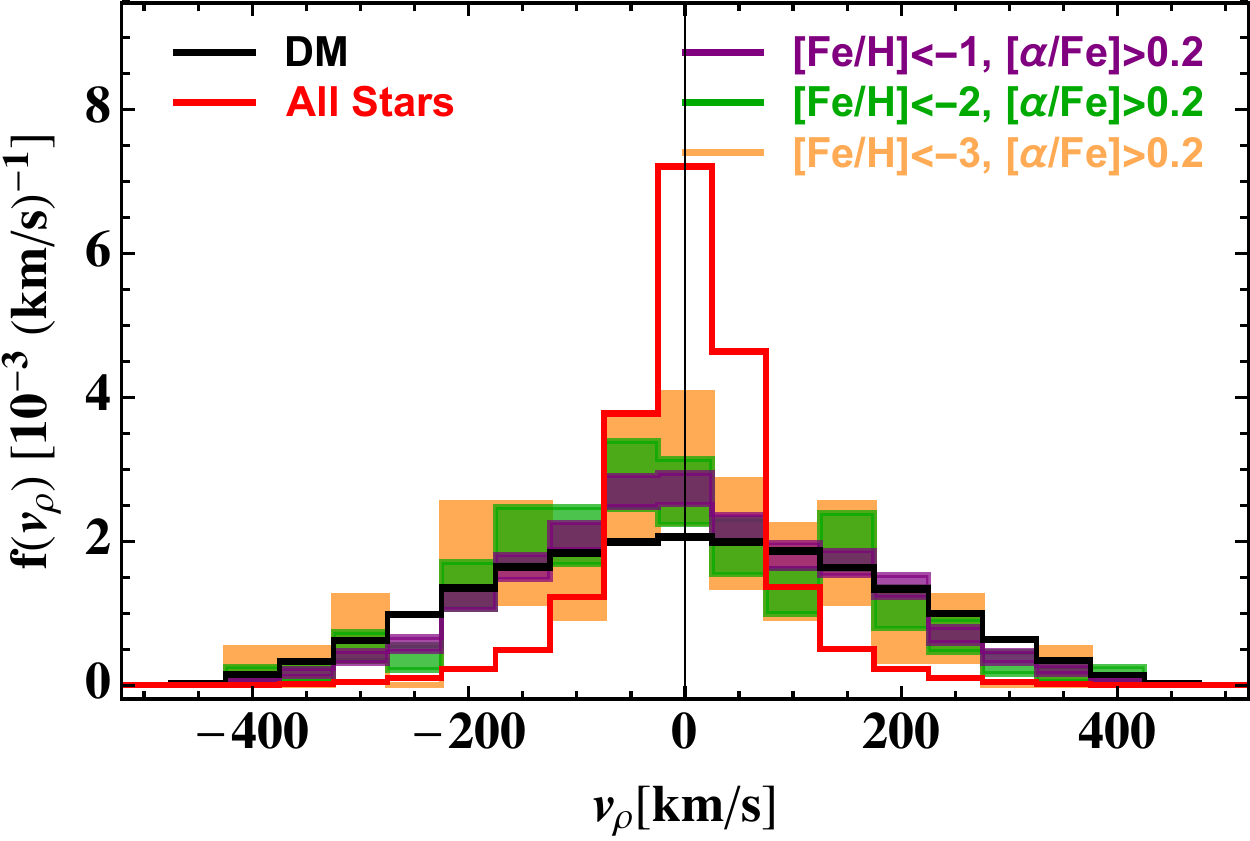}
   \includegraphics[width=0.31\textwidth]{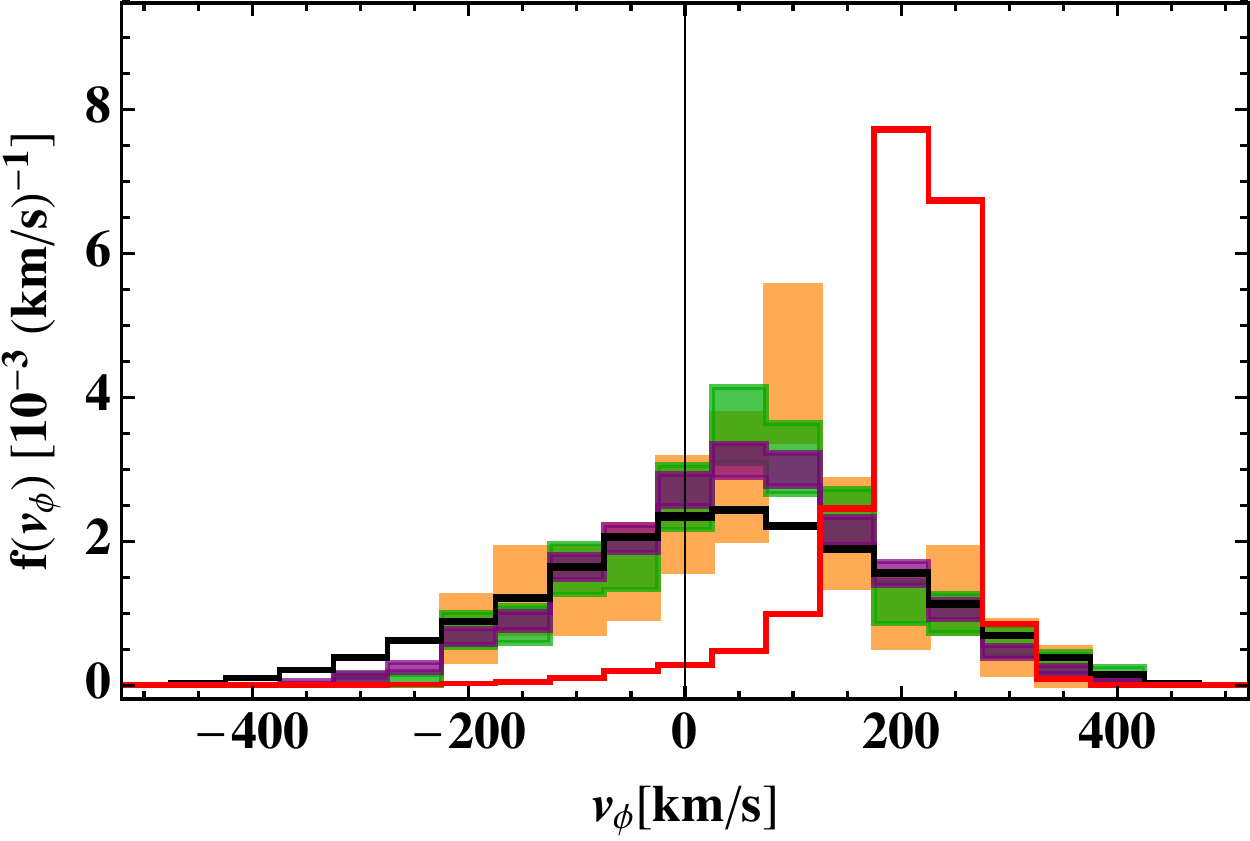}
   \includegraphics[width=0.31\textwidth]{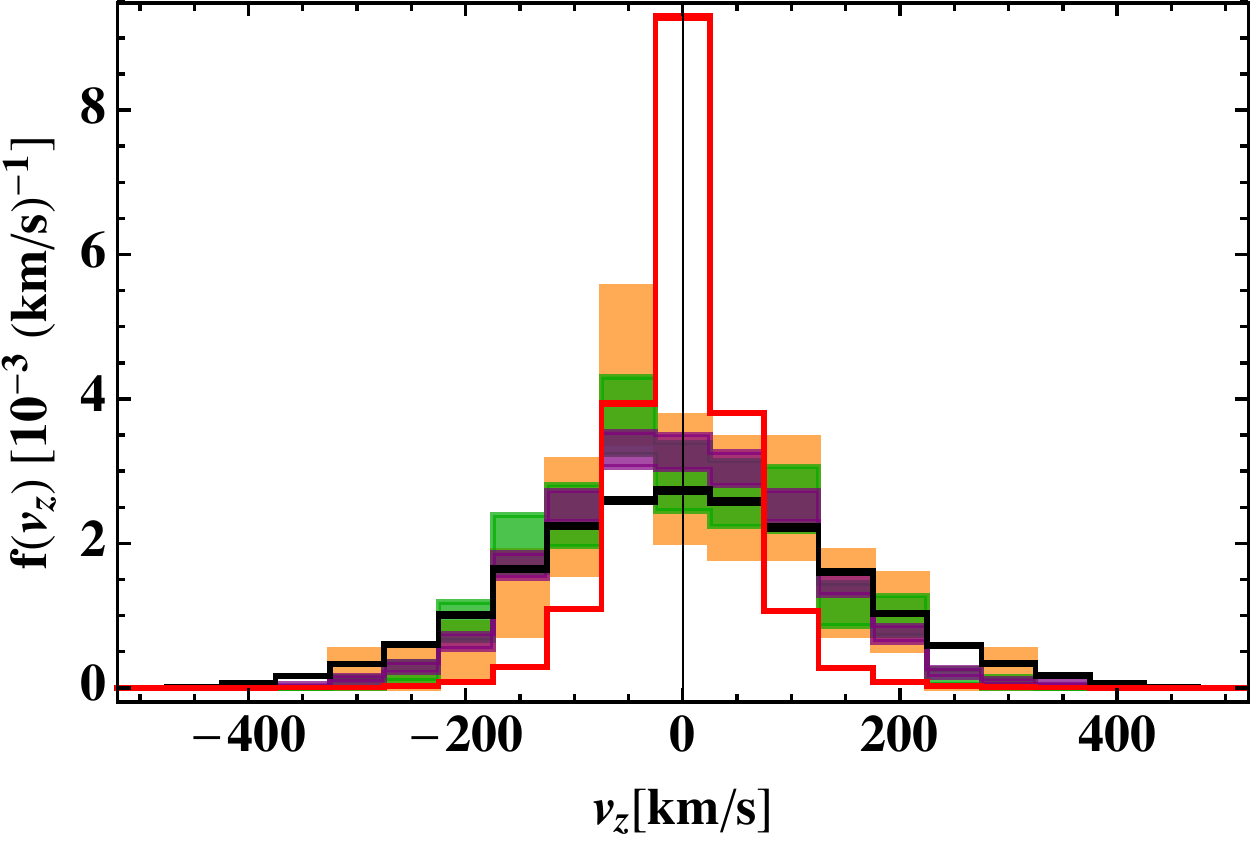}\\
   \includegraphics[width=0.31\textwidth]{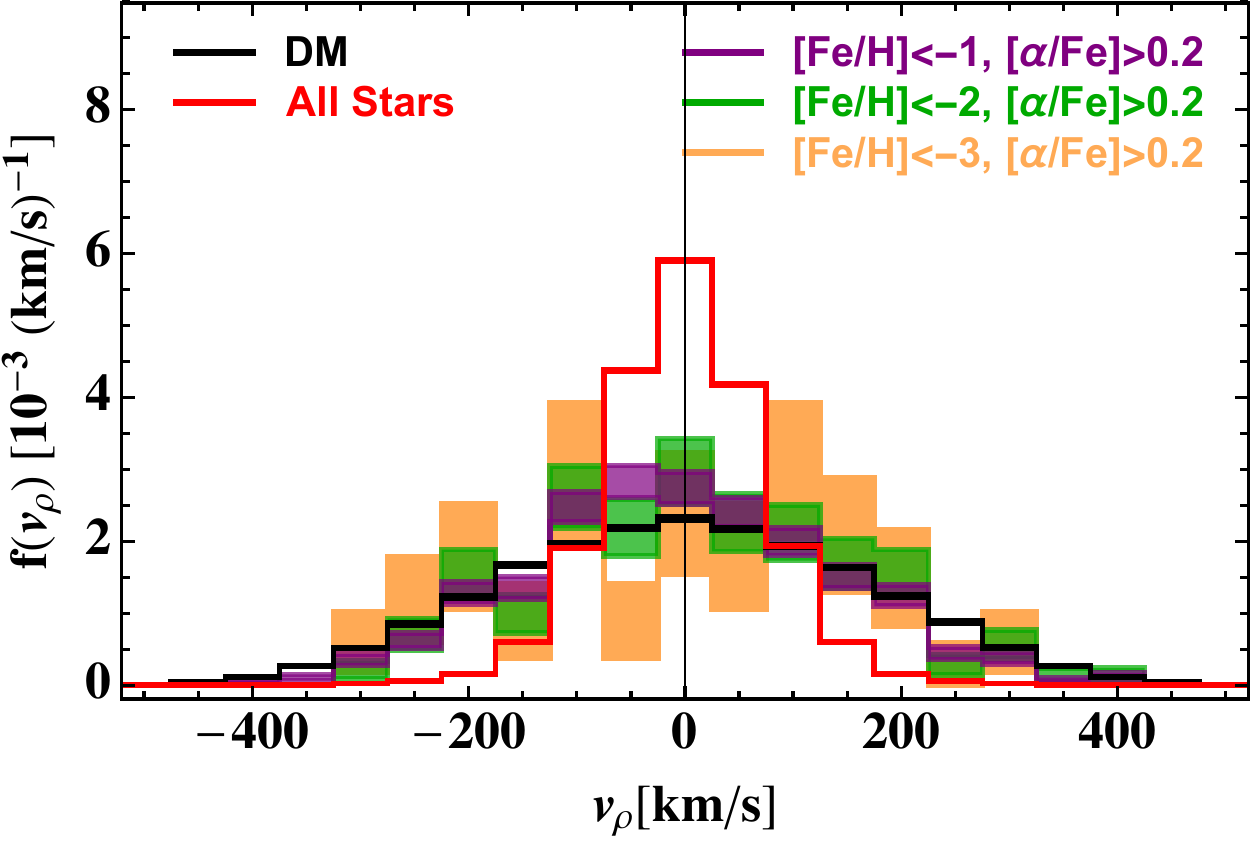}
   \includegraphics[width=0.31\textwidth]{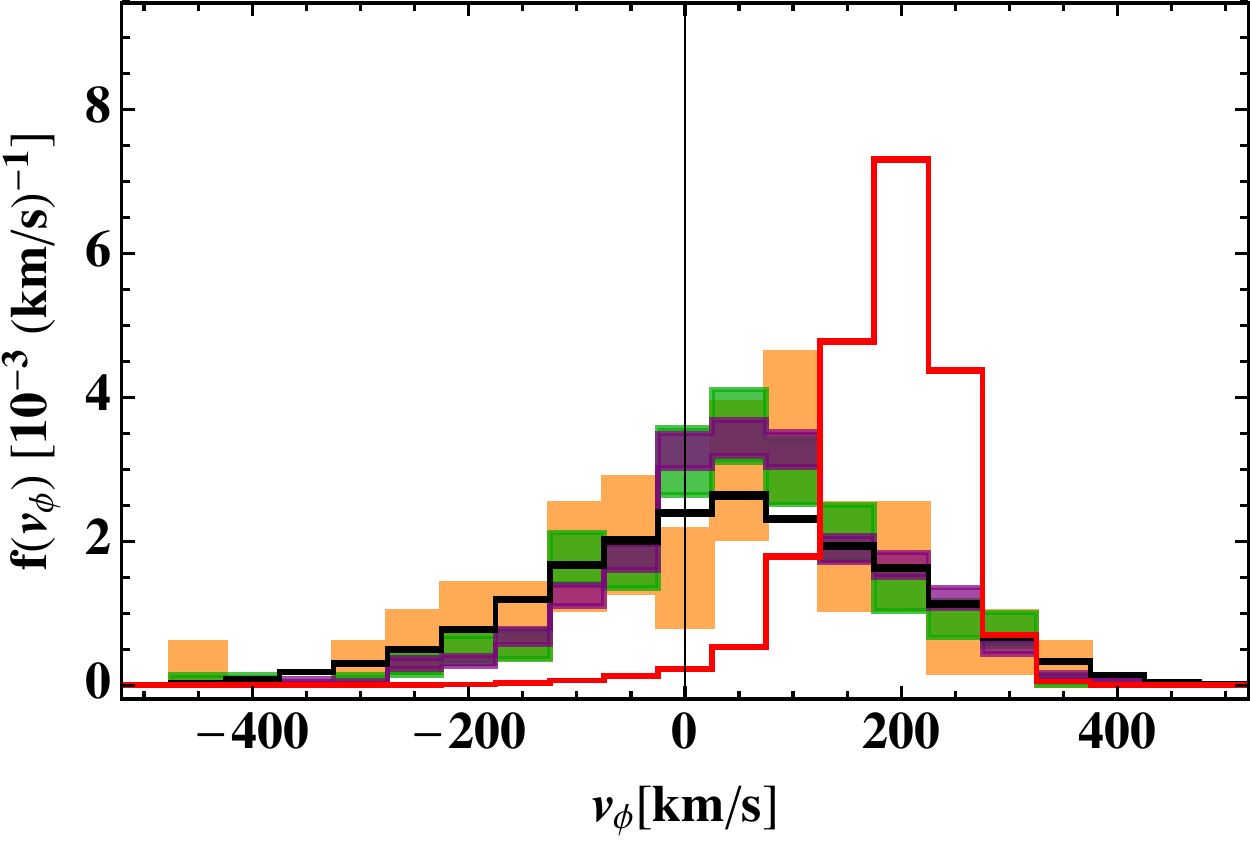}
   \includegraphics[width=0.31\textwidth]{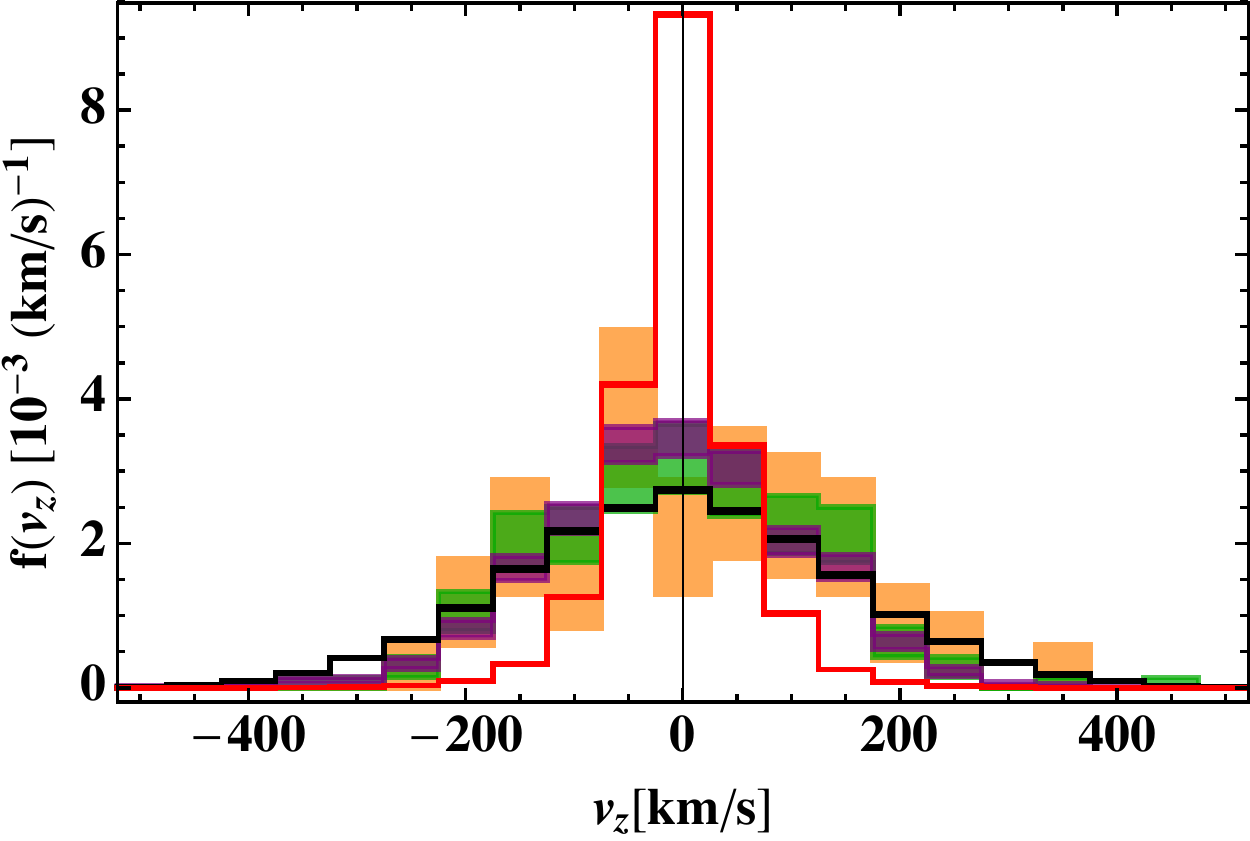}\\
   \includegraphics[width=0.31\textwidth]{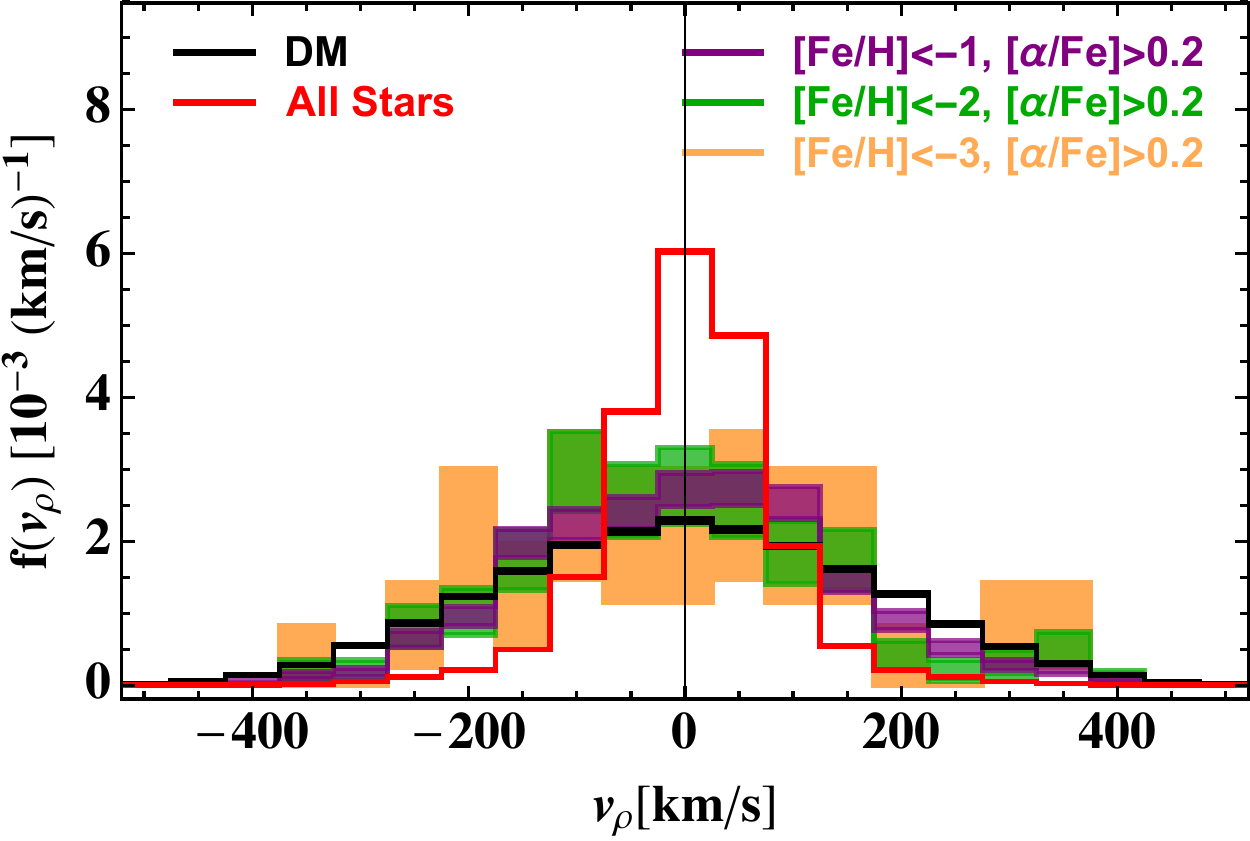}
   \includegraphics[width=0.31\textwidth]{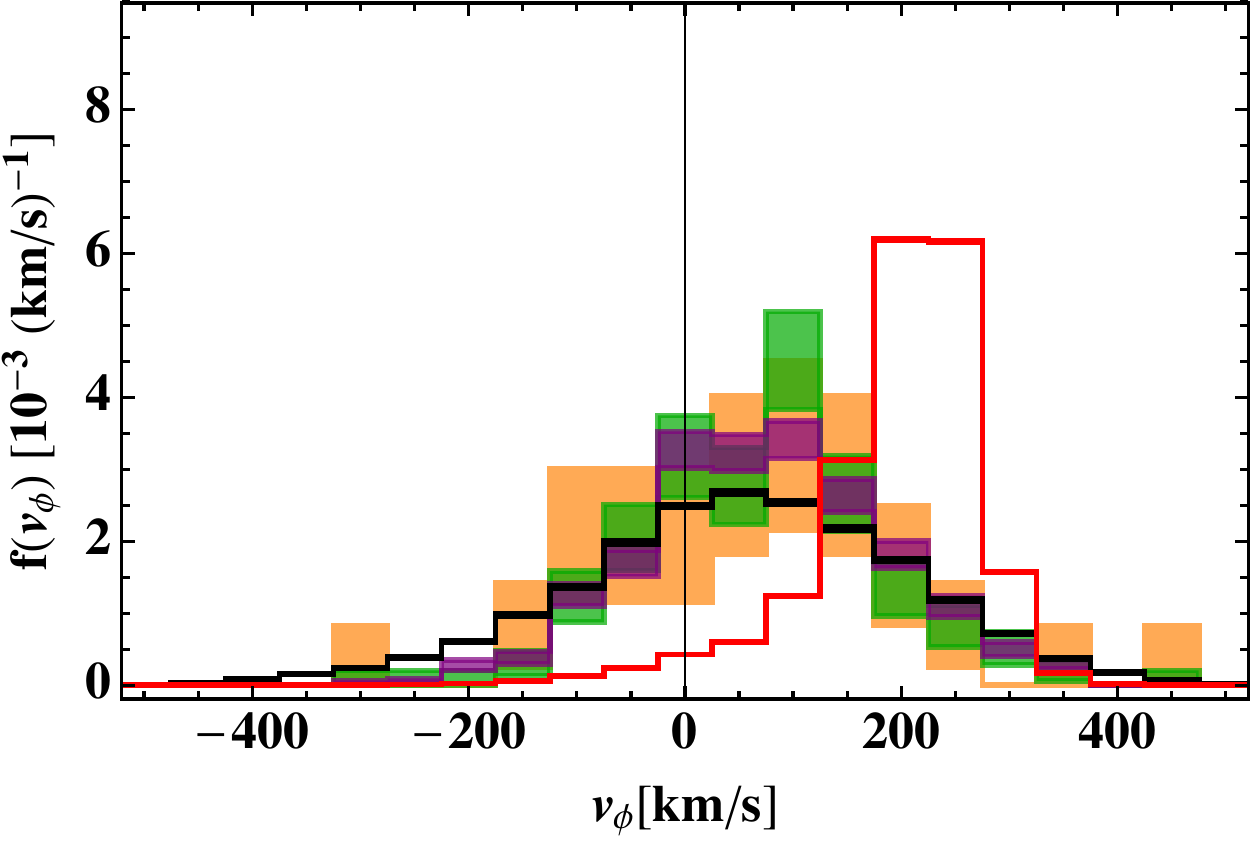}
   \includegraphics[width=0.31\textwidth]{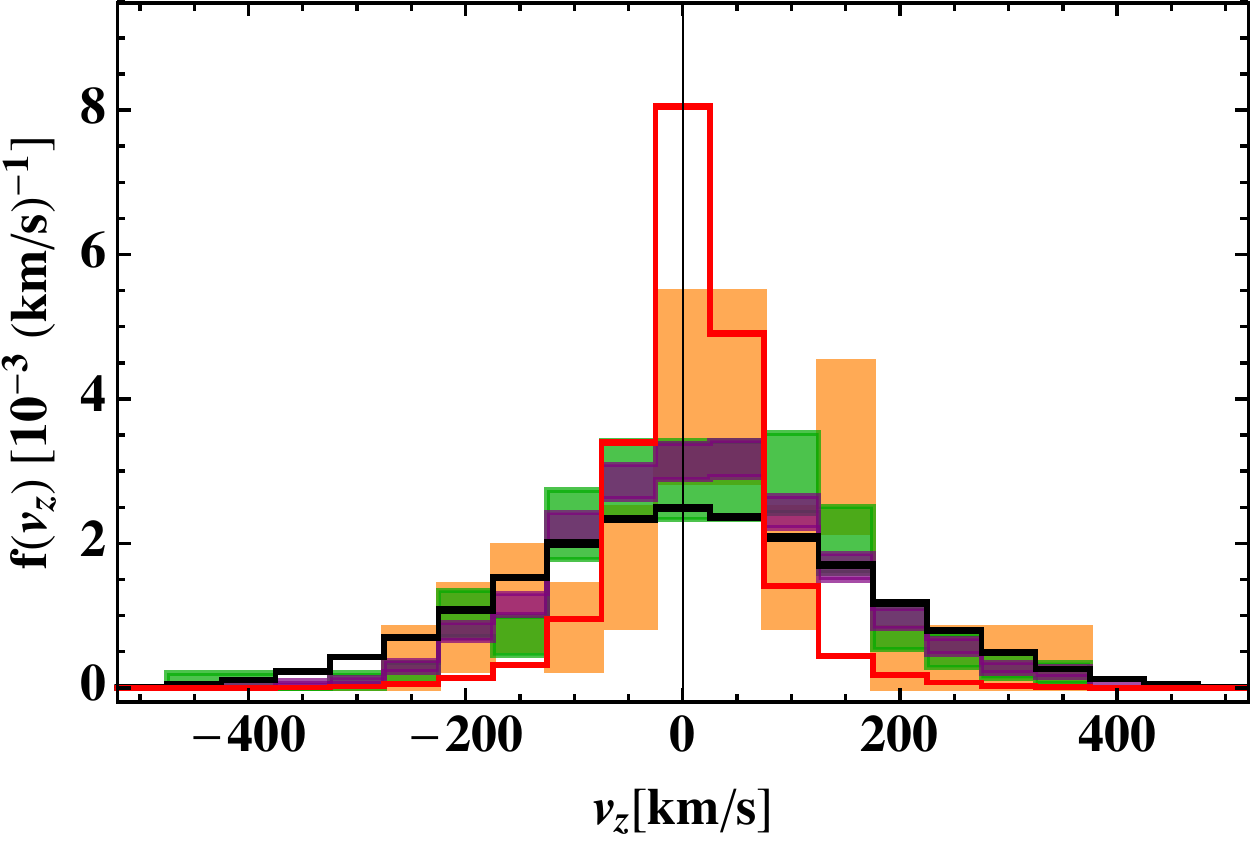}\\
   \includegraphics[width=0.31\textwidth]{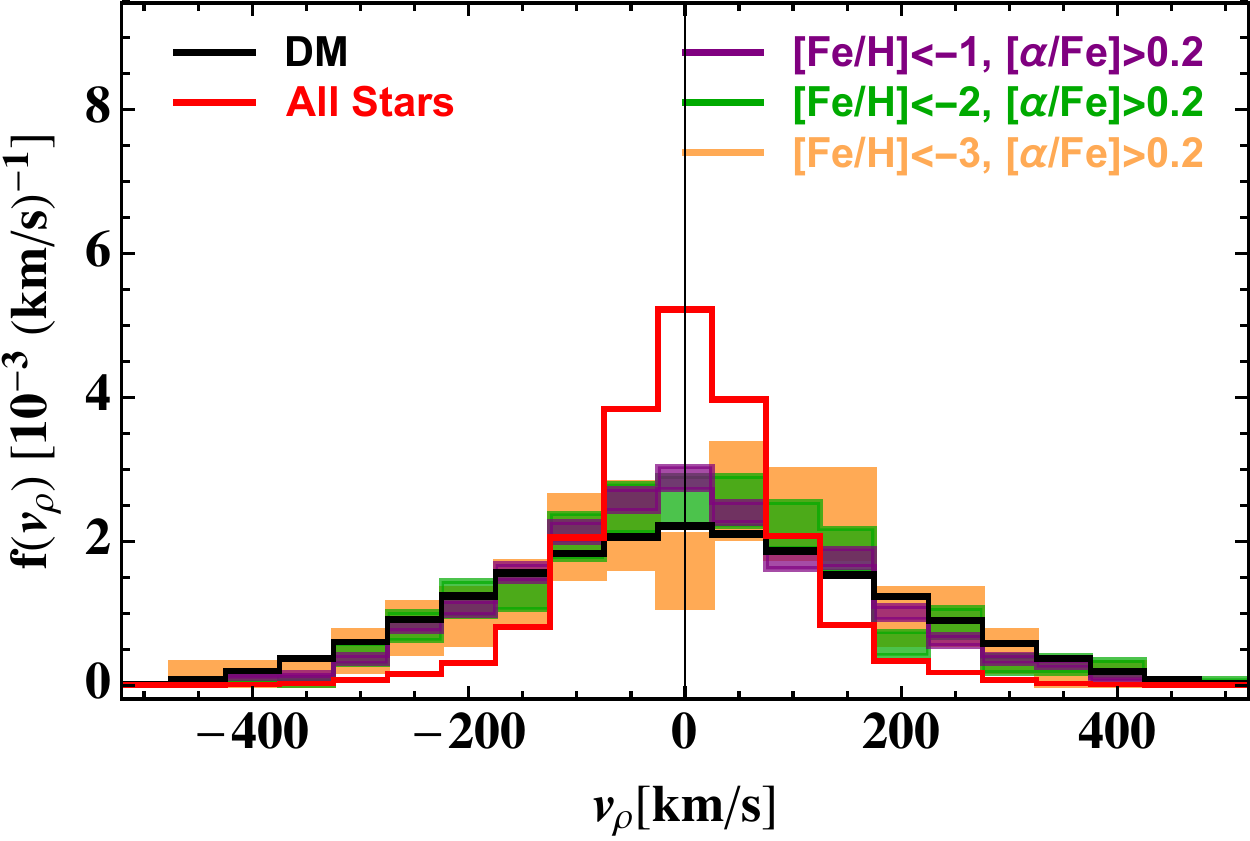}
   \includegraphics[width=0.31\textwidth]{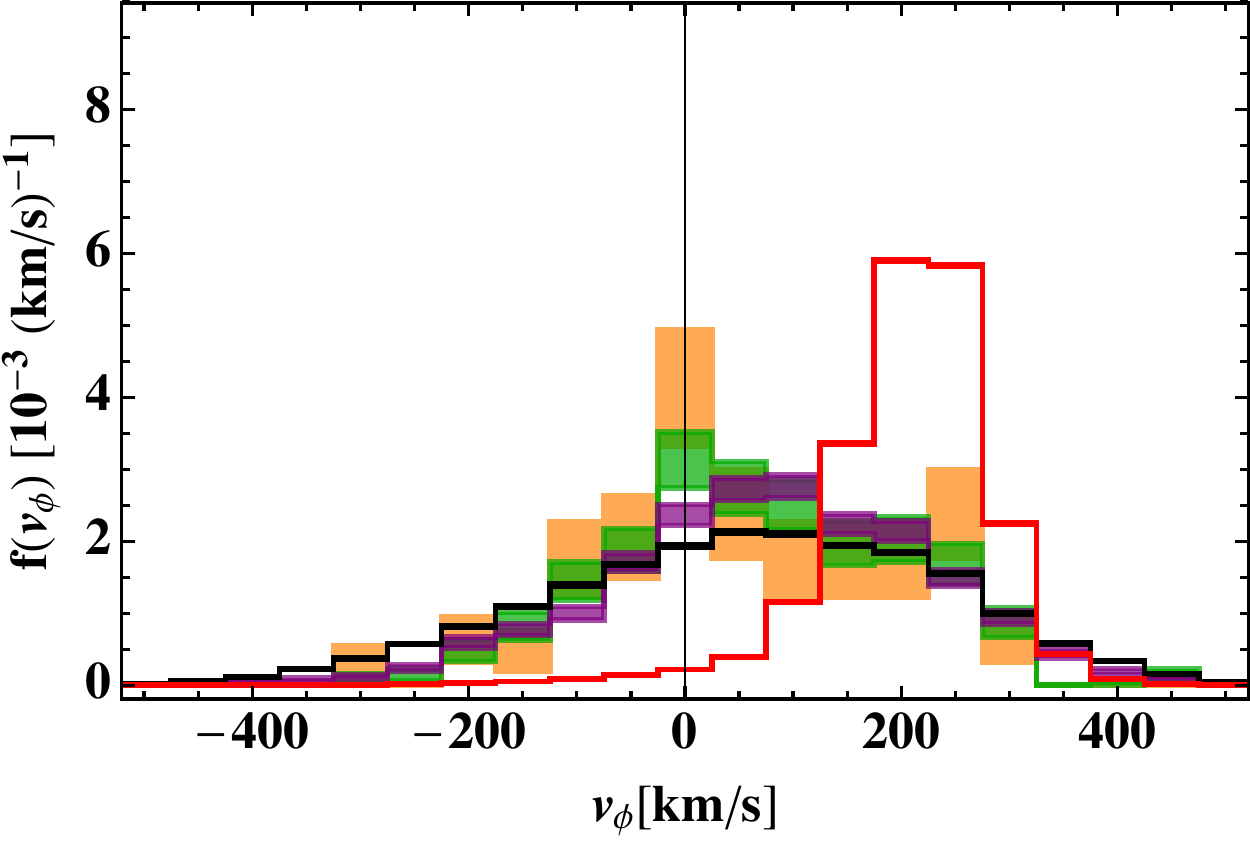}
   \includegraphics[width=0.31\textwidth]{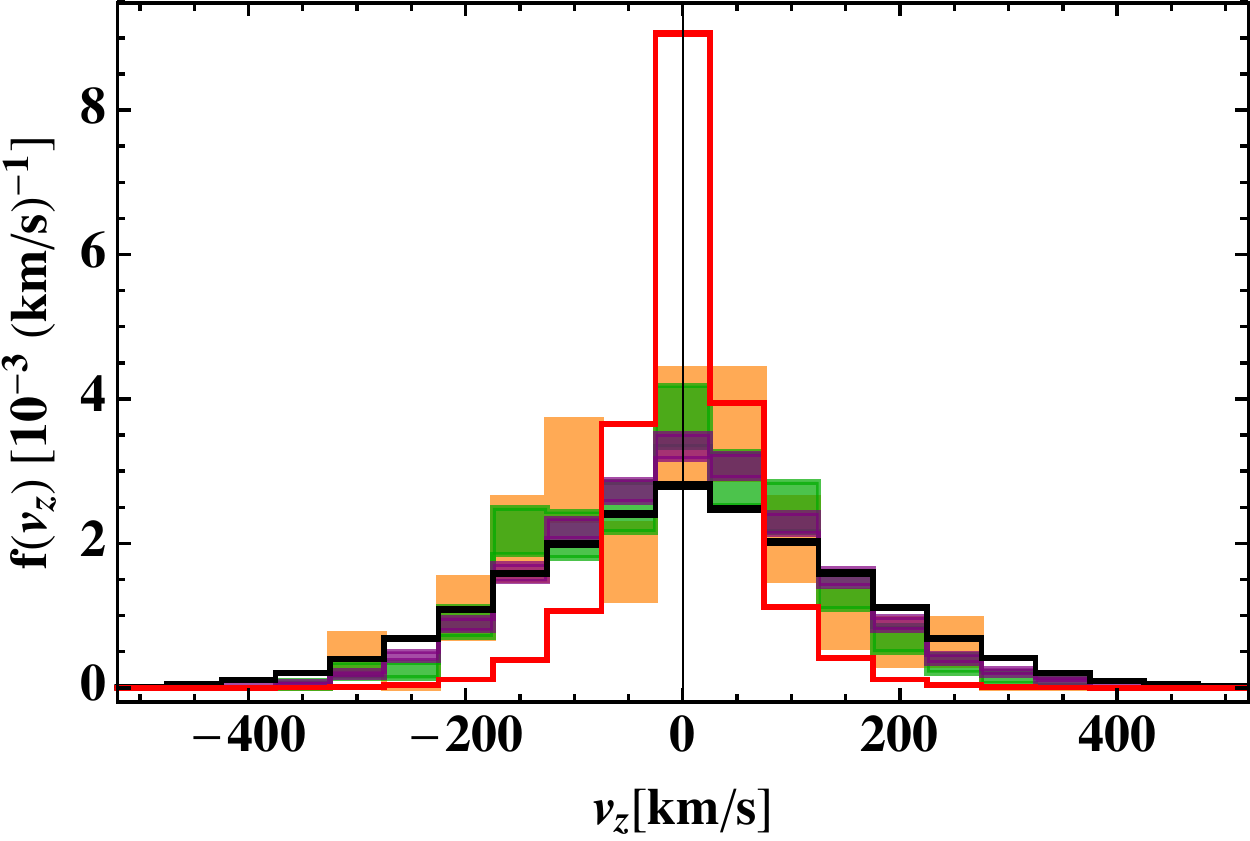}\\
   \includegraphics[width=0.31\textwidth]{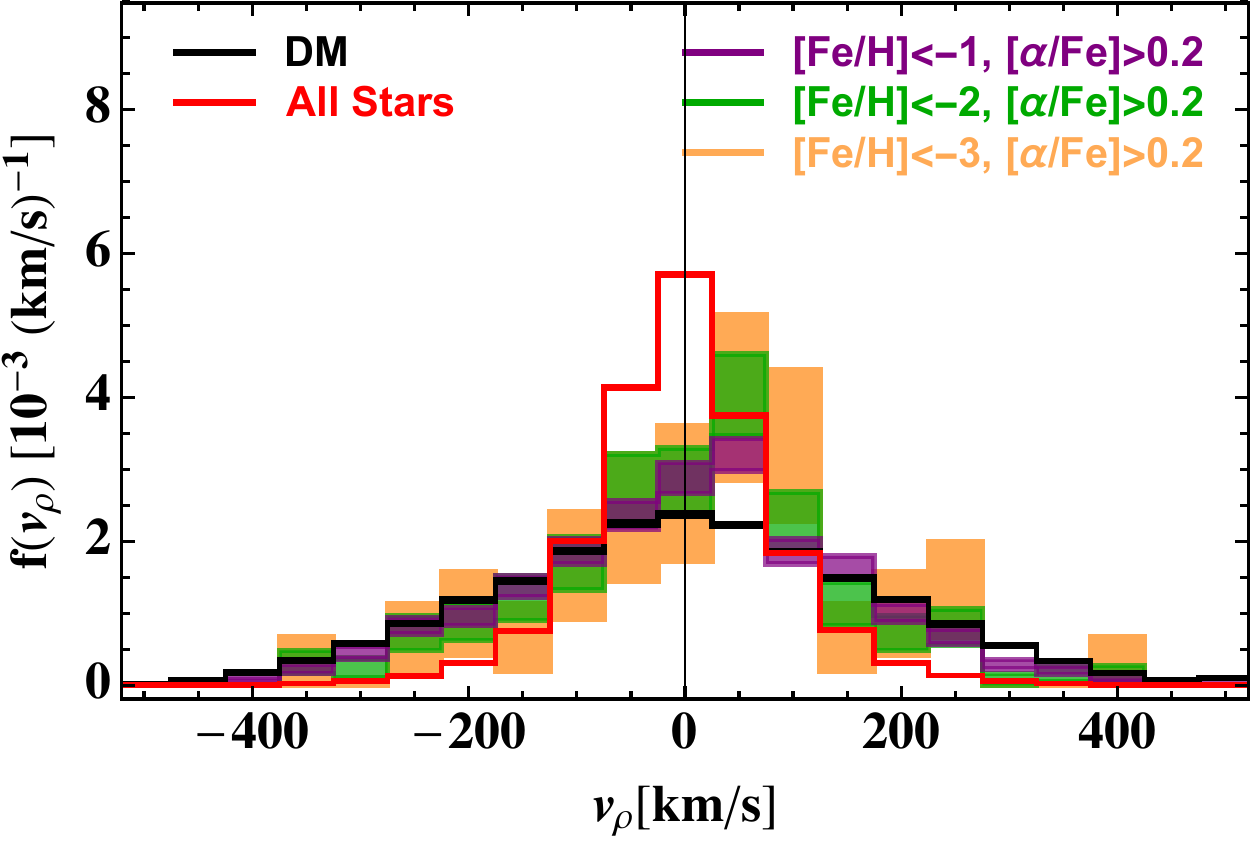}
   \includegraphics[width=0.31\textwidth]{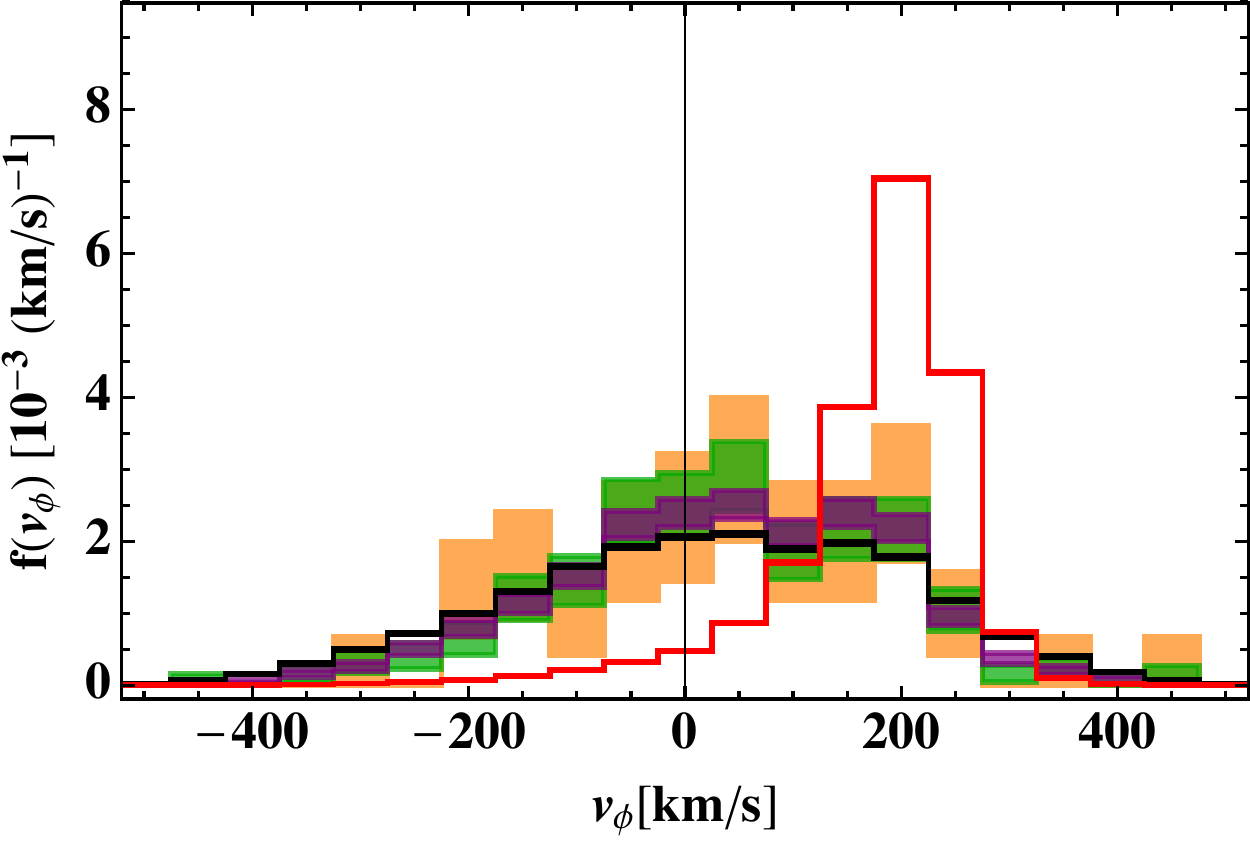}
   \includegraphics[width=0.31\textwidth]{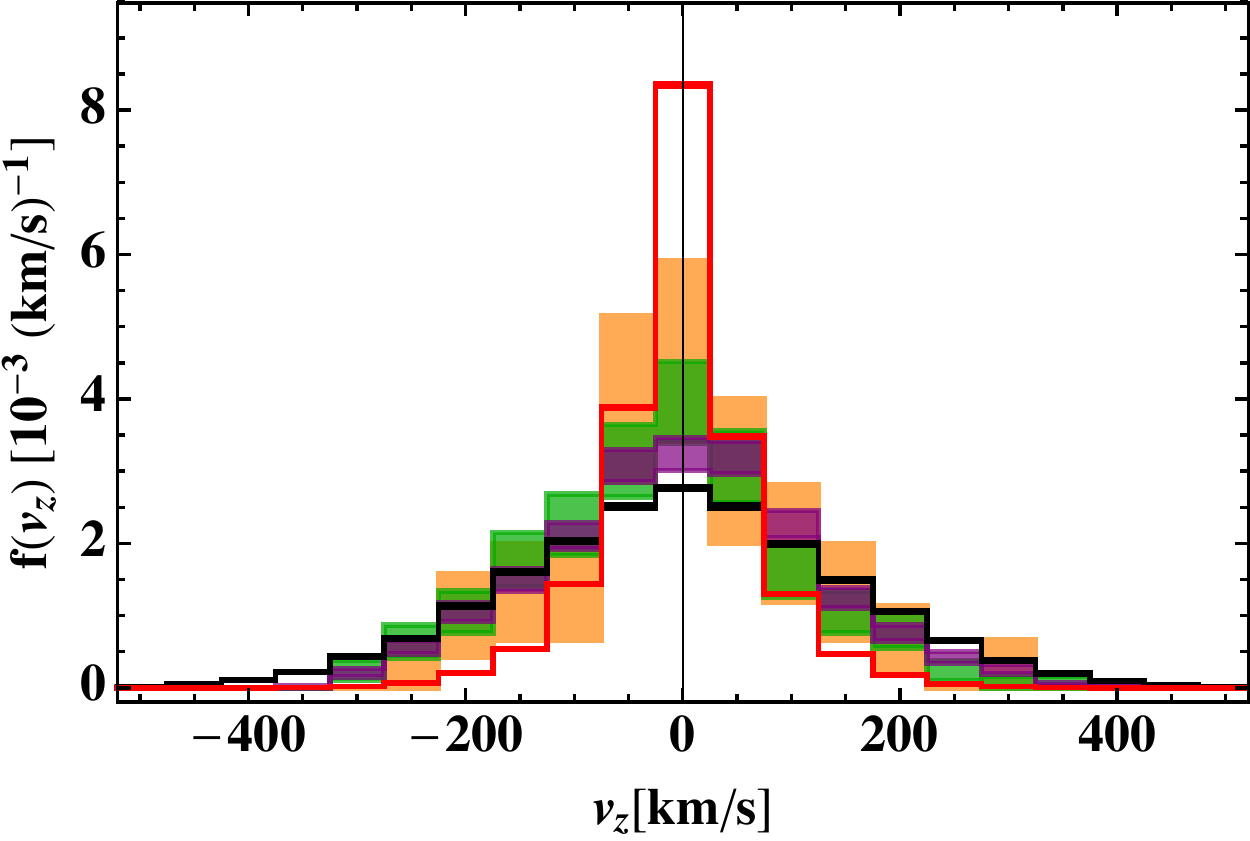}\\
   \includegraphics[width=0.31\textwidth]{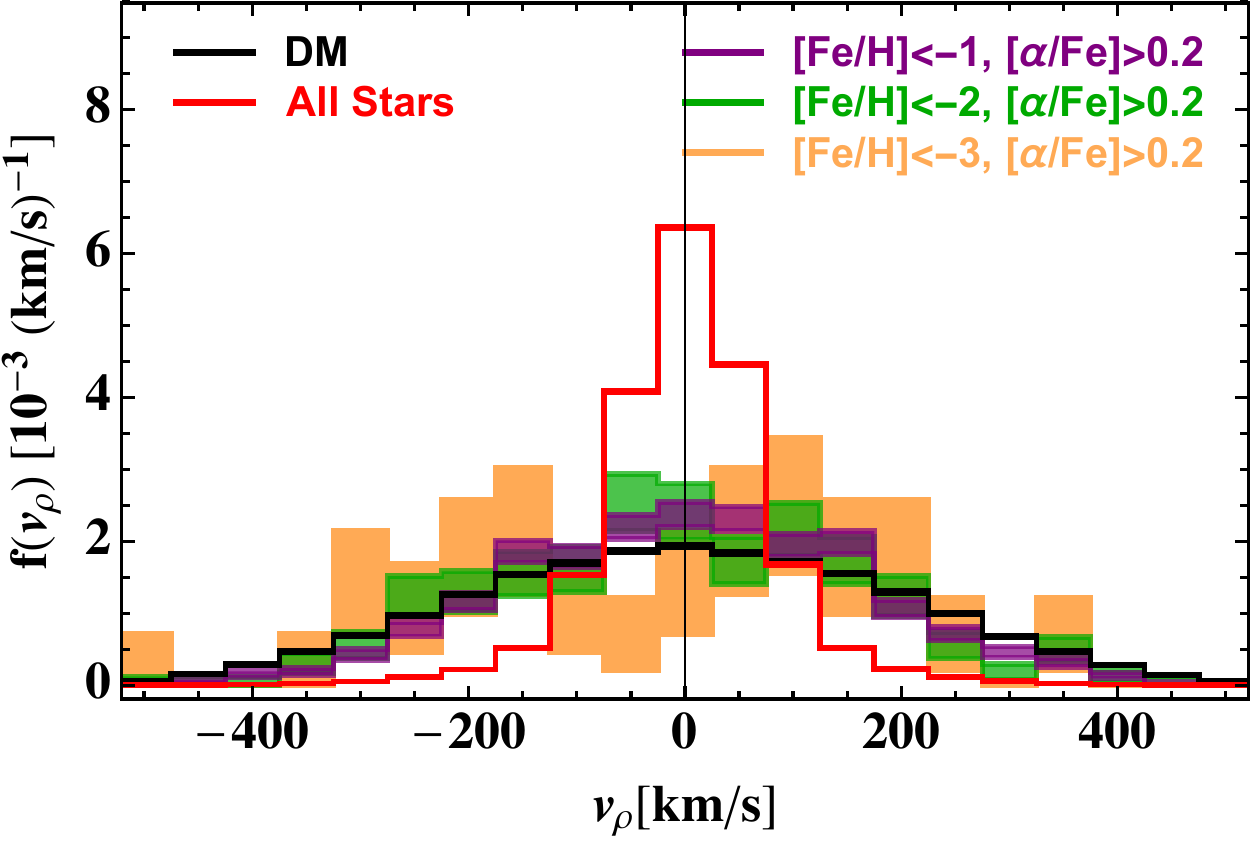}
   \includegraphics[width=0.31\textwidth]{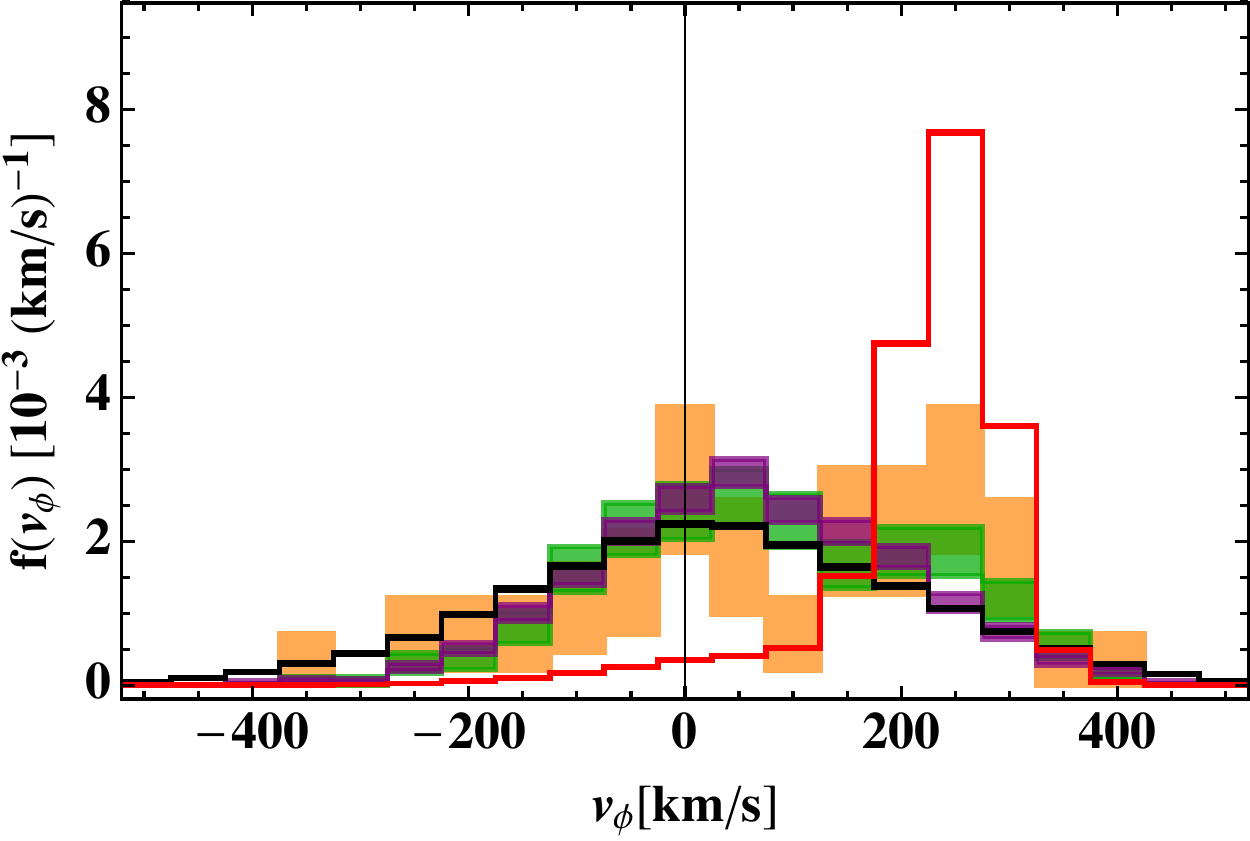}
   \includegraphics[width=0.31\textwidth]{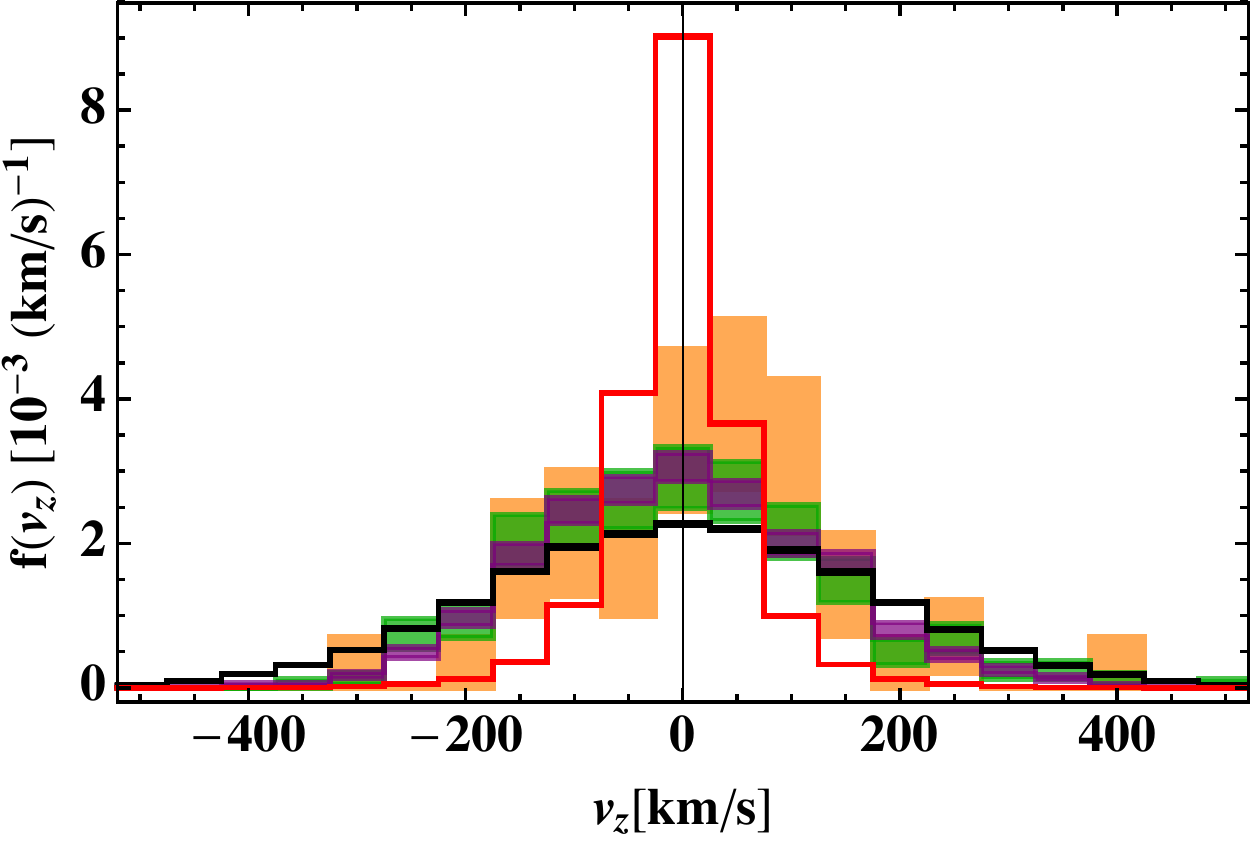}
\caption{Same as figure \ref{fig:TCuts}, but showing the stellar velocity distributions with various metallicity cuts: \FeHone~(purple),  \FeHtwo~(green), and  \FeHthree~(orange), with an additional metallicity cut \aFetwo~in all cases.}
\label{fig:FeHalphaHeCuts}
\end{center}
\end{figure}

In figure~\ref{fig:FeHalphaHeCuts}, we show how setting different cuts on the metallicities \FeH~and \aFe~of stars changes their velocity distributions. In particular we consider three metallicity cuts: \FeHone, \FeHtwo, and \FeHthree, all with \aFetwo. From figure~\ref{fig:FeHalphaHeCuts}, we can see that in some simulated galaxies the velocity distribution of very metal-poor stars (with \FeHthree~and \aFetwo) show some similarities to the DM velocity distribution. These apparent correlations are due to the larger Poisson errors in the distribution of metal-poor stars. As we place severe cuts on the metallicity, we lose statistics. The number of stars in the Solar region with \FeHone, \FeHtwo, and \FeHthree~all with the additional \aFetwo~cut is $[1.0 - 2.2] \times 10^3$, $[2.0 - 4.2] \times 10^2$, and $[0.48 - 1.3]  \times 10^2$, respectively, and depending on the halo.

We will gain statistics if we set cuts only on \FeH~without any cuts on \aFe. This can be seen from the bottom panels of figure~\ref{fig:MetalAge} where setting the additional cut of \aFetwo~significantly reduces the number of stars in the Solar neighbourhood. In appendix~\ref{sec:Diffmetal} we show how the results differ when the cut of \aFetwo~is not included. The comparison between the velocity distributions of DM and metal-poor stars is qualitatively similar with and without the \aFetwo~cut. Notice that setting an even more stringent cut on \aFe, such as $[\alpha/{\rm Fe}]>0.4$, results in even poorer statistics. In this case, the velocity distributions have such large Poisson error bars that we cannot draw any strong conclusions.

In appendix~\ref{sec:volume} we show how the results differ if we do not constrain the metal-poor stars to the disc, i.e.~considering the stars in a cylindrical shell with radius $7 \leq \rho \leq 9$~kpc, and within a maximum Galactocentric radius of $r_{\rm max}=15$~kpc (hence with an implicit cut on the vertical distance, $|z| \leq \sqrt{r_{\rm max}^2 - \rho^2} \sim 12 \kpc$). In this case, the results will remain qualitatively similar to figure \ref{fig:FeHalphaHeCuts}, but with smaller Poisson error bars due to larger statistics. Moreover, we have checked that considering the same cylindrical shell region, but removing stars from the disc does not change the main results in figure \ref{fig:FeHalphaHeCuts}.

To quantify the correlation between the velocity distribution of metal-poor stars and DM, we apply the two-sided Kolmogorov-Smirnov (KS) test to the distributions. In table~\ref{tab:pvalues}, we show the $p$-values of the KS test for the radial, azimuthal, and vertical velocity distributions. For stars with \FeHtwo~and \aFetwo, the $p$-values are always smaller than 0.05 for the azimuthal component of the velocity distribution in all halos, and the null hypothesis that DM and those populations of metal-poor stars share the same distribution is rejected at 95\% CL. The $p$-values can, however, become larger than 0.05 for the radial and vertical components of the velocity distribution in some halos, reaching the largest value of 0.14 for the vertical velocity distribution in halo Au16. When placing the more extreme cut of \FeHthree~on the metallicity, we lose statistics. Hence, the stellar distributions with \FeHthree~and \aFetwo~show some correlations with the DM distributions, solely due to the large Poisson errors. This can be observed from the $p$-values shown in the second column of table~\ref{tab:pvalues}, where they become larger than 0.05 for the three components of the velocity distribution in almost all halos. Notice however that in order to draw any conclusions regarding DM direct detection event rates for standard DM-nucleus interactions, the appropriate function to study is the so-called \emph{halo integral} which the direct detection event rate is proportional to. We present a comparison of the the DM and stellar halo integrals in section \ref{sec:halointegral}.

 \bigskip 
  
  \begin{table}[h!]
    \centering
    \begin{tabular}{|c|c c c|c c c|}
      \hline
         & \multicolumn{3}{|c|}{\FeHtwo, \aFetwo} & \multicolumn{3}{|c|}{\FeHthree, \aFetwo} \\
       \hline
       Halo Name  & $\rho$ & $\phi$  & $z$ & $\rho$ & $\phi$  & $z$\\
       \hline
       Au6 & $1.1 \times 10^{-1}$ & $3.8 \times 10^{-3}$ & $1.2 \times 10^{-1}$ & $3.2 \times 10^{-1}$ & $1.4 \times 10^{-1}$ & $5.3 \times 10^{-1}$\\
       Au16 & $1.3 \times 10^{-1}$ & $1.4 \times 10^{-2}$ & $1.4 \times 10^{-1}$ & $9.9 \times 10^{-1}$ & $9.2 \times 10^{-1}$ & $4.0 \times 10^{-1}$\\
       Au21 & $1.0 \times 10^{-1}$ & $1.6 \times 10^{-2}$ & $5.1 \times 10^{-2}$ & $7.6 \times 10^{-1}$ & $9.2 \times 10^{-1}$ & $7.0 \times 10^{-2}$\\
       Au23 & $5.6 \times 10^{-2}$ & $1.4 \times 10^{-3}$ & $8.4 \times 10^{-3}$ & $2.3 \times 10^{-1}$ & $1.1 \times 10^{-1}$ & $1.3 \times 10^{-1}$\\
       Au24  & $4.3 \times 10^{-2}$ & $4.5 \times 10^{-2}$ & $4.9 \times 10^{-3}$ & $2.3 \times 10^{-1}$ & $4.0 \times 10^{-1}$ & $3.2 \times 10^{-1}$\\
       Au27 & $6.8 \times 10^{-2}$ & $1.6 \times 10^{-5}$ & $7.0 \times 10^{-3}$ & $6.4 \times 10^{-1}$ & $2.0 \times 10^{-2}$ & $1.2 \times 10^{-1}$\\
      \hline
    \end{tabular}
\caption{$p$-values for the KS test to check the correlation between the radial, azimuthal, and vertical distributions of DM and metal-poor stars with \FeHtwo~(left column) and \FeHthree~(right column), both with an additional cut of \aFetwo~for the six Auriga halos.}
    \label{tab:pvalues}
  \end{table}

\bigskip

\begin{figure}[t!]
\begin{center}
  \includegraphics[width=0.32\textwidth]{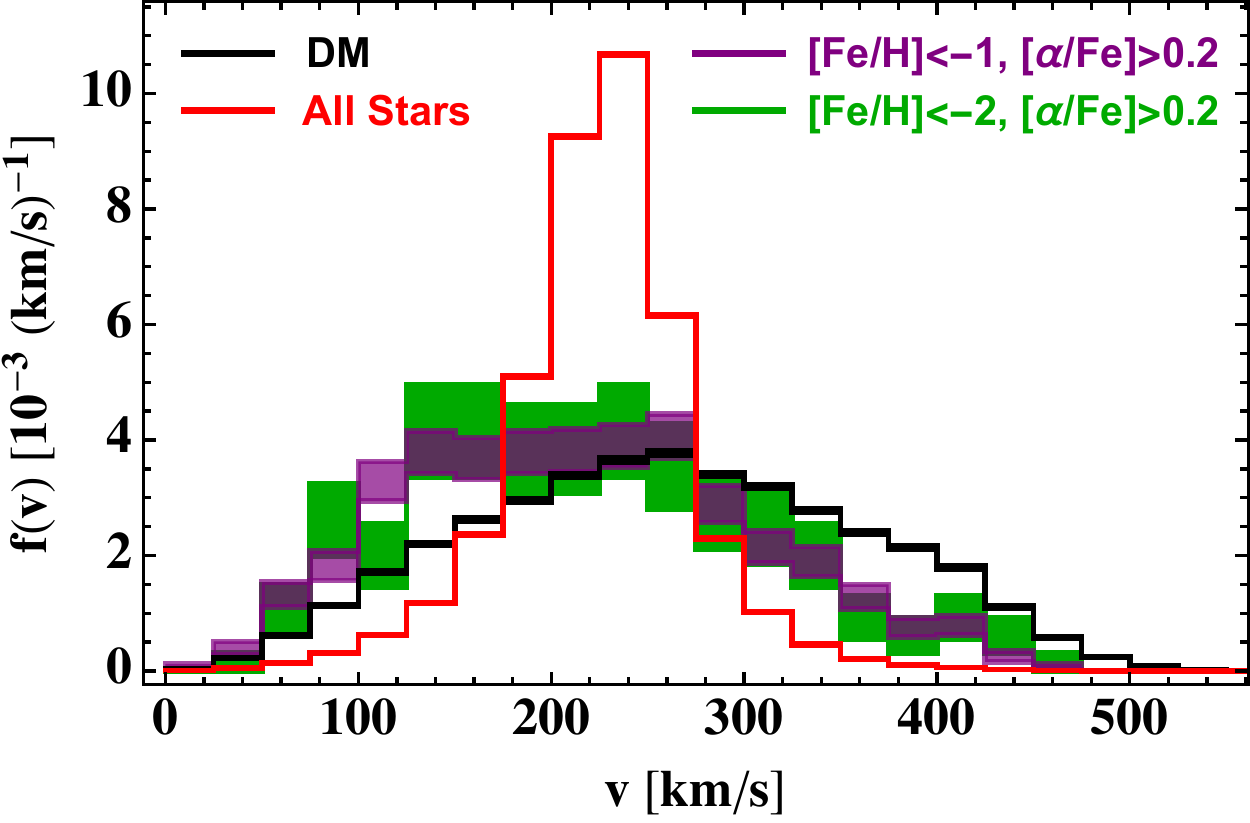}
  \includegraphics[width=0.32\textwidth]{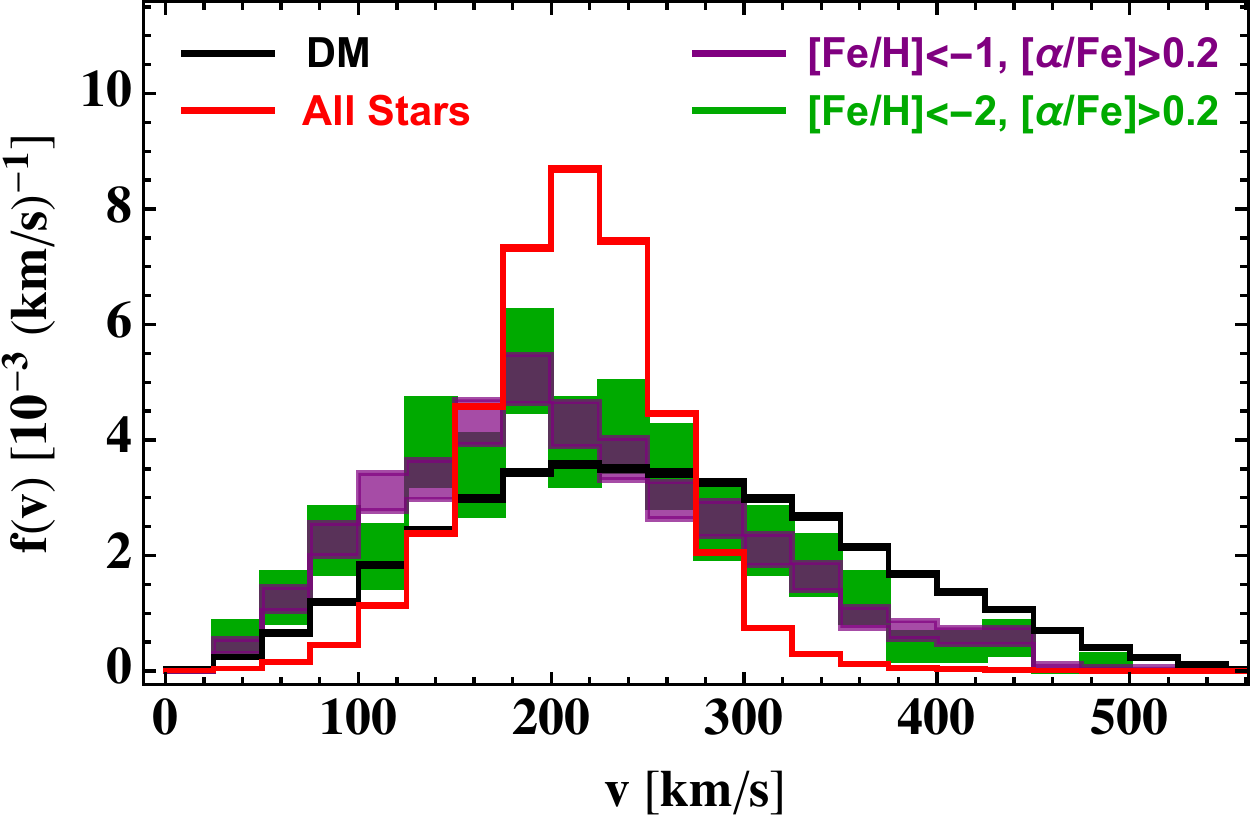}
  \includegraphics[width=0.32\textwidth]{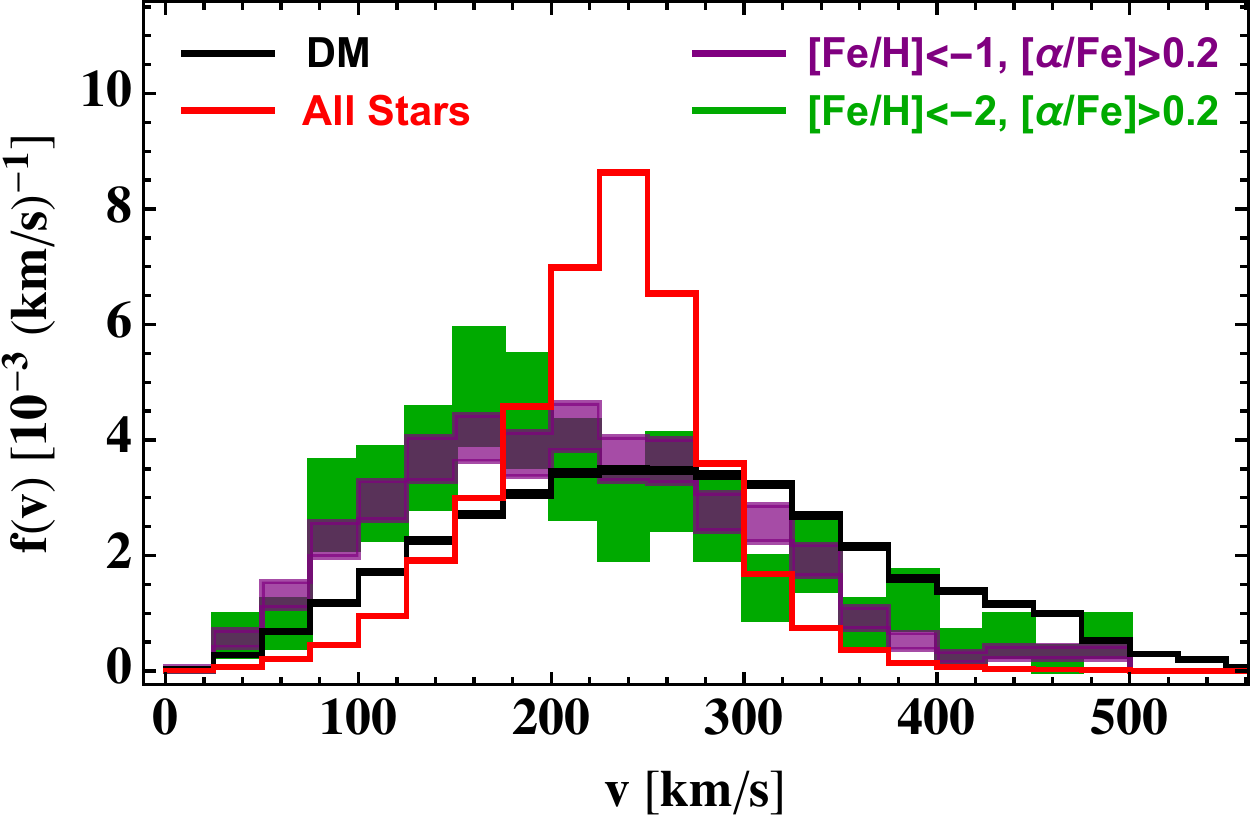}\\
  \includegraphics[width=0.32\textwidth]{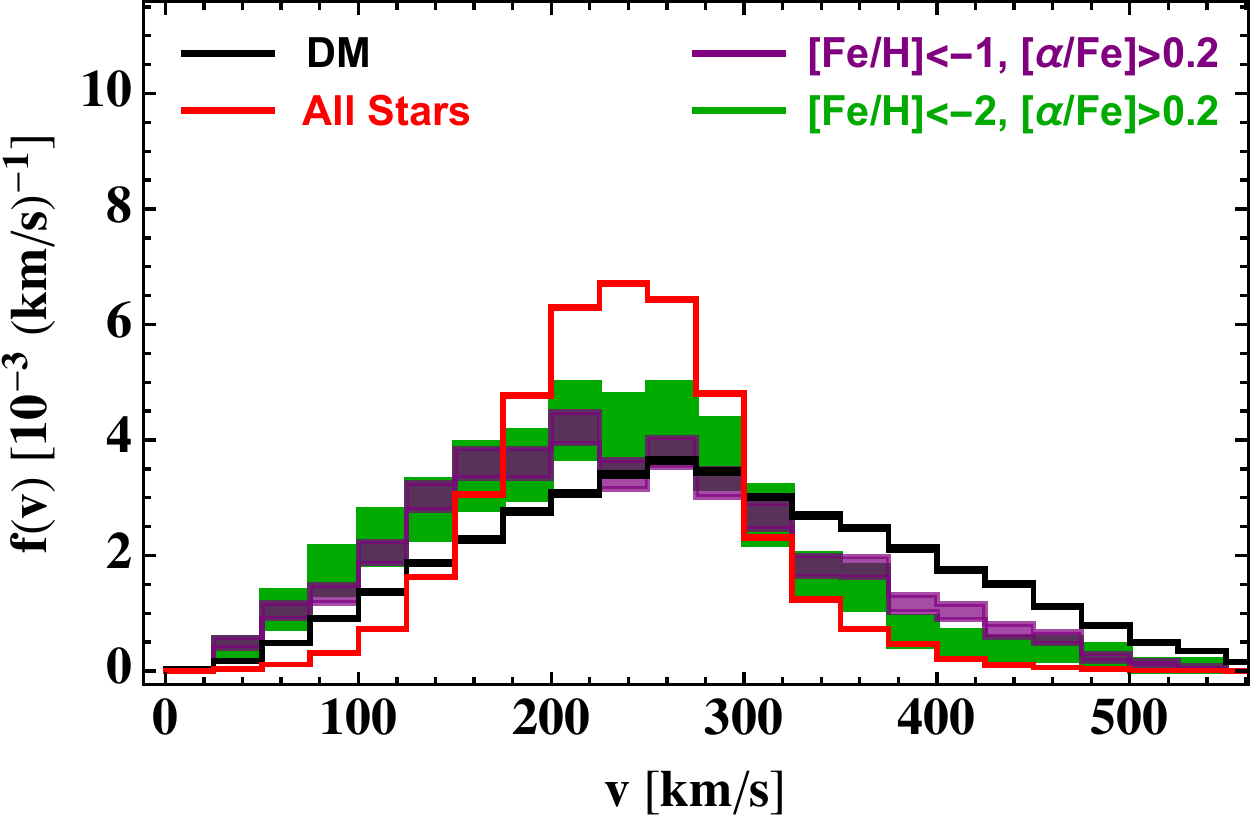}
  \includegraphics[width=0.32\textwidth]{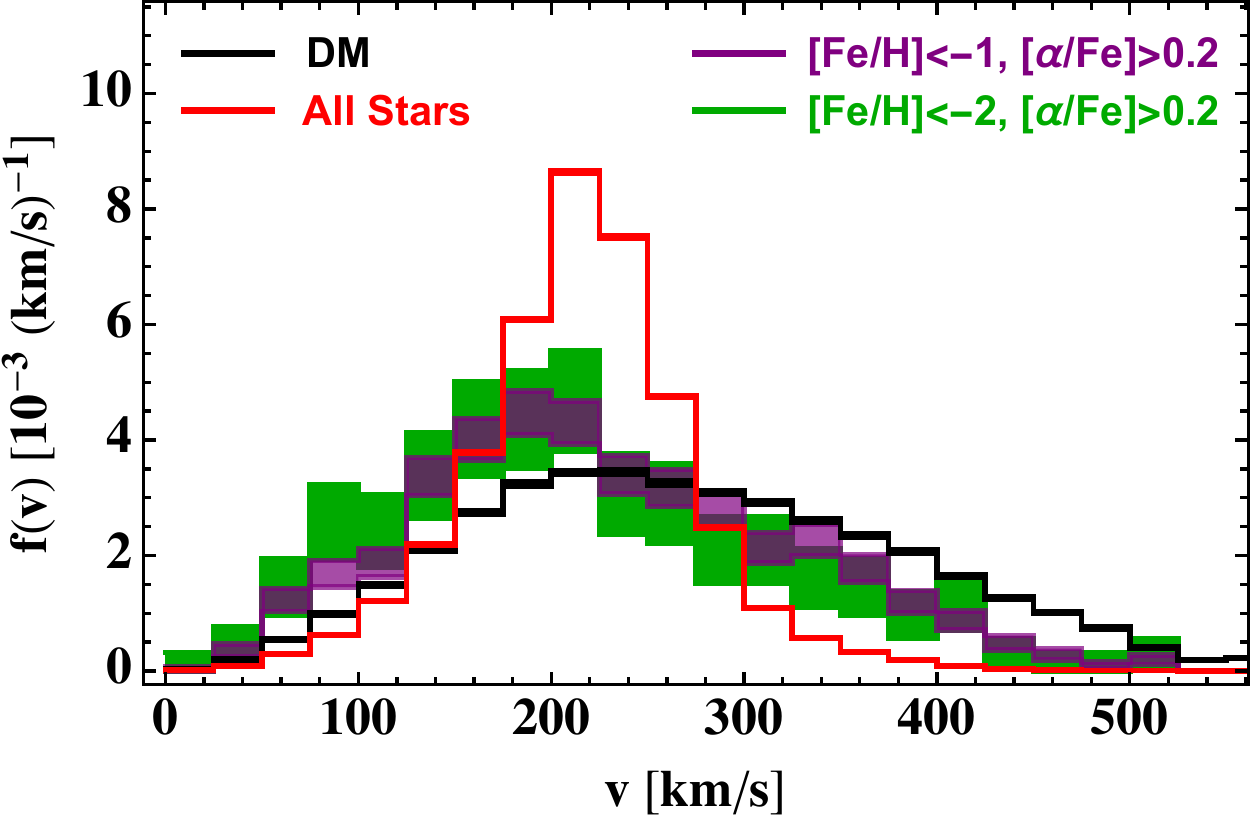}
  \includegraphics[width=0.32\textwidth]{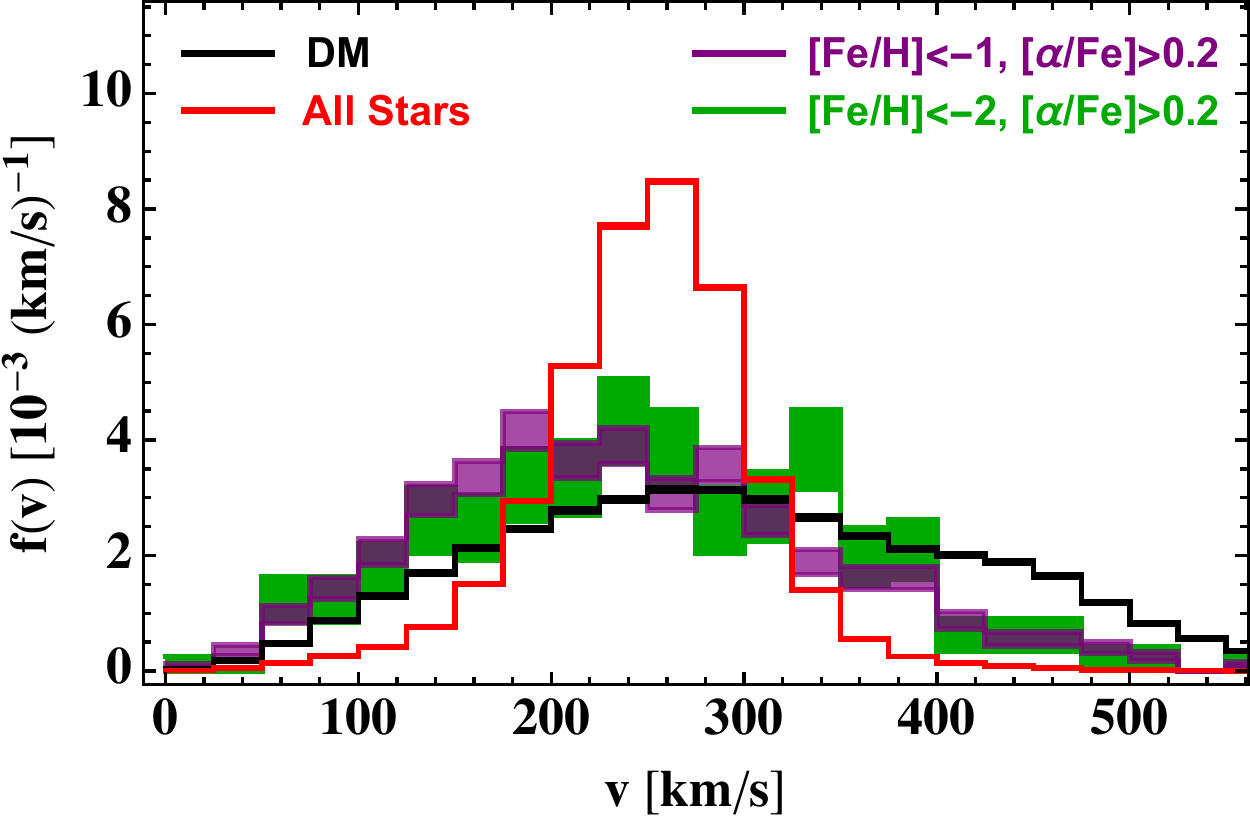}
\caption{The Galactic frame speed distributions of DM (black) and all stars (red), as well as stars with \FeHone~(purple) and \FeHtwo~(green) and \aFetwo~in the Solar neighbourhood for halos Au6 (top left), Au16 (top middle), Au21 (top right), Au23 (bottom left), Au24 (bottom middle), and Au27 (top right).}
\label{fig:fvMod}
\end{center}
\end{figure}

In figure~\ref{fig:fvMod}, we show the speed distributions of DM and all stars, as well as stars with metallicity cut \FeHone~and \FeHtwo, both with \aFetwo, in the Solar neighbourhood for the six halos. We can see that the local DM speed distribution has a peak speed which is always larger than the peak of the speed distribution of metal-poor stars. This is consistent with what we find for the density profiles of DM and stars (see section~\ref{sec:density}). There are also substantial differences in the tails of the speed distributions of DM and metal-poor stars, a region that affects the search for low mass DM in direct detection experiments (see, e.g., \cite{Bottino:2005qj}). Note that we do not show the results of \FeHthree~since the error bars are very large in that case, so it is difficult to perform any meaningful comparison with the DM distribution.
Also, as we have already argued, such low metallicities are not consistent with observations in the MW.

\section{Halo integrals}
\label{sec:halointegral}

To study the implications of possible correlations between the DM and stellar velocity distributions for DM direct detection, we need to study  the DM and stellar \emph{halo integrals}. Consider the scattering of a DM particle $\chi$ with mass $m_\chi$ off a target nucleus with mass $m_T$, in an underground direct detection experiment. The differential event rate is given by
\begin{equation}
\frac{dR}{dE_R} = \frac{\rho_\chi}{m_\chi}\frac{1}{m_T} \int_{v>v_{\rm min}} d^3 v ~\frac{d\sigma_T}{d E_R}~v~f_{\rm det}({\bf v}, t),
\end{equation}
where $E_R$ is the recoil energy of the target nucleus, $d\sigma_T/dE_R$ is the differential DM-nucleus scattering cross section, $\rho_\chi$ is the local DM density, $f_{\rm det}({\bf v}, t)$ is the local DM velocity distribution in the detector rest frame, and ${\bf v}$ is the relative velocity between the DM and the nucleus, with $v \equiv |{\bf v}|$. For elastic DM-nucleus scattering, the minimum speed required for the DM particle to deposit a recoil energy $E_R$ in the detector is given by,
\begin{equation}
v_{\rm min} = \sqrt{\frac{m_T E_R}{2 \mu^2_{\chi T}}},
\end{equation}
where $\mu_{\chi T}$ is the DM-nucleus reduced mass. 

For the standard spin-independent and spin-dependent interactions, $d\sigma_T/dE_R \propto 1/v^2$, and the differential event rate becomes proportional to the halo integral,
\begin{equation}
\frac{dR}{dE_R} \propto \eta(v_{\rm min}, t) \equiv \int_{v>v_{\rm min}} d^3 ~ \frac{f_{\rm det}({\bf v}, t)}{v}.
\end{equation}
The astrophysical dependence of the direct detection event rate is contained in the halo integral, $\eta(v_{\rm min}, t)$, together with the local DM density.

\begin{figure}[t!]
\begin{center}
  \includegraphics[width=0.32\textwidth]{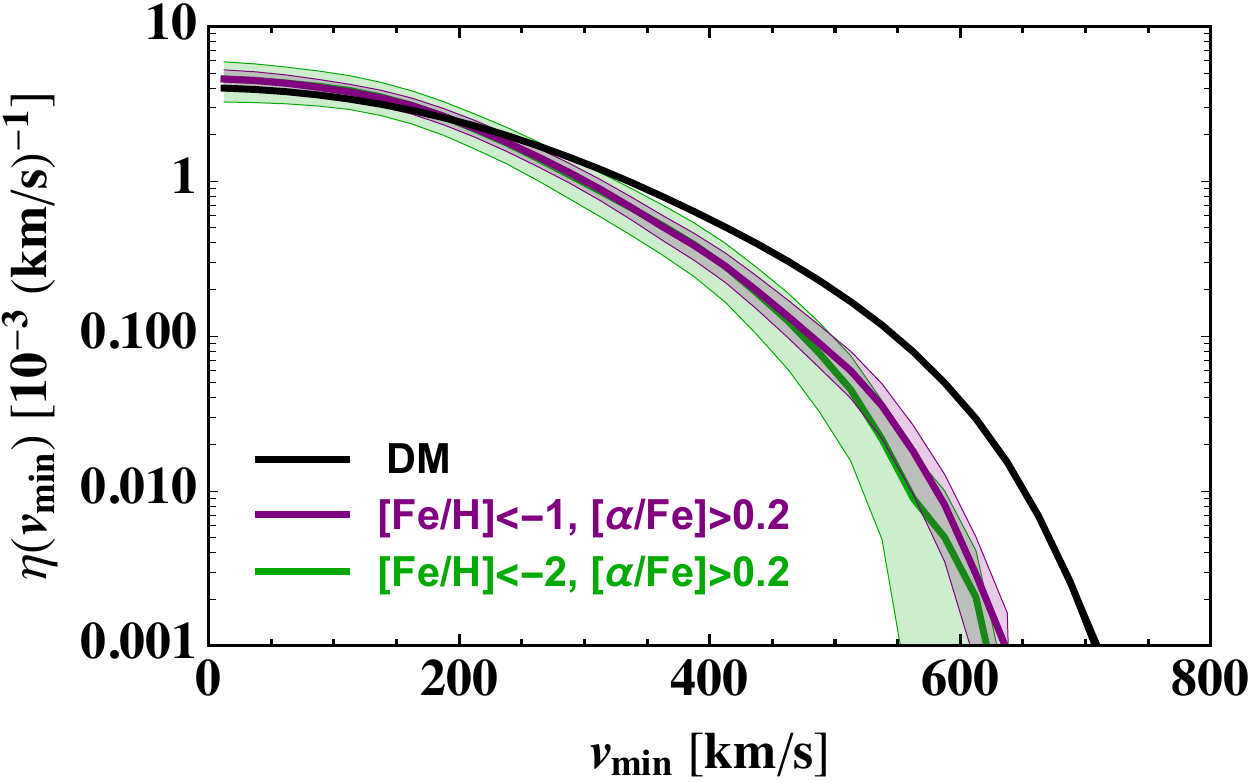}
  \includegraphics[width=0.32\textwidth]{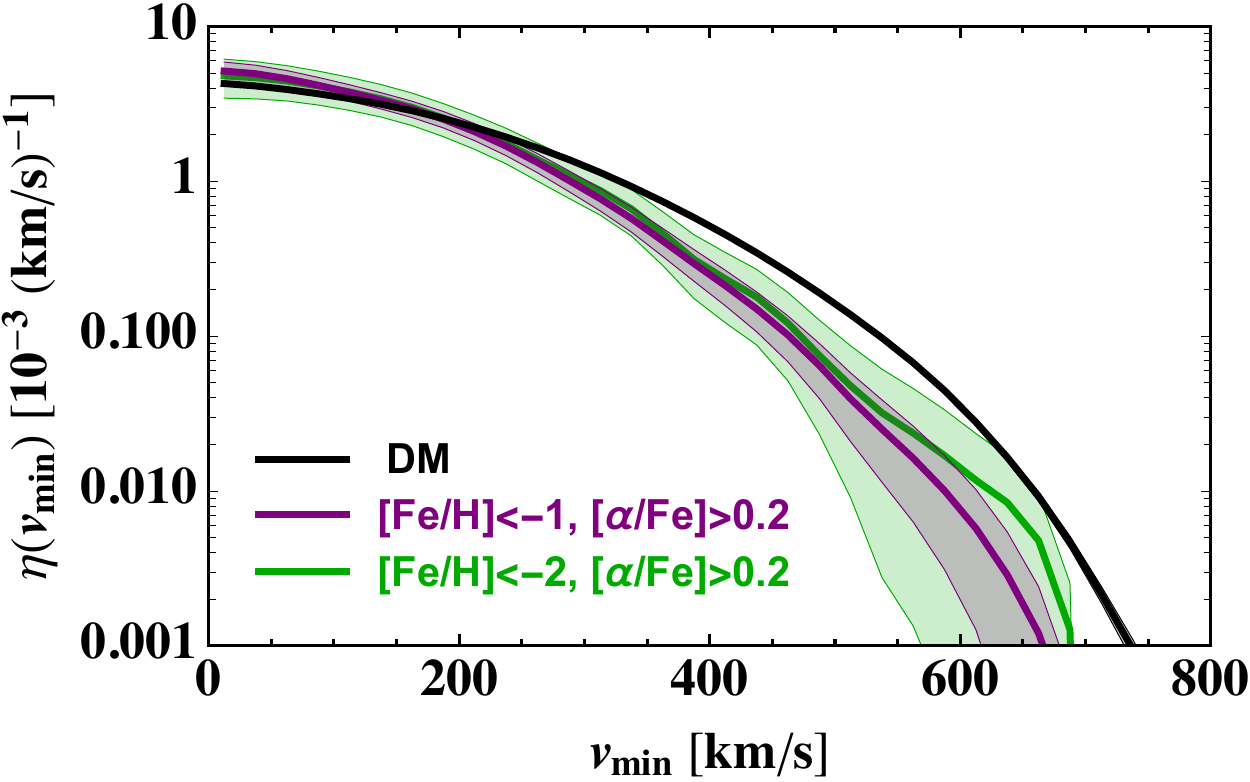}
  \includegraphics[width=0.32\textwidth]{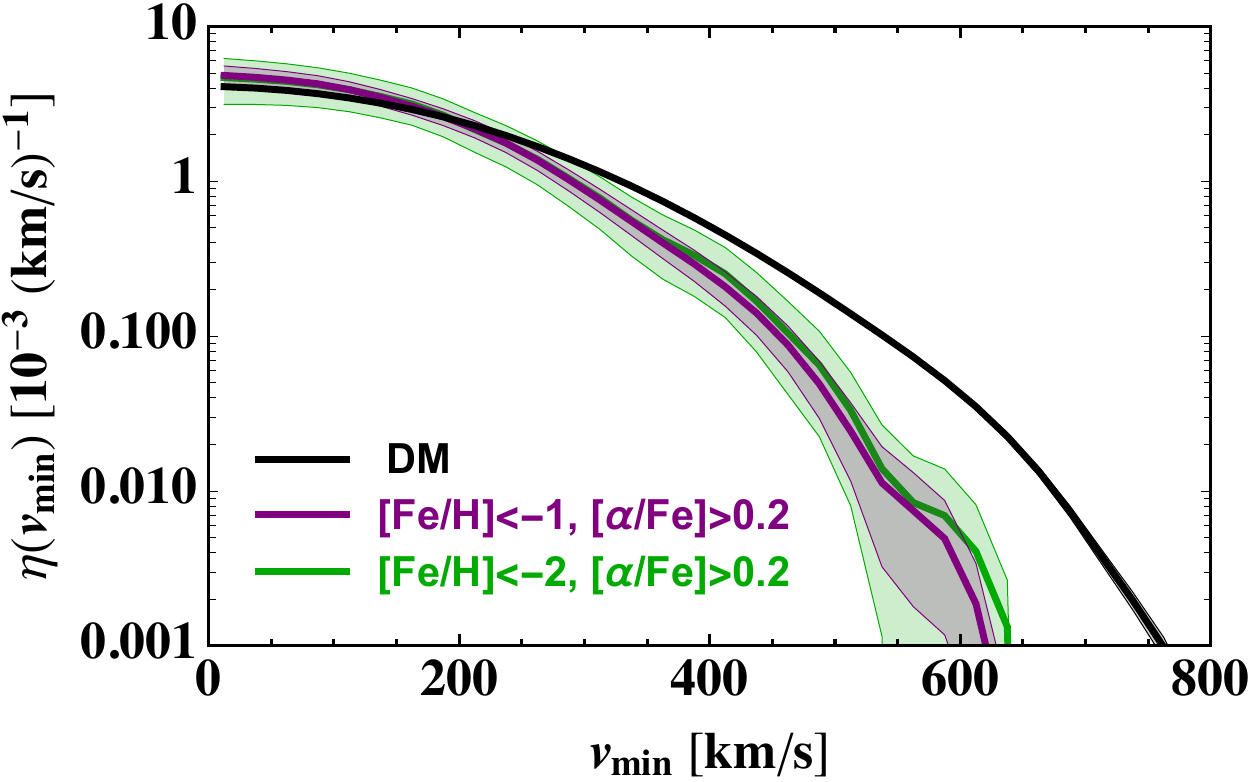}\\
  \includegraphics[width=0.32\textwidth]{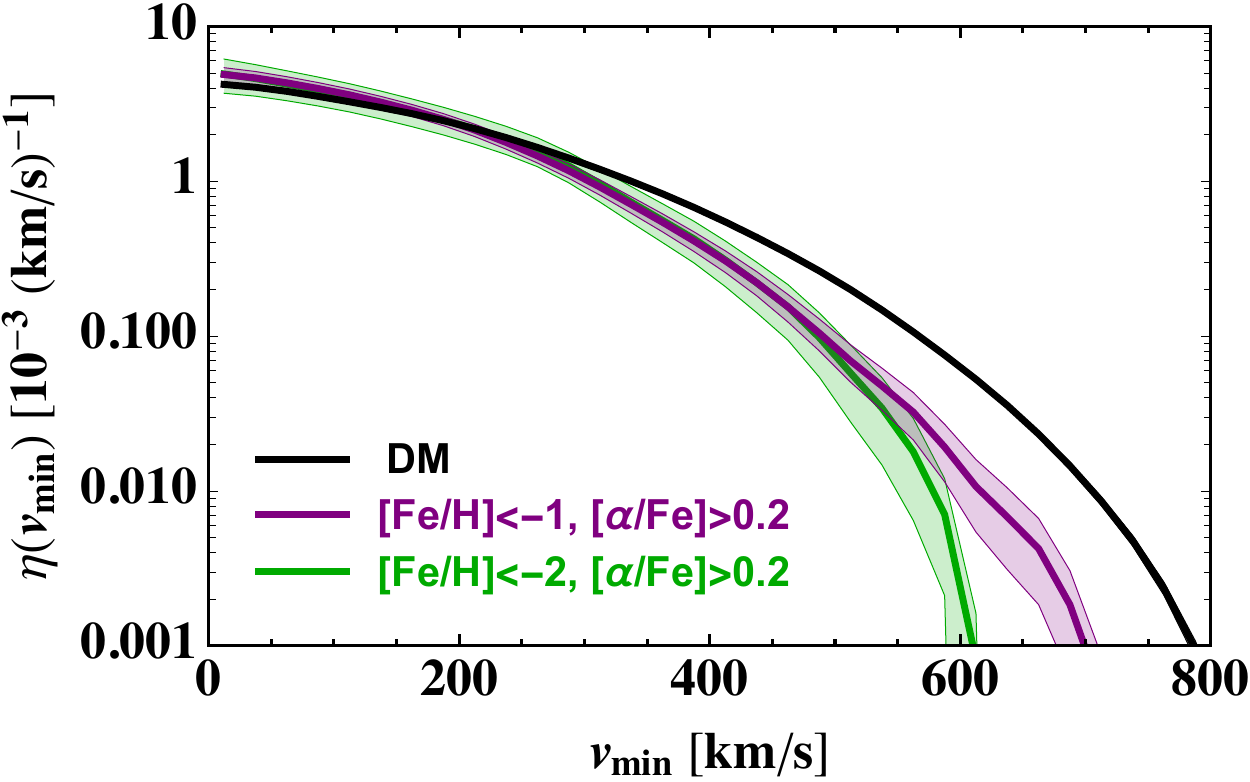}
  \includegraphics[width=0.32\textwidth]{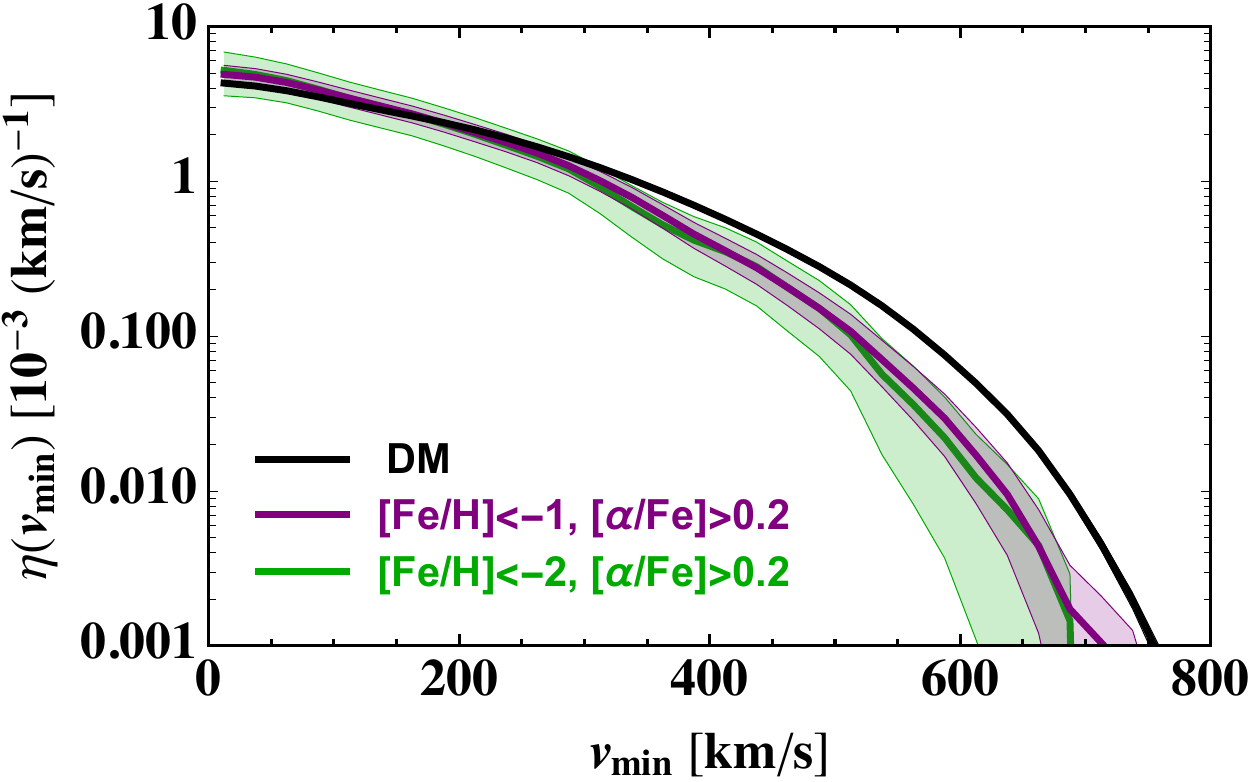}
  \includegraphics[width=0.32\textwidth]{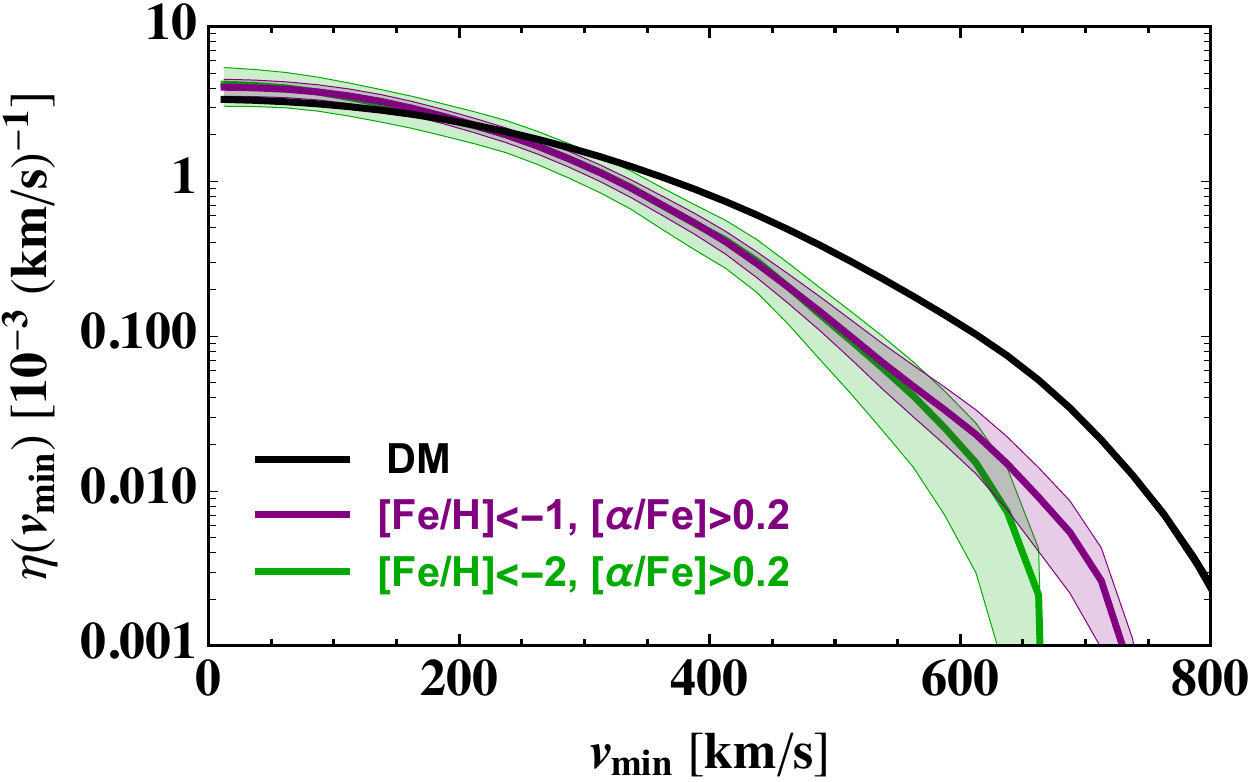}
\caption{The time-averaged halo integrals for DM (black) and stars with \FeHone~(purple) and \FeHtwo~(green) and \aFetwo~in the Solar neighbourhood for halos Au6 (top left), Au16 (top middle), Au21 (top right), Au23 (bottom left), Au24 (bottom middle), and Au27 (top right). The solid coloured lines and the shaded bands correspond to the halo integrals obtained from the mean velocity distributions and the velocity distributions at $\pm 1 \sigma$ from the mean, respectively.}
\label{fig:eta}
\end{center}
\end{figure}

In figure~\ref{fig:eta} we show the time-averaged halo integrals as a function of the minimum speed, $v_{\rm min}$, for  DM and metal-poor stars with  \FeHone~and  \FeHtwo, both with  \aFetwo~in the Solar neighbourhood for the six Auriga halos. The solid lines specify the halo integrals computed from the mean  velocity distributions, while the shaded bands are obtained by adding and subtracting one standard deviation to the mean velocity distributions. It is clear that the DM halo integrals do not trace the halo integrals of the metal-poor stars, especially at large $v_{\rm min}$. This is consistent with the speed distributions presented in figure~\ref{fig:fvMod} where there is a large discrepancy between the tails of the DM and stellar speed distributions. As in the case of the speed distributions, in figure~\ref{fig:eta} we do not show the stellar halo integrals for \FeHthree~due to the large Poisson errors which make it difficult to perform meaningful statistical comparisons. Notice however that in most cases, the halo integrals for stars with \FeHtwo~and \aFetwo~are even further from the DM halo integrals, compared to the halo integrals of stars with \FeHone~and \aFetwo. Due to this systematic trend we do not expect that even with better statistics, setting more stringent cuts on the metallicity of the stars (i.e.~ \FeHthree) would result in a better agreement with the DM halo integrals.

\section{Density profiles}
\label{sec:density}

\begin{figure}[t]
\begin{center}
  \includegraphics[width=0.32\textwidth]{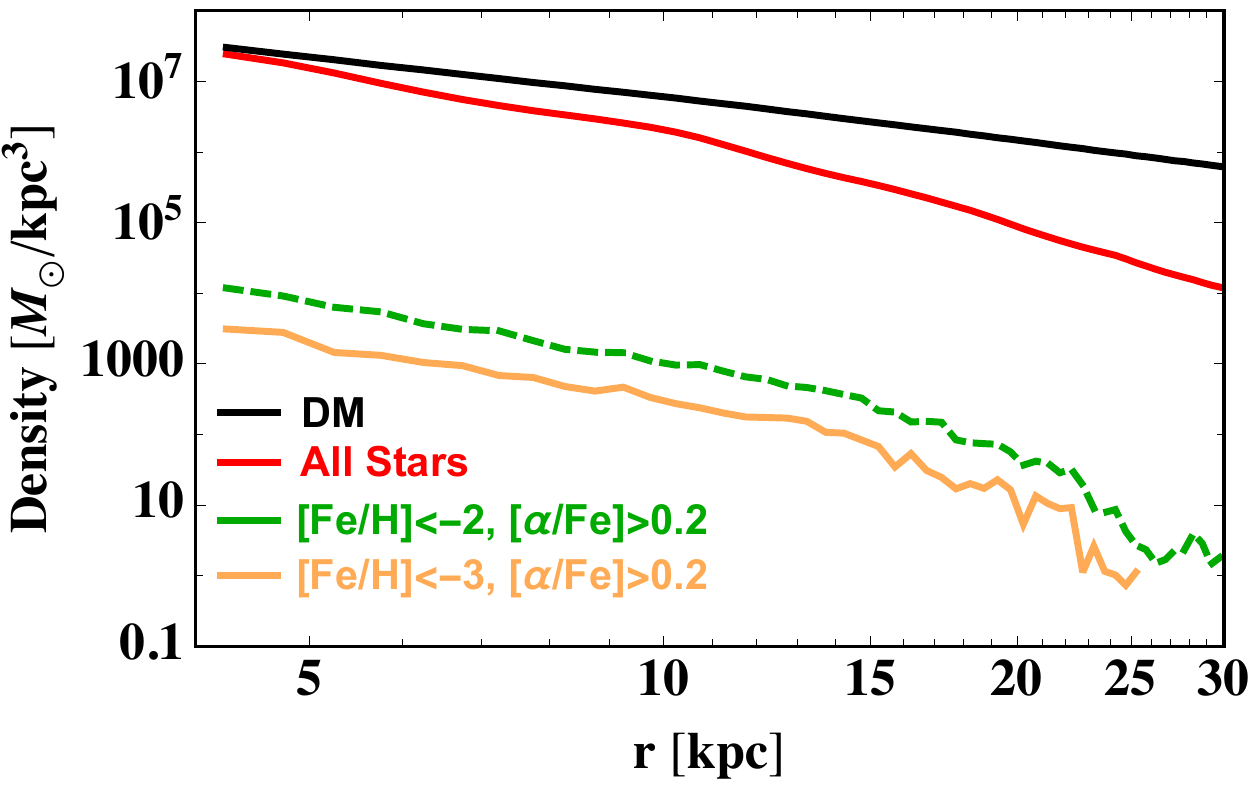}
 \includegraphics[width=0.32\textwidth]{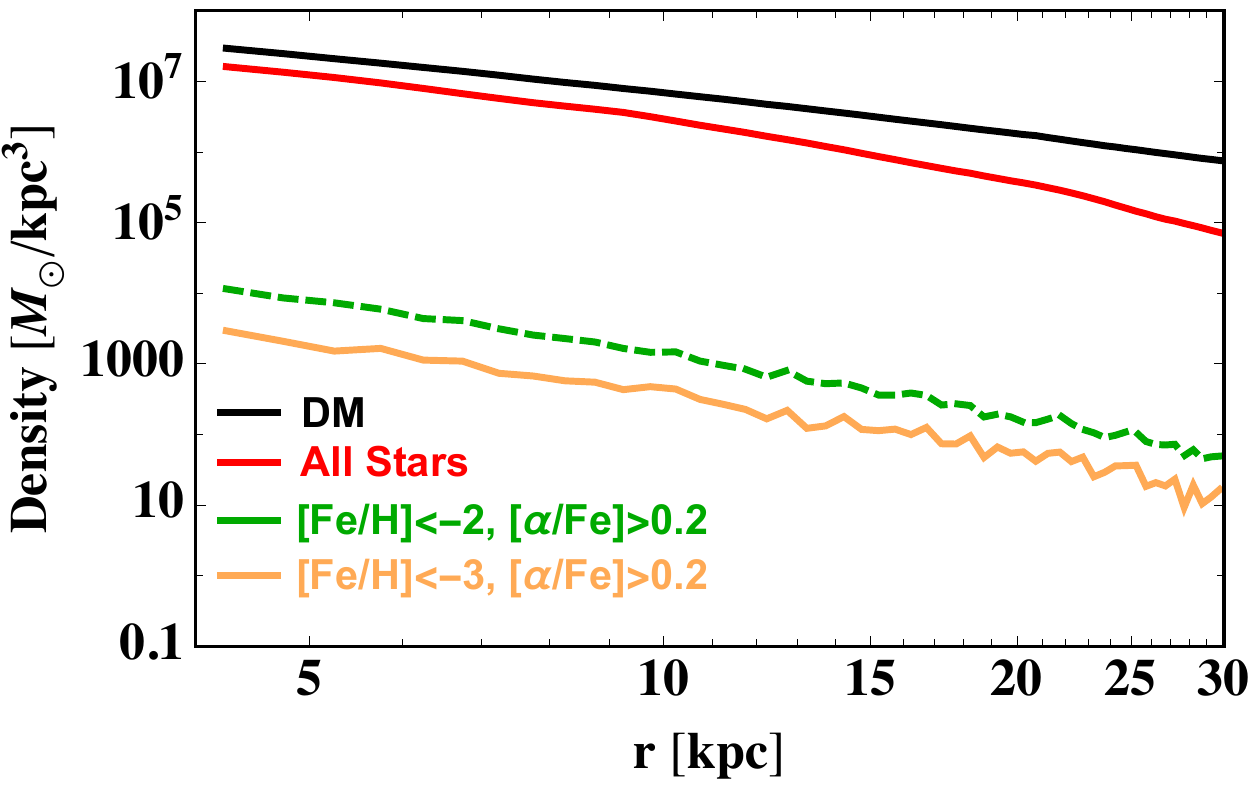}
 \includegraphics[width=0.32\textwidth]{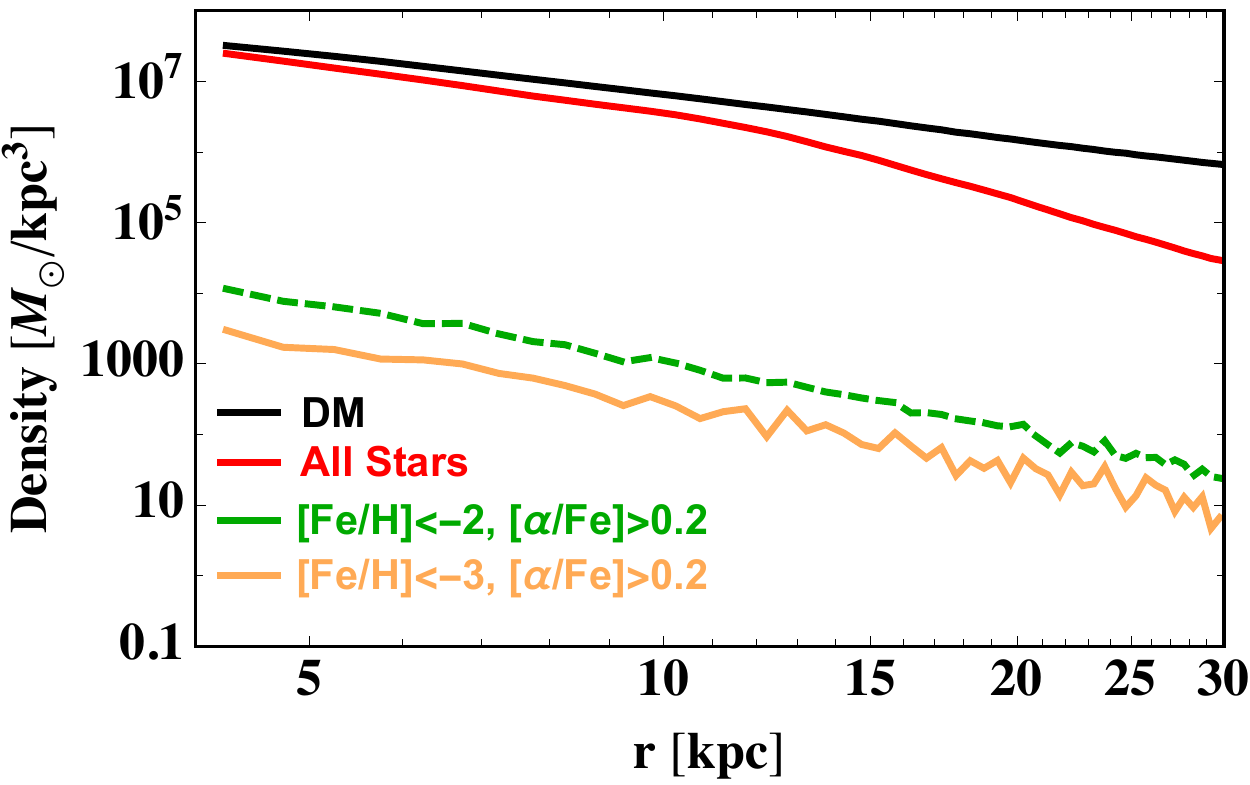}\\
\hspace{7pt}\includegraphics[width=0.305\textwidth]{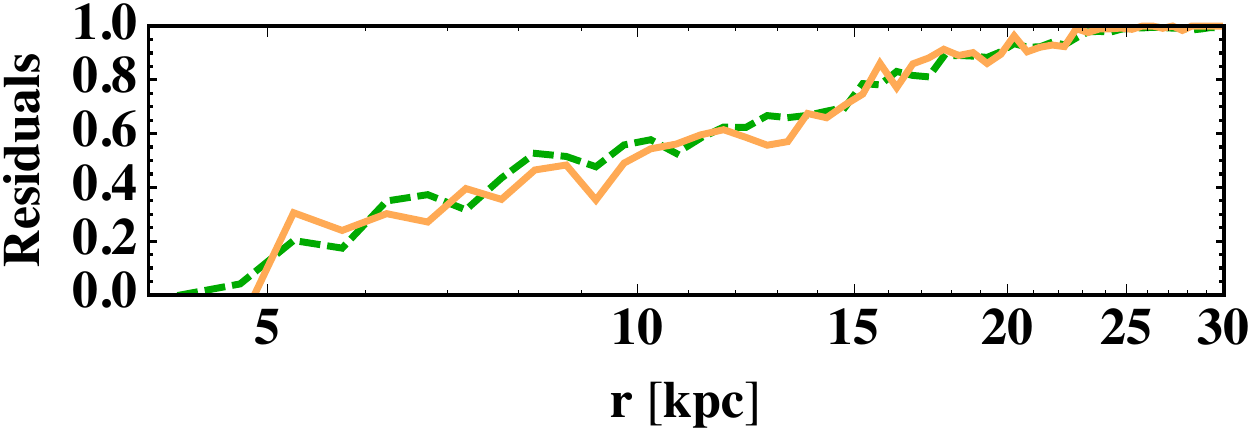}
\hspace{7pt}\includegraphics[width=0.305\textwidth]{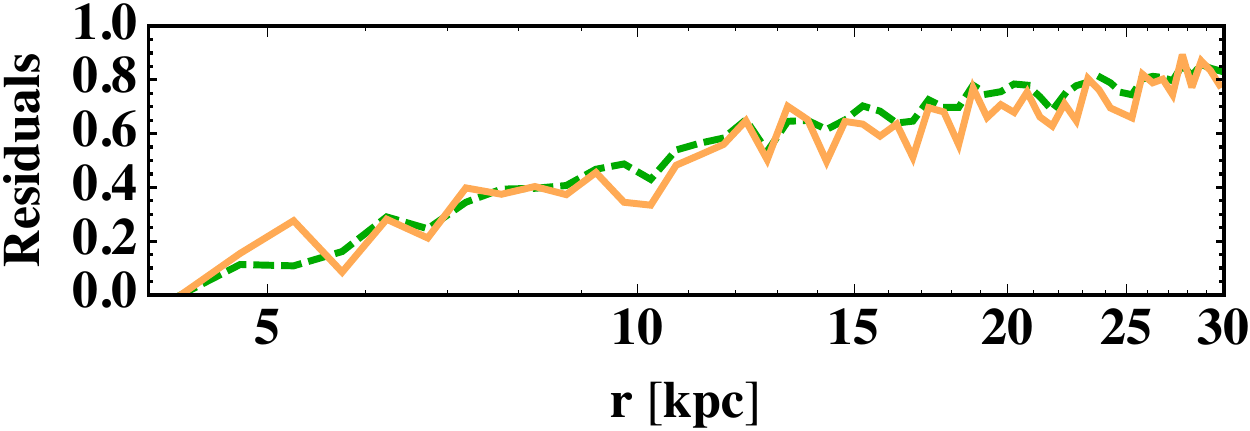}
\hspace{7pt}\includegraphics[width=0.305\textwidth]{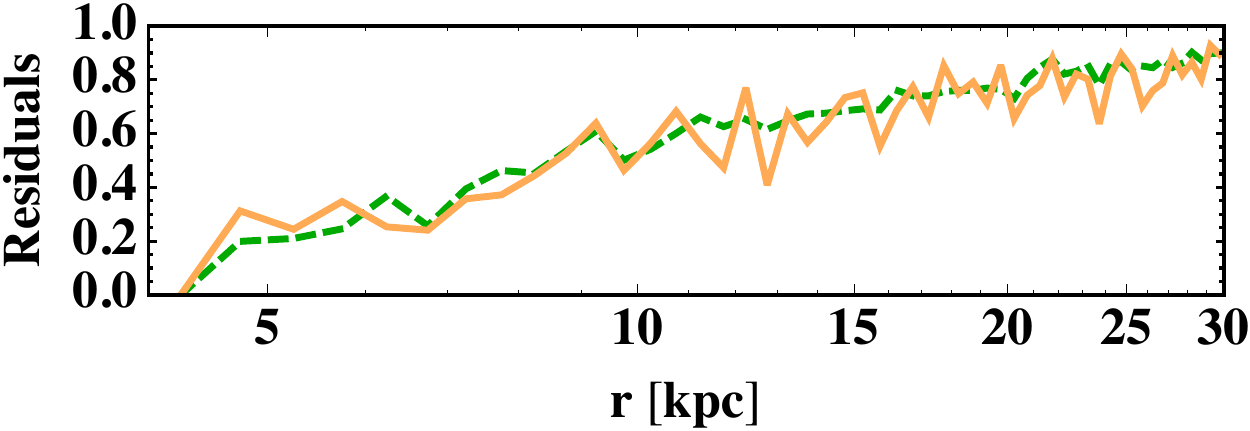}\\
  \vspace{15pt}
  \includegraphics[width=0.32\textwidth]{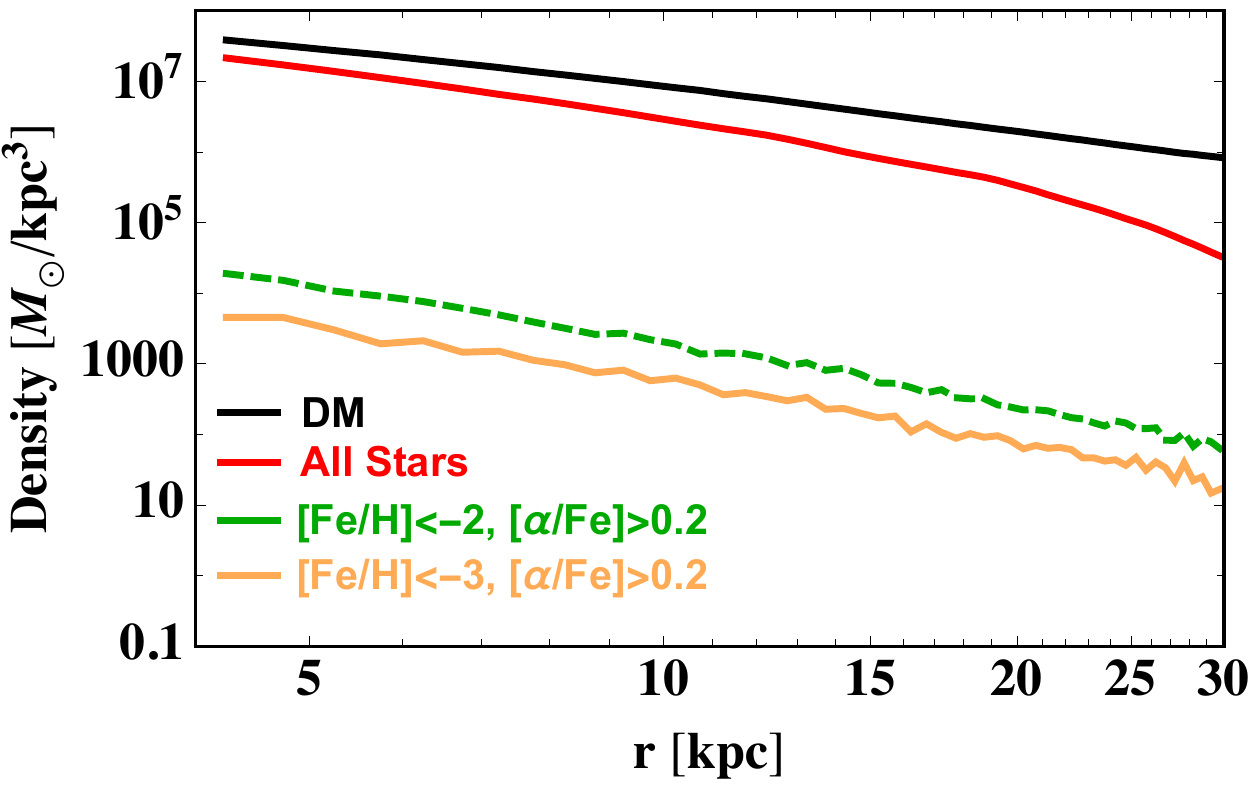}
  \includegraphics[width=0.32\textwidth]{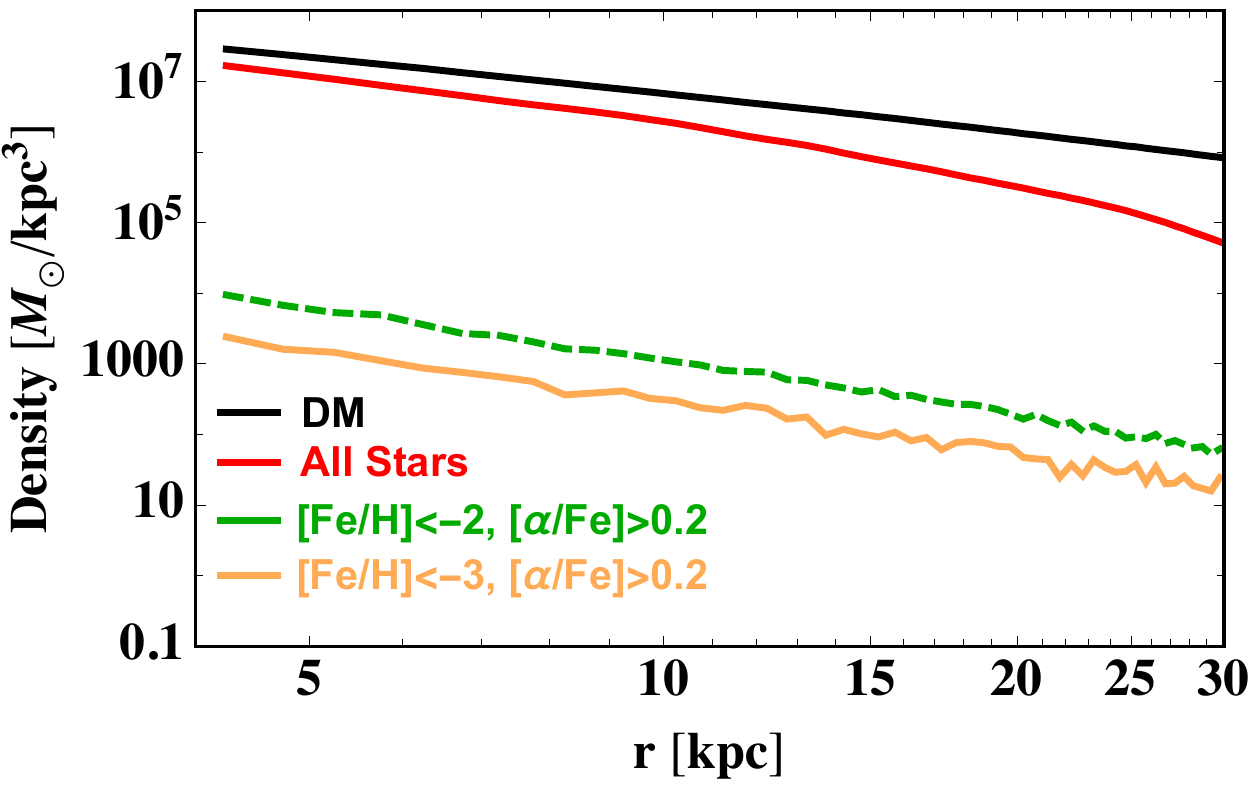}
  \includegraphics[width=0.32\textwidth]{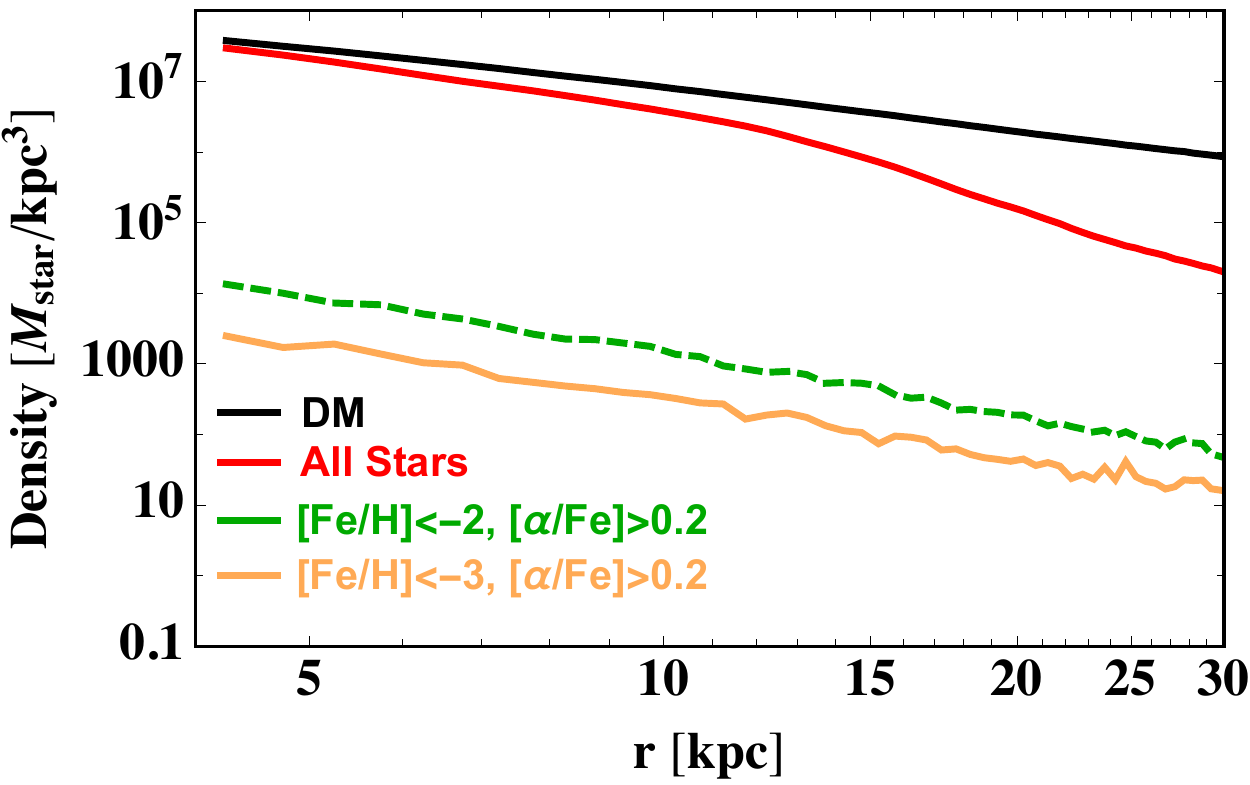}\\
  \hspace{7pt}\includegraphics[width=0.305\textwidth]{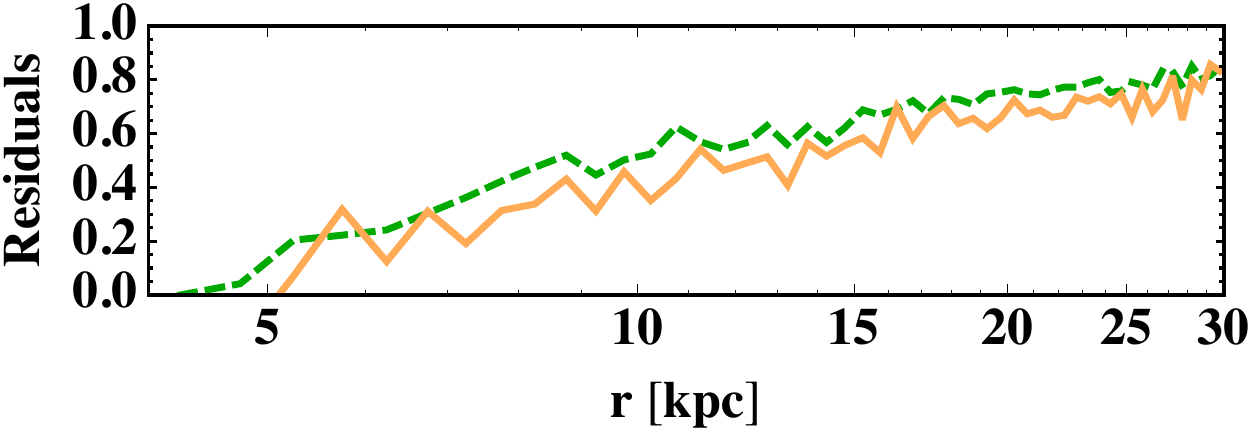}
  \hspace{7pt}\includegraphics[width=0.305\textwidth]{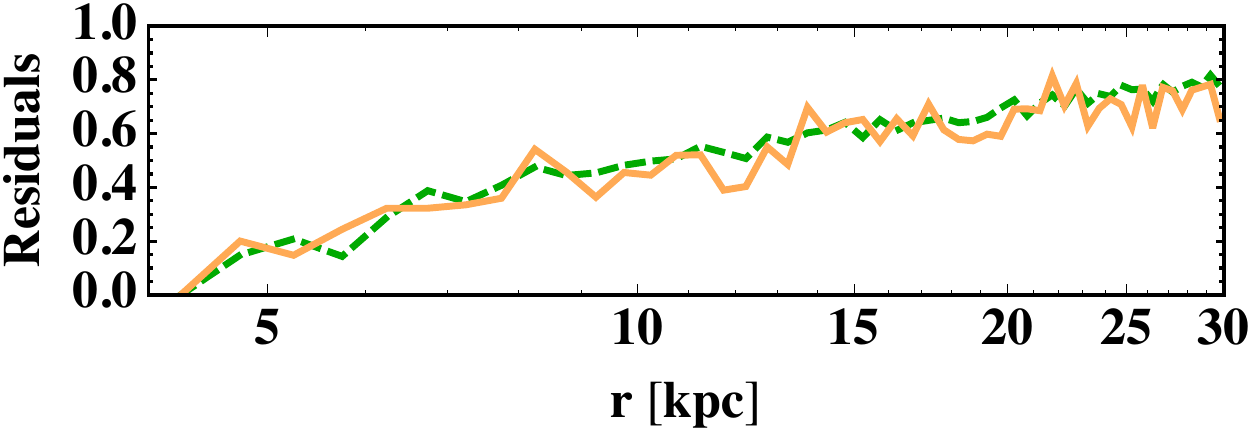}
  \hspace{7pt}\includegraphics[width=0.305\textwidth]{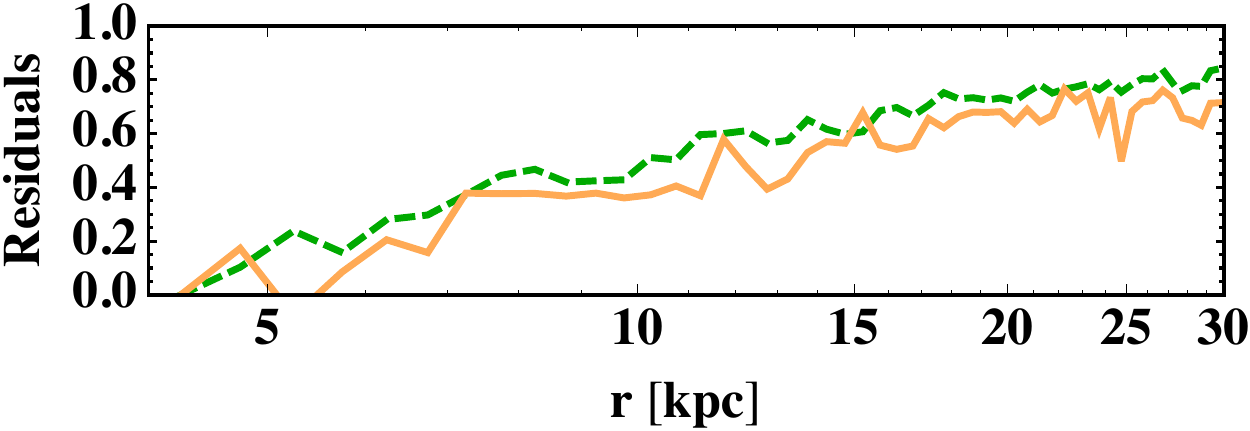}\\
\caption{Spherically averaged density profiles as a function of Galactocentric distance, for DM (black) and all stars (red), as well as stars with \FeHtwo~(green) and \FeHthree~(orange) both with \aFetwo~for halos Au6 (top left), Au16 (top middle), Au21 (top right), Au23 (bottom left), A24 (bottom middle), and Au27 (top right). The smaller panels below each density plot show the residuals, $(\rho_{\rm DM}-\rho_{\rm star}^{\rm scaled})/\rho_{\rm DM}$, where $\rho_{\rm DM}$ is the DM density and $\rho_{\rm star}^{\rm scaled}$ is the density of stars scaled to the DM density at 4 kpc, with \FeHtwo~(green) and \FeHthree~(orange), both with \aFetwo.}
\label{fig:density}
\end{center}
\end{figure}

Figure~\ref{fig:density} shows the density profiles of DM, all stars, as well as stars with metallicity cut \FeHtwo~and \FeHthree, both with \aFetwo, as a function of Galactocentric distance, $r$. The density profiles are averaged in spherical shells with radial width of 0.5 kpc. The smaller panels at the bottom of the density plots show the residuals, $(\rho_{\rm DM}-\rho_{\rm star}^{\rm scaled})/\rho_{\rm DM}$, where $\rho_{\rm DM}$ is the DM density and $\rho_{\rm star}^{\rm scaled}$ is the density of metal-poor stars with \FeHtwo~or \FeHthree~(both with \aFetwo) scaled to the DM density at 4 kpc. It is clear from the positive residuals that the density of metal-poor stars fall faster with Galactocentric distance than the DM density. 

To assess the slope of the density profiles in the Solar neighbourhood, we fit a power-law $\rho(r) \propto r^{-a}$ to each density profile in the range of $6 \leq r \leq 10$~kpc. We choose this larger range of radius in order to have enough bins to perform the fit. The values of the slope, $a$, with their standard error are given in table~\ref{tab:slopes}. One can see that the slope of the density profiles of metal-poor stars and DM are not similar, with the slope of stars with \FeHtwo~and \aFetwo~greater than the DM slope by 32\% to 54\%, depending on the halo. Notice that for \FeHthree, the error bars become large again due to poor statistics. 

Since in the Solar neighbourhood, the density profiles of any population of stars (metal-poor or metal-rich) fall faster with Galactocentric radius compared to the DM density profiles, there are more high speed DM particles compared to stars in the Solar neighbourhood. This results in the peak speed of the local DM distributions to be larger than the stellar speed distributions as shown in figure~\ref{fig:fvMod}. This was also discussed in detail in ref.~\cite{Evans:2018bqy}. 

\begin{table}[th!]
    \centering
    \begin{tabular}{|c|c|c|c|c|}
      \hline
       Halo Name  & All Stars  & \FeHtwo & \FeHthree & DM\\
       \hline
       Au6 & $2.52 \pm 0.004$ & $2.75 \pm 0.18$  & $2.53 \pm 0.34$ & $1.85 \pm 0.008$ \\
       Au16 & $2.01 \pm 0.005$ & $2.55 \pm 0.17$ & $2.24 \pm 0.33$ & $1.75 \pm 0.008$ \\
       Au21 & $2.29 \pm 0.004$ & $3.01 \pm 0.20$ & $3.33 \pm 0.37$ & $1.95 \pm 0.008$ \\
       Au23 & $2.45 \pm 0.004$  & $2.84 \pm 0.15$ & $2.69 \pm 0.27$ & $1.85 \pm 0.007$ \\
       Au24 & $2.08 \pm 0.004$ & $2.38 \pm 0.20$ & $2.29 \pm 0.38$  & $1.73 \pm 0.008$ \\
       Au27 & $2.43 \pm 0.003$ & $2.44 \pm 0.17$ & $2.45 \pm 0.37$  & $1.85 \pm 0.007$ \\
      \hline
    \end{tabular}
    \caption{The slope (and its standard error) of the density profiles of all stars, stars with \FeHtwo~and \FeHthree~(both with \aFetwo), and DM at the Solar neighbourhood, $6 \leq r \leq 10$~kpc, for the six Auriga halos.}
    \label{tab:slopes}
  \end{table}


\section{Conclusions}
\label{sec:conclusions}

The local dark matter velocity distribution is an important input in the calculation of dark matter direct detection event rates and the interpretation of current and future experimental results. If there were a population of stars that trace the dark matter velocity distribution in the Solar neighbourhood, one would be able to use observations of those stellar velocities to infer the dark matter velocity distribution directly from data.

In this work, we have studied whether the velocity distributions of stars and dark matter in the Solar neighbourhood are correlated in six Milky Way-like galaxies in the Auriga hydrodynamic simulations. In particular, we set various cuts on the formation time and metallicity of stars in a cylindrical shell with radius $7-9$ kpc and height $|z| \leq 2$~kpc from the Galactic plane, and compared their velocity distributions with that of dark matter in the same Solar neighbourhood region. Our main findings are listed below:

\begin{itemize}
\item The velocity distributions of old stars formed less than 1 Gyr or 3 Gyr after the Big Bang show no correlation with the dark matter velocity distribution in the Solar neighbourhood.

\item The velocity distributions of metal-poor stars with metallicity cuts \FeHone~and \FeHtwo~do not show strong correlations with the dark matter distribution, with $p$-values for the KS test remaining smaller than 0.05 for most halos. Setting an additional cut on \aFe~does not change this conclusion, with $p$-values always smaller than 0.05 for at least one component of the velocity distribution in all halos. Setting a stronger cut of \FeHthree~results in large Poisson errors due to the low number of stars with such low metallicities in the simulations. As a result of these large errors, one cannot statistically distinguish the velocity distribution of stars with \FeHthree~and dark matter in the Auriga halos. 

\item The local dark matter speed distributions have peak speeds which are systematically larger than the peak speeds of the distributions of metal-poor stars. 

\item The halo integrals of metal-poor stars do not show any correlations with the dark matter halo integrals, especially at their high speed tails.

\item The density profiles of stars with various metallicity cuts drop faster than that of dark matter. As expected, in the Solar neighbourhood, the slopes of the density profiles of metal-poor stars and dark matter are not similar, with the  slope of stars with \FeHtwo~and \aFetwo~greater than the  dark matter slope by 32\% to 54\%, depending on the halo.
\end{itemize}

Hence, the results of our work do not confirm the conclusions of ref.~\cite{Herzog-Arbeitman:2017fte} which found excellent correlation between the local velocity distributions of metal-poor stars and dark matter in one simulated galaxy with an especially quiet merger history, and proposed that observations of metal-poor
stars in the Solar neighbourhood could be used to empirically determine the local dark matter velocity distribution. The difference in the local velocity distribution of dark matter and stellar halo is not surprising; the dark matter halo is built up not only by the accretion of dwarf galaxies, but also through smooth accretion and accretion of dark substructures. Ref.~\cite{Wang2011} shows that a non-negligible fraction of mass in Milky Way size halos of the Aquarius project \citep{Springel2008} comes from smooth accretion. Moreover, the steep shape of the stellar mass-halo mass in $\Lambda$CDM predicts that almost all low mass substructures are dark \citep{Sawala2015}, which can contribute to the dark matter mass due to their large abundance \citep{Springel2008}. These points have also been discussed in refs.~\cite{Necib:2018iwb} and ~\cite{Necib:2018igl}. Dark matter direct detection experiments, however, are sensitive to all dark matter in the Solar neighbourhood (as considered in this work and in ref.~\cite{Herzog-Arbeitman:2017fte}), regardless of its accretion history.

Before the submission of this work, ref.~\cite{Necib:2018igl} appeared where the authors find strong correlations between the local velocity distribution of stars and dark matter accreted from luminous satellites in two Milky Way-mass halos from the Latte suite of FIRE-2 simulations. In particular, ref.~\cite{Necib:2018igl} demonstrates how to infer the local velocity distribution of dark matter accreted from luminous satellites when it is dominated by a relaxed population and debris  flow. The correlation between the velocity distributions of stars and dark matter accreted from luminous satellites is expected due to the similar origin of those two populations. However, it needs to be tested in multiple simulations to ensure that the conclusions are not sensitive to the specific baryonic feedback, resolution, or merger history of a particular simulation. We leave the study of this correlation and possibility of other tracers of the local dark matter distribution in Auriga halos to future work.

We finally note that one obvious limitation of this and similar work is the poor statistics of metal-poor stars in both simulations and observations. The availability of higher resolution simulations and better observations could significantly improve the results of such analyses in the future.

\subsection*{Acknowledgements}
We thank  Alis Deason, Christopher McCabe, Mariangela Lisanti, and Piero Madau for useful discussions on the results of this work. NB is grateful to the Institute for Research in Fundamental Sciences in Tehran for their hospitality during her visit. NB and AF acknowledge support by the European Union COFUND/Durham Junior Research Fellowship (under EU grant agreement no. 609412). NB has received support from the European Union's Horizon 2020 research and innovation programme under the Marie Sklodowska-Curie grant agreement No 690575. This work was supported by the Science and Technology Facilities Council 
(STFC) consolidated grant ST/P000541/1. CSF acknowledges support by the 
European Research Council (ERC) through Advanced Investigator grant DMIDAS 
(GA 786910). FAG acknowledges financial support from CONICYT through the project FONDECYT Regular Nr.~1181264, and funding from the Max Planck Society through a Partner Group grant. This work used the DiRAC Data Centric system at Durham 
University, operated by the Institute for Computational Cosmology on 
behalf of the STFC DiRAC HPC Facility (\url{www.dirac.ac.uk}). This 
equipment was funded by BIS National E-infrastructure capital grant 
ST/K00042X/1, STFC capital grant ST/H008519/1, and STFC DiRAC Operations 
grant ST/K003267/1 and Durham University. DiRAC is part of the National 
E-Infrastructure.

\appendix


\section{Sensitivity to metallicity cut}
\label{sec:Diffmetal}

\begin{figure}[h!]
\begin{center}
\includegraphics[width=0.31\textwidth]{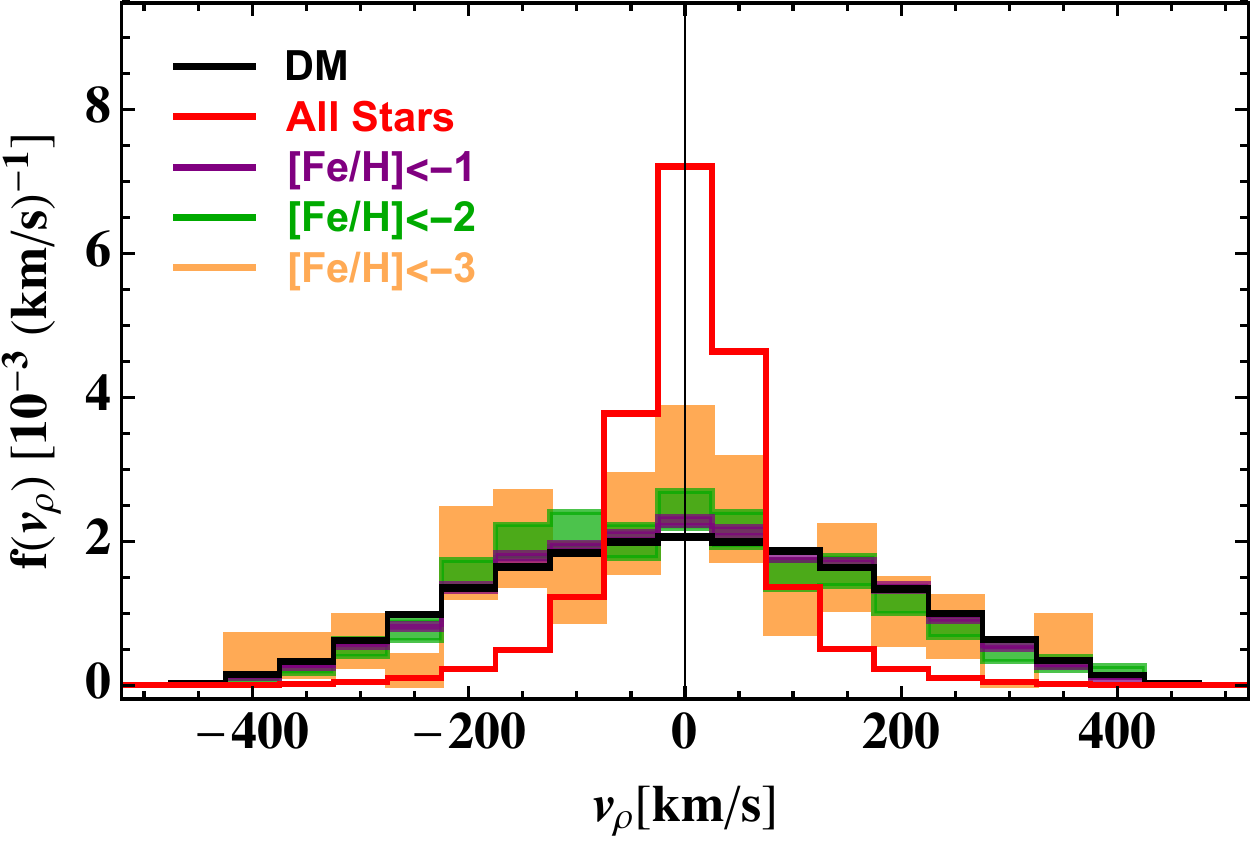}
  \includegraphics[width=0.31\textwidth]{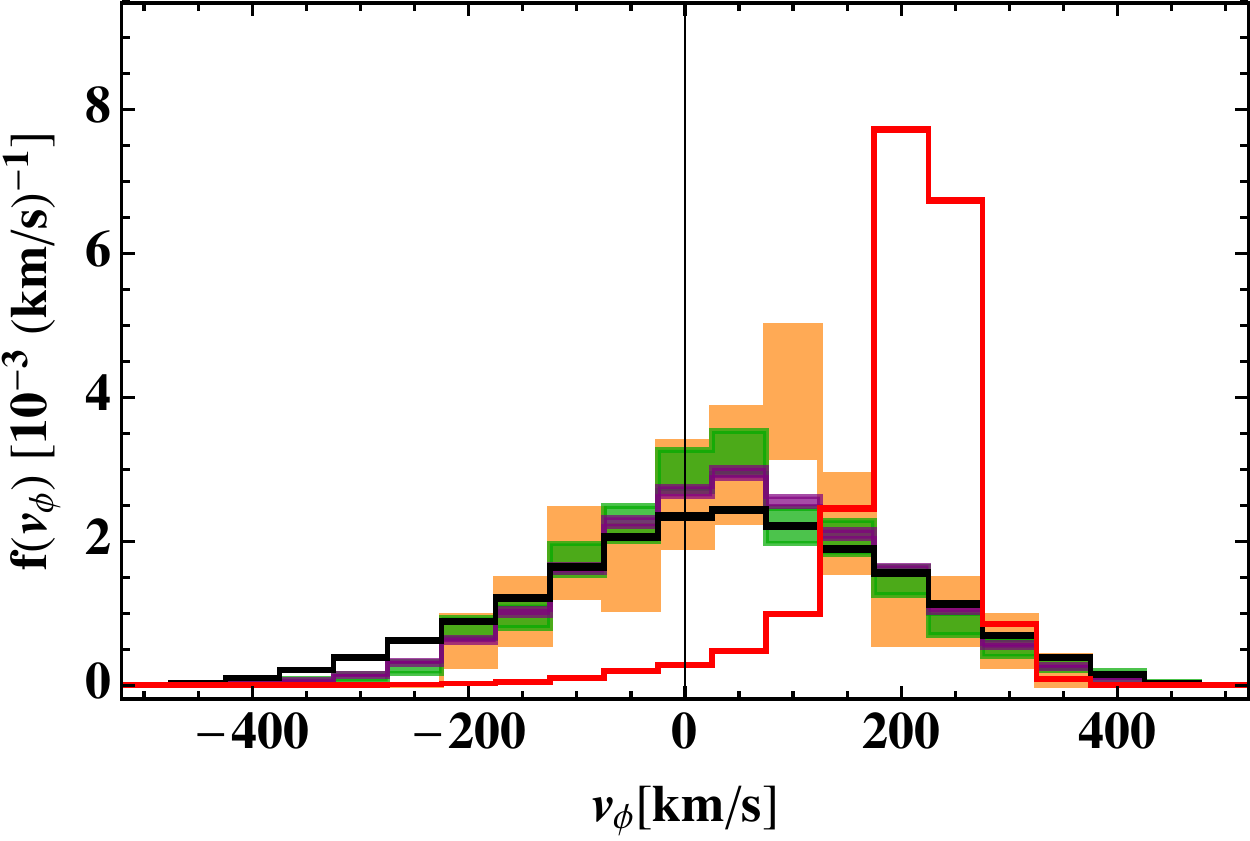}
  \includegraphics[width=0.31\textwidth]{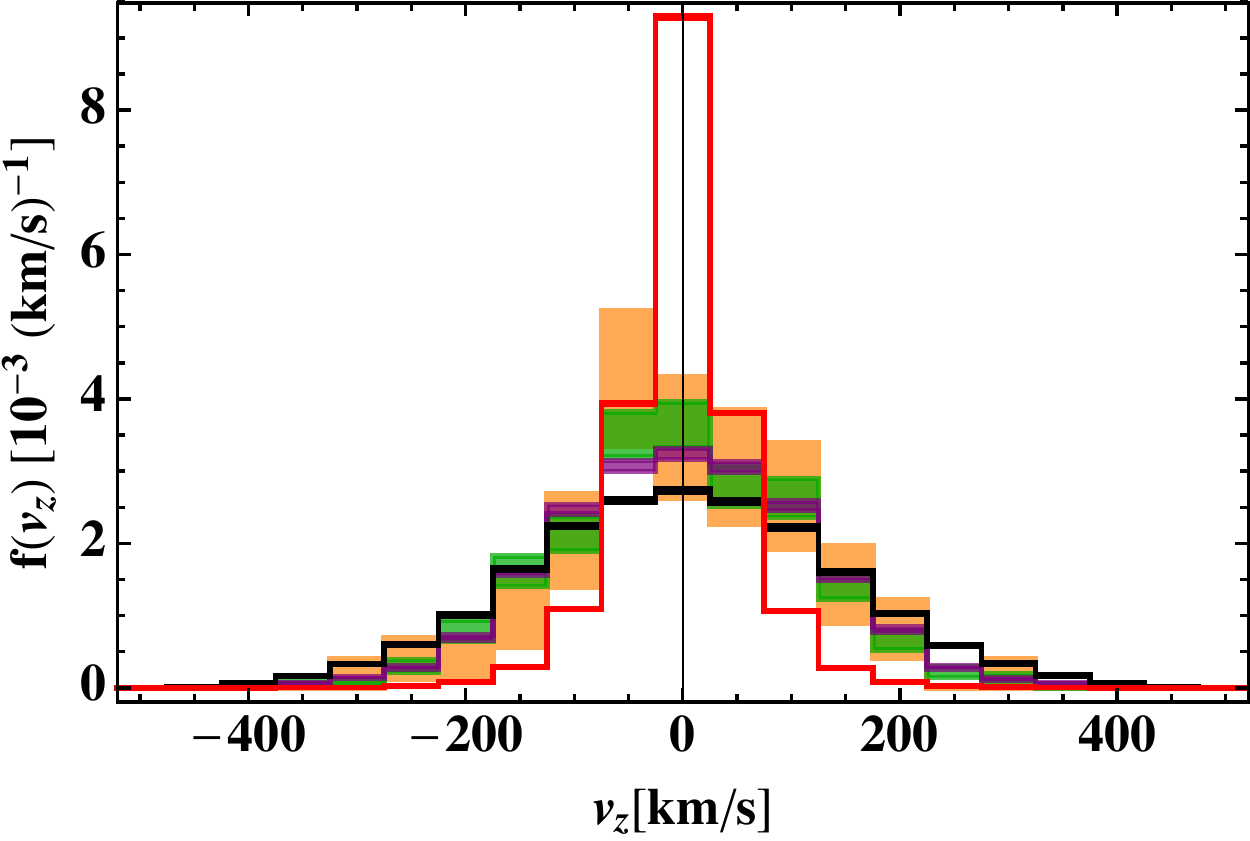}\\
  \includegraphics[width=0.31\textwidth]{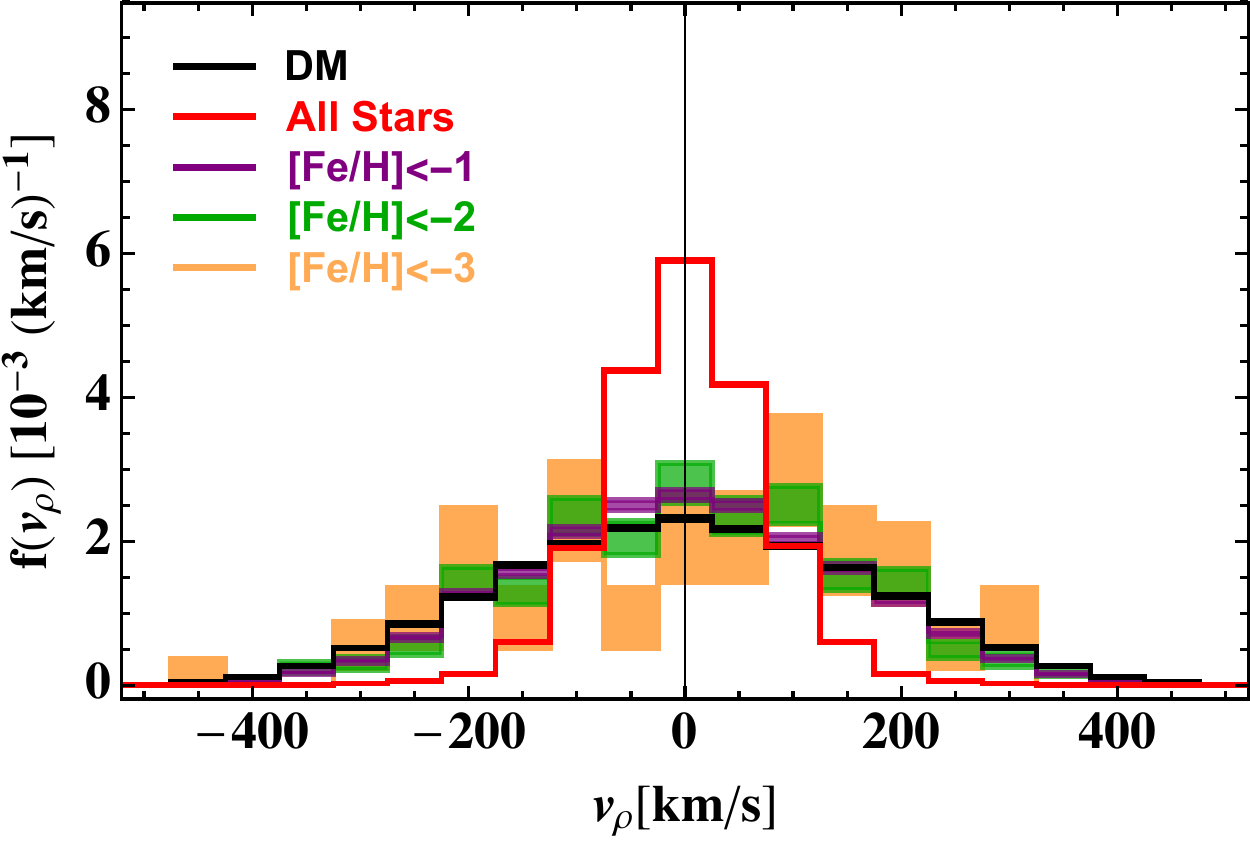}
  \includegraphics[width=0.31\textwidth]{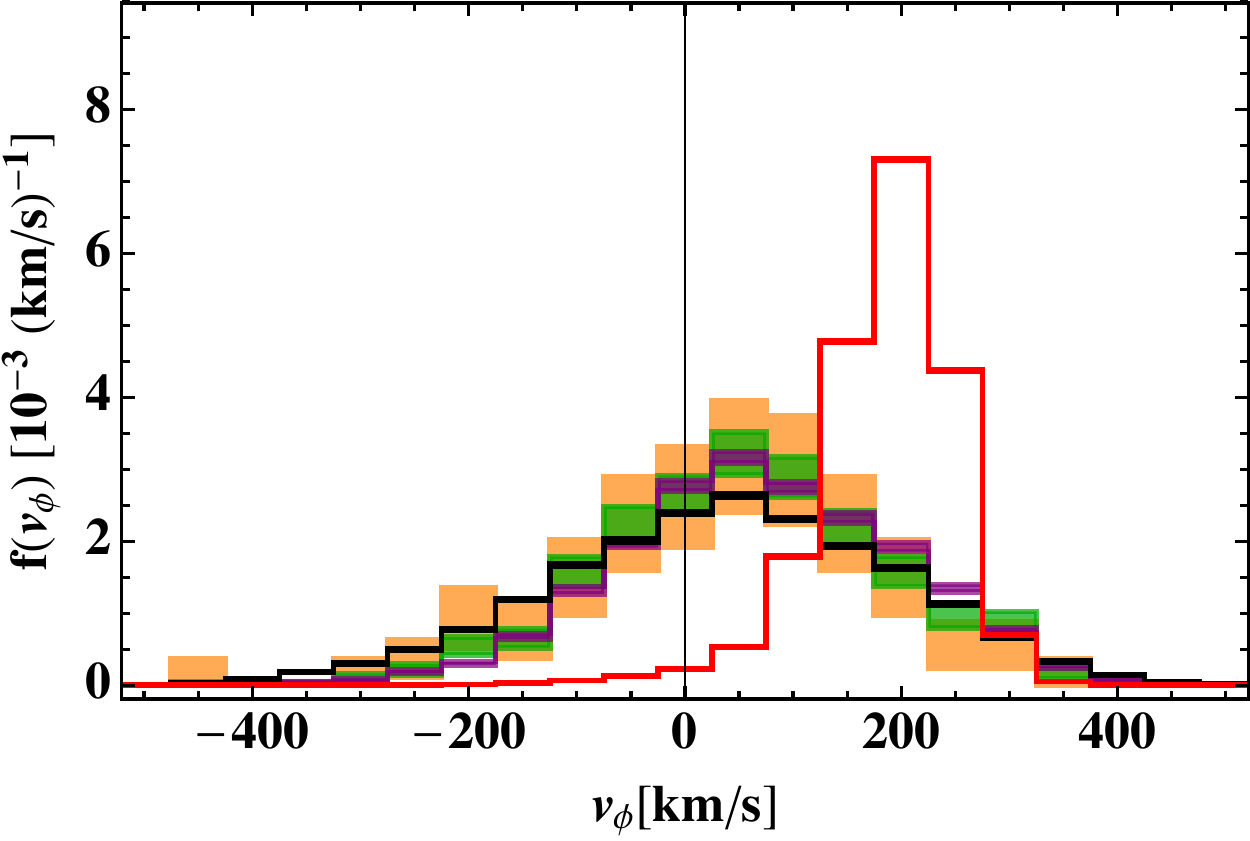}
  \includegraphics[width=0.31\textwidth]{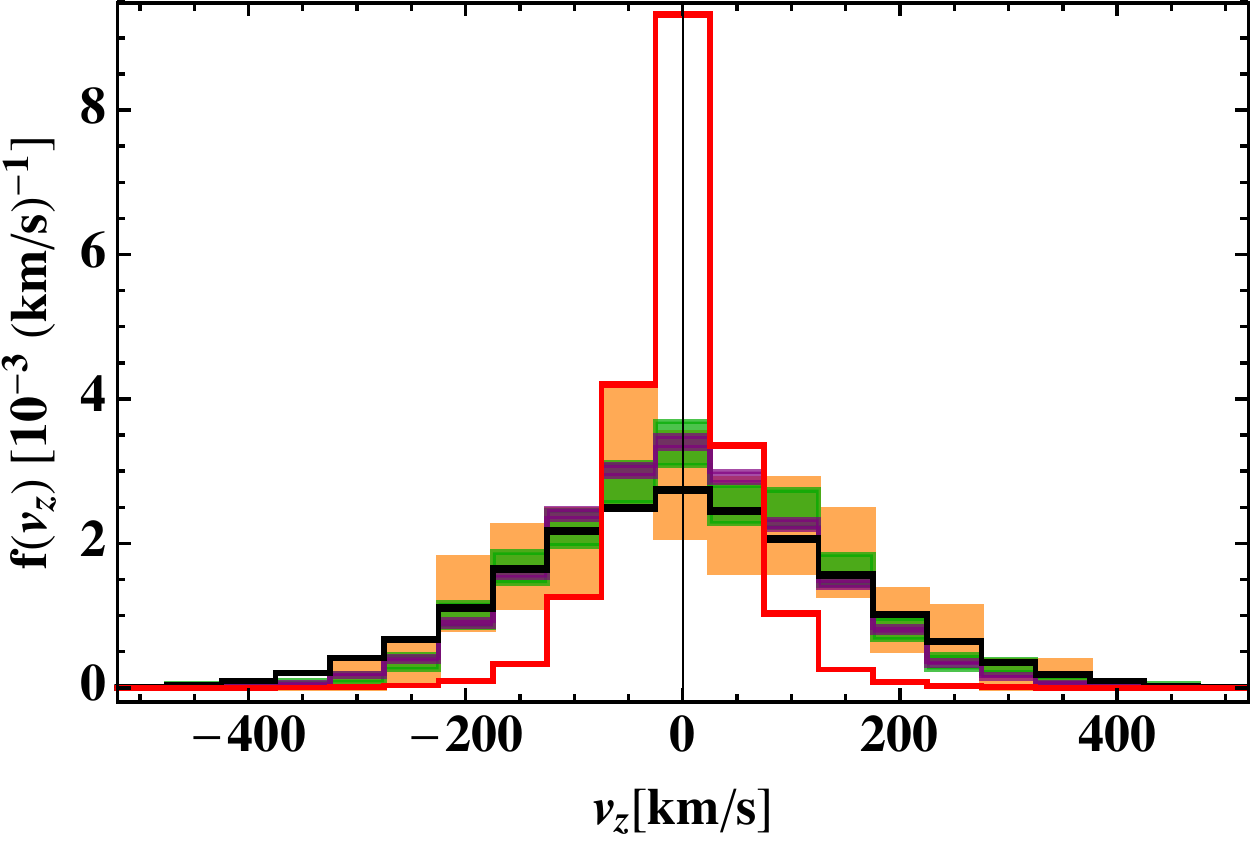}\\
  \includegraphics[width=0.31\textwidth]{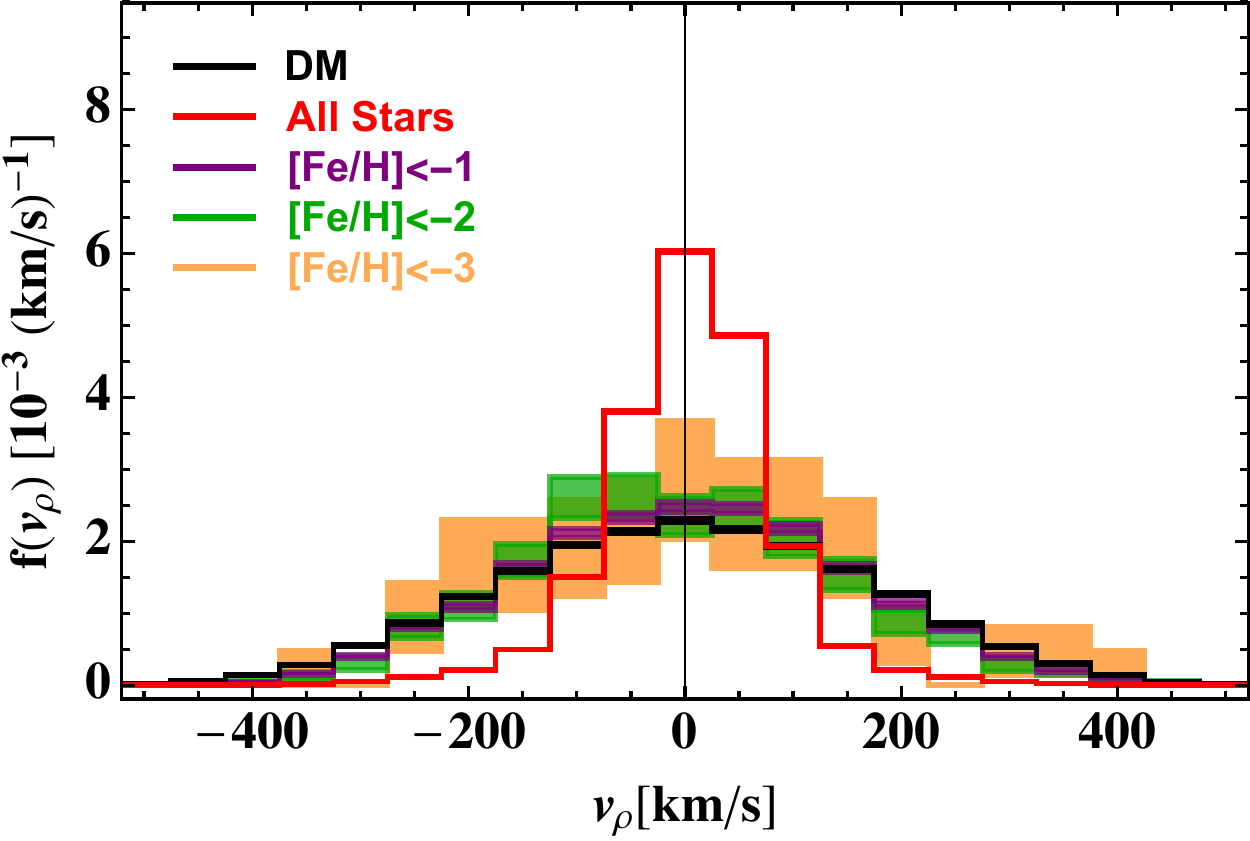}
  \includegraphics[width=0.31\textwidth]{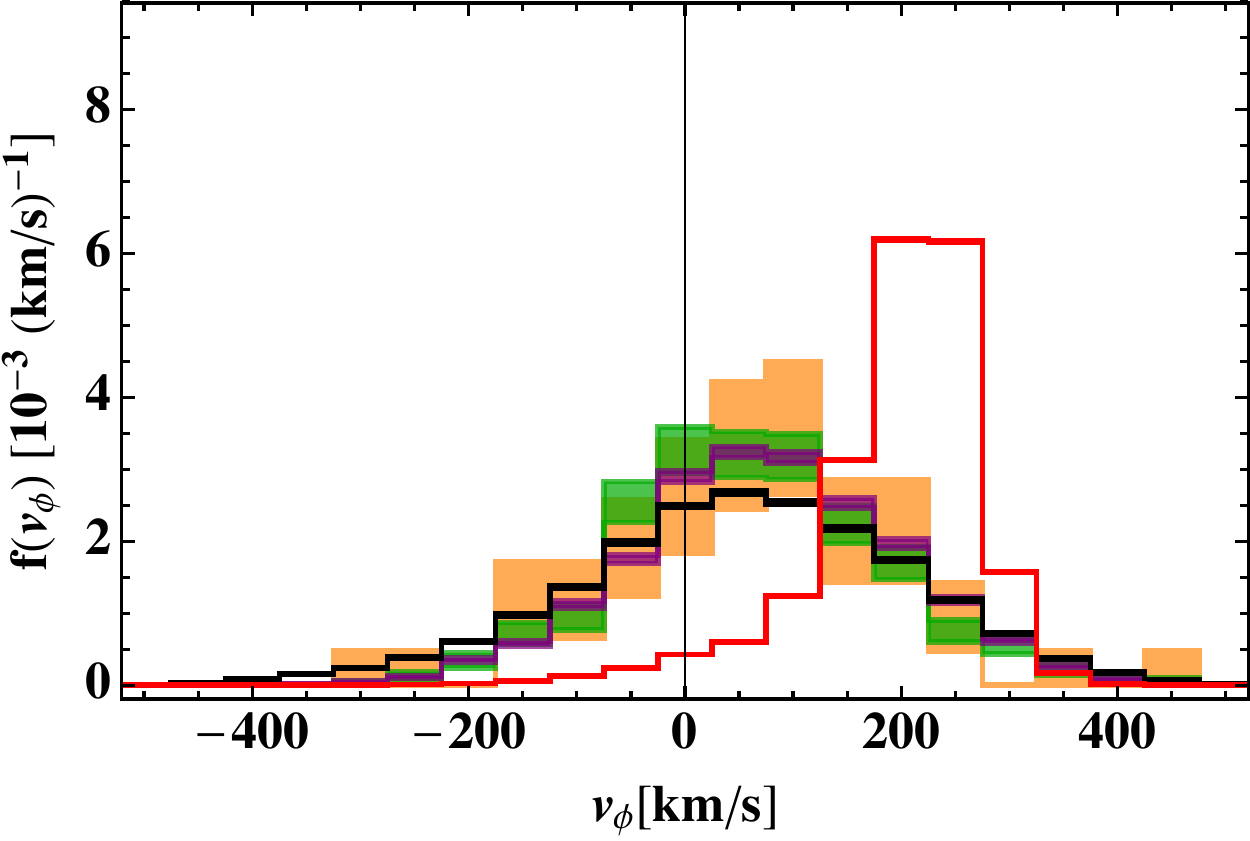}
  \includegraphics[width=0.31\textwidth]{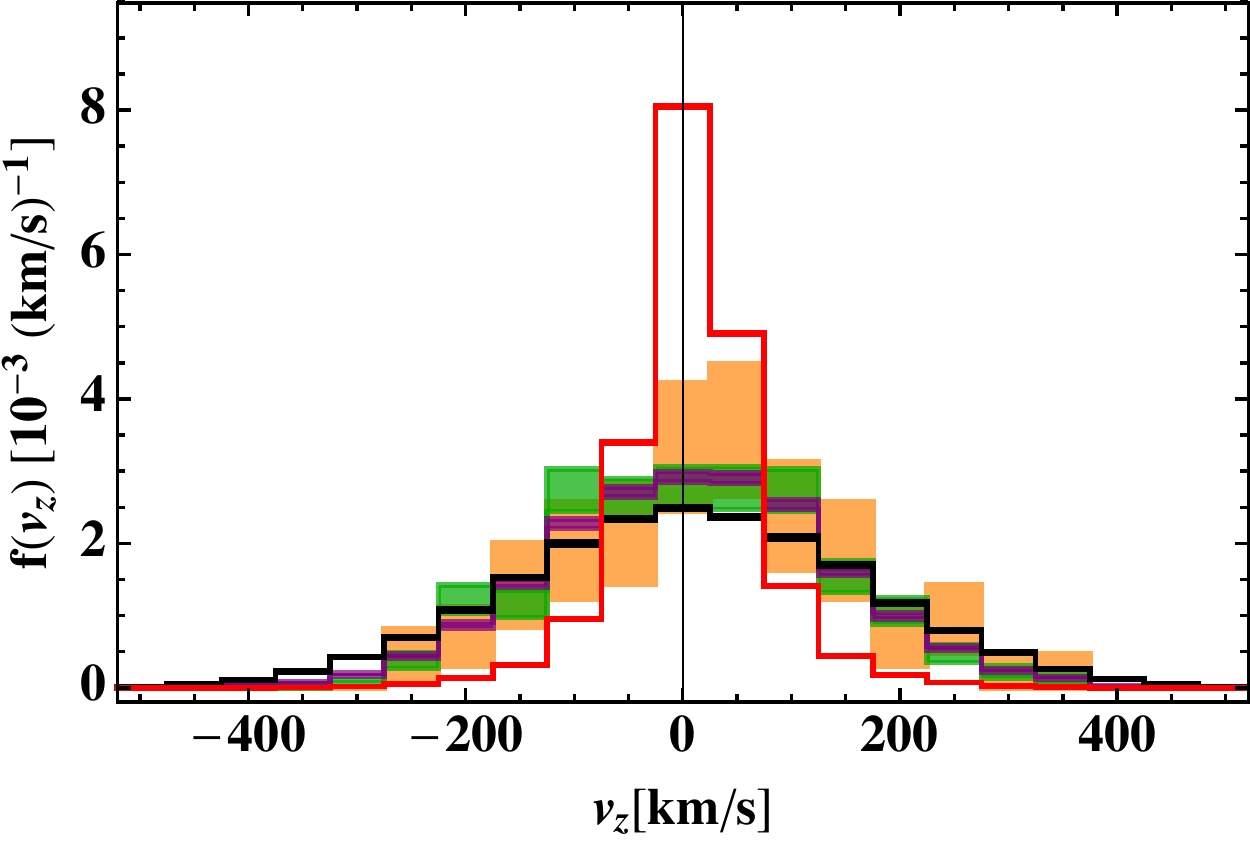}\\
  \includegraphics[width=0.31\textwidth]{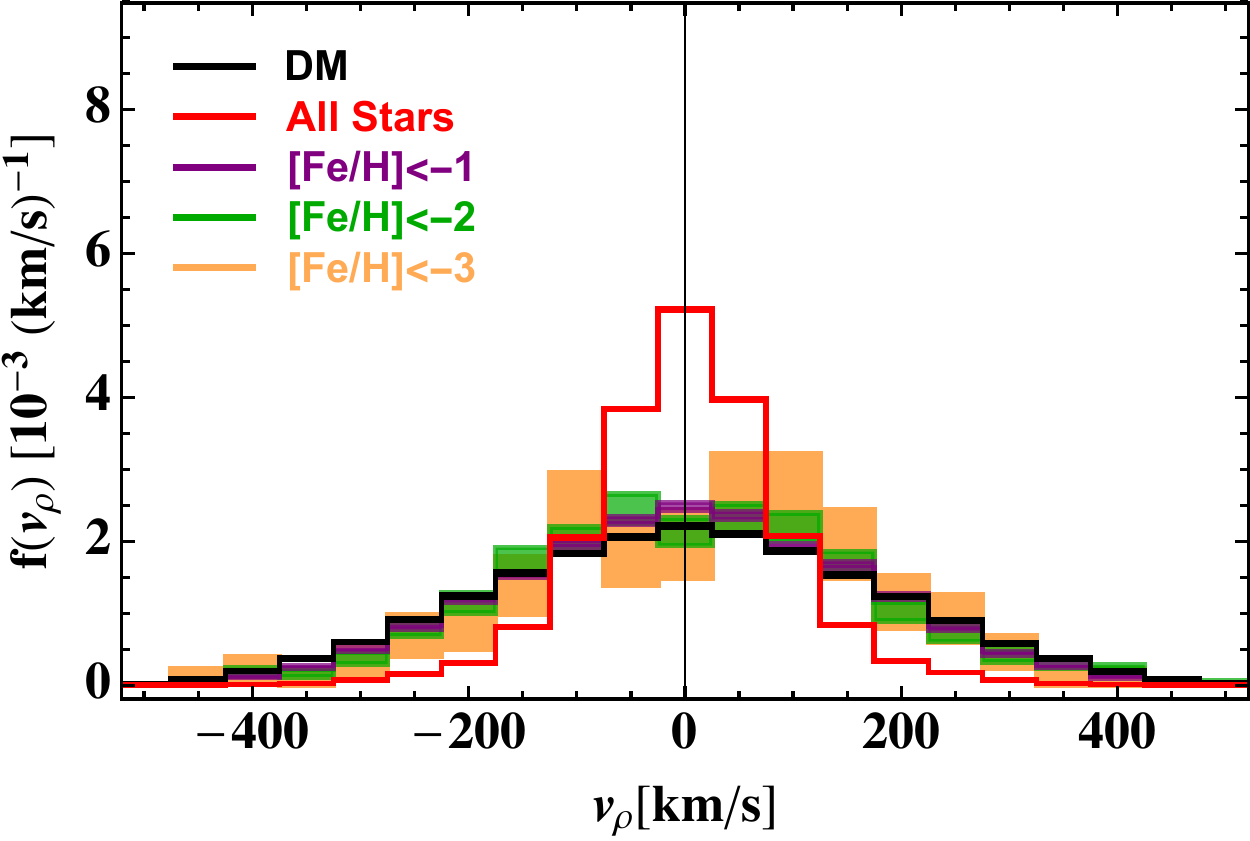}
  \includegraphics[width=0.31\textwidth]{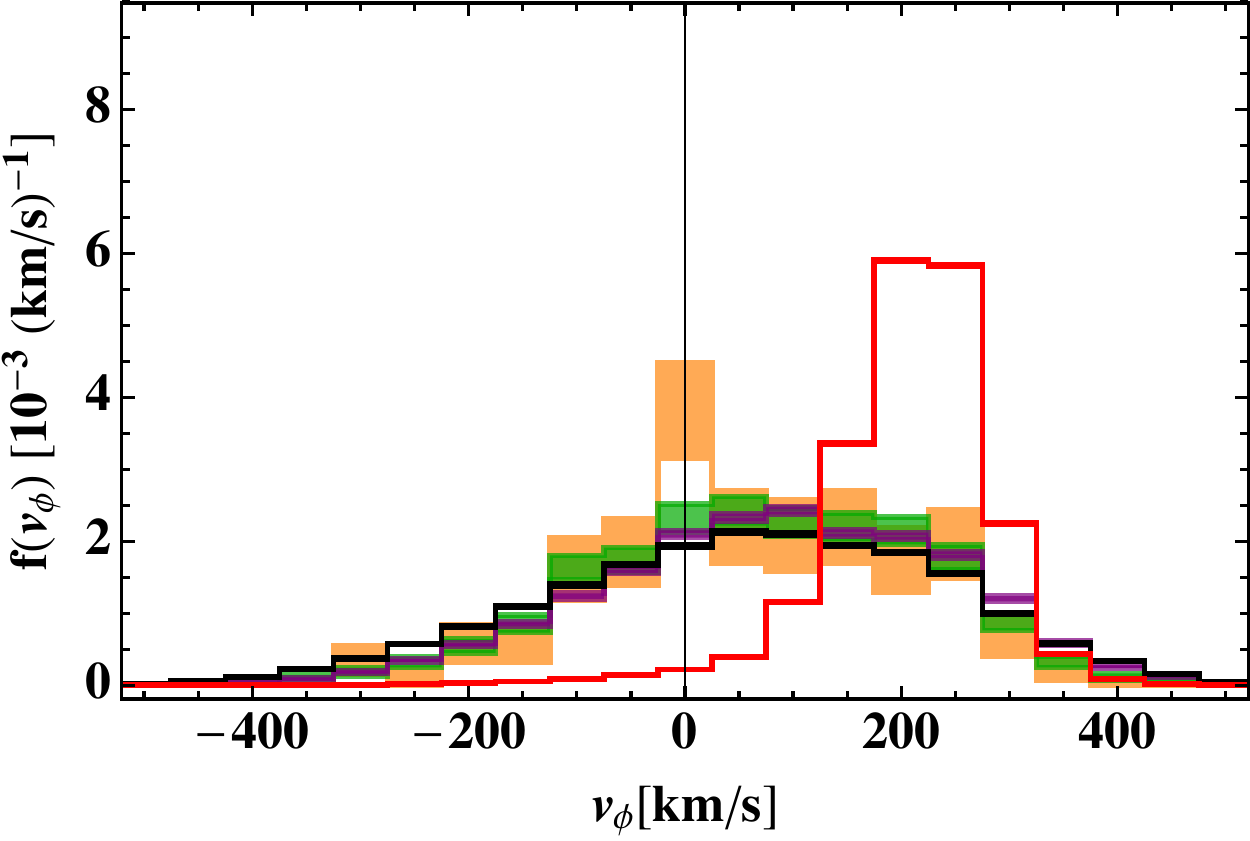}
  \includegraphics[width=0.31\textwidth]{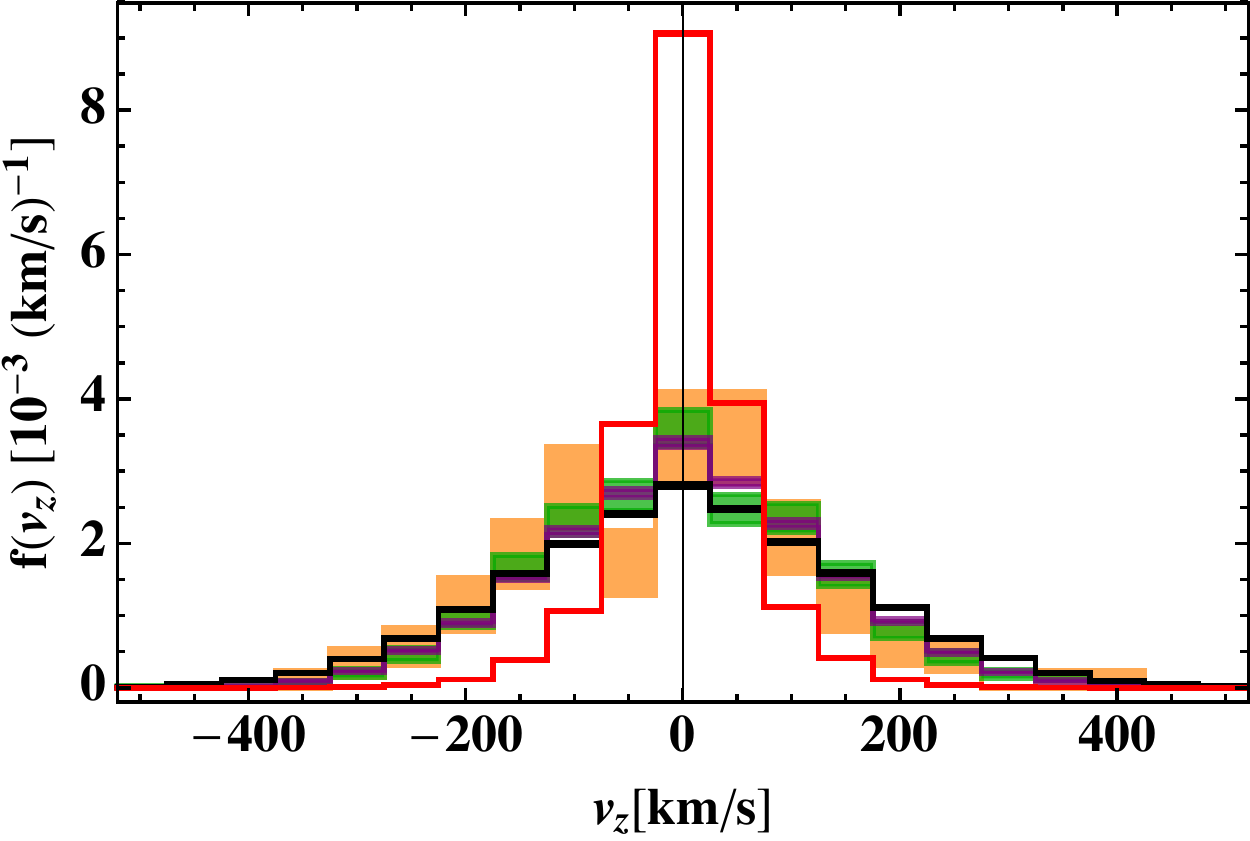}\\
  \includegraphics[width=0.31\textwidth]{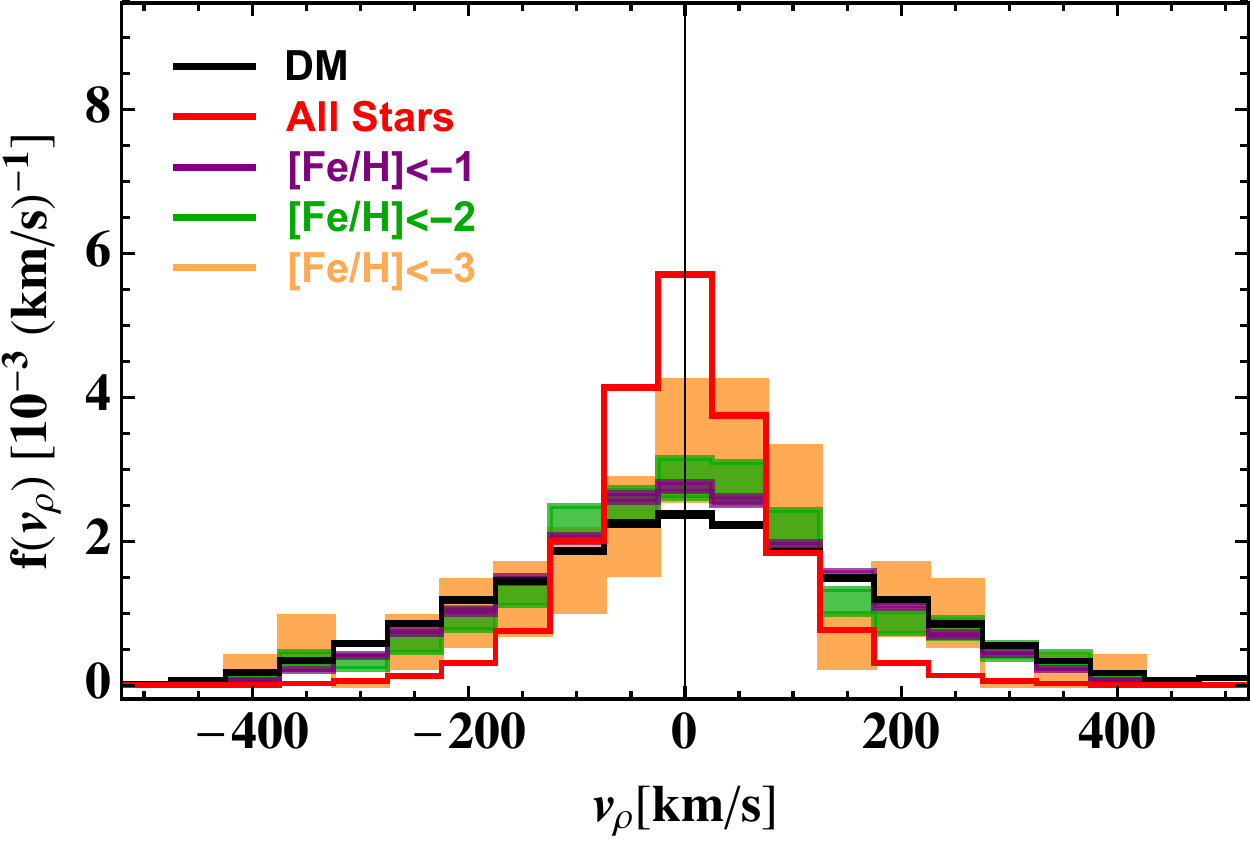}
  \includegraphics[width=0.31\textwidth]{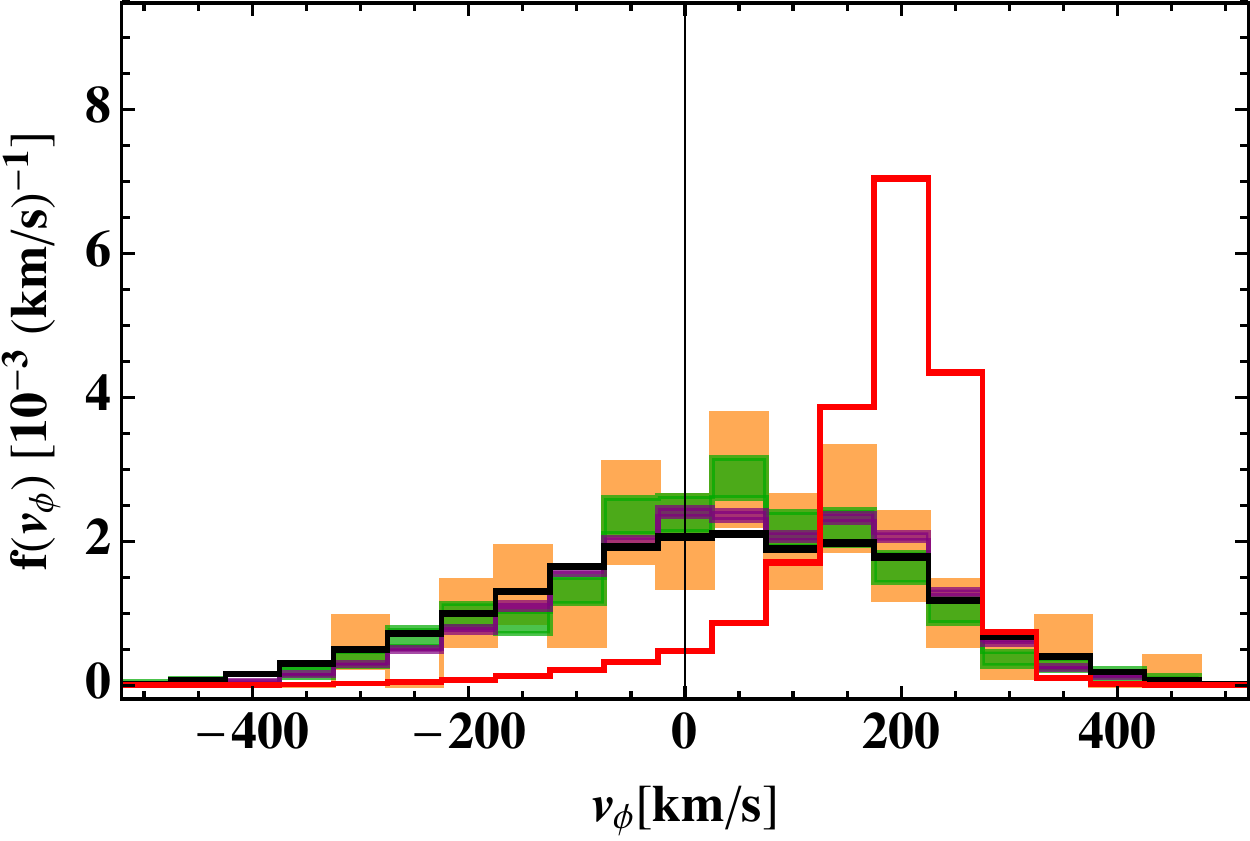}
  \includegraphics[width=0.31\textwidth]{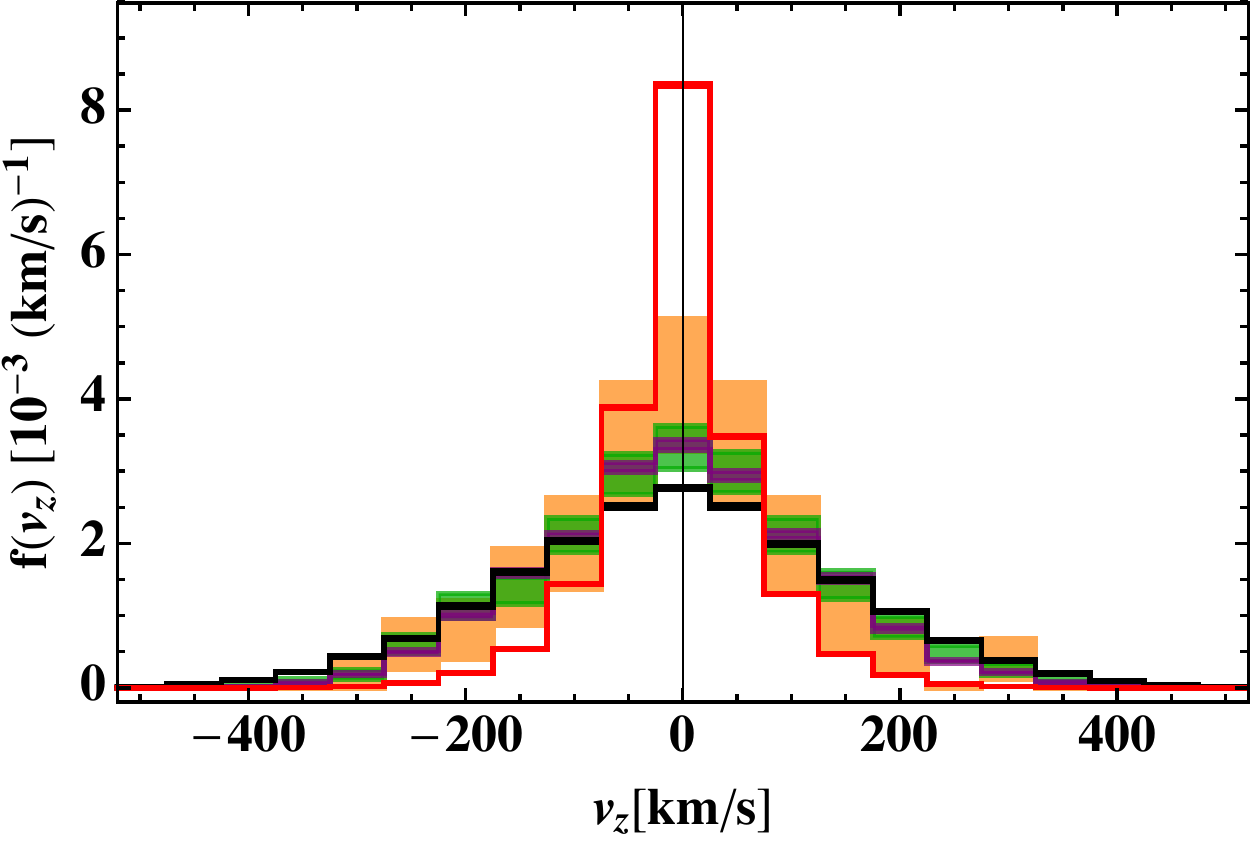}\\
  \includegraphics[width=0.31\textwidth]{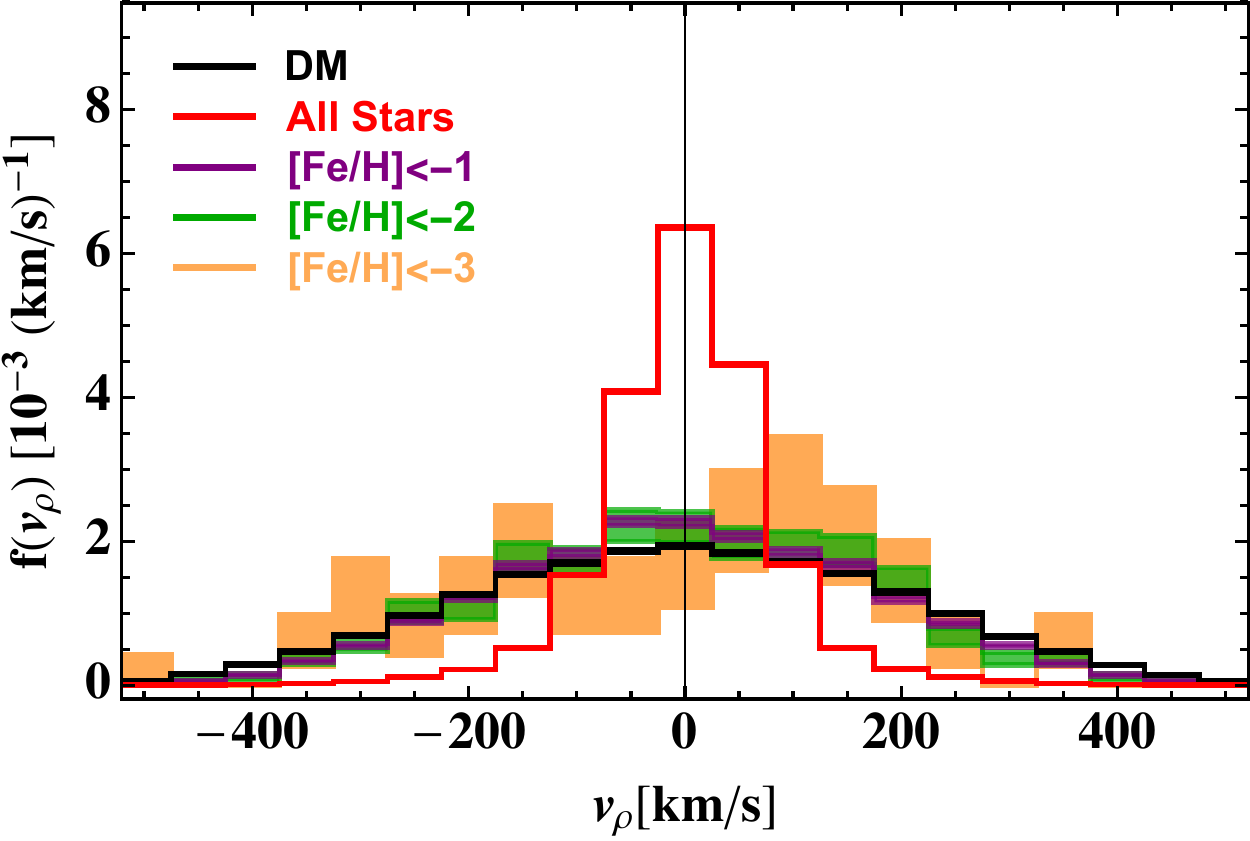}
  \includegraphics[width=0.31\textwidth]{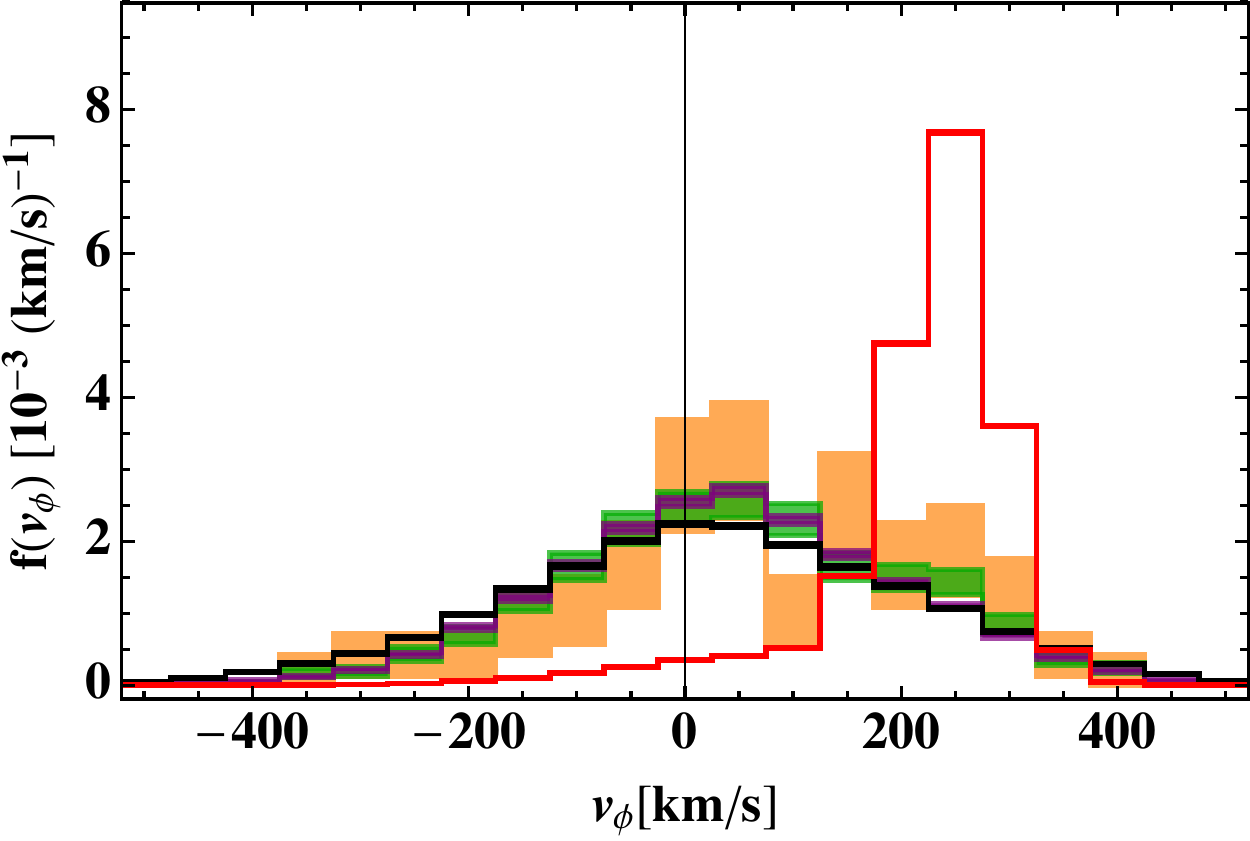}
  \includegraphics[width=0.31\textwidth]{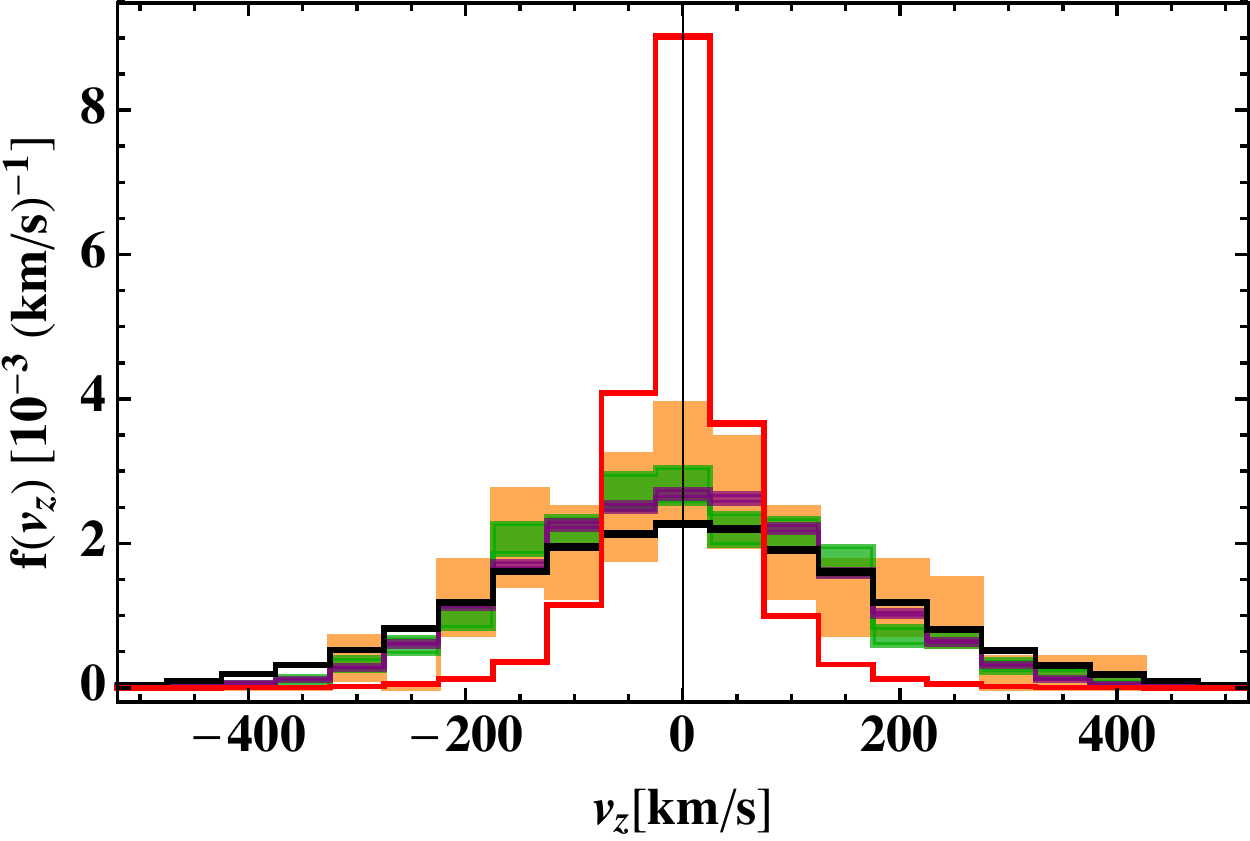}
\caption{Same as figure \ref{fig:FeHalphaHeCuts}, but showing the stellar velocity distributions with cuts on \FeH~only.}
\label{fig:FeHCuts}
\end{center}
\end{figure}

In this appendix we show how the correlations between the local DM velocity distribution and the distribution of metal-poor stars change when we only consider the cut on \FeH~without including any cut on \aFe. In figure~\ref{fig:FeHCuts} we show the velocity distributions of DM, all stars, and stars with different cuts on \FeH, in the Solar neighbourhood.

\bigskip

\begin{table}[H]
    \centering
    \begin{tabular}{|c|c c c|c c c|}
      \hline
         & \multicolumn{3}{|c|}{\FeHtwo} & \multicolumn{3}{|c|}{\FeHthree} \\
       \hline
       Halo Name  & $\rho$ & $\phi$  & $z$ & $\rho$ & $\phi$  & $z$\\
       \hline
       Au6 & $4.1 \times 10^{-2}$ & $2.8 \times 10^{-3}$  & $1.7 \times 10^{-3}$ & $3.0 \times 10^{-1}$ & $1.6 \times 10^{-1}$ & $1.6 \times 10^{-1}$\\
       Au16 & $9.2 \times 10^{-2}$ & $7.1 \times 10^{-5}$ & $1.5 \times 10^{-2}$ & $5.2 \times 10^{-1}$ & $4.7 \times 10^{-1}$  & $2.8 \times 10^{-1}$ \\
       Au21 & $4.1 \times 10^{-2}$  & $1.7 \times 10^{-4}$  &  $9.3 \times 10^{-3}$ & $4.9 \times 10^{-1}$ & $5.4 \times 10^{-1}$ & $2.8 \times 10^{-1}$ \\
       Au23 & $4.1 \times 10^{-2}$ & $2.0 \times 10^{-4}$  & $6.4 \times 10^{-4}$  & $1.9 \times 10^{-1}$ & $8.7 \times 10^{-2}$  & $8.7 \times 10^{-2}$ \\
       Au24  & $5.2 \times 10^{-3}$ & $1.0 \times 10^{-3}$  & $5.5 \times 10^{-2}$  & $3.6 \times 10^{-1}$& $1.9 \times 10^{-1}$ & $1.3 \times 10^{-1}$ \\
       Au27 & $4.0 \times 10^{-3}$ & $2.4 \times 10^{-4}$ & $2.3 \times 10^{-3}$  & $3.6 \times 10^{-1}$  & $1.4 \times 10^{-2}$ & $4.0 \times 10^{-1}$ \\
      \hline
    \end{tabular}
    \caption{$p$-values for the KS test to check the correlation between the radial, azimuthal, and vertical distributions of DM and metal-poor stars with \FeHtwo~(left column) and \FeHthree~(right column) for the six Auriga halos.}
    \label{tab:pvalues-app}
  \end{table}

In table \ref{tab:pvalues-app} we present the $p$-values for the KS test to check the correlations of DM and metal-poor stars with \FeHtwo~or \FeHthree. The $p$-values are in general smaller compared to the case of setting the cut \aFetwo~on the stars (given in table \ref{tab:pvalues}), but the conclusions remain the same. Namely, for \FeHtwo, the $p$-values are always smaller than 0.05 for most halos (other than for Au16 in the radial direction, and for Au24 in the vertical direction), and no strong correlation with the DM velocity distribution is present. Only for \FeHthree, there are some correlations present with the DM distribution, due to the low number of these very metal-poor stars in the Solar neighbourhood in the simulated halos.

\section{Sensitivity to the Solar neighbourhood region}
\label{sec:volume}

In this appendix we investigate the effect of choosing a larger Solar neighbourhood region for the stars. As done in ref.~\cite{Herzog-Arbeitman:2017fte}, we define the Solar neighbourhood region for extracting the stellar velocity distributions, as a cylindrical shell with radius $7 \leq \rho \leq 9$~kpc, and considering all particles within a maximum Galactocentric radius of $r_{\rm max}=15$~kpc (hence there is an implicit cut on the vertical distance, $|z| \leq \sqrt{r_{\rm max}^2 - \rho^2} \sim 12~\kpc$). Since old or metal-poor stars are mostly distributed in the stellar halo, we do not restrict the Solar neighbourhood stellar distributions to the disc. For the DM velocity distribution, however, we still require that $|z| \leq 2$~kpc, such that the particles are constrained to the disc.

\begin{figure}[t]
\begin{center}
   \includegraphics[width=0.31\textwidth]{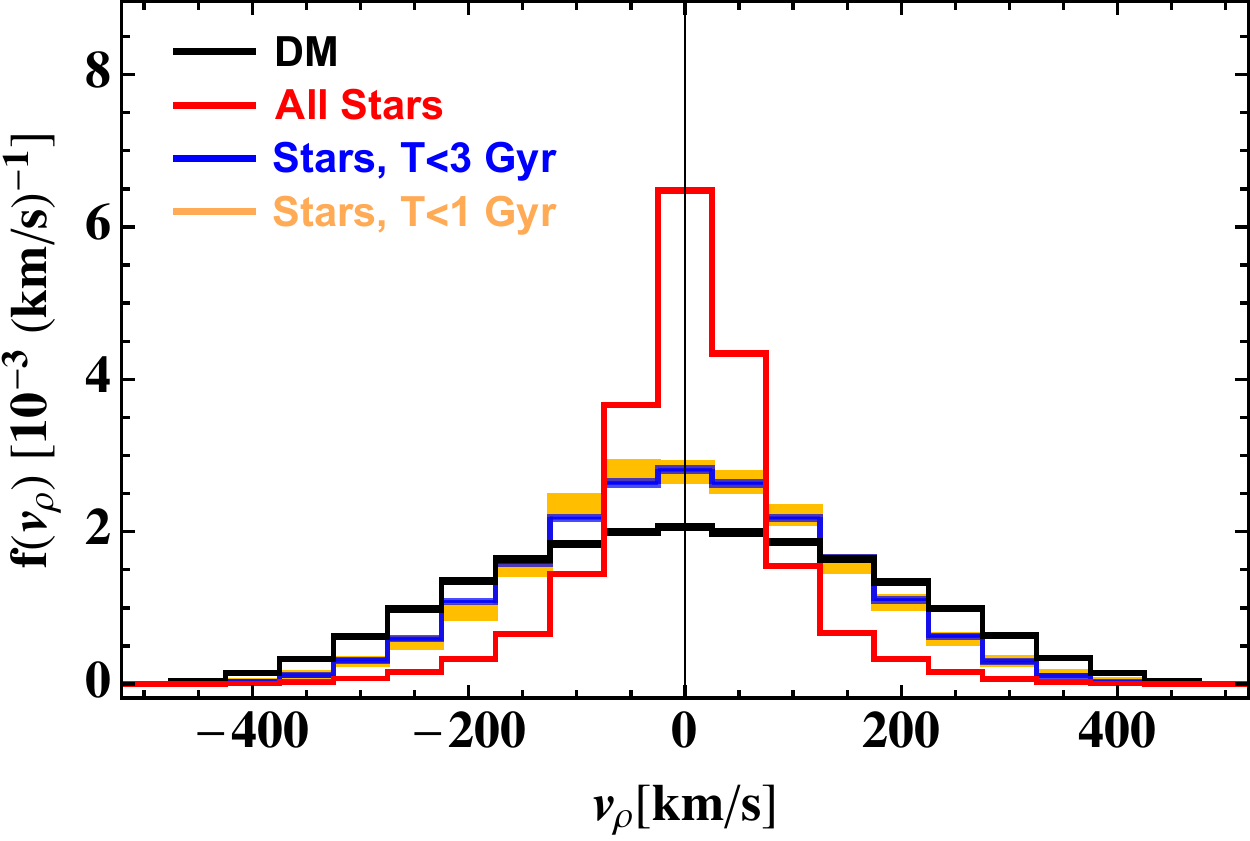}
   \includegraphics[width=0.31\textwidth]{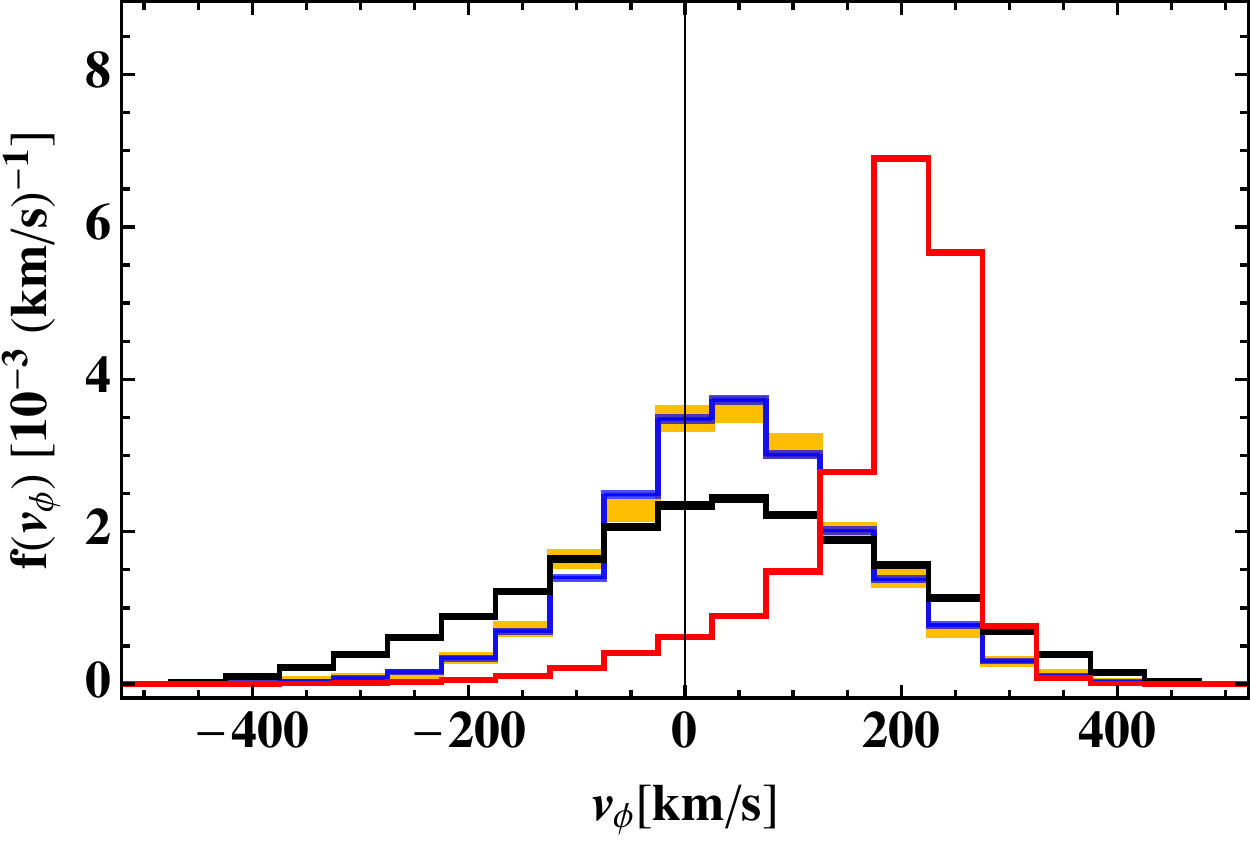}
   \includegraphics[width=0.31\textwidth]{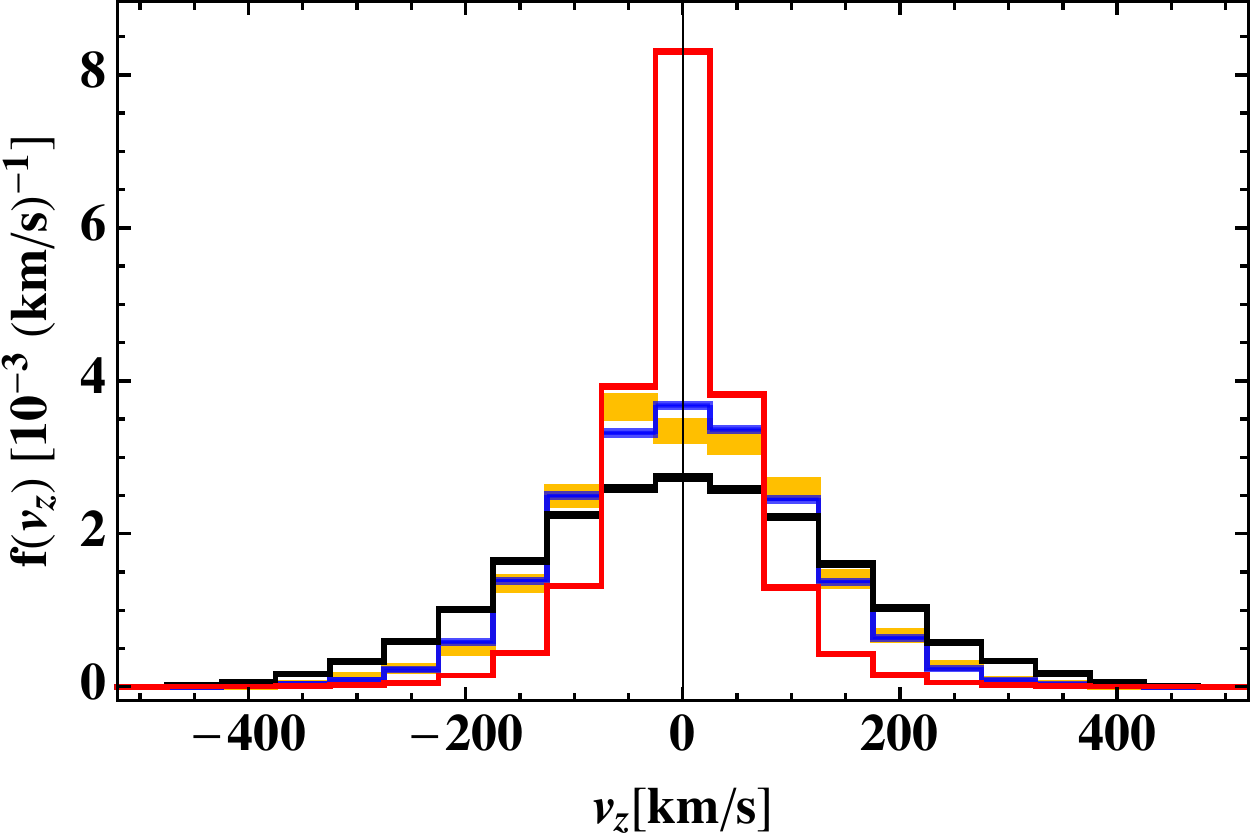}\\
   \includegraphics[width=0.31\textwidth]{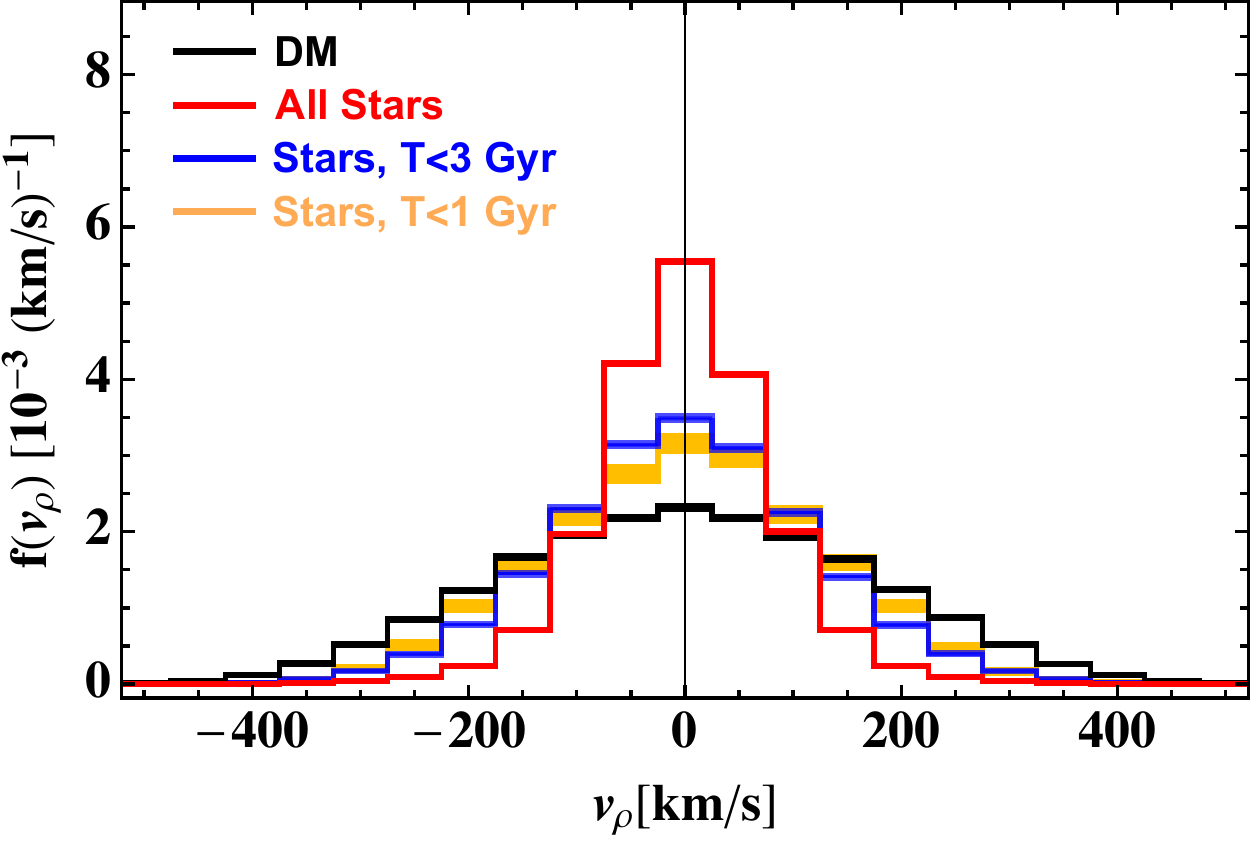}
   \includegraphics[width=0.31\textwidth]{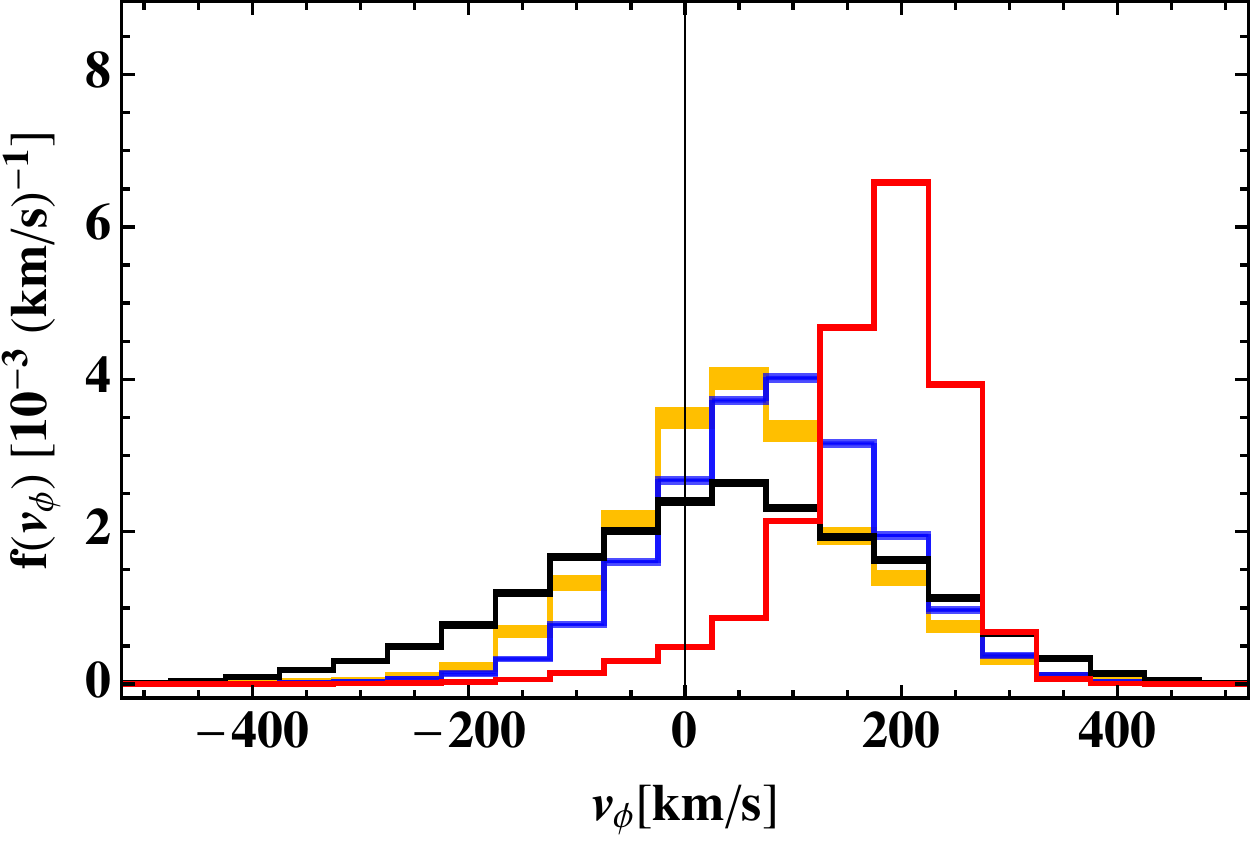}
   \includegraphics[width=0.31\textwidth]{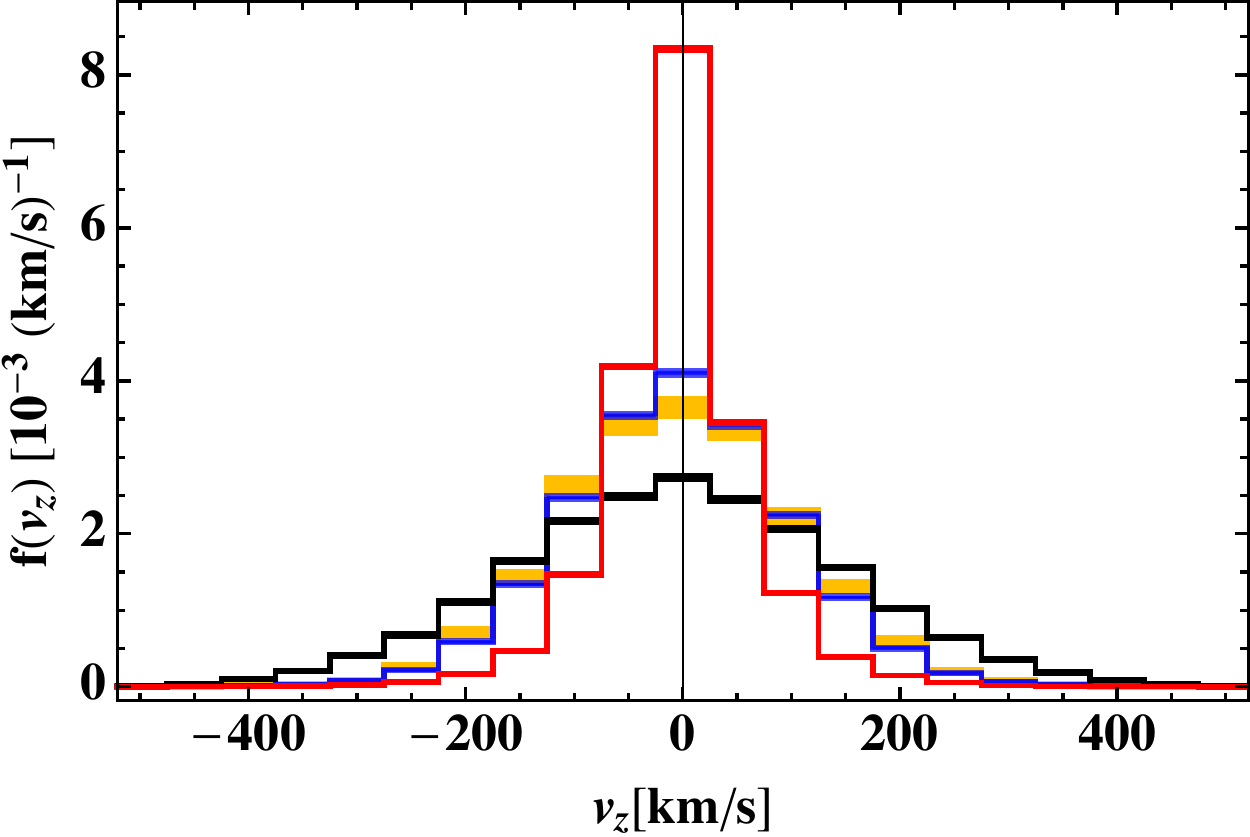}\\
   \includegraphics[width=0.31\textwidth]{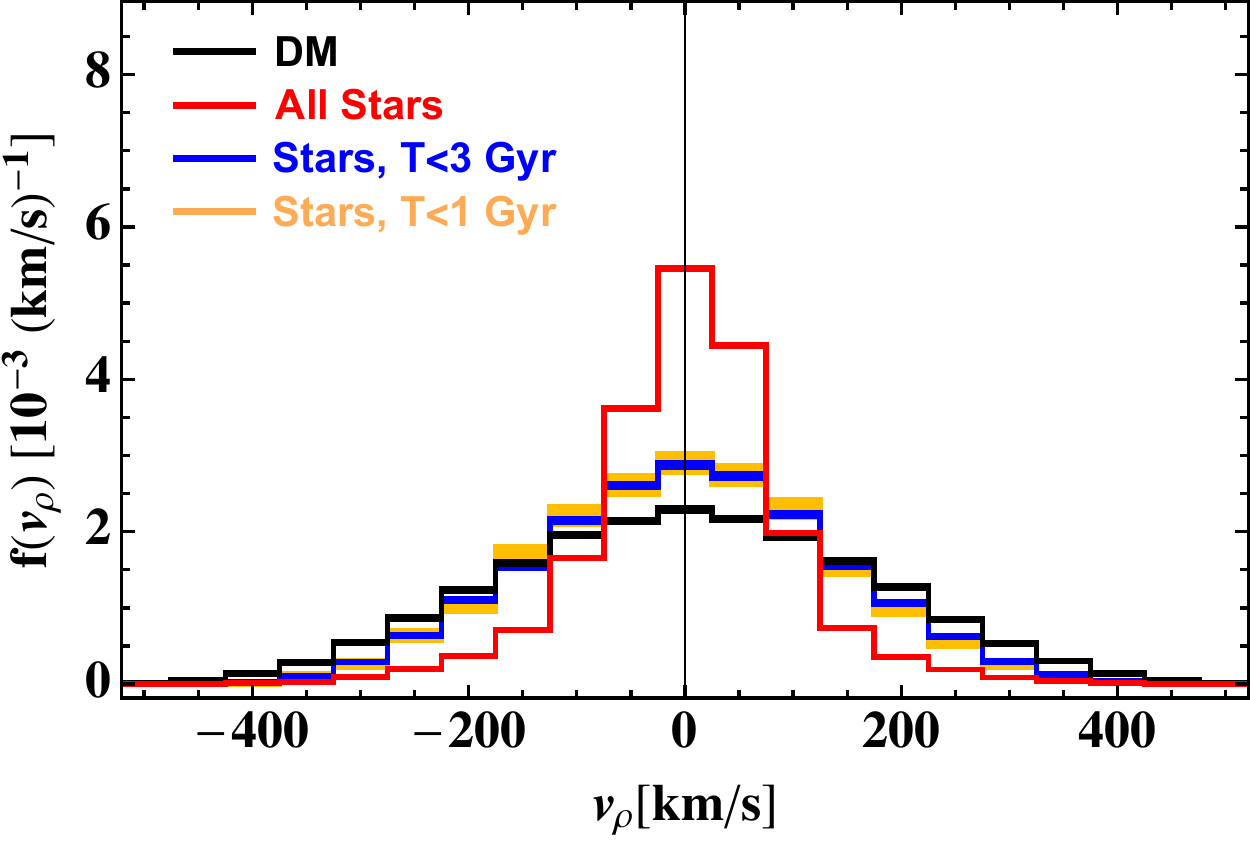}
   \includegraphics[width=0.31\textwidth]{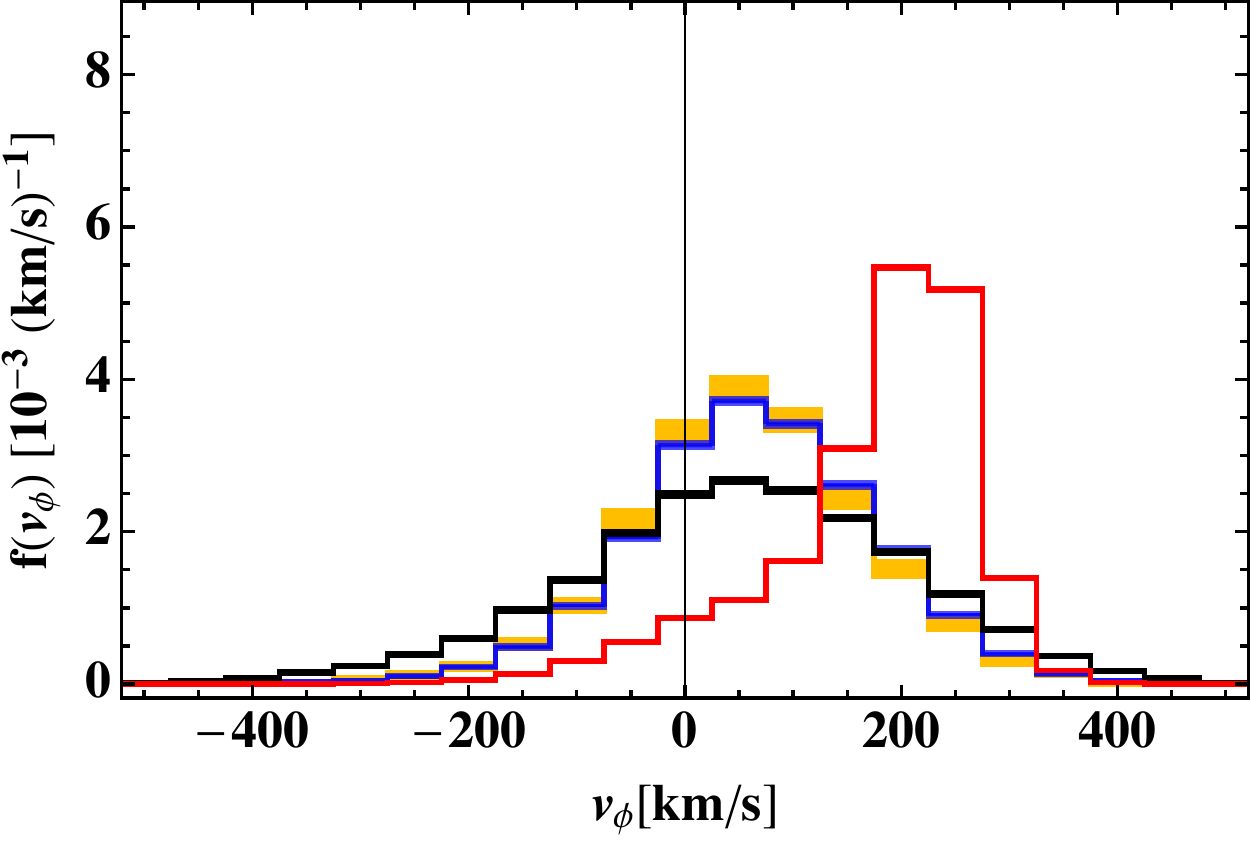}
   \includegraphics[width=0.31\textwidth]{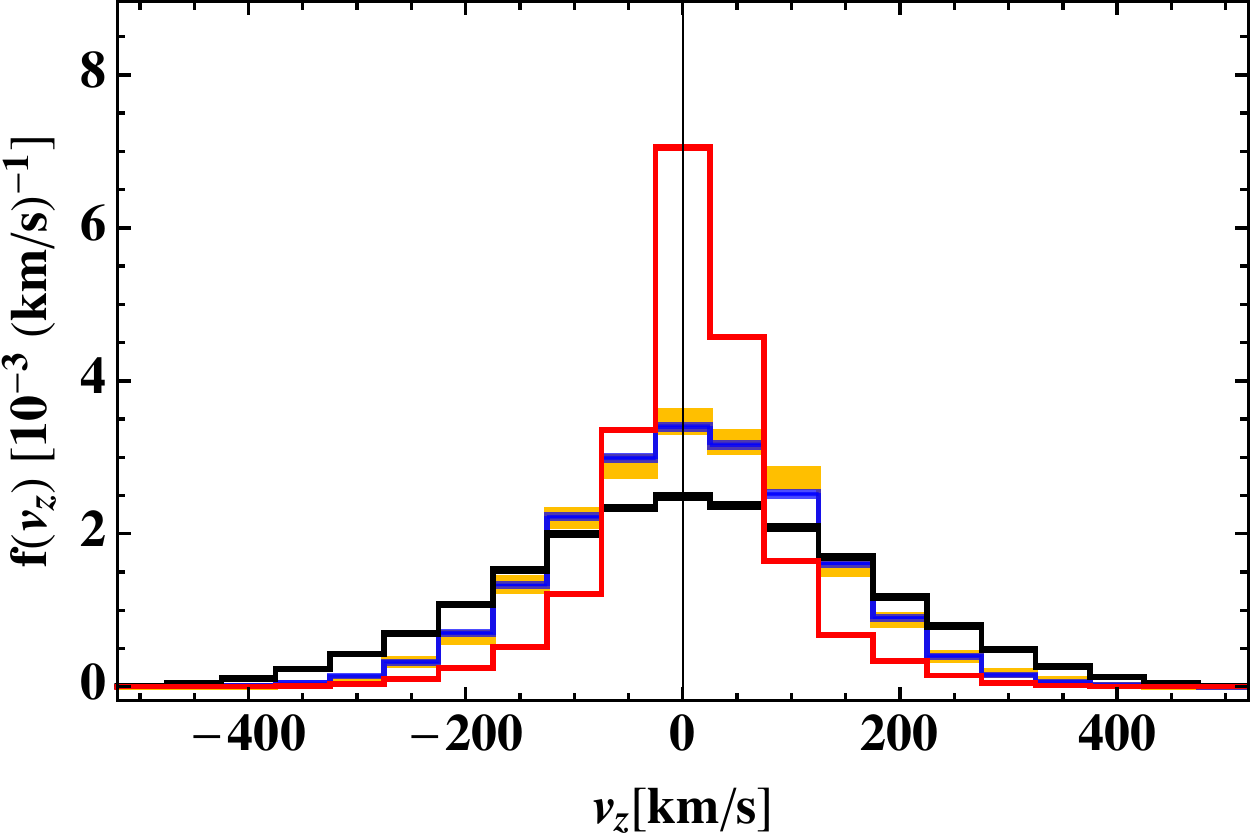}\\
   \includegraphics[width=0.31\textwidth]{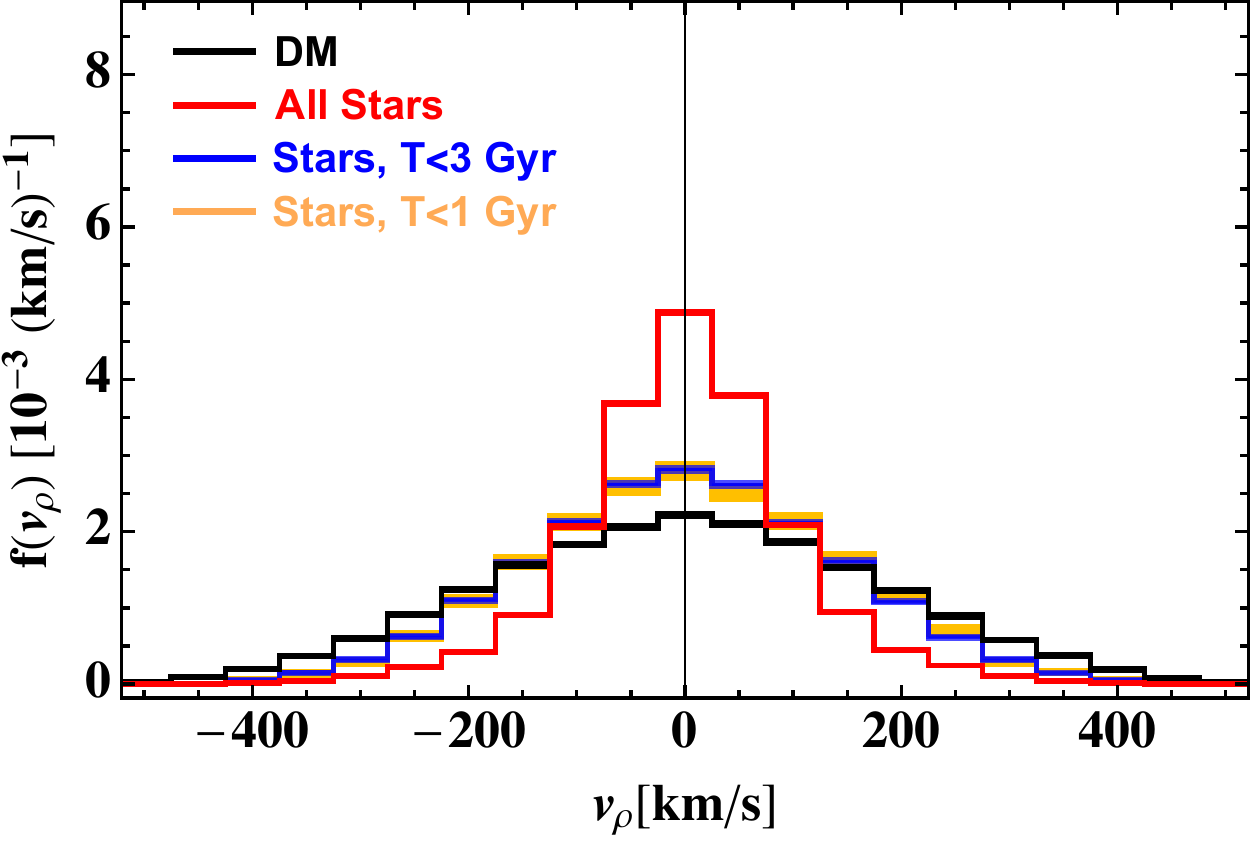}
   \includegraphics[width=0.31\textwidth]{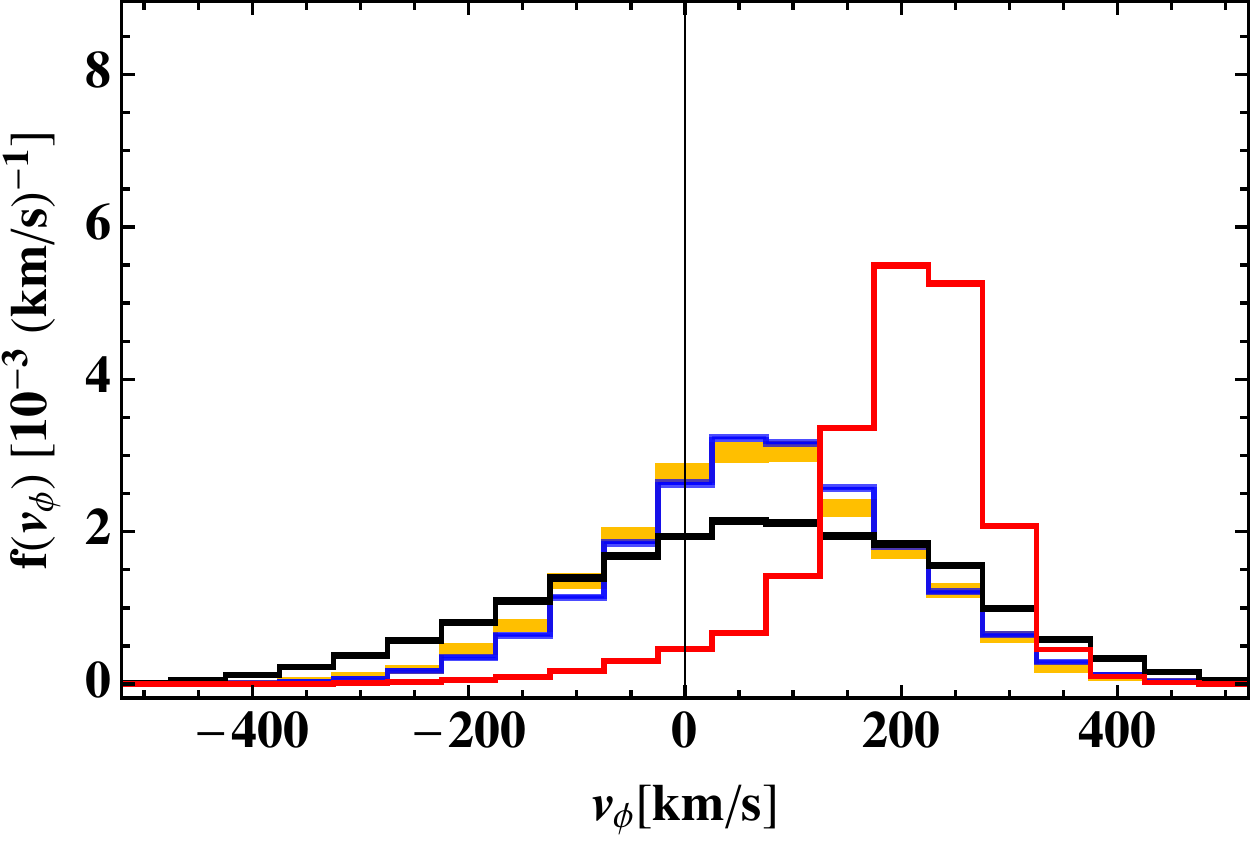}
   \includegraphics[width=0.31\textwidth]{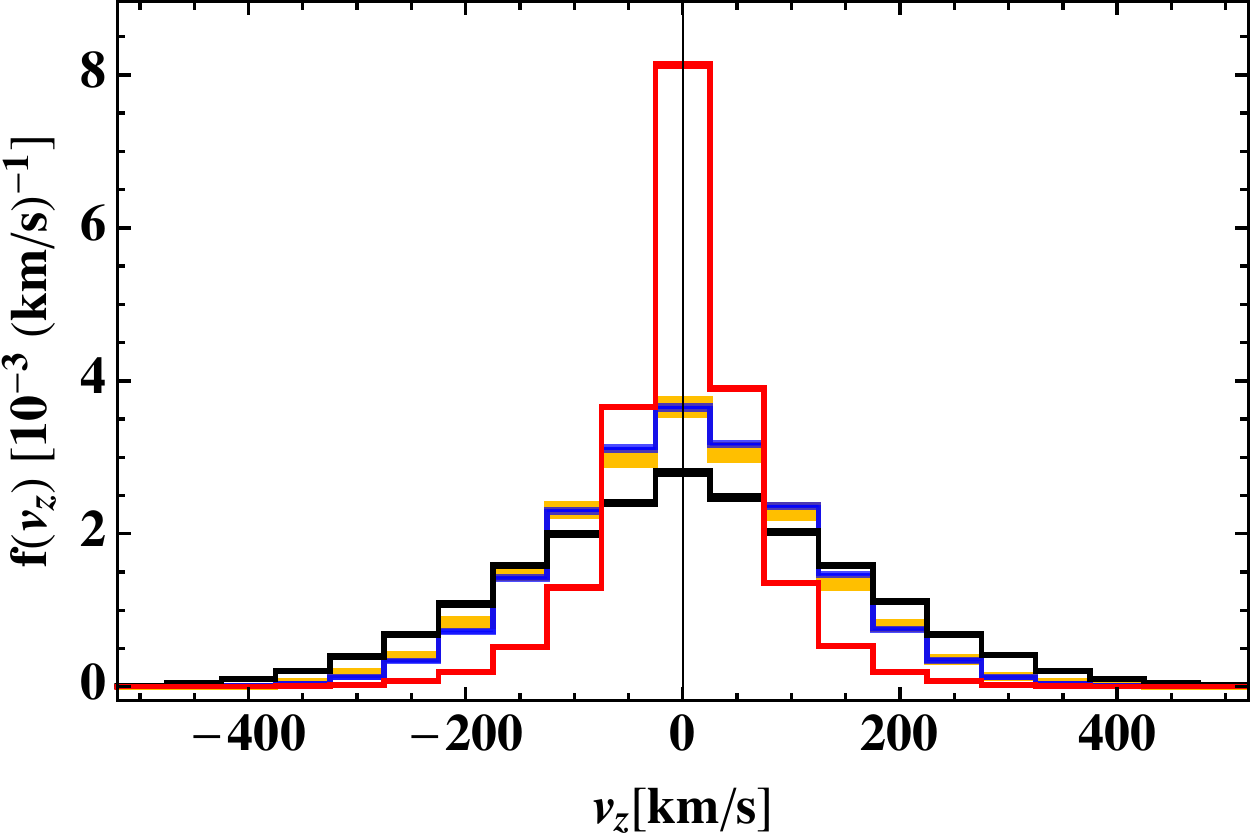}\\
   \includegraphics[width=0.31\textwidth]{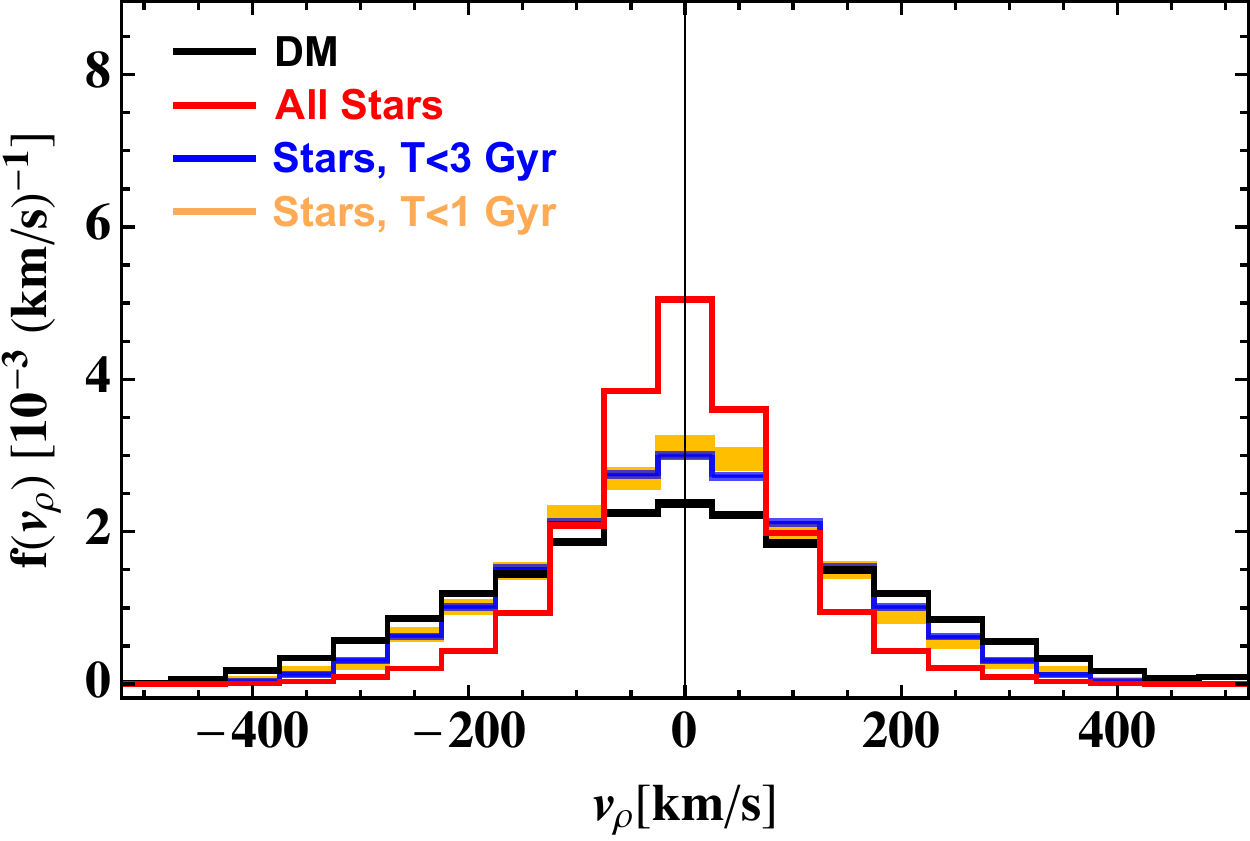}
   \includegraphics[width=0.31\textwidth]{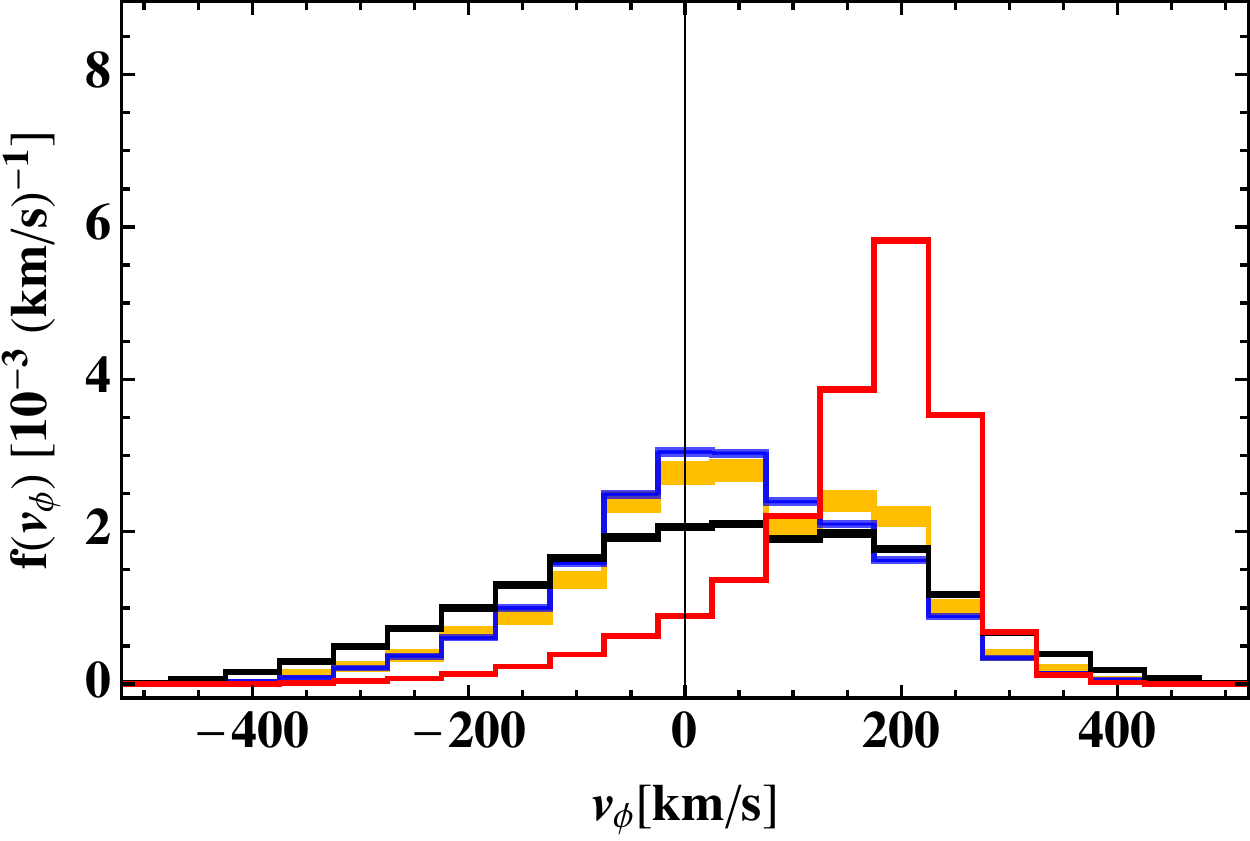}
   \includegraphics[width=0.31\textwidth]{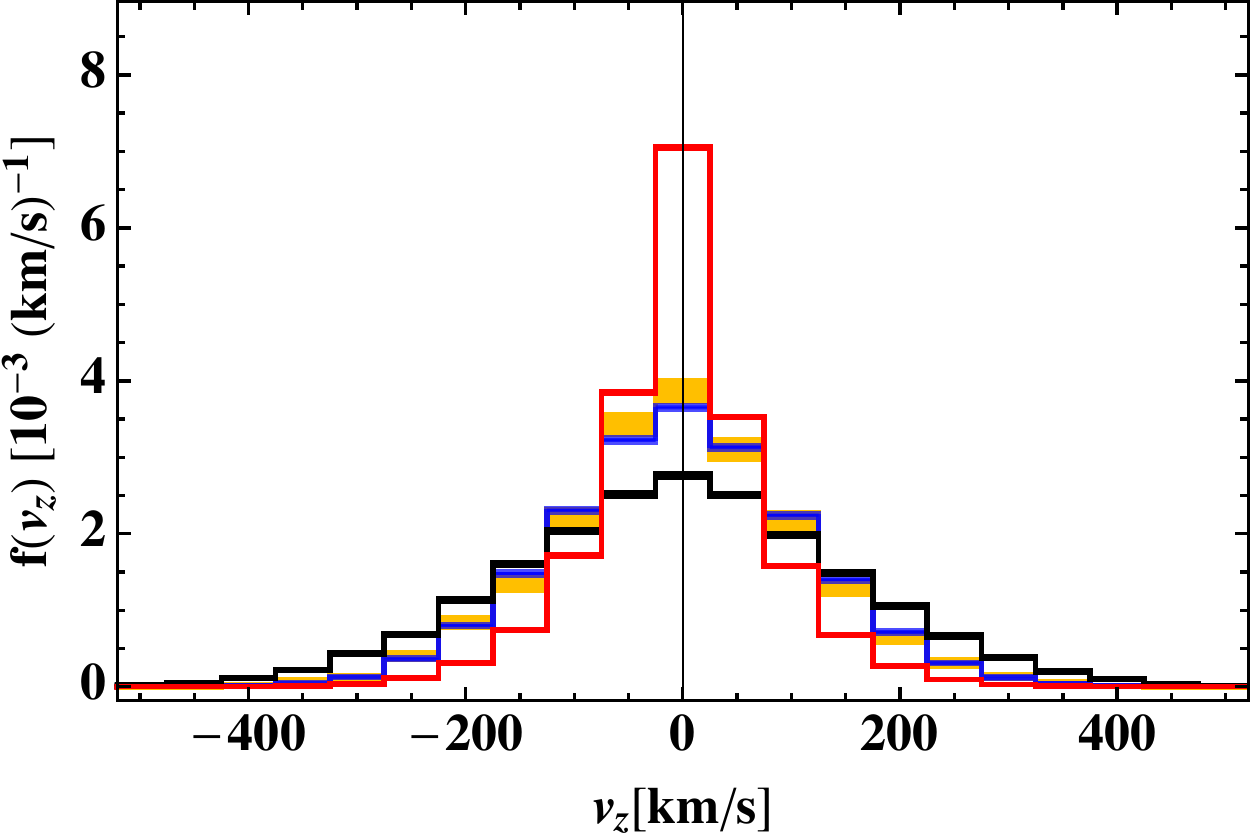}\\
   \includegraphics[width=0.31\textwidth]{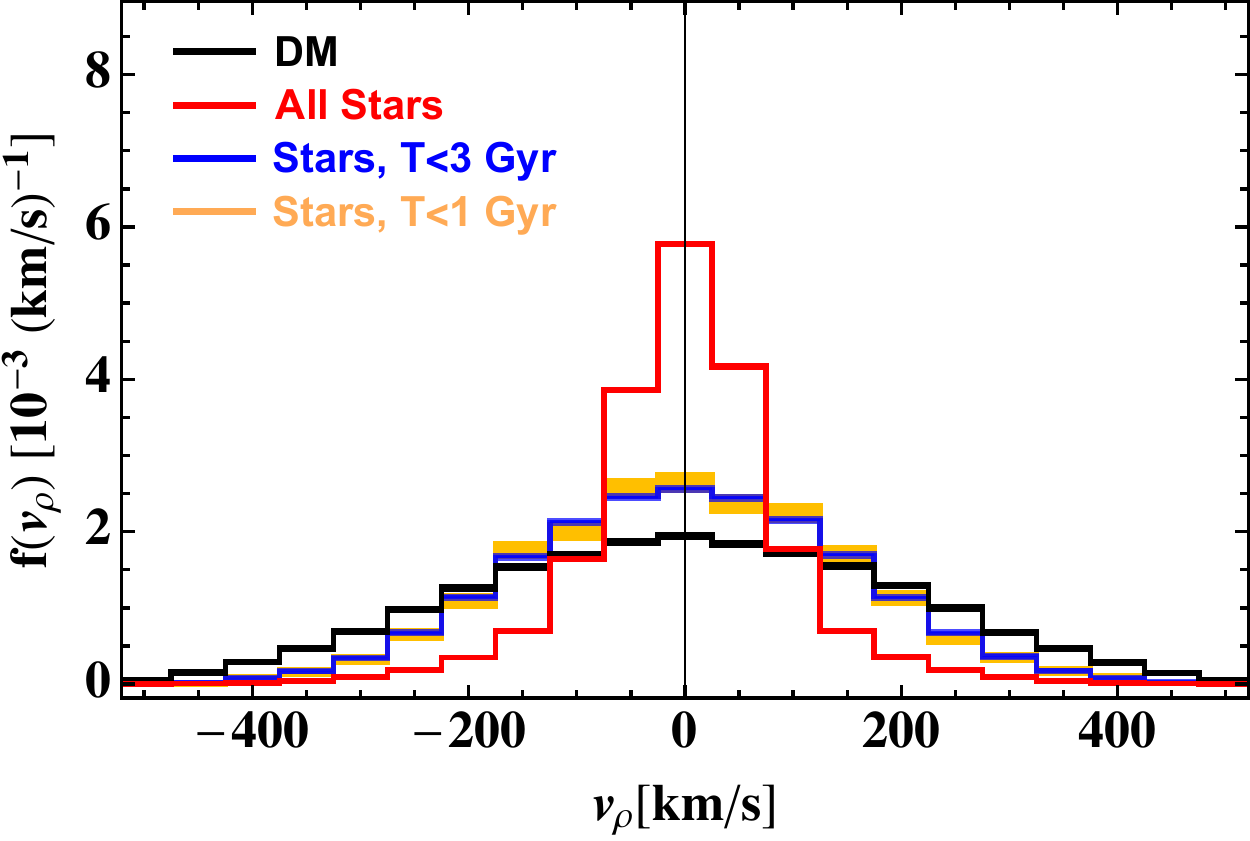}
   \includegraphics[width=0.31\textwidth]{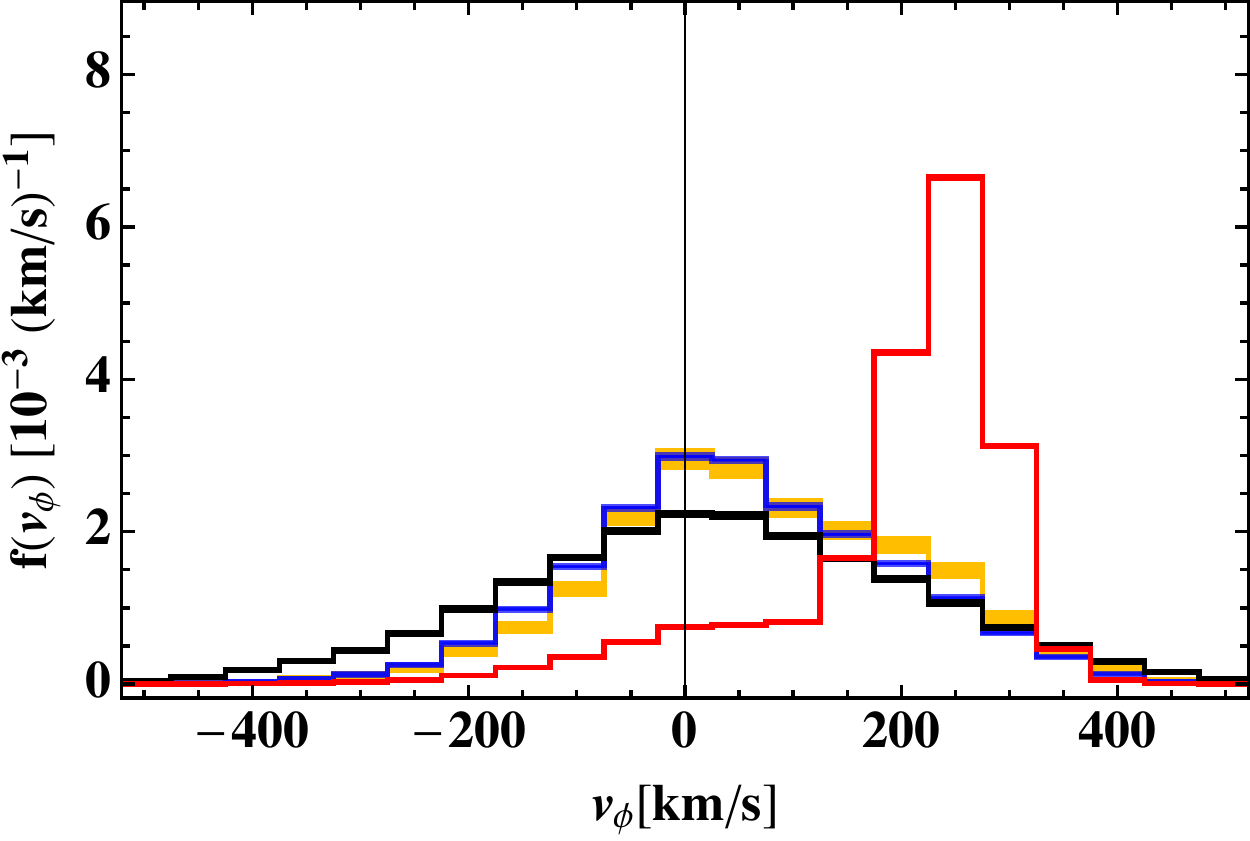}
   \includegraphics[width=0.31\textwidth]{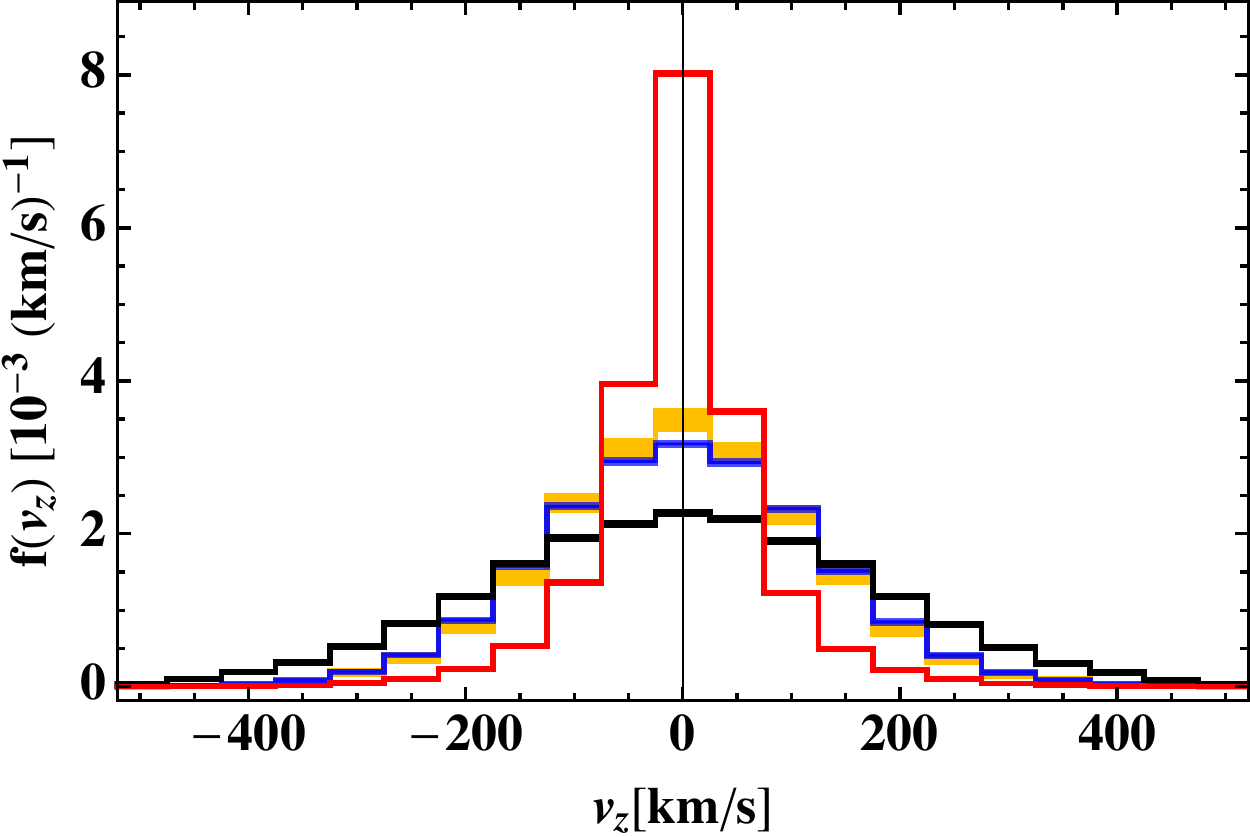}
\caption{Same as figure \ref{fig:TCuts} but without constraining the stellar velocity distributions to the disc.}
\label{fig:TCutsShell}
\end{center}
\end{figure}

\begin{figure}[t]
\begin{center}
   \includegraphics[width=0.31\textwidth]{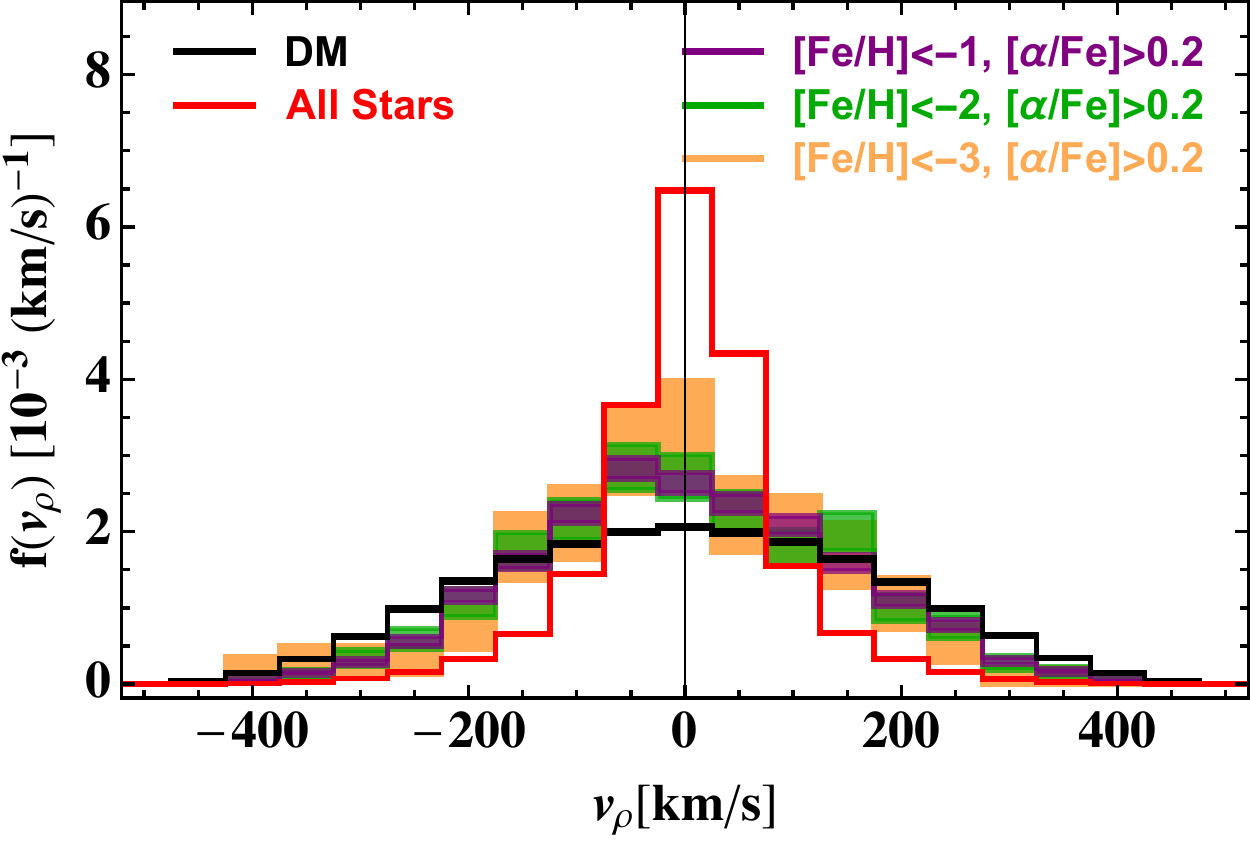}
   \includegraphics[width=0.31\textwidth]{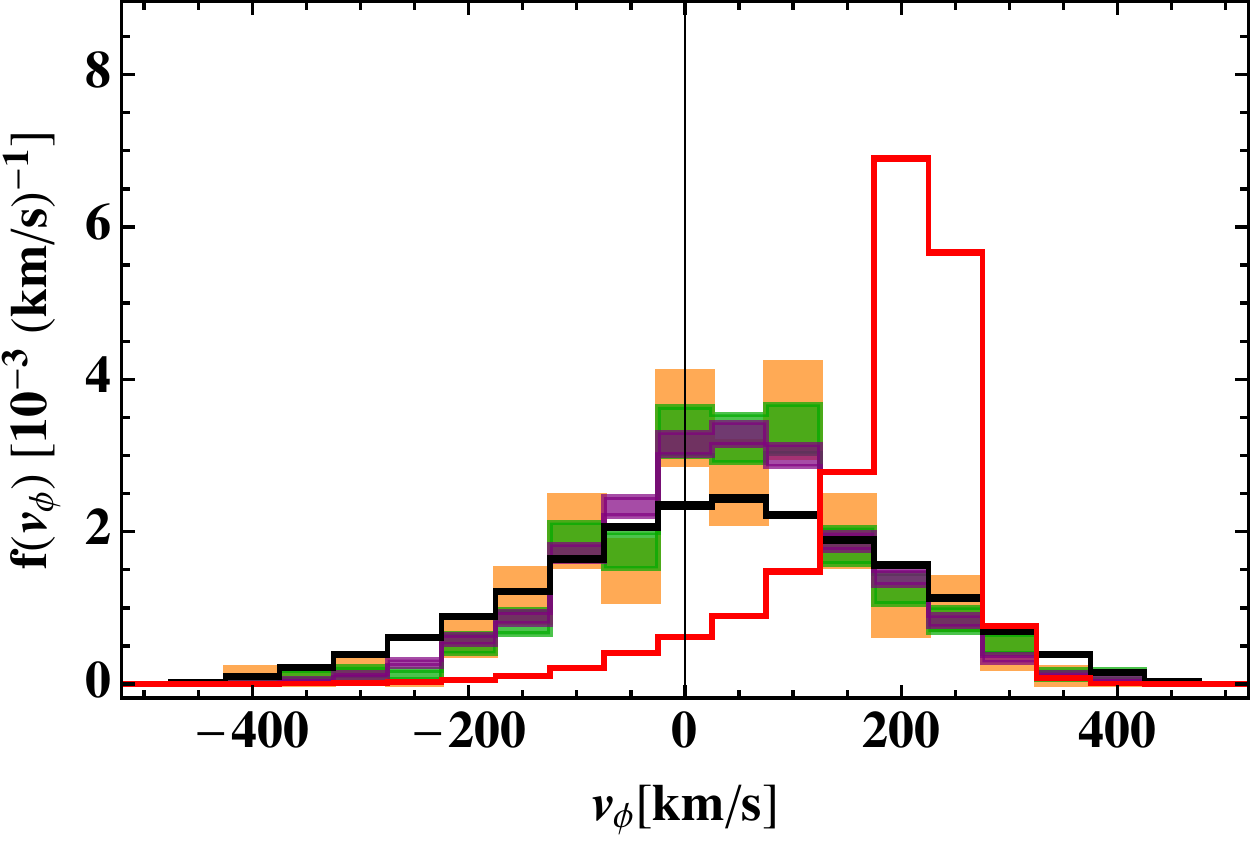}
   \includegraphics[width=0.31\textwidth]{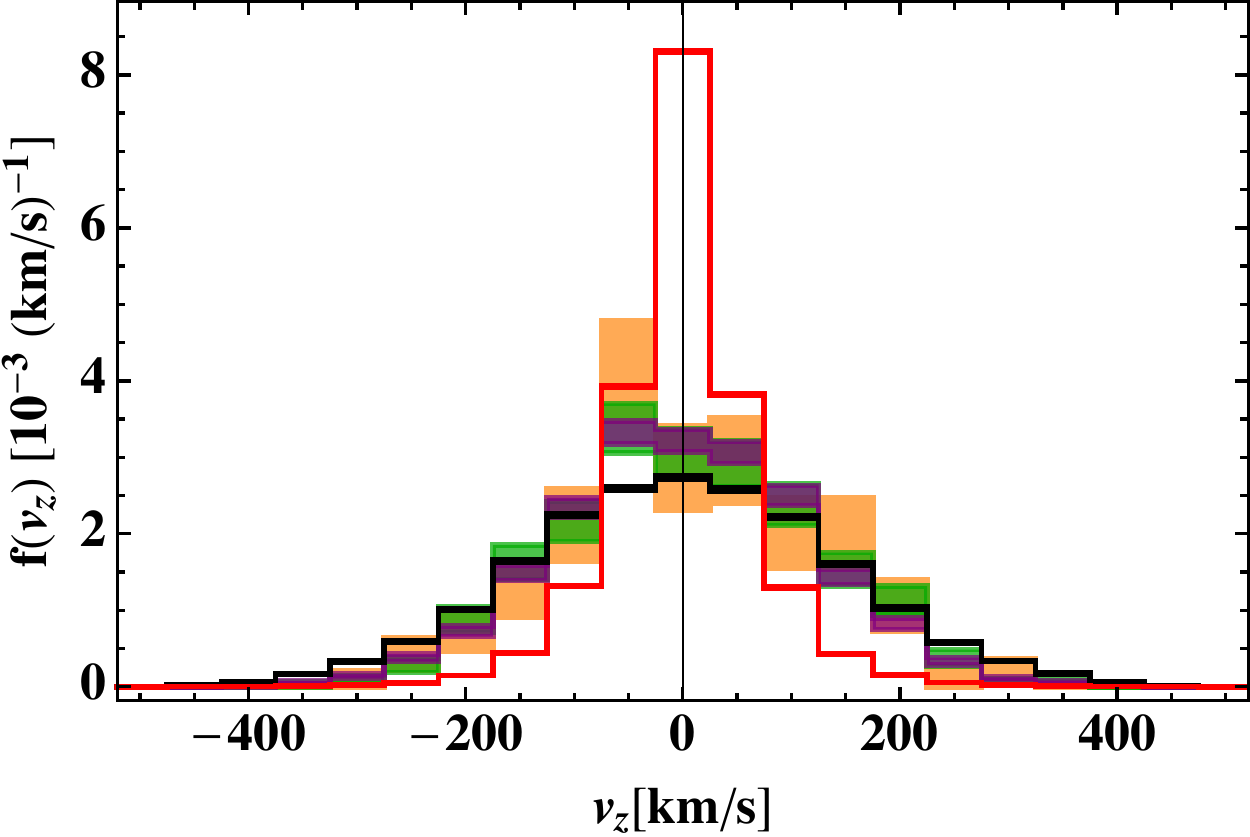}\\
   \includegraphics[width=0.31\textwidth]{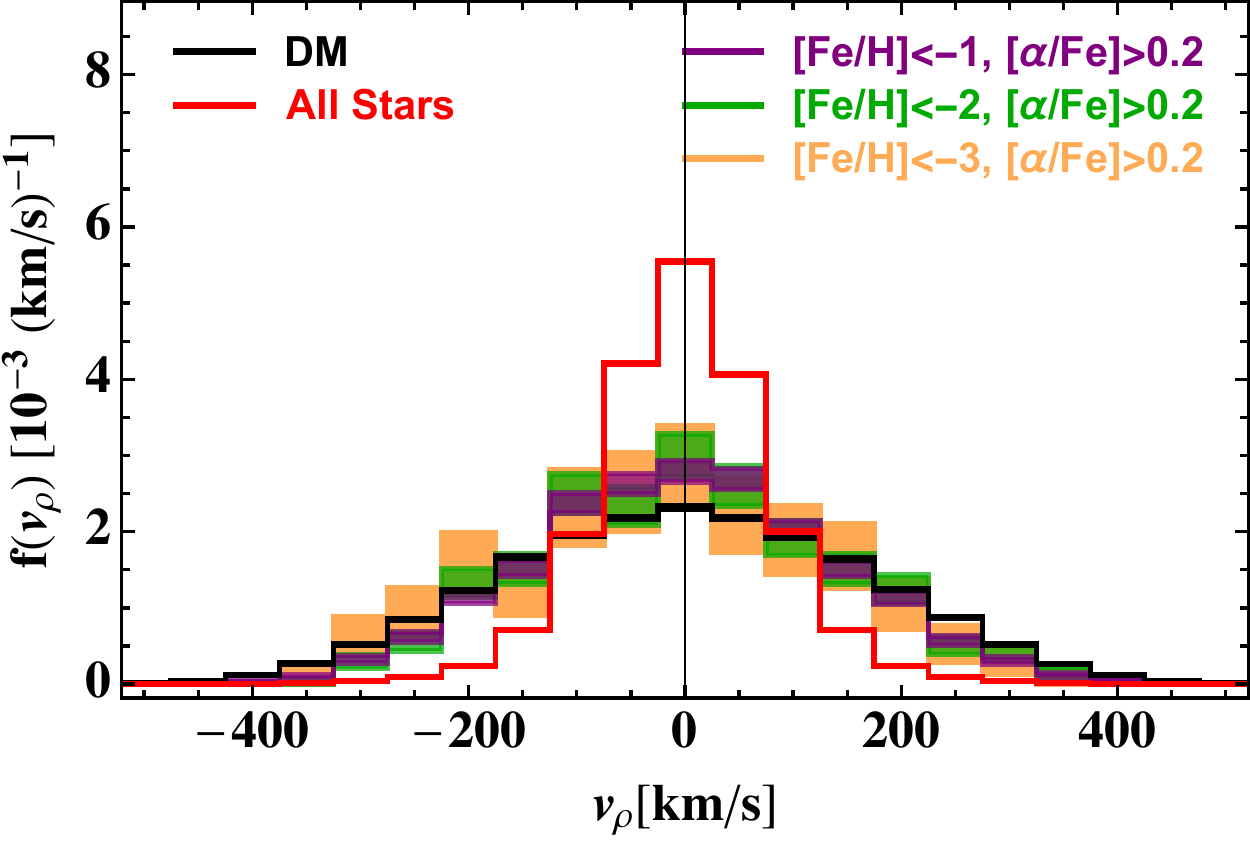}
   \includegraphics[width=0.31\textwidth]{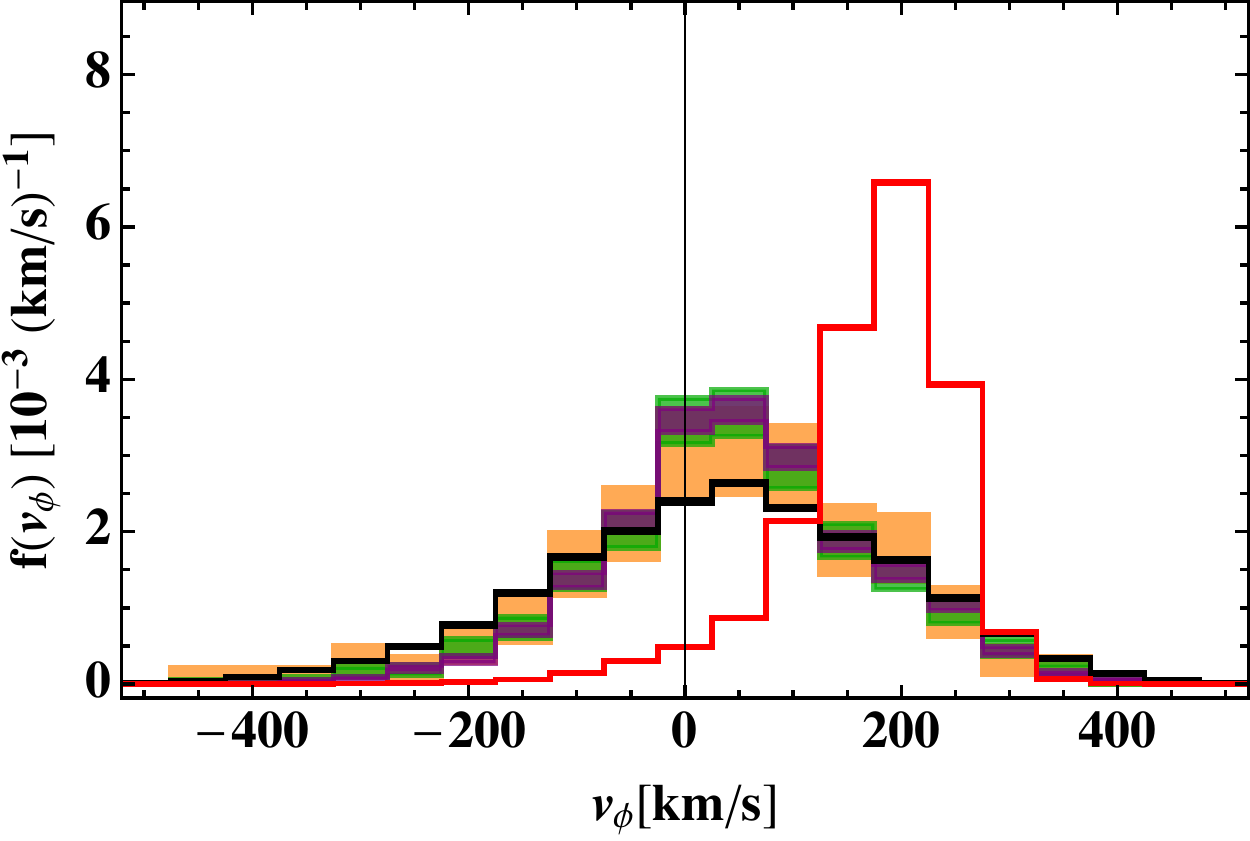}
   \includegraphics[width=0.31\textwidth]{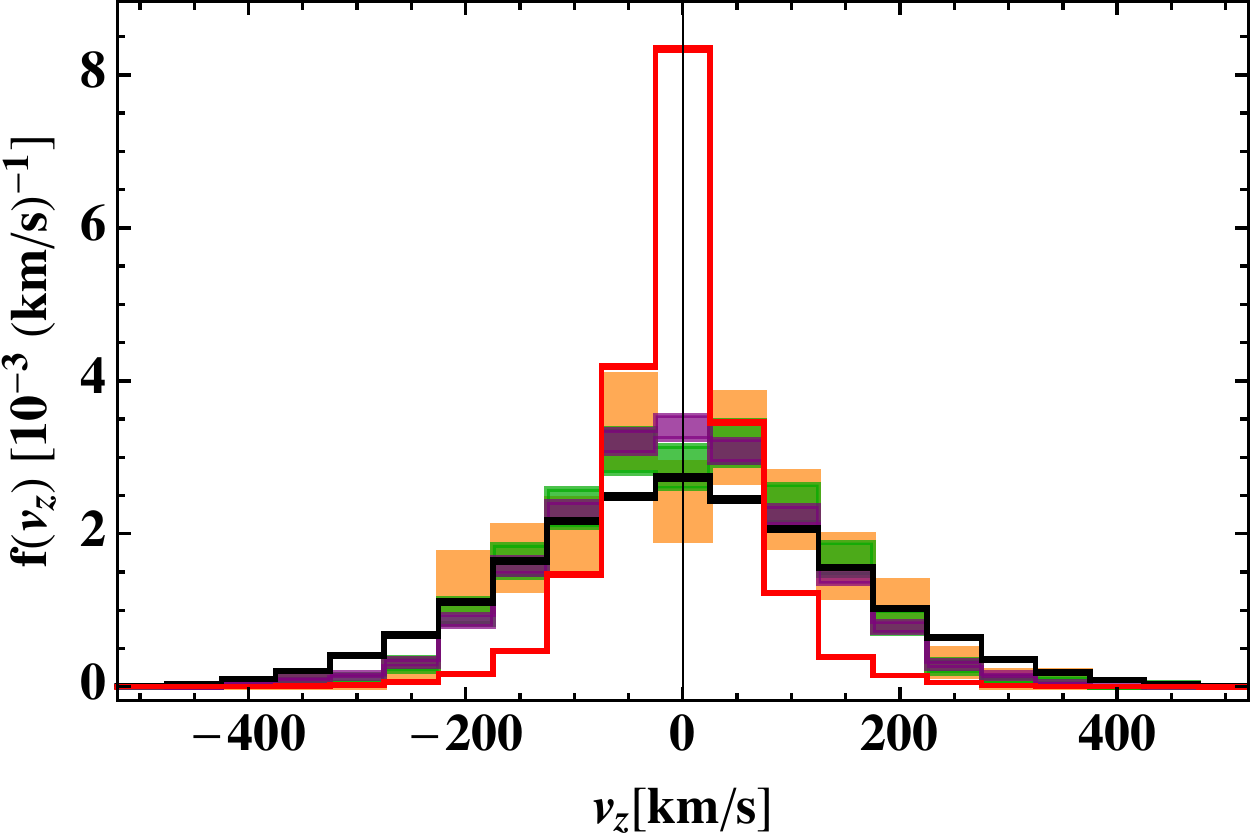}\\
   \includegraphics[width=0.31\textwidth]{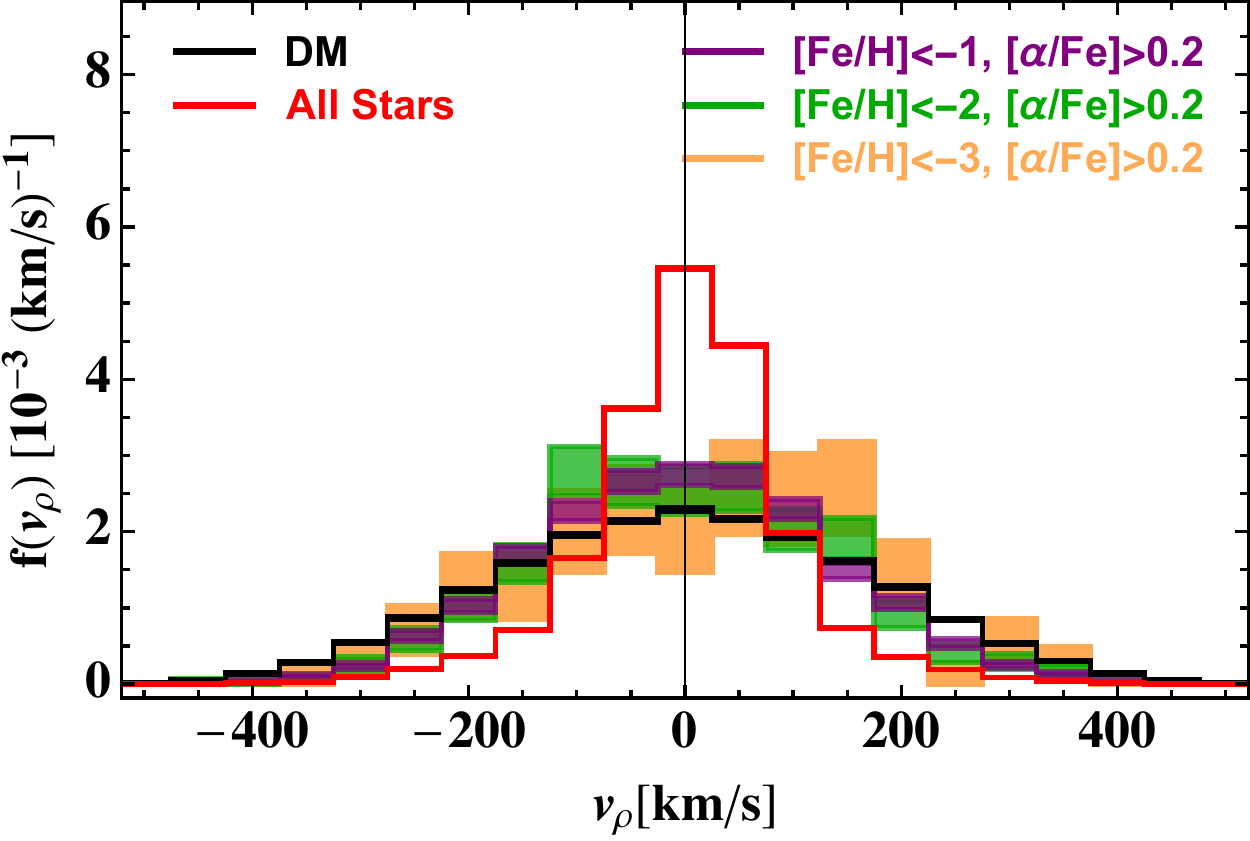}
   \includegraphics[width=0.31\textwidth]{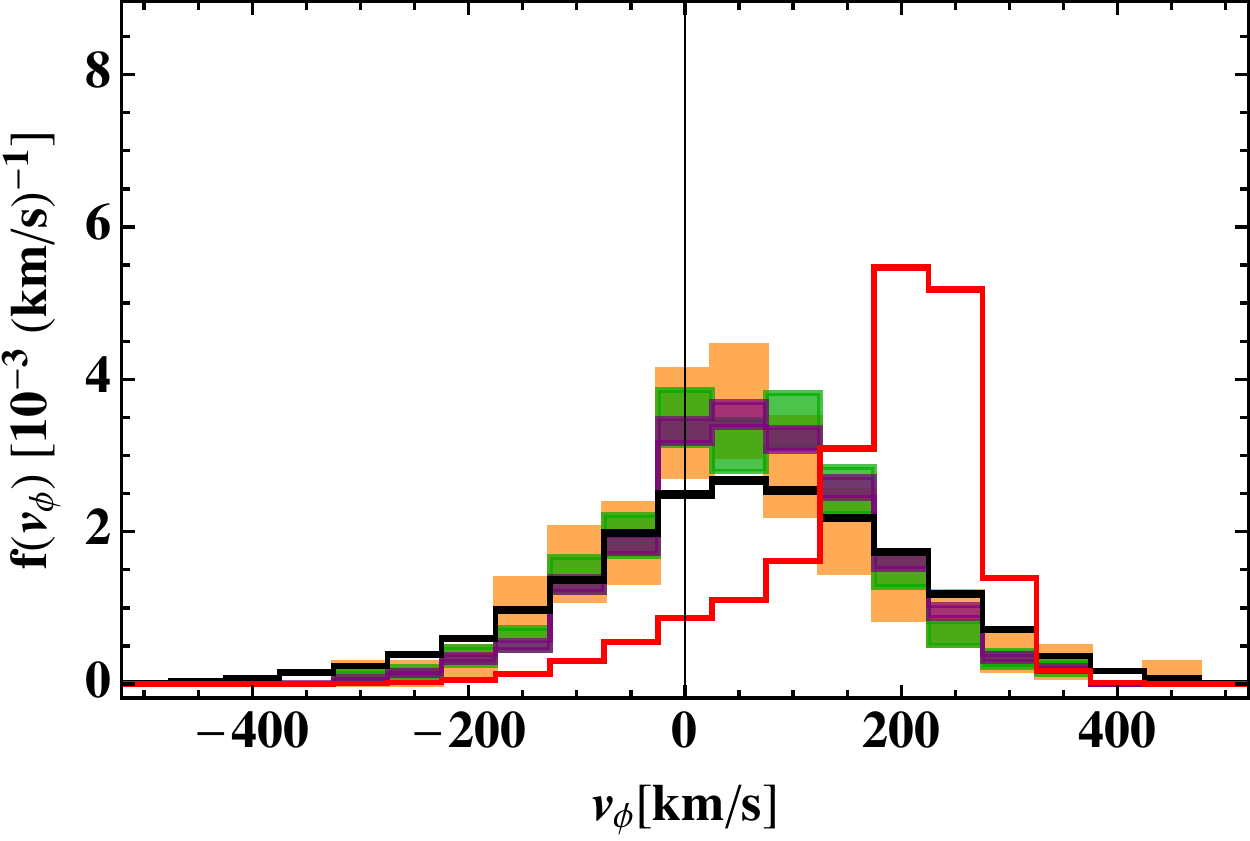}
   \includegraphics[width=0.31\textwidth]{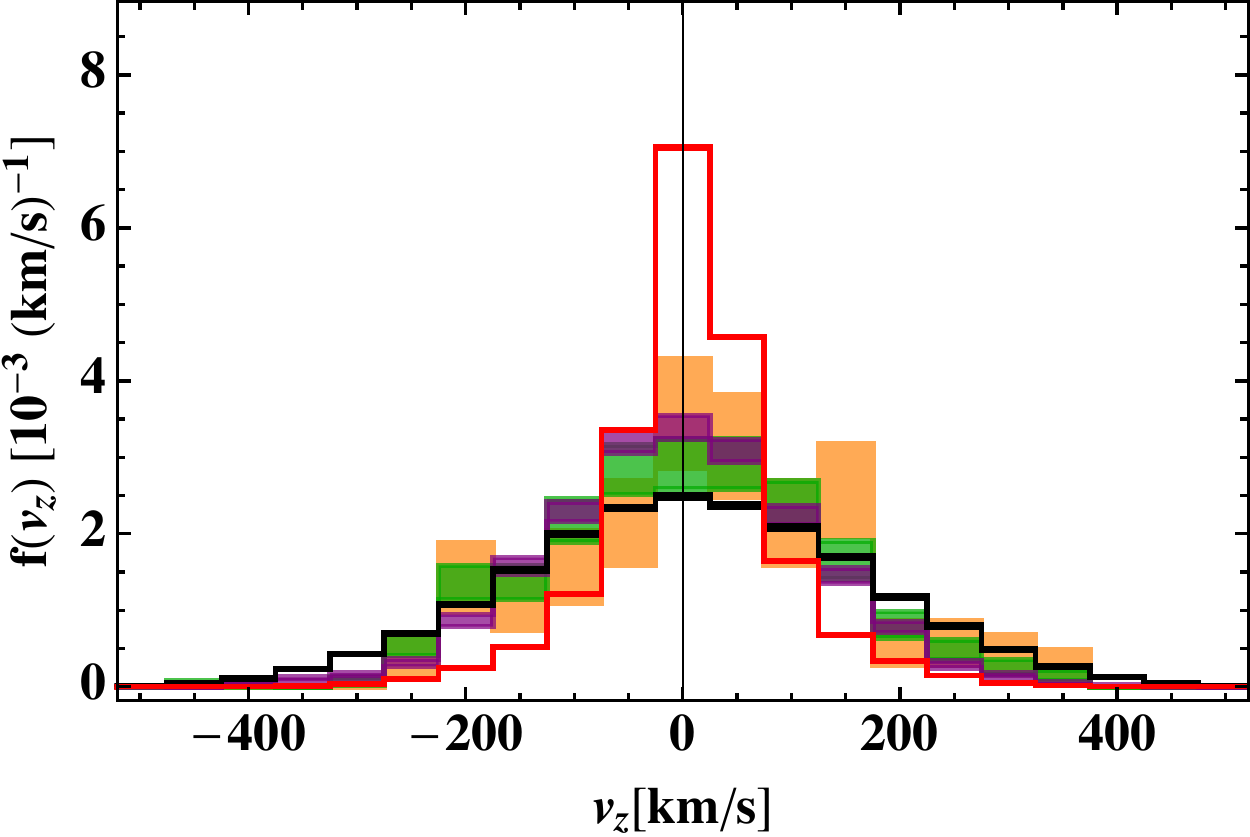}\\
   \includegraphics[width=0.31\textwidth]{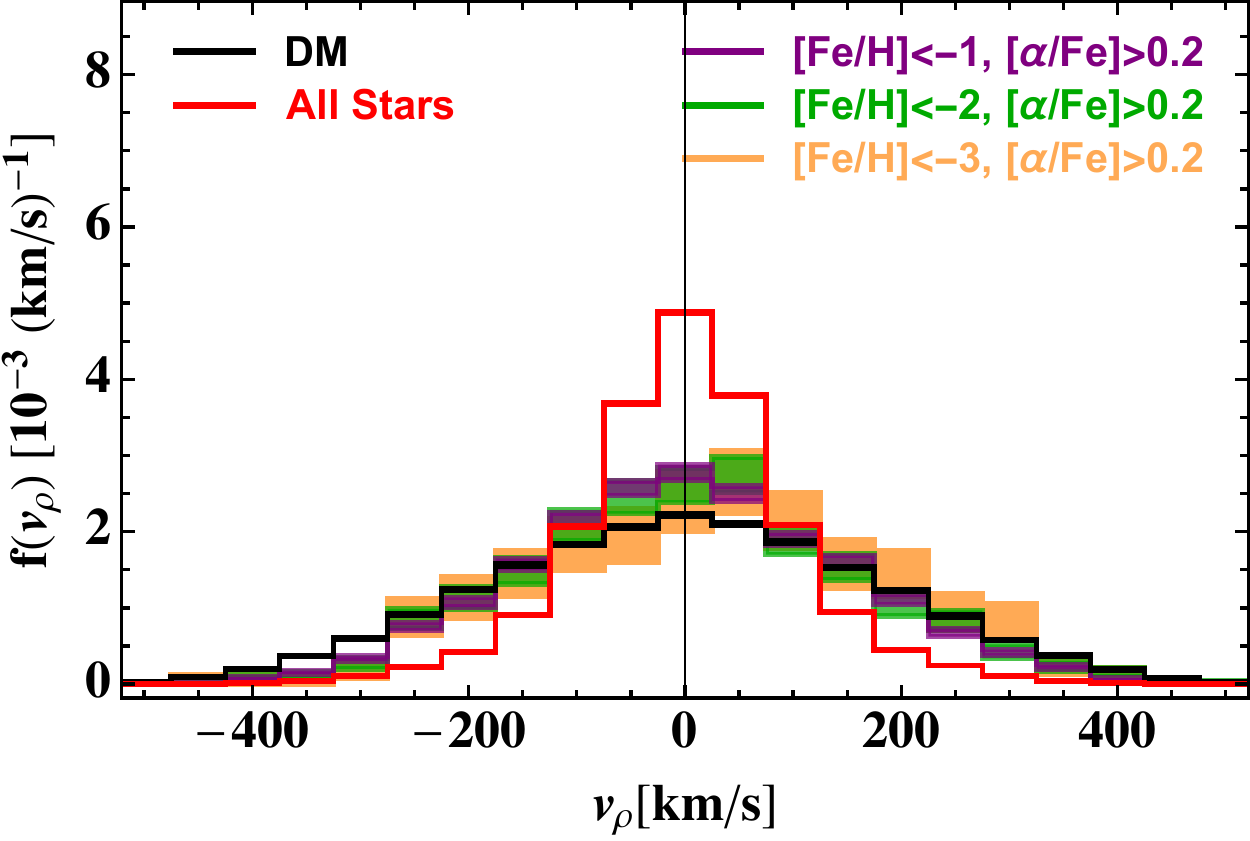}
   \includegraphics[width=0.31\textwidth]{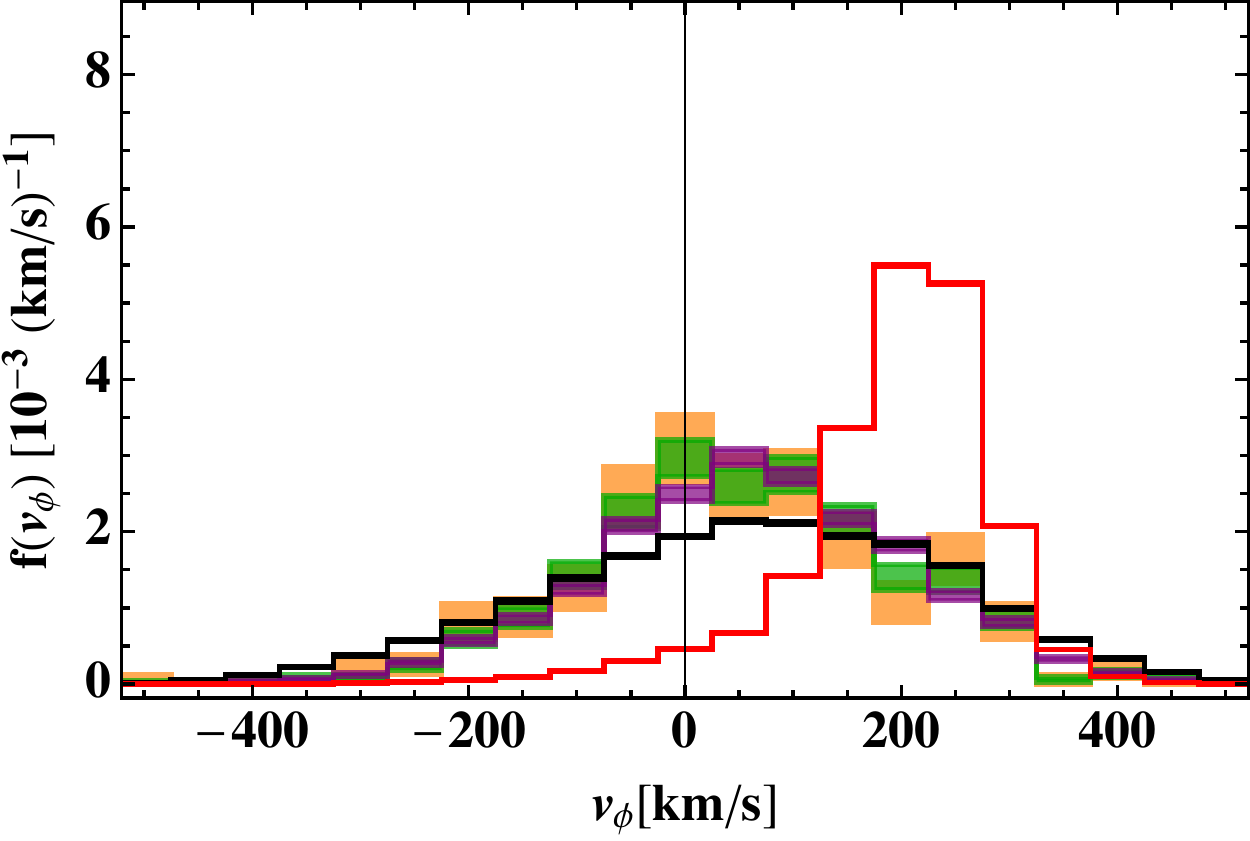}
   \includegraphics[width=0.31\textwidth]{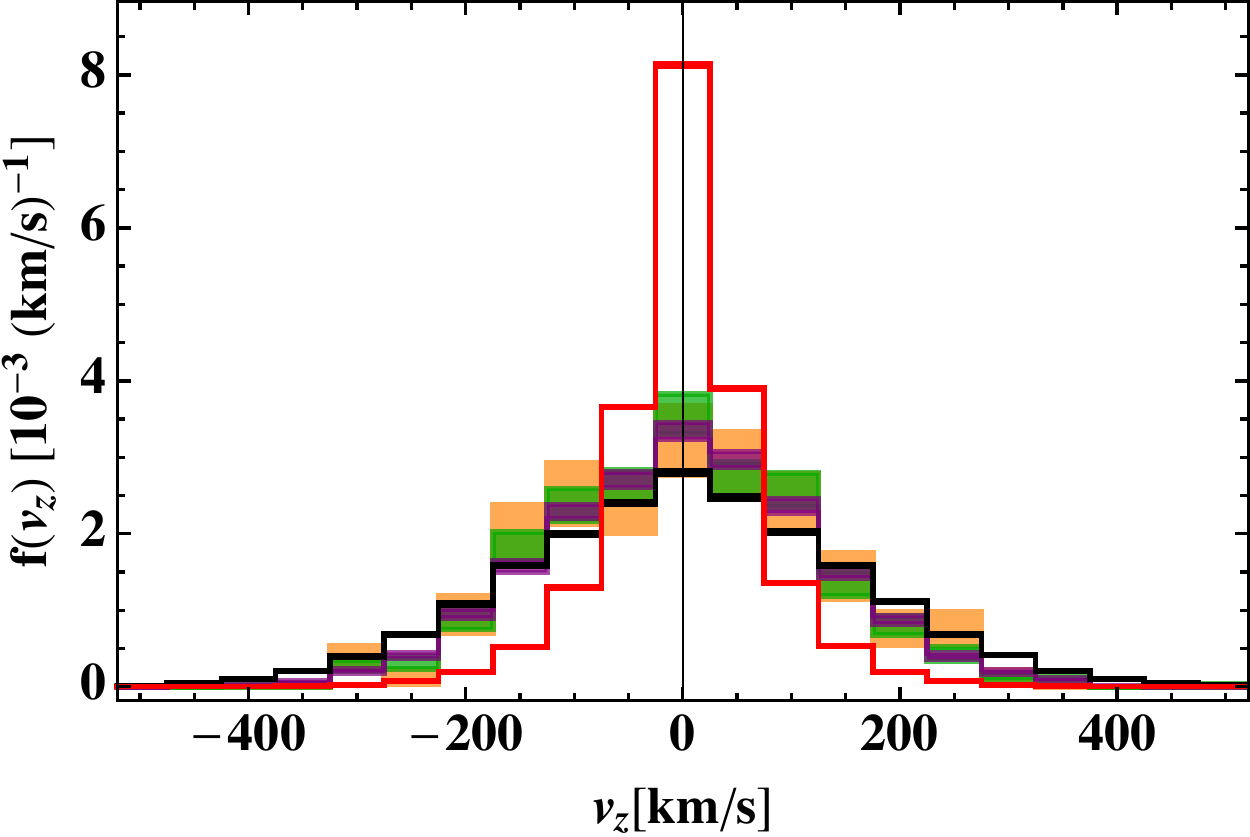}\\
   \includegraphics[width=0.31\textwidth]{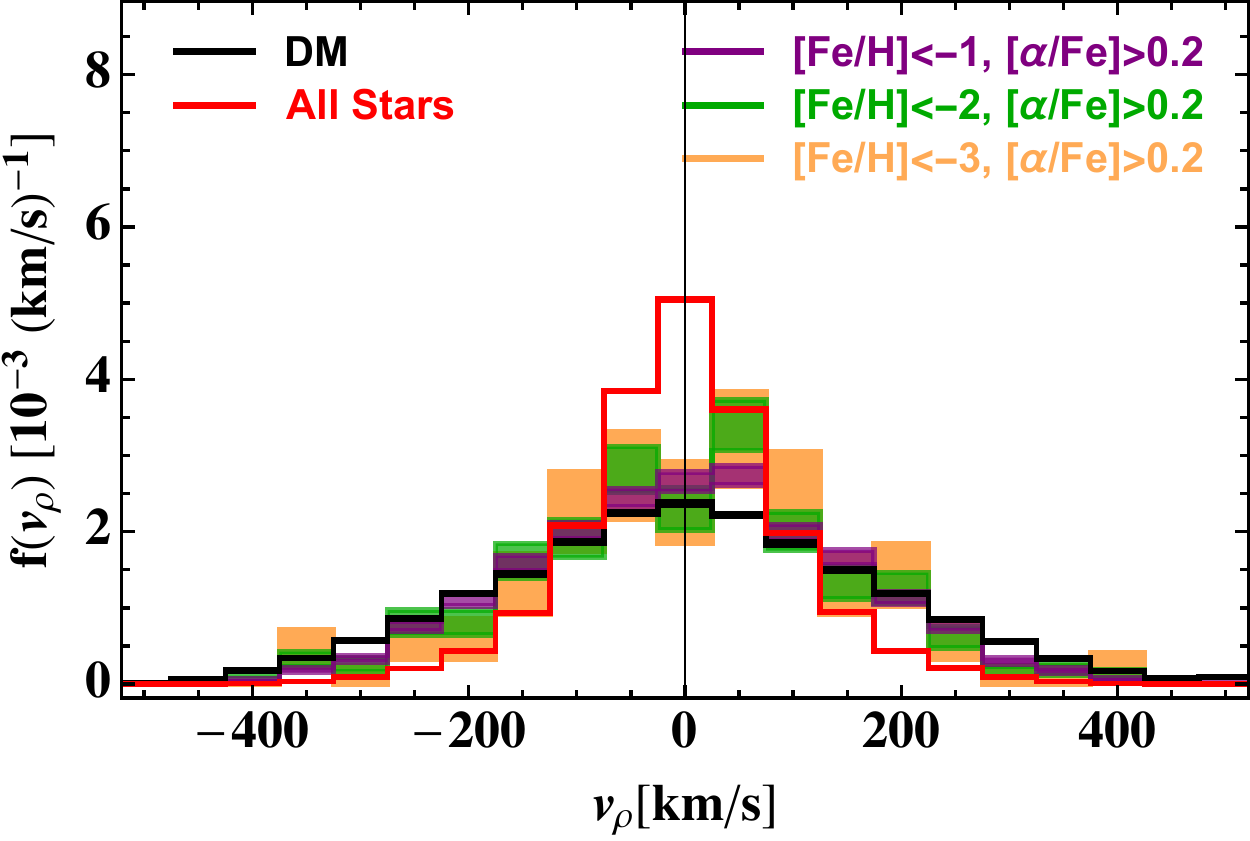}
   \includegraphics[width=0.31\textwidth]{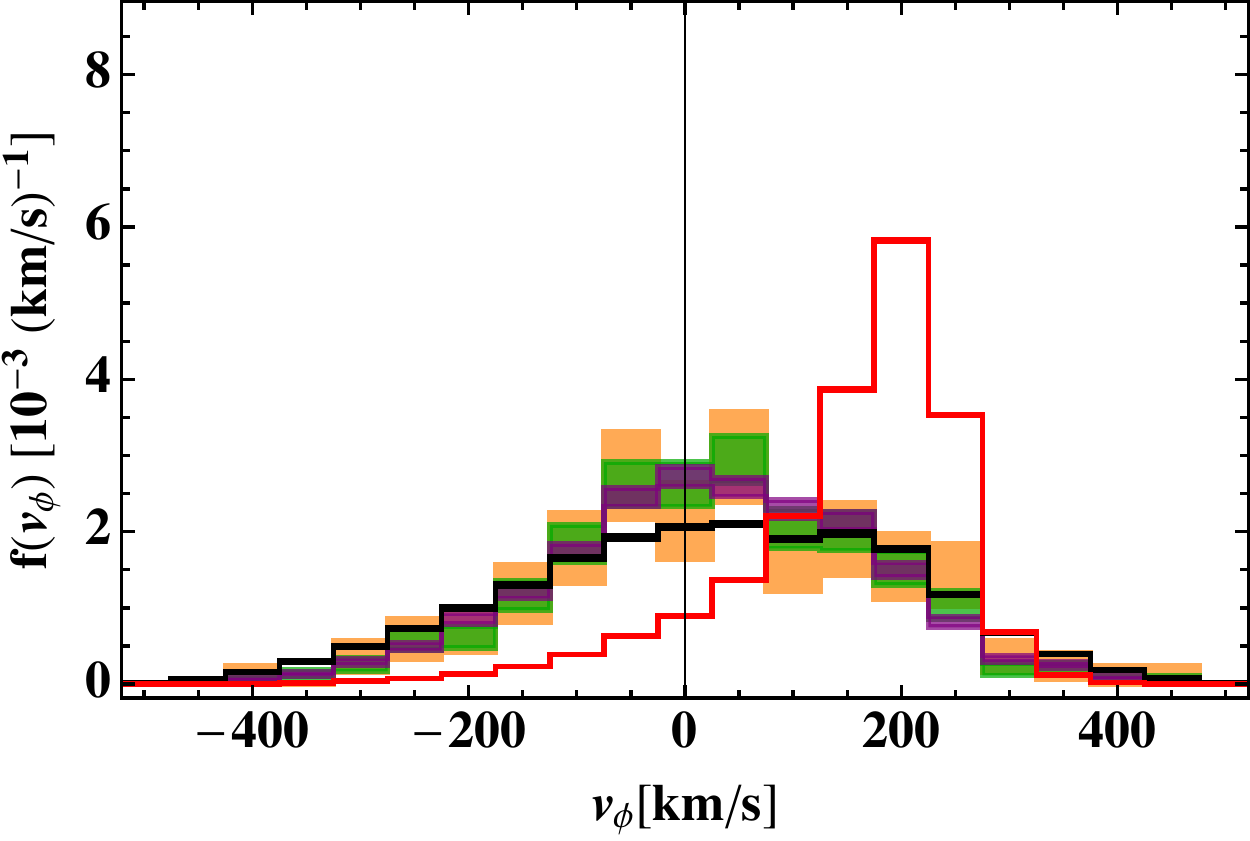}
   \includegraphics[width=0.31\textwidth]{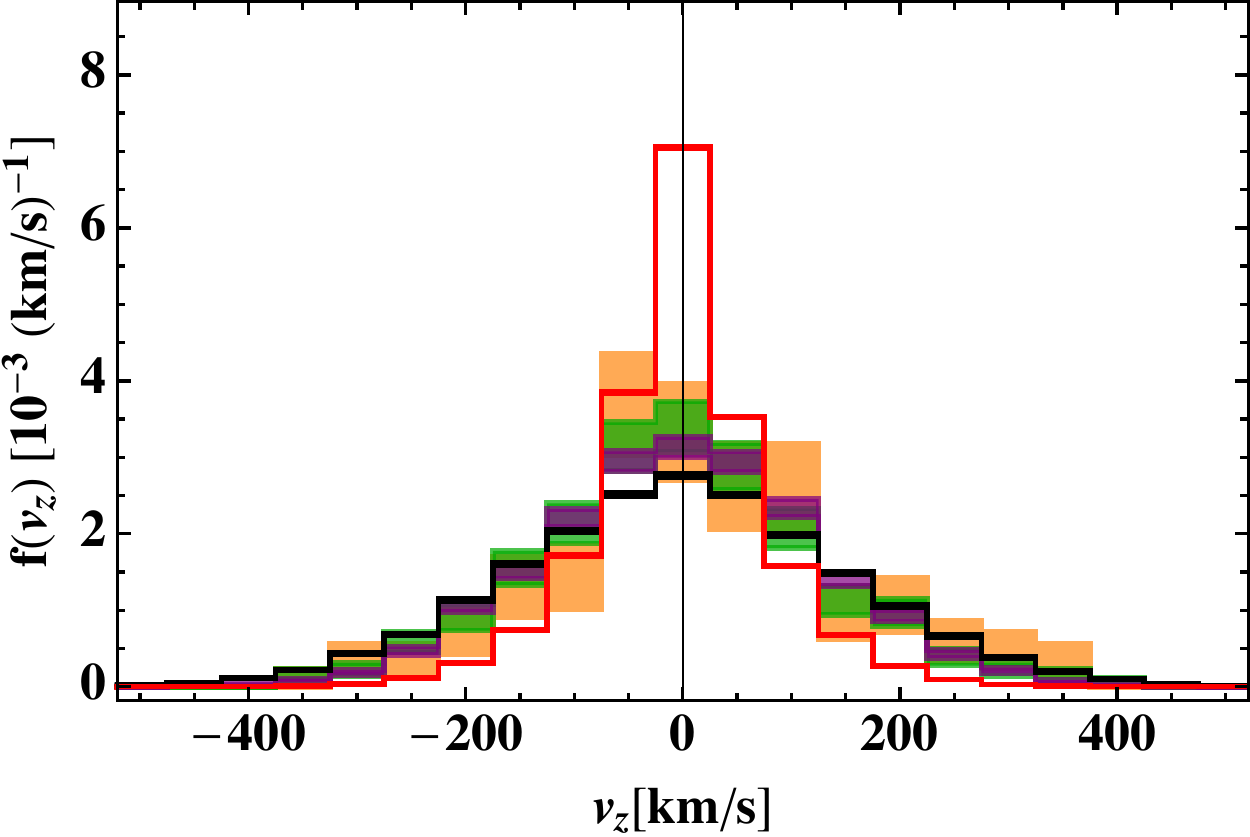}\\
   \includegraphics[width=0.31\textwidth]{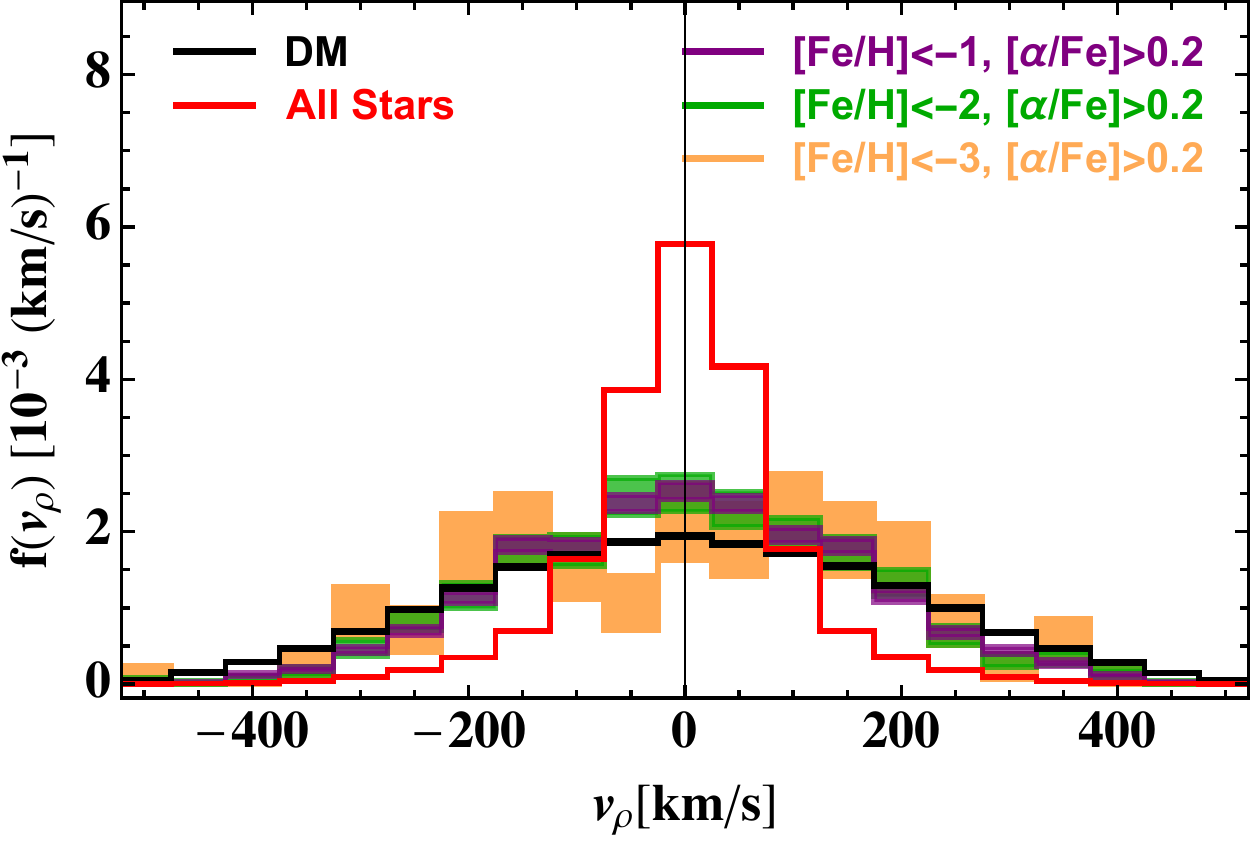}
   \includegraphics[width=0.31\textwidth]{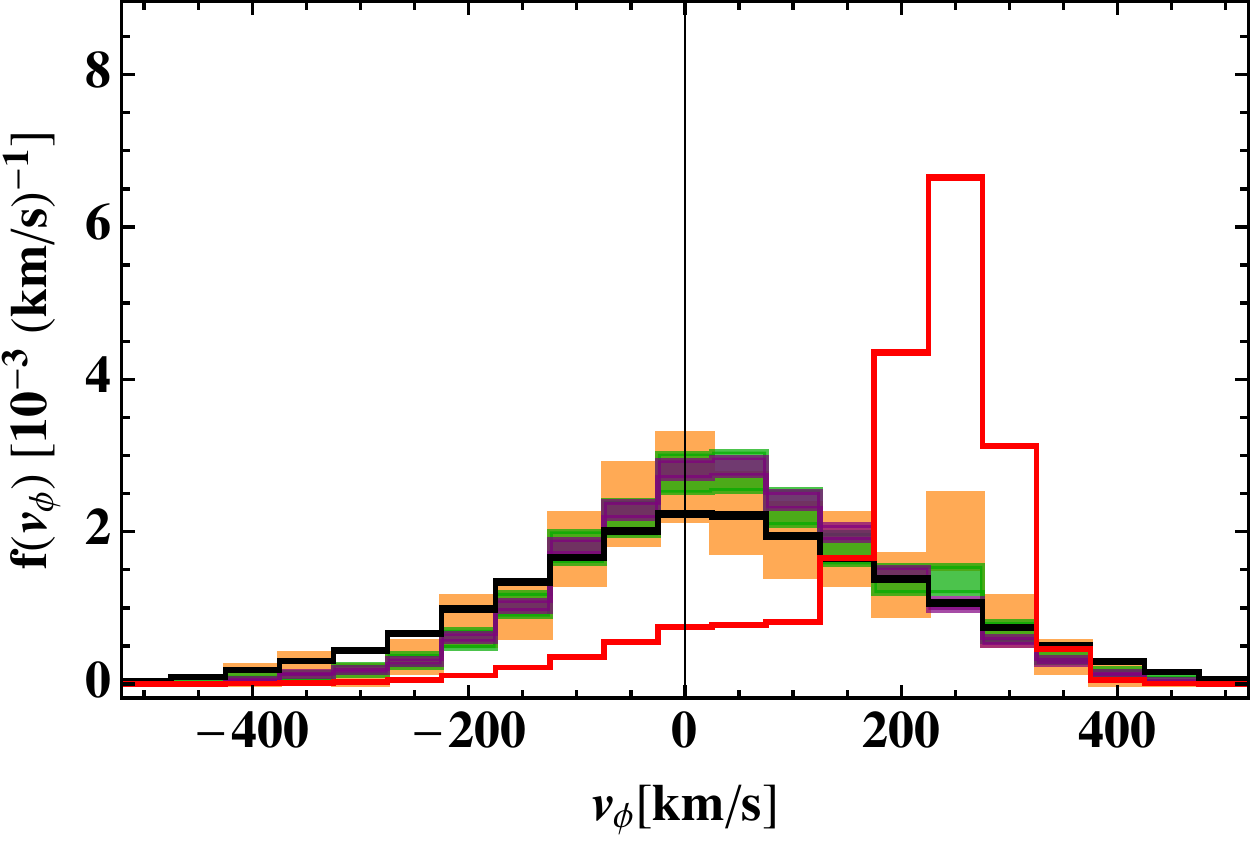}
   \includegraphics[width=0.31\textwidth]{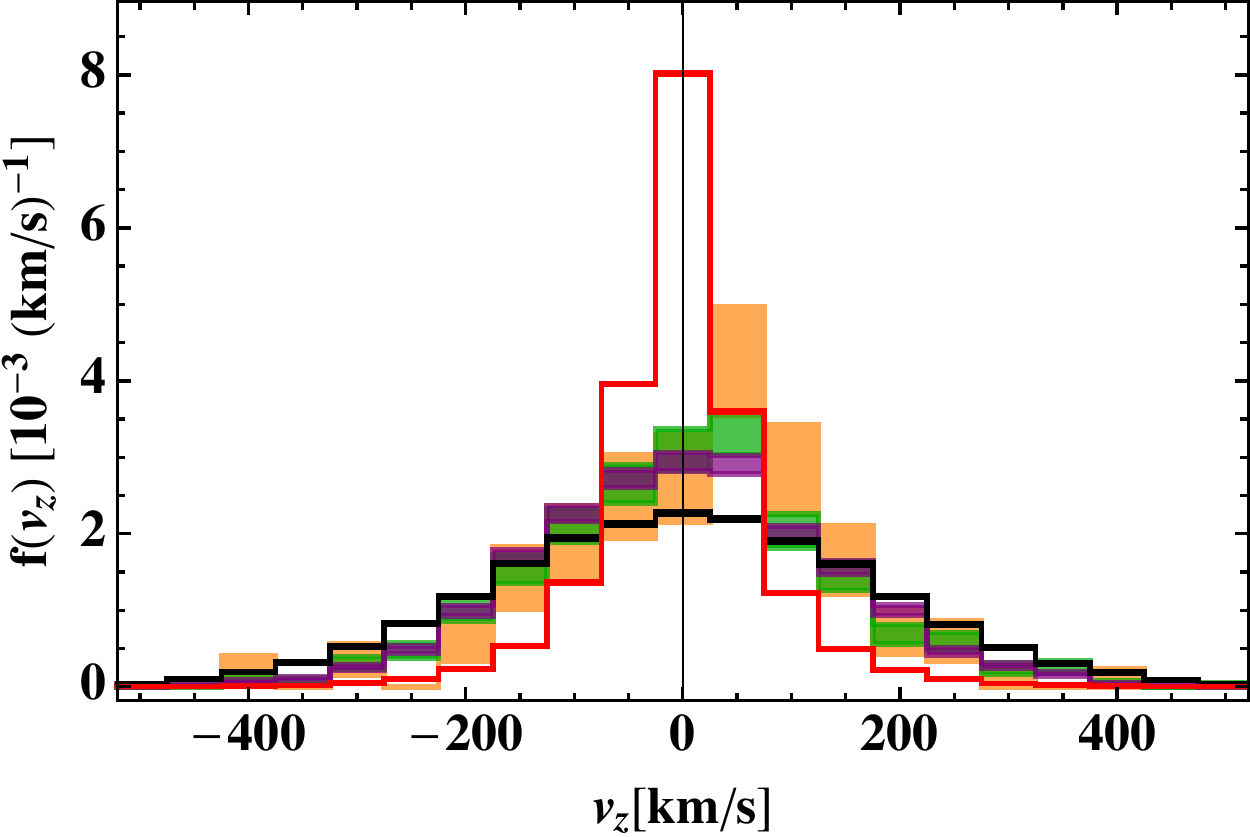}
\caption{Same as figure \ref{fig:FeHalphaHeCuts} but without constraining the stellar velocity distributions to the disc.}
\label{fig:FeHalphaHeCutsShell}
\end{center}
\end{figure}

\begin{figure}[t!]
\begin{center}
  \includegraphics[width=0.32\textwidth]{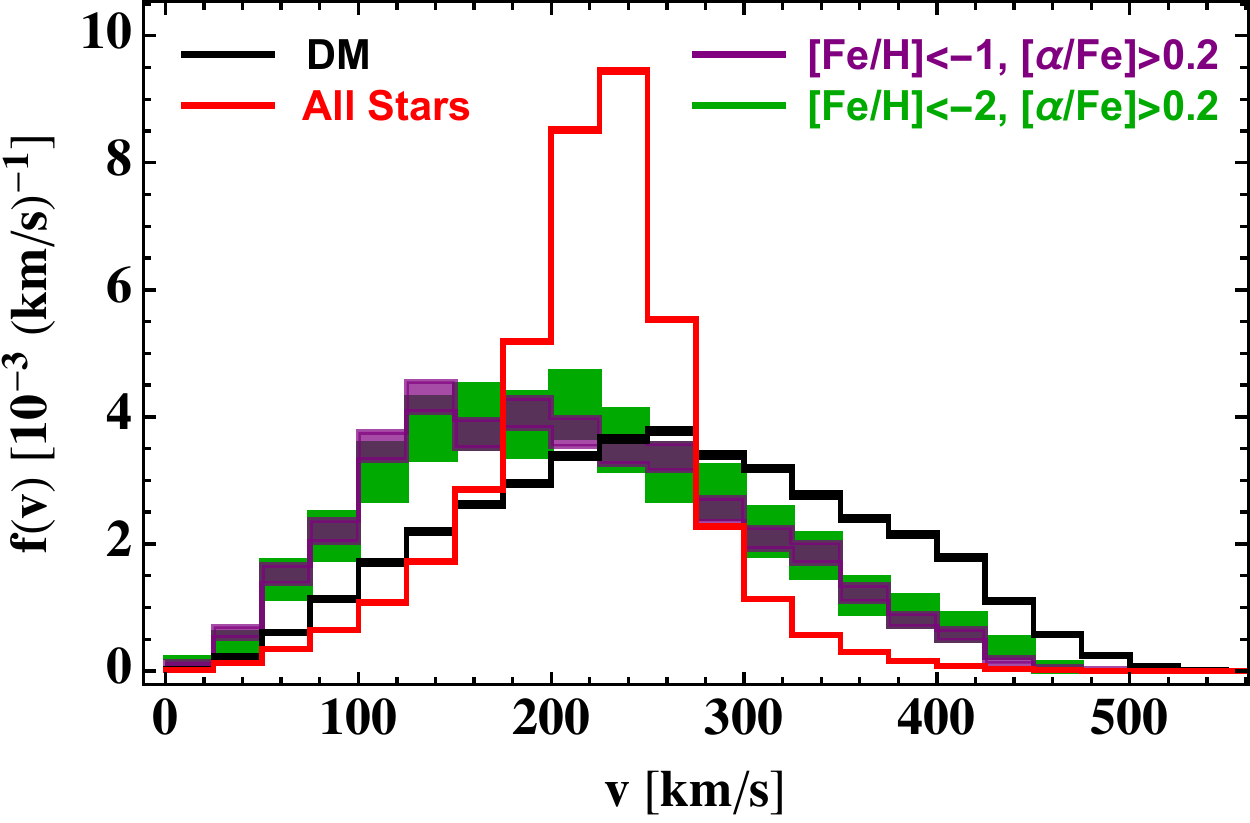}
  \includegraphics[width=0.32\textwidth]{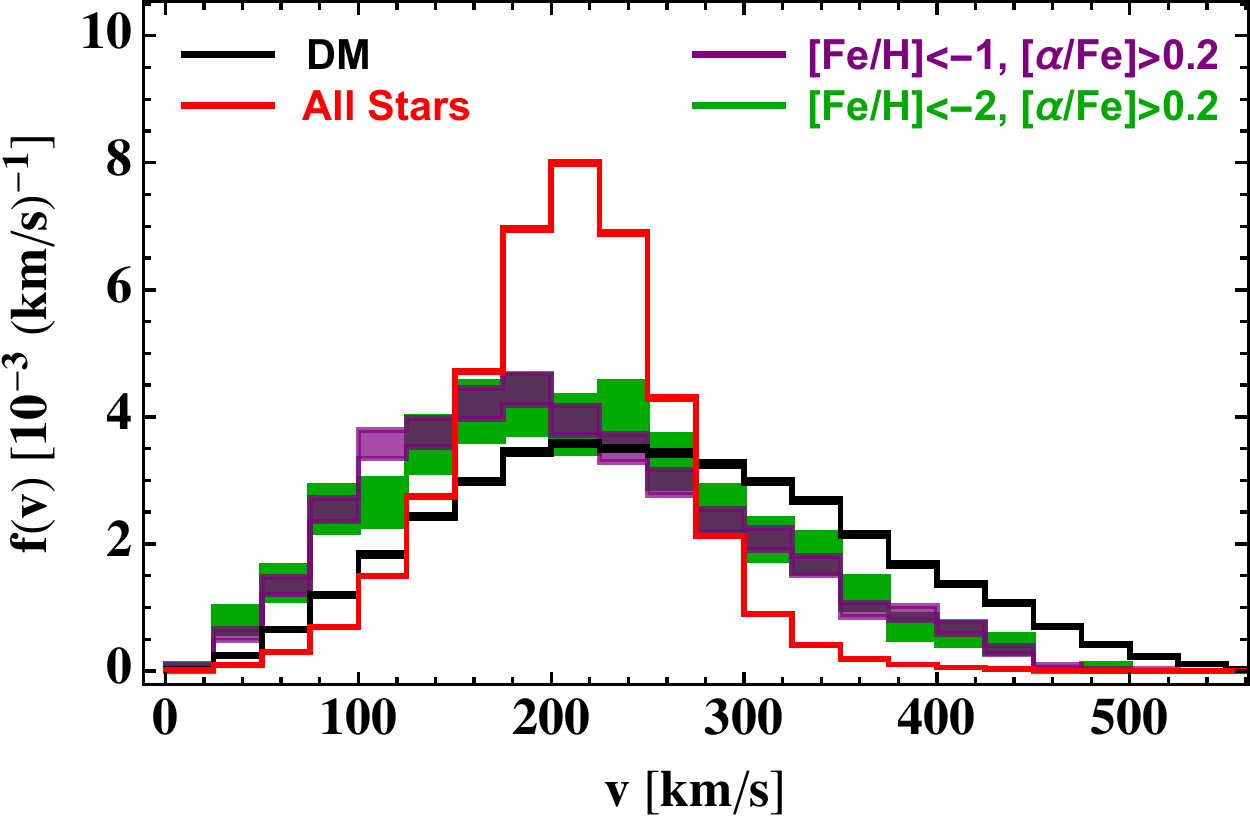}
  \includegraphics[width=0.32\textwidth]{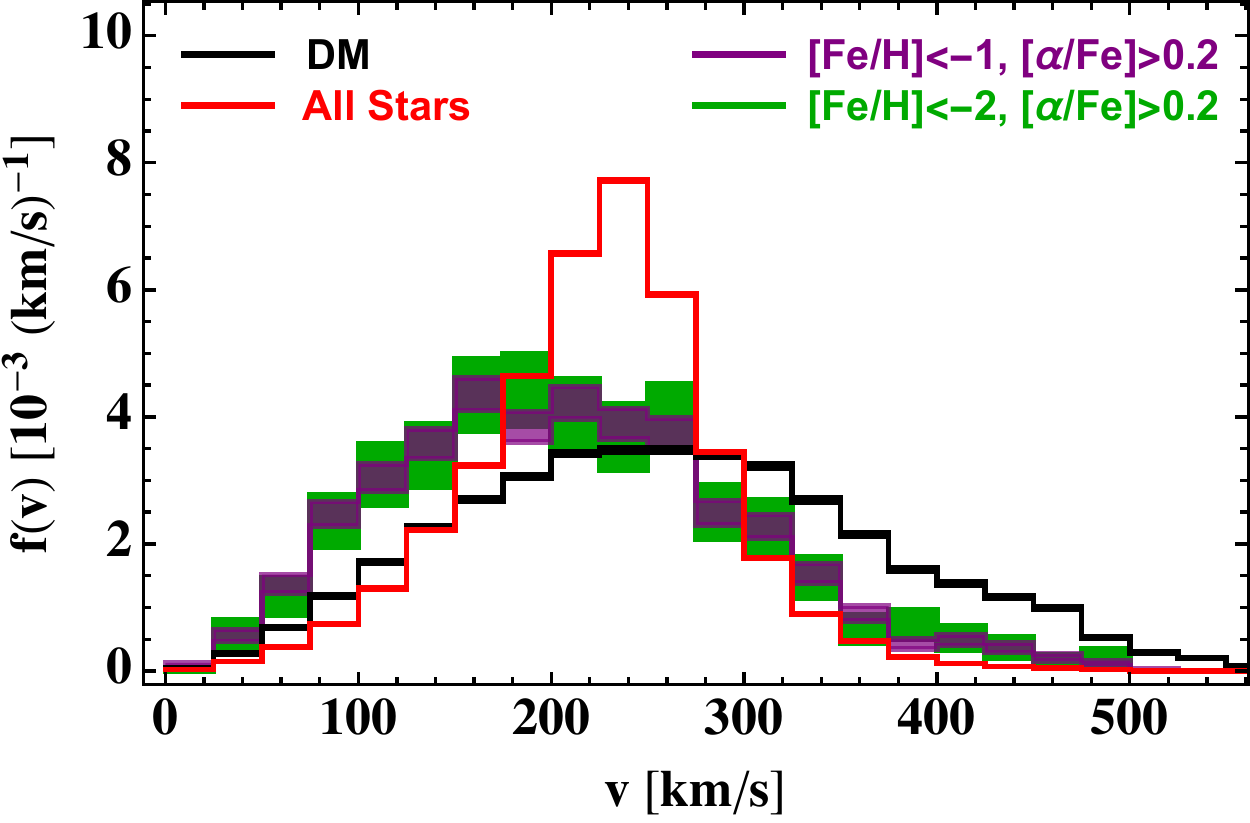}\\
  \includegraphics[width=0.32\textwidth]{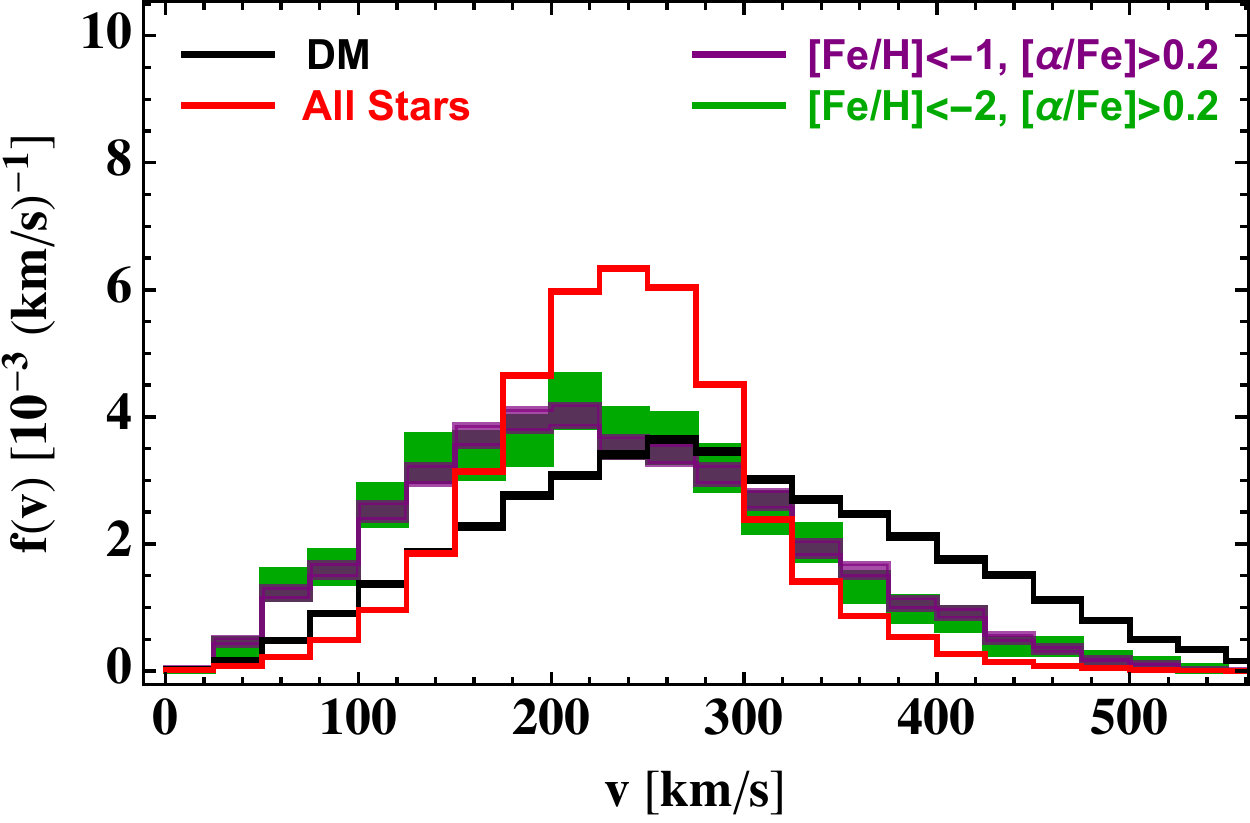}
  \includegraphics[width=0.32\textwidth]{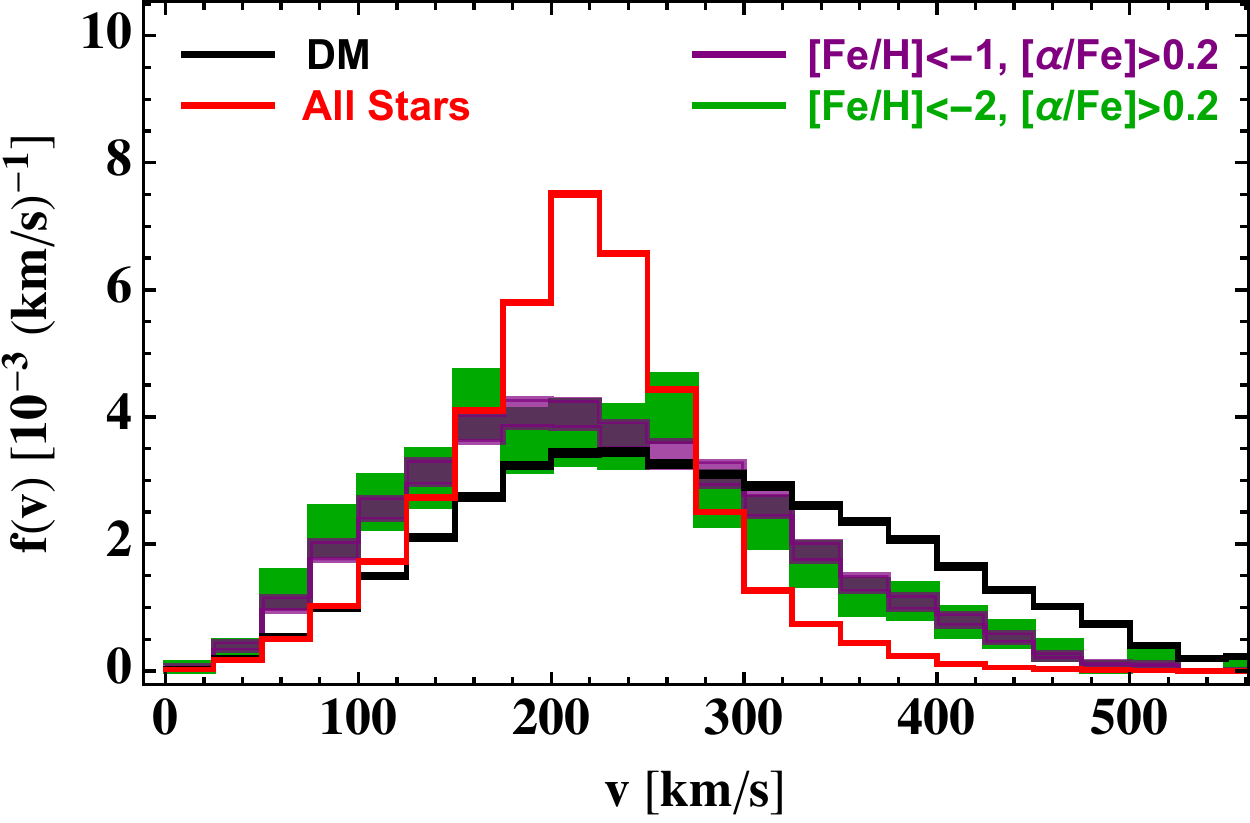}
  \includegraphics[width=0.32\textwidth]{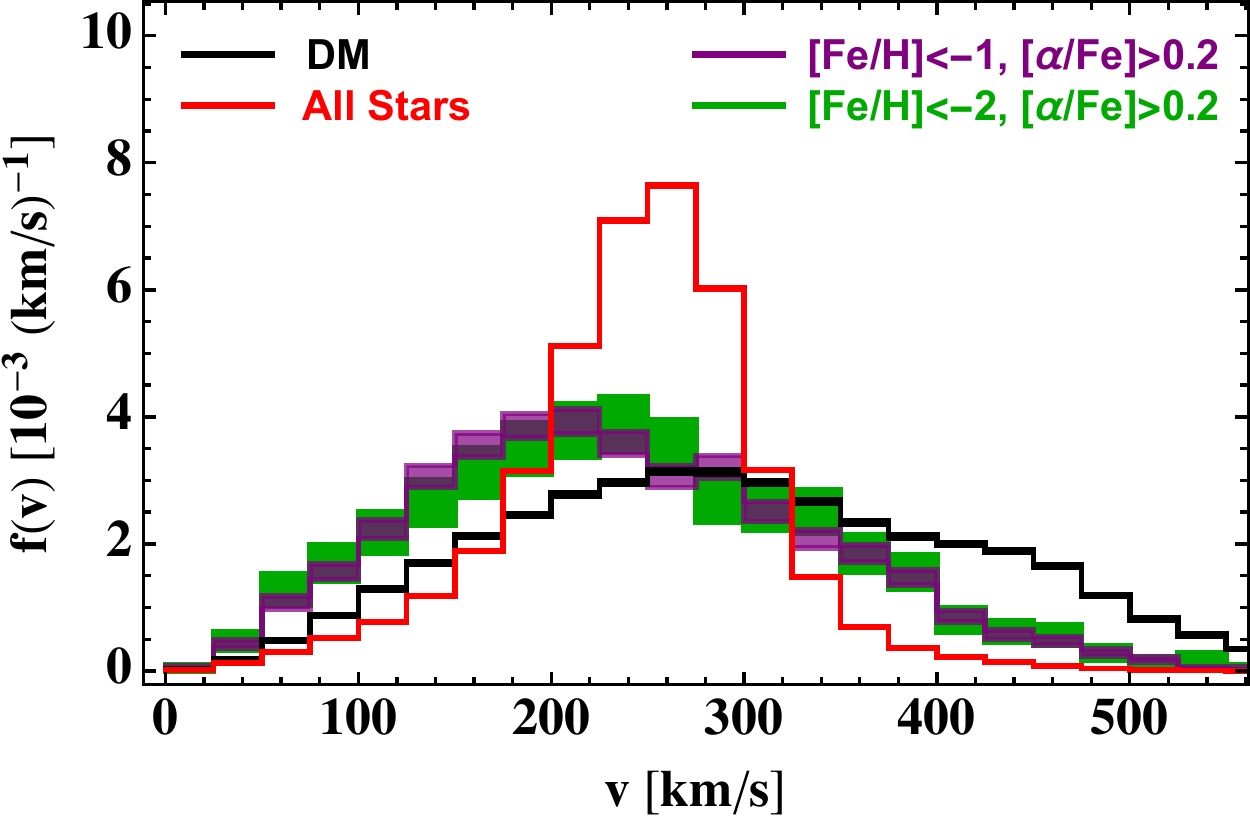}
\caption{Same as figure \ref{fig:fvMod} but without constraining the stellar velocity distributions to the disc.}
\label{fig:fvModShell}
\end{center}
\end{figure}

\begin{figure}[h!]
\begin{center}
  \includegraphics[width=0.32\textwidth]{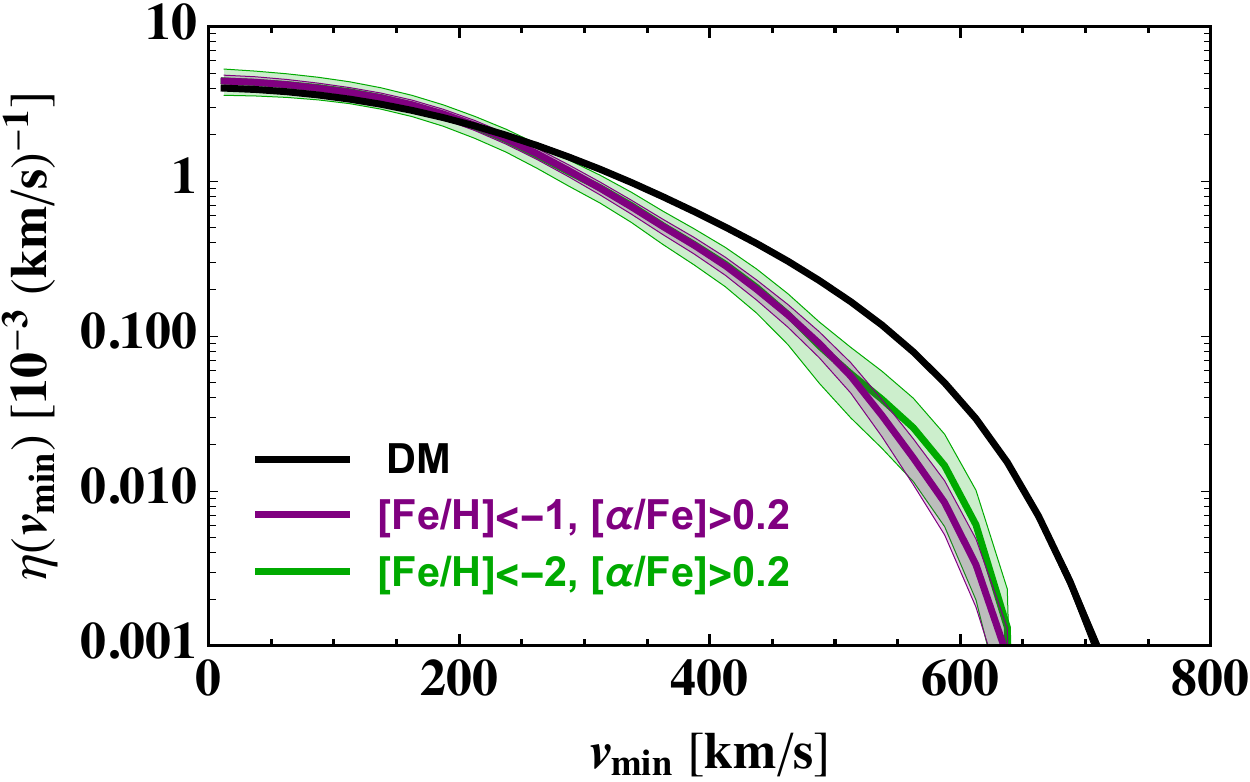}
  \includegraphics[width=0.32\textwidth]{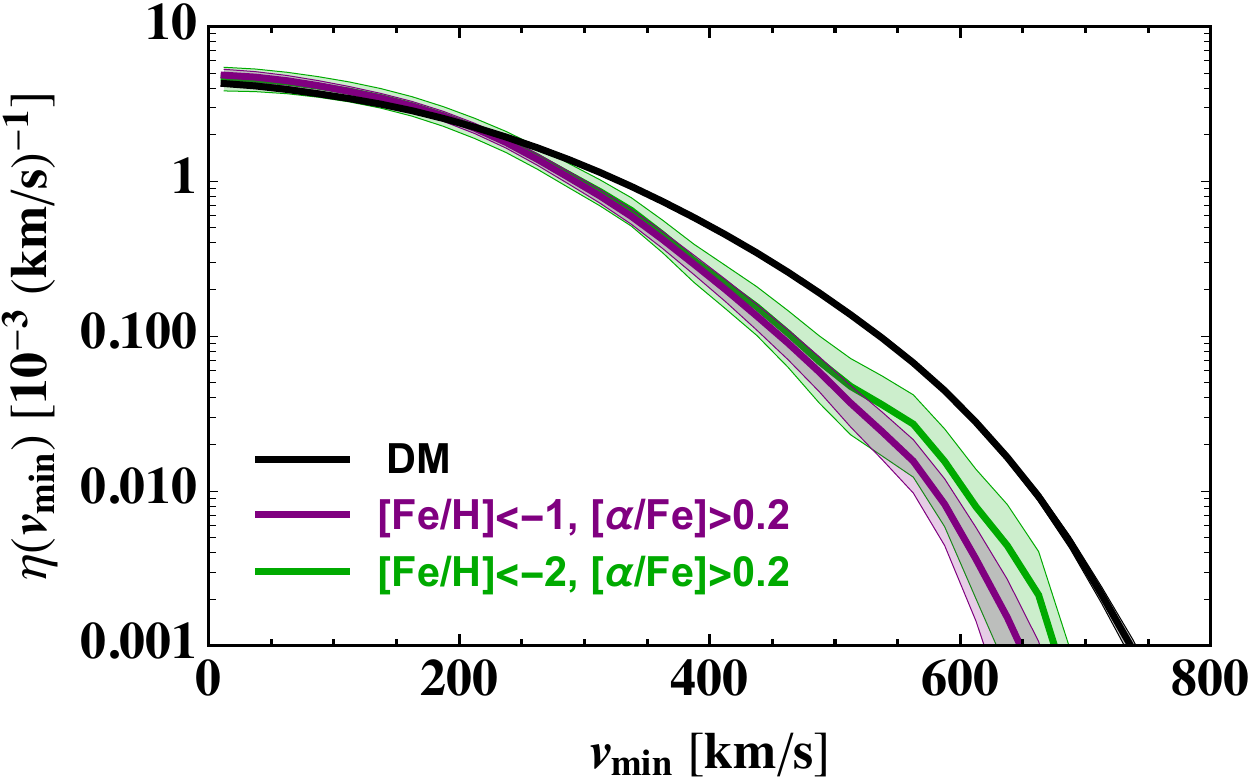}
  \includegraphics[width=0.32\textwidth]{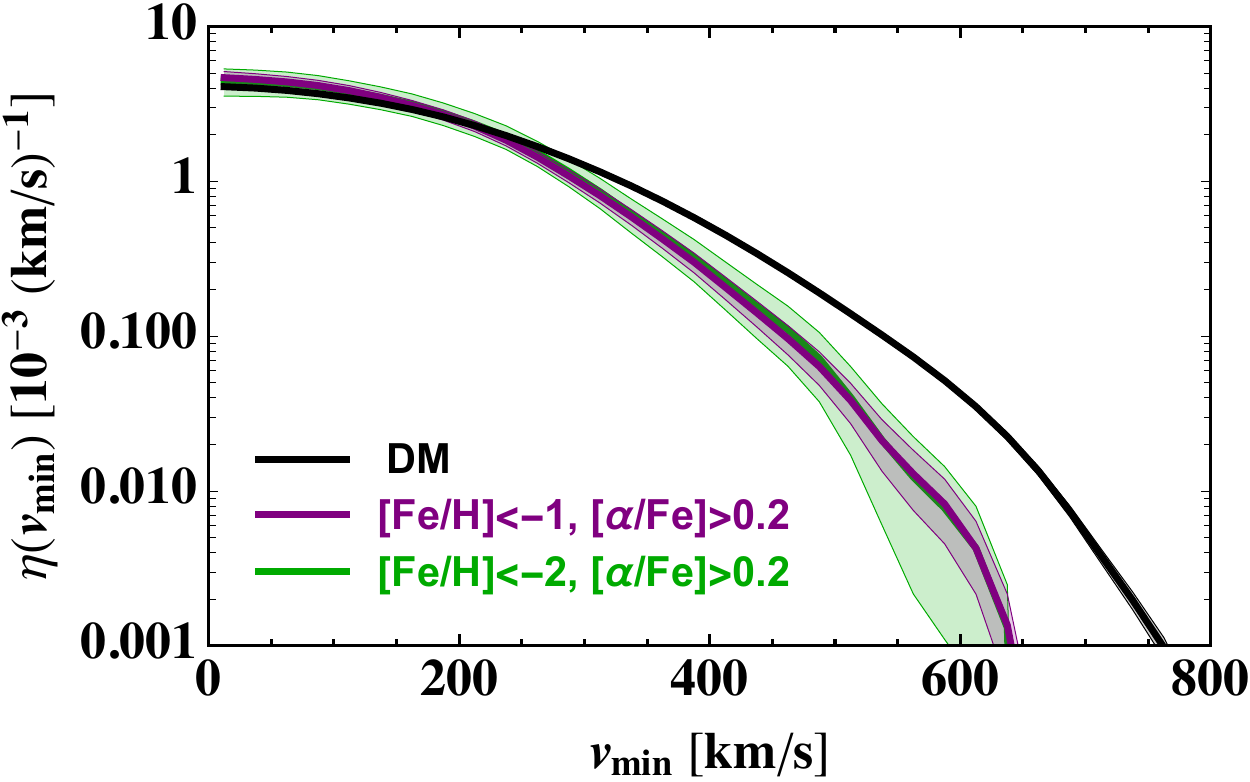}\\
  \includegraphics[width=0.32\textwidth]{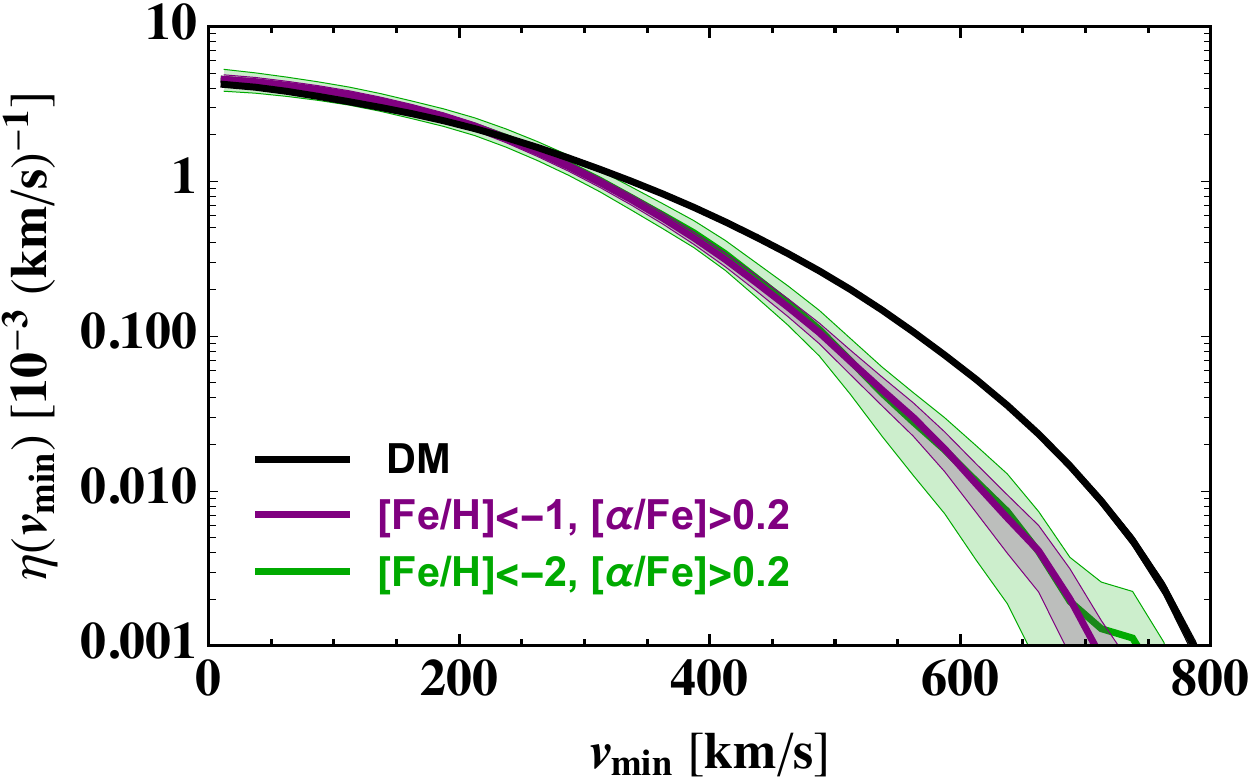}
  \includegraphics[width=0.32\textwidth]{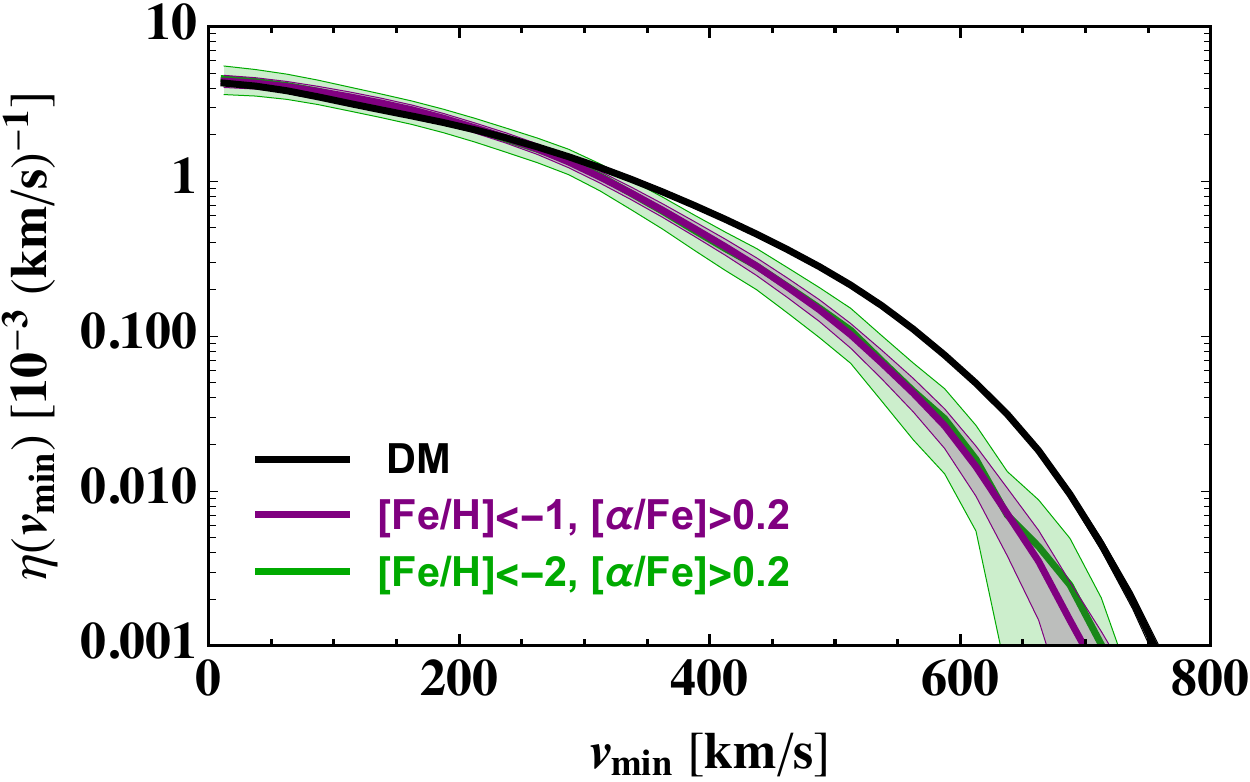}
  \includegraphics[width=0.32\textwidth]{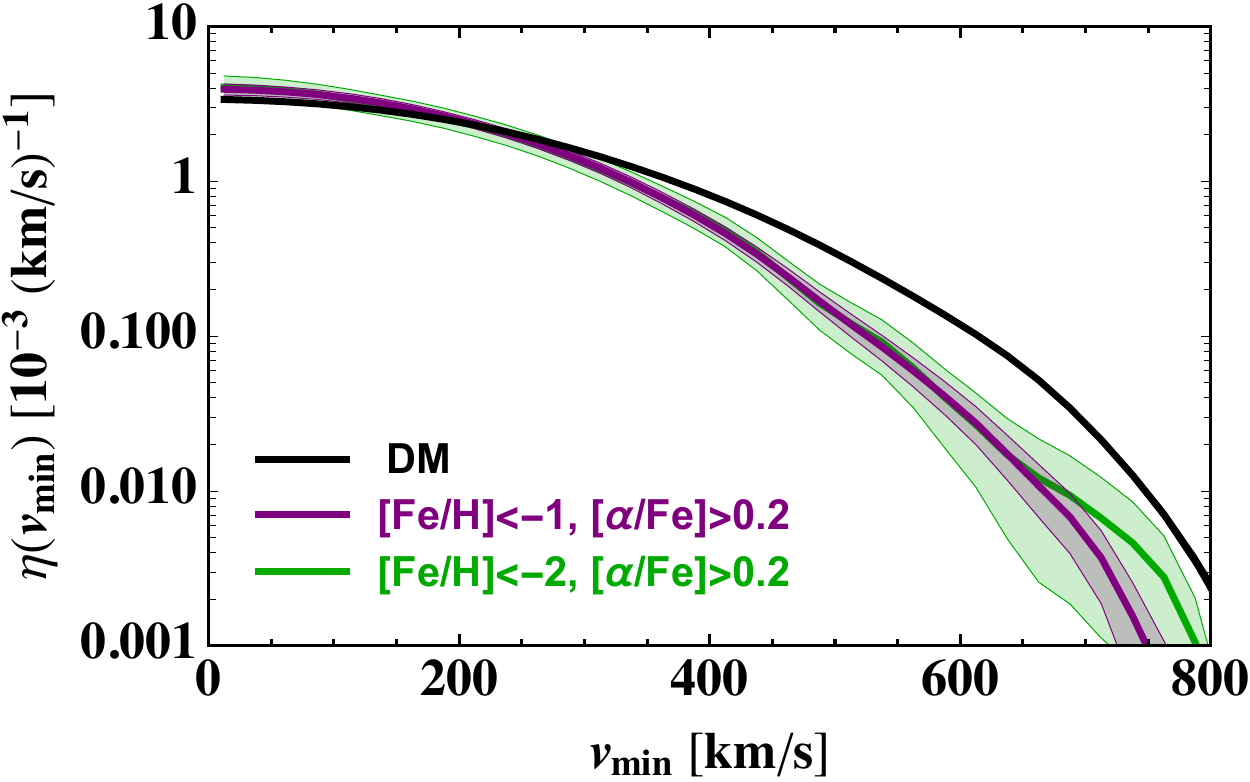}
\caption{Same as figure \ref{fig:eta} but without constraining the stellar velocity distributions to the disc.}
\label{fig:etaShell}
\end{center}
\end{figure}

In figures \ref{fig:TCutsShell}, \ref{fig:FeHalphaHeCutsShell}, \ref{fig:fvModShell}, and \ref{fig:etaShell} we present the equivalents of figures \ref{fig:TCuts}, \ref{fig:FeHalphaHeCuts}, \ref{fig:fvMod}, and \ref{fig:eta}, respectively, when the stellar velocity distributions are not constrained to the disc. In the Solar neighbourhood region with $7 \leq \rho \leq 9$~kpc, the total number of stars is in the range of $[1.1 - 2.2] \times 10^6$, depending on the halo. The number of stars with formation time $T<3$~Gyr and $T<1$~Gyr, is $[7.8 - 17] \times 10^4$ and $[3.6 - 6.6] \times 10^3$, respectively. The number of stars with \FeHone, \FeHtwo, and \FeHthree~all with the additional \aFetwo~cut is $[2.7 - 5.4] \times 10^3$, $[5.4 - 10.3] \times 10^2$, and $[1.4 - 3.2] \times 10^2$, respectively, and depending on the halo.

The $p$-values for the KS test to check the correlation between the velocity distributions of DM (constrained to the disc with $|z| \leq 2$~kpc) and metal-poor stars (not constrained to the disc) in the Solar circle with \FeHtwo~(left column) and \FeHthree~(right column), both with an additional cut of \aFetwo~are given in table \ref{tab:pvalues-Shell} for the six Auriga halos.

As it can be seen from figures \ref{fig:TCutsShell}, \ref{fig:FeHalphaHeCutsShell}, \ref{fig:fvModShell}, and \ref{fig:etaShell}, not constraining the stars to the disc, does not change the results qualitatively. The main difference from before is the smaller Poisson error bars due to larger statistics which leads to even weaker correlations between the velocity distributions of DM and metal-poor stars. This is clear from the $p$-values presented in table \ref{tab:pvalues-Shell} which are never larger than 0.05 (other than for the vertical velocity distribution  of Au24 where $p=0.063$) for metal-poor stars with \FeHtwo~and  \aFetwo.

\bigskip

\begin{table}[H]
    \centering
    \begin{tabular}{|c|c c c|c c c|}
      \hline
         & \multicolumn{3}{|c|}{\FeHtwo, \aFetwo} & \multicolumn{3}{|c|}{\FeHthree, \aFetwo} \\
       \hline
       Halo Name  & $\rho$ & $\phi$  & $z$ & $\rho$ & $\phi$  & $z$\\
       \hline
       Au6 & $5.2 \times 10^{-3}$ & $1.1 \times 10^{-4}$ & $2.2 \times 10^{-2}$ & $4.2 \times 10^{-2}$ & $4.6 \times 10^{-2}$ & $2.3 \times 10^{-1}$\\
       Au16 & $1.6 \times 10^{-2}$ & $6.6 \times 10^{-6}$ & $9.3 \times 10^{-3}$ & $3.6 \times 10^{-1}$ & $2.8 \times 10^{-1}$ & $2.2 \times 10^{-1}$\\
       Au21 & $1.6 \times 10^{-2}$ & $1.2 \times 10^{-3}$ & $8.6 \times 10^{-3}$ & $3.1 \times 10^{-1}$ & $2.0 \times 10^{-1}$ & $5.5 \times 10^{-2}$\\
       Au23 & $2.0 \times 10^{-3}$ & $4.9 \times 10^{-7}$ & $7.1 \times 10^{-5}$ & $3.0 \times 10^{-2}$ & $1.1 \times 10^{-3}$ & $1.4 \times 10^{-1}$\\
       Au24  & $2.8 \times 10^{-2}$ & $4.4 \times 10^{-3}$ & $6.3 \times 10^{-2}$ & $1.1 \times 10^{-1}$ & $5.1 \times 10^{-1}$ & $4.5 \times 10^{-2}$\\
       Au27 & $3.6 \times 10^{-4}$ & $3.4 \times 10^{-6}$ & $4.5 \times 10^{-5}$ & $6.0 \times 10^{-1}$ & $1.6 \times 10^{-1}$ & $7.6 \times 10^{-3}$\\
      \hline    \end{tabular}
    \caption{Same as table \ref{tab:pvalues} but for the metal-poor stars in a cylindrical shell at the Solar circle with $7 \leq \rho \leq 9$~kpc.}
    \label{tab:pvalues-Shell}
  \end{table}

\pagebreak


\clearpage
\bibliographystyle{JHEP}
\bibliography{./refs}

\end{document}